\documentclass[12pt]{article}
\usepackage[letterpaper,margin=1in]{geometry}
\usepackage[T1]{fontenc}
\usepackage{lmodern,microtype,amsmath,amssymb,amsthm,mathtools}
\usepackage{booktabs,longtable,array,graphicx,xcolor}
\usepackage[round,authoryear,longnamesfirst]{natbib}
\setcitestyle{citesep={,}}
\usepackage{enumitem}
\usepackage{algorithm,algpseudocode}
\usepackage[colorlinks=true,linkcolor=blue!40!black,citecolor=blue!40!black,urlcolor=blue!40!black]{hyperref}
\hypersetup{pdftitle={Calibrated Transfer Learning for Survival Prediction and Treatment Inference},pdfauthor={Shuoyang Wang, Wanyu Zhang, Lingli Yang, Kexuan Li}}
\setlist{nosep}
\newtheorem{theorem}{Theorem}[section]
\newtheorem{proposition}[theorem]{Proposition}
\newtheorem{lemma}[theorem]{Lemma}
\newtheorem{corollary}[theorem]{Corollary}

\newcommand{\E}{\mathbb E}
\newcommand{\Pp}{\mathbb P}
\newcommand{\J}{\mathcal J}
\newcommand{\op}{o_{\Pp}}
\newcommand{\Op}{O_{\Pp}}
\newcommand{\norm}[1]{\lVert #1\rVert}

\newcommand{\ind}{\mathbf 1}
\newcommand{\cnt}{\mathcal N}%

\usepackage{float}
\newcommand{\setstretch}[1]{\renewcommand{\baselinestretch}{#1}\selectfont}

\newcommand{\papertitle}{Calibrated Transfer Learning for Survival Prediction and Treatment Inference}
\newcommand{\codeurl}{\url{https://github.com/ShuoyangWang/ctcox}}
\makeatletter\g@addto@macro\@parboxrestore{\parfillskip=0pt plus 0.8\textwidth\relax}\makeatother

\newcommand{\OakGlobalPh}{0.009}
\newcommand{\OakTimeHr}{0.709 (0.596, 0.844)}
\newcommand{\GbsgClinHr}{0.651 (0.501, 0.853)}
\newcommand{\GbsgNeuralHr}{0.651 (0.499, 0.851)}
\newcommand{\GbsgProjectedHr}{0.651 (0.488, 0.871)}
\newcommand{\GbsgClinIbs}{0.1658}
\newcommand{\GbsgClinSmallIbs}{0.1816}
\newcommand{\GbsgClinSmallCal}{0.64}
\newcommand{\GbsgCurrentSplineIbs}{0.1644}
\newcommand{\GbsgCurrentSplineSmallIbs}{0.1970}
\newcommand{\GbsgTargetNeuralIbs}{0.1694}
\newcommand{\GbsgNeuralIbs}{0.1657}
\newcommand{\GbsgSplineIbs}{0.1649}
\newcommand{\GbsgFreeIbs}{0.1667}
\newcommand{\GbsgFreeSmallIbs}{0.1728}
\newcommand{\GbsgAdaptiveSmallIbs}{0.1728}
\newcommand{\GbsgAdaptiveSmallCal}{0.95}
\newcommand{\GbsgAdaptiveSplineIbs}{0.1640}
\newcommand{\GbsgGlobalPh}{0.006}
\newcommand{\GbsgTimeHr}{0.644 (0.495, 0.838)}
\newcommand{\OakRmst}{1.45}
\newcommand{\OakRmstLow}{0.49}
\newcommand{\OakRmstHigh}{2.33}
\newcommand{\OakThreePartitionRange}{0.689 to 0.755}
\newcommand{\OakFivePartitionRange}{0.696 to 0.731}
\newcommand{\GbsgNonlinearPercent}{0.48}
\newcommand{\GbsgThreePartitionRange}{0.615 to 0.690}
\newcommand{\GbsgFivePartitionRange}{0.628 to 0.672}
\newcommand{\OakThreePartitionRangeFinal}{0.689 to 0.754}
\newcommand{\OakFivePartitionRangeFinal}{0.696 to 0.731}
\newcommand{\OakFullOsCtFinal}{0.702}
\newcommand{\OakFullOsPtFinal}{0.726}
\newcommand{\OakFullOsClin}{0.703}

\usepackage[letterpaper,margin=1in]{geometry}
\usepackage[T1]{fontenc}
\usepackage{lmodern,microtype,amsmath,amssymb,amsthm,mathtools}
\usepackage{booktabs,longtable,array,graphicx,xcolor}
\usepackage[round,authoryear,longnamesfirst]{natbib}
\setcitestyle{citesep={,}}
\usepackage{enumitem}
\usepackage{algorithm,algpseudocode}
\usepackage[colorlinks=true,linkcolor=blue!40!black,citecolor=blue!40!black,urlcolor=blue!40!black]{hyperref}
\hypersetup{pdftitle={Calibrated Transfer Learning for Survival Prediction and Treatment Inference},pdfauthor={Shuoyang Wang, Wanyu Zhang, Lingli Yang, Kexuan Li}}
\setlist{nosep}
\usepackage{float}

\makeatletter\g@addto@macro\@parboxrestore{\parfillskip=0pt plus 0.8\textwidth\relax}\makeatother

\makeatletter
\renewcommand{\@seccntformat}[1]{\ifnum\pdfstrcmp{#1}{section}=0 Appendix \csname the#1\endcsname.\quad\else\csname the#1\endcsname\quad\fi}
\makeatother

\renewcommand{\thetable}{S\arabic{table}}

\title{\papertitle}
\author{Shuoyang Wang$^{1,*}$, Wanyu Zhang$^{2,3,*}$, Lingli Yang$^{4}$, and Kexuan Li$^{5,\dagger}$\\[2ex]
\parbox{\dimexpr\textwidth-2\tabcolsep\relax}{\centering\normalsize
$^{1}$Department of Bioinformatics and Biostatistics, University of Louisville, Louisville, KY, USA\\
$^{2}$Division of Gastroenterology, Hepatology and Nutrition, University of Louisville School of Medicine, Louisville, KY, USA\\
$^{3}$Louisville Clinical and Translational Research Center, University of Louisville, Louisville, KY, USA\\
$^{4}$Takeda Pharmaceuticals, Cambridge, MA, USA\\
$^{5}$Data \& Quantitative Sciences, Bristol Myers Squibb, Cambridge, MA 02141, USA\\[2ex]
$^{*}$These authors contributed equally to this work as co-first authors.\\
$^{\dagger}$Corresponding author: Kexuan Li, Data \& Quantitative Sciences, Bristol Myers Squibb, Cambridge, MA 02141, USA. Email: \texttt{kexuan.li.77@gmail.com}}}
\date{}
\begin{document}
\maketitle
\clearpage

\begin{abstract}\normalsize
Historical survival data can improve prediction in a cohort with few events, but prognostic differences can make borrowing harmful. To borrow while adapting, we propose calibrated transfer learning for the Cox model, in which a historical risk function enters the target analysis as a fixed offset with a prespecified correction for changing covariate effects and a separate baseline hazard. Calibration removes historical error in the correction space exactly, so only the remaining error governs the prediction rate, for any bounded learner. Accurate prediction does not, however, ensure valid treatment inference. In randomized trials, the remaining error biases the conditional treatment coefficient, because a treatment effect changes the prognostic composition of later risk sets. We characterize this bias and its sharp worst-case direction, and we introduce a cross-fitted projected score that removes it at first order and leaves only a second-order bias when a fixed error lies in its dictionary. A historical learning bound separates approximation, sampling, and optimization error. In simulations, calibration improves prediction when the historical risk is well learned, and the projected score restores interval coverage under drift outside the correction, also at a hazard ratio of 0.7. In a retrospective transfer between the POPLAR and OAK lung cancer trials, selected corrections reduce the integrated Brier score by 4.3\% relative to ridge-penalized clinical Cox with 100 training patients, without improving treatment precision.
\end{abstract}

\clearpage

\section*{Introduction}\label{m:introduction}
Survival prediction is often needed where few events are available, for example in a new clinical cohort, a later recruitment period, or a randomized trial whose comparator arm resembles patients in an earlier study. An earlier cohort can supply prognostic information, but its patients may differ in case mix, measurement, and care, so its model rarely transports unchanged. Classical updating of Cox proportional hazards (PH) models \citep{cox1972} keeps an external prognostic index and re-estimates the baseline hazard, a calibration slope, or selected covariate effects \citep{vanhouwelingen2000,steyerberg2004,janssen2008,Royston2013}. Flexible learners such as neural Cox models \citep{faraggi1995,katzman2018} make the external index richer but its errors harder to characterize. It is then unclear which historical errors such updating removes, and what the remaining error does to prediction and, in a trial, to treatment inference.

In transfer learning, shared structure improves target estimation in high-dimensional linear regression \citep{licaili2022} and in nonparametric regression with deep networks \citep{ren2026}. For survival outcomes, \citet{li2023} allow heterogeneous Cox coefficients and baselines, \citet{gu2026} transfer source predictions through a penalty in a target transformation model, and \citet{craig2026} adapt a source offset with the lasso. These methods target prediction or coefficient estimation. In randomized trials, external prognostic scores increase the efficiency of tests and marginal summaries \citep{lu2008,schuler2022,ye2024,zhou2026}, and historical control methods borrow outcomes directly \citep{schmidli2014}. A conditional Cox treatment coefficient raises a further difficulty. Randomization balances prognosis at enrollment, but a treatment effect changes who remains at risk, so treatment and prognosis become associated within later risk sets \citep{hernan2010,martinussen2013,aalen2015}. An error in a borrowed prognostic score can then bias the treatment coefficient even when it barely affects prediction. To our knowledge, this bias has not been characterized when prognosis is transferred from another cohort.

To make the balance between borrowing and adaptation explicit, we propose a two-stage procedure. A bounded historical log-risk function, estimated by any learner, is held fixed as an offset in the target partial likelihood, and the target analysis estimates a prespecified correction for covariate effects that may change, a separate baseline hazard, and, in a trial, a conditional treatment coefficient. Historical patients never enter target risk sets, and the correction space determines how much is borrowed. Our central finding is that the calibration that makes prediction accurate does not by itself make treatment inference valid. First, we show that calibration removes historical error in the correction space exactly, so only the remaining error enters the survival prediction rate, also after data-dependent selection among prespecified corrections. Second, we show that the remaining error enters the treatment coefficient through one linear functional with an explicit covariate representer and a sharp local bound, so errors of equal prediction size can produce very different treatment biases. Third, to remove this functional we introduce a cross-fitted score projected onto functions of the covariates under risk-set weighting. Its unrestricted limit is the efficient score of the target-only partially linear Cox model \citep{huang1999,zhong2022}, and cross-fitting follows \citet{chernozhukov2018}, but both estimate the nuisance from target data. We derive what they leave open for a transported offset calibrated in a restricted space, namely the bias of the direct fit, a finite-dictionary trade-off whose nuisance rate is set by the historical error, and limits under a fixed misspecification, where the projected score keeps only a second-order bias when the error lies in its dictionary. Fourth, we derive a historical partial likelihood bound that accounts for dependence among risk sets and separates approximation, sampling, and optimization error. We also give a label-blind randomization test that remains valid when the transported prognosis is wrong.

Simulations and two retrospective applications clarify the role of each analysis. Calibrated transfer improves prediction when target information is limited and the historical risk is well learned, including against penalized target-only and coefficient-transfer comparators, and can harm prediction when the historical sample is small. Drift outside the correction can make direct treatment intervals undercover, also at a hazard ratio of 0.7 in trials of 2000 or 3000 patients, and the projected score repairs this. In both applications the benefit comes mainly from external clinical weighting with parsimonious calibration, and treatment precision does not improve.

\section*{Model and estimands}\label{m:modelsection}

\subsection*{Survival prediction in a target cohort}
We observe independent historical and target samples of sizes $N$ and $n$. A target record is $O=(U,\Delta,X)$, with covariates $X\in\mathbb R^d$, event time $T$, censoring time $C$, fixed horizon $\tau>0$, $U=\min(T,C,\tau)$, and $\Delta=\ind(T\le C,T\le\tau)$, where $\ind(\cdot)$ is the indicator function. Historical records $O^H=(U^H,\Delta^H,X^H)$ are defined in the same way. Records are independent and identically distributed within each cohort, and censoring is independent of the event time given the measured covariates.

Let $f_0$ denote the historical prognostic log risk and let $z(x)\in\mathbb R^p$ be a deterministic correction vector of fixed dimension $p$. For $\gamma_0\in\mathbb R^p$, write $g_0=f_0+\gamma_0^\top z$. We propose the model
\begin{equation}\label{m:predictionmodel}
\lambda_H(t\mid x)=\lambda_{H,0}(t)e^{f_0(x)},\qquad
\lambda_T(t\mid x)=\lambda_{T,0}(t)e^{g_0(x)},
\end{equation}
where the baseline hazards $\lambda_{H,0}$ and $\lambda_{T,0}$ are unrelated and have cumulative hazards $\Lambda_{H,0}$ and $\Lambda_{T,0}$. The prediction target is $S(t\mid x)=\exp\{-\Lambda_{T,0}(t)e^{g_0(x)}\}$. Model~\eqref{m:predictionmodel} requires no treatment variable and permits different covariate distributions, event rates, and censoring laws in the two cohorts, but it assumes that the changes in log risk lie in the correction space. We fix $z$ independently of target outcomes, exclude an intercept, and center $z$ under a fixed reference measure. Transferring a historical prediction rate also requires a bounded ratio of target to historical covariate densities, and comparable endpoints, time origins, and covariate definitions.

\subsection*{Treatment inference in a randomized target trial}
In a trial, the target record is $O=(U,\Delta,A,X)$, where $A=1$ denotes the investigational treatment and $A=0$ the background treatment that historical patients received. We assume $A\sim\operatorname{Bernoulli}(p_A)$ independently of $X$, with $0<p_A<1$, censoring independent of the event time given $(A,X)$, and
\begin{equation}\label{m:model}
\lambda_T(t\mid a,x)=\lambda_{T,0}(t)e^{\theta_0a+g_0(x)},
\qquad g_0=f_0+\gamma_0^\top z.
\end{equation}
We estimate the conditional log hazard ratio $\theta_0$ and the arm-specific survival function $S_a(t\mid x)=\Pr(T>t\mid A=a,X=x)$. Model~\eqref{m:model} excludes time-varying effects and interactions between treatment and covariates, and $\theta_0$ is a conditional rather than a marginal hazard ratio \citep{martinussen2013}. For testing we also consider the observed-data null hypothesis $H_0^{\rm obs}:A\perp(X,U,\Delta)$, which randomization implies when the arms have equal conditional event distributions and a common censoring law. Norms $\|\cdot\|_2$ are Euclidean for vectors and target $L_2$ for functions, $\E$ denotes target expectation, and $O_\Pp$ and $o_\Pp$ denote stochastic orders. Appendix A lists the remaining notation.

\section*{Calibrated prediction and treatment analysis}\label{m:methodssection}
\subsection*{Historical learning and direct calibration}
Let $\mathcal F_N$ be a deterministic separable class of functions with $|f|\le B_f$. Any learner with bounded output can be used, and our implementations use bounded neural networks and fixed-knot splines. To learn the historical prognosis, we estimate $\widehat f\in\mathcal F_N$ from the historical records by maximizing the partial likelihood
\begin{equation}\label{m:fitH}
\ell_H(f)=N^{-1}\sum_j\Delta_j^H\left[f(X_j^H)-\log\left\{\sum_{l:U_l^H\ge U_j^H}e^{f(X_l^H)}\right\}\right],
\end{equation}
allowing regularization and validation stopping, and then hold $\widehat f$ fixed. Set $W=z(X)$ and $\beta=\gamma$ for prediction, or $W=(A,z^\top)^\top$ and $\beta=(\theta,\gamma^\top)^\top$ in a trial, and put $\eta_\beta=\widehat f(X)+\beta^\top W$. We then propose direct calibrated transfer (CT), which maximizes the concave criterion
\begin{equation}\label{m:fit}
\ell_n(\beta)=n^{-1}\sum_i\Delta_i
 \left[\eta_{\beta,i}-\log\left\{\sum_{j:U_j\ge U_i}e^{\eta_{\beta,j}}\right\}\right].
\end{equation}
With maximizer $\widehat\beta$, components $\widehat\gamma$ and $\widehat\theta$, $\widehat g=\widehat f+\widehat\gamma^\top z$, and $\widehat\eta=\eta_{\widehat\beta}$, the Breslow estimate of the target cumulative baseline hazard and the survival estimates are
\[
\widehat\Lambda_{T,0}(t)=\sum_{i:U_i\le t}\frac{\Delta_i}{\sum_{j:U_j\ge U_i}e^{\widehat\eta_j}},\qquad
\widehat S_a(t\mid x)=\exp\{-\widehat\Lambda_{T,0}(t)e^{\widehat\theta a+\widehat g(x)}\},
\]
with $\widehat S(t\mid x)=\exp\{-\widehat\Lambda_{T,0}(t)e^{\widehat g(x)}\}$ in the prediction setting. Frozen transfer omits $z$. Free score recalibration adds $\widehat f$ itself as a correction coordinate, and given the independent historical fit it is the special case $z=\widehat f$ of the analysis below.

For a supplied index $\eta$ and function $v$, let $Y_i(t)=\ind(U_i\ge t)$, $\cnt_i(t)=\ind(U_i\le t,\Delta_i=1)$, $\nu_n=n^{-1}\sum_i\cnt_i$, $s_{n,\eta}(t)=n^{-1}\sum_iY_i(t)e^{\eta_i}$, and $\bar v_{n,\eta}(t)=\sum_iY_i(t)e^{\eta_i}v_i/\sum_iY_i(t)e^{\eta_i}$, with ratios set to zero on empty risk sets. Define the score and the empirical event-weighted covariance
\begin{align*}
\mathcal U_n(\eta,v)&=n^{-1}\sum_i\int(v_i-\bar v_{n,\eta})\,d\cnt_i,\qquad
\widehat\J_n(\eta,u,v)=\int
 (\overline{uv^\top}_{n,\eta}-\bar u_{n,\eta}\bar v_{n,\eta}^\top)\,d\nu_n,
\end{align*}
so that \eqref{m:fit} has gradient $\mathcal U_n(\eta_\beta,W)$ and negative Hessian $\widehat\J_n(\eta_\beta,W,W)$. We estimate uncertainty with the robust sandwich covariance \citep{linwei1989}. Its patient influence contributions $\widehat\varphi_i(\eta,v)$ include the correction for estimated risk-set means (Appendix B.2), and in a trial $\widehat{\operatorname{se}}_{\rm sw}$ denotes its standard error (SE) for the treatment coefficient.

\subsection*{A finite-dictionary projected score}
For treatment inference under \eqref{m:model}, let $Y(t)=\ind(U\ge t)$, $\eta_0=\theta_0 A+g_0(X)$, $s_0(t)=\E\{Y(t)e^{\eta_0}\}$, and $Q_t(v)=\E(Ye^{\eta_0}v)/s_0$, and define the event-weighted covariance form
\begin{equation*}
\J(u,v)=\int s_0(t)\{Q_t(uv^\top)-Q_t(u)Q_t(v)^\top\}\,d\Lambda_{T,0}(t).
\end{equation*}
Under a treatment alternative, patients with poorer prognosis leave the risk set at different rates in the two arms, so $A$ and $X$ become associated under $Q_t$ although they are independent at enrollment. Subtracting $p_A$ from $A$ does not remove this association. To remove it, we introduce a projected score that subtracts from $A$ its regression on functions of $X$, weighted as in the risk sets.

Let $B(X)\in\mathbb R^M$ be a deterministic dictionary of bounded functions whose span contains $z$ modulo constants, centered under a fixed reference measure with bounded density ratio. Partition the trial indices, independently of the data, into $K\ge2$ folds $\mathcal I_k$ of sizes $m_k$, and let $w_k=m_k/n$. On each training complement, calibrate as above to obtain $\widetilde\theta_k$, $\widetilde\gamma_k$, $\widetilde g_k=\widehat f+\widetilde\gamma_k^\top z$, and $\widetilde\eta_k=\widetilde\theta_k A+\widetilde g_k$. Let $\widehat\J_{{\rm tr},k}$ be the empirical covariance on that complement, let $\lambda_k\in[0,\lambda_n]$ be a ridge parameter with prespecified bound $\lambda_n$, and compute
\[
\widehat\mu_k=
 \{\widehat\J_{{\rm tr},k}(\widetilde\eta_k,B,B)+\lambda_k\mathrm{Id}_M\}^{-1}
 \widehat\J_{{\rm tr},k}(\widetilde\eta_k,B,A),
\]
where $\mathrm{Id}_M$ is the $M\times M$ identity. Set $\widehat q_k=A-\widehat h_k$ with $\widehat h_k=\operatorname{clip}_{[-L_h,L_h]}(\widehat\mu_k^\top B)$ for a prespecified bound $L_h$. Using only fold $k$, compute the score $\mathcal U_k=\mathcal U_{m_k}(\widetilde\eta_k,\widehat q_k)$, the derivative $D_k=\widehat\J_k(\widetilde\eta_k,\widehat q_k,A)$, and the influence variance $V_k=m_k^{-1}\sum_{i\in\mathcal I_k}\widehat\varphi_{i,k}(\widetilde\eta_k,\widehat q_k)^2$, where $\widehat\J_k$ and $\widehat\varphi_{i,k}$ are evaluated on fold $k$. The projected estimate and its SE are
\begin{equation}\label{m:projected}
\widehat\theta^{\rm score}=\sum_k w_k(\widetilde\theta_k+\mathcal U_k/D_k),\qquad
\widehat{\operatorname{se}}_{\rm score}^2=\sum_k\frac{w_k^2 V_k}{m_kD_k^2}.
\end{equation}
Cross-fitting makes the score evaluation independent of the nuisance training, and projection reduces the sensitivity to prognostic error. Survival predictions always come from the direct fit.

\subsection*{Computation, correction choice, and resampling}\label{m:selectionsection}
We implement both computations in Algorithms S1 and S2 (Appendix B), including Newton steps, interiority and information checks, and failure rules, and return a projected estimate only when every fold succeeds. All tuning constants are prespecified. For intervals we also implement a patient bootstrap that holds $\widehat f$ fixed and refits every target step, resampling within the original folds for the projected score, with 999 resamples (Appendix N). It targets the working coefficient, so it covers $\theta_0$ only when the bias of that coefficient is negligible.

We treat the choice of correction as part of the prediction procedure. We prespecify a small set of candidates, for example a clinical core, the full linear clinical vector, a spline basis, and free rescaling of $\widehat f$, select among them within the target training data, and evaluate the complete procedure on held-out target patients against strong target-only fits and conventional recalibration, using prediction error, calibration, and concordance \citep{graf1999,Royston2013}. Corollary~\ref{m:selection} shows that selection preserves the prediction rate. Our results do not cover treatment intervals after selection, so we fix the correction for treatment analysis.

To test $H_0^{\rm obs}$ when the transported prognosis may be wrong, we fit the correction with $\theta=0$, using only $(X,U,\Delta)$ and $\widehat f$, and obtain $\widehat g_{\rm null}$. With $\widehat p_A$ the treated fraction and $\mathbf a=(a_1,\ldots,a_n)$ an assignment vector, define
\[
\widehat R_{{\rm null},i}=\Delta_i-\int\frac{Y_i e^{\widehat g_{\rm null}(X_i)}}{s_{n,\widehat g_{\rm null}}}\,d\nu_n,\qquad
T_{\rm rnd}(\mathbf a)=\Big|\sum_i(a_i-\widehat p_A)\widehat R_{{\rm null},i}\Big|.
\]
Comparing the statistic at the observed assignment $\mathbf A=(A_1,\ldots,A_n)$ with its values at $B_{\rm rnd}$ assignments $\mathbf A^{(b)}$ drawn uniformly with the observed number treated gives the calibrated randomization (CR) $p$ value
\begin{equation}\label{m:random}
p_{\rm CR}=\frac{1+\sum_{b=1}^{B_{\rm rnd}}\ind\{T_{\rm rnd}(\mathbf A^{(b)})\ge T_{\rm rnd}(\mathbf A)\}}{B_{\rm rnd}+1}.
\end{equation}
No fitting step uses the treatment labels, so the test is valid conditionally on the number treated \citep{hemerik2018}. Other allocation designs require their actual randomization law.

\section*{Theoretical properties}\label{m:theorysection}
We consider joint limits of $N$ and $n$ and condition on the independent historical fit. Appendix C restates each result with complete conditions.

\subsection*{Prediction in a general target cohort}\label{m:generalprediction}
For model~\eqref{m:predictionmodel} we assume three conditions. P1 requires independent cohorts, censoring independent of the event time given $X$, continuous event times, and $0<\Lambda_{T,0}(\tau)<\infty$. P2 requires uniformly bounded true and fitted risks, correction, and candidate indices, compact coefficient sets with interior optima, and $\Pr(U\ge\tau\mid X)\ge c_{\rm obs}>0$. P3 requires, for $p>0$, a locally positive definite population calibration Hessian and an interior score root with numerical residual $\xi_n$. Decompose $\widehat f-f_0=c_N+b_N^\top z+e_N$ by target $L_2$ projection onto constants and $z$, so that $\E e_N=\E(ze_N)=0$, and put $\rho_N=\|e_N\|_2$ and $\gamma_N=\gamma_0-b_N$.

\begin{proposition}[Calibration invariance]\label{m:invariance}
Replacing a supplied offset $f$ by $f+c+b^\top z$ leaves the unpenalized fitted survival curves, and in a trial the fitted treatment coefficient, unchanged whenever the coefficient shifts are feasible. The correction becomes $\gamma-b$ and the Breslow baseline is multiplied by $e^{-c}$.
\end{proposition}
Historical errors in $\operatorname{span}\{1,z\}$ are therefore harmless in finite samples, whatever their size. A correction penalty or a binding coefficient constraint breaks this exact property.

\begin{theorem}[Survival prediction without treatment]\label{m:generalpredictiontheorem}
Under P1 through P3, let $r_{n,N}=n^{-1/2}+\rho_N+\xi_n=o_\Pp(1)$. The direct calibrated fit satisfies
\begin{multline*}
\|\widehat\gamma-\gamma_N\|_2+\|\widehat g-g_0-c_N\|_2
+\sup_{t\le\tau}|e^{c_N}\widehat\Lambda_{T,0}(t)-\Lambda_{T,0}(t)|\\
+\Big\|\sup_{t\le\tau}|\widehat S(t\mid\cdot)-S(t\mid\cdot)|\Big\|_2
=O_\Pp(r_{n,N}).
\end{multline*}
For $p=0$, omit the coefficient term and P3, and set $\xi_n=0$.
\end{theorem}
The bound needs no treatment variable and does not guarantee improvement over a target-only fit, which is why we evaluate the complete procedure on held-out patients. Drift outside $\operatorname{span}\{1,z\}$ adds to $e_N$, and $\rho_N$ then has a nonzero limit.

\begin{corollary}[Selection among prespecified corrections]\label{m:selection}
Let $z^{(1)},\ldots,z^{(L)}$ be fixed corrections that each satisfy the conditions of Theorem~\ref{m:generalpredictiontheorem}, with rates $r_{n,N}^{(l)}=o_\Pp(1)$ whose $\rho_N^{(l)}$ includes any drift outside the span of $z^{(l)}$. For any data-dependent choice $\widehat l$, possibly based on target outcomes, the survival bound of Theorem~\ref{m:generalpredictiontheorem} holds for $\widehat S^{(\widehat l)}$ with rate $\max_lr_{n,N}^{(l)}$.
\end{corollary}

\subsection*{Conditions for randomized trial inference}
For model~\eqref{m:model}, conditions R1 to R3 extend P1 to P3 to the trial index $\theta A+g(X)$ and the design $W=(A,z^\top)^\top$, with censoring independent given $(A,X)$, $\Pr(U\ge\tau\mid A,X)\ge c_{\rm obs}$, and positive definite $I_W=\J(W,W)$. Condition R4, for projected scores, requires fixed $K$, fold fractions bounded away from zero, uniformly bounded fitted projections, and a clipping interval that contains the centered limiting projection. Under these conditions, which Appendix D discusses, the projection $h_*$ of $A$ onto square-integrable functions of $X$ in the seminorm $\J(u,u)^{1/2}$ exists modulo constants, with positive information $I_X=\J(A-h_*,A-h_*)$ (Lemma C.2).

\subsection*{Calibration, prediction, and direct treatment inference}\label{m:directsection}
Use the decomposition of Section~\ref{m:generalprediction} under the trial covariate law, with comparison coefficient $\beta_N=(\theta_0,(\gamma_0-b_N)^\top)^\top$. Let $q_z=A-\J(A,z)\J(z,z)^{-1}z$ and $I_z=\J(q_z,q_z)$. For the counting process $\cnt(t)=\ind(U\le t,\Delta=1)$, let $\mathcal M(t)=\cnt(t)-\int_0^tY(s)e^{\eta_0}\,d\Lambda_{T,0}(s)$ be its martingale and $\phi_q=\int\{q-Q_t(q)\}\,d\mathcal M$ for a fixed score direction $q$.

\begin{theorem}[Direct calibration and prediction]\label{m:directtheorem}
Under conditions R1 through R3 and $\rho_N=o_\Pp(1)$, exact roots of \eqref{m:fit} satisfy
\begin{align*}
\|\widehat\beta-\beta_N\|_2+\|\widehat g-g_0-c_N\|_2
 &=O_\Pp(n^{-1/2}+\rho_N),\\
\widehat\theta-\theta_0
 &=I_z^{-1}\{n^{-1}\textstyle\sum_i\phi_{q_z}(O_i)-\J(q_z,e_N)\}
 +O_\Pp(n^{-1}+\rho_N^2+\rho_Nn^{-1/2}).
\end{align*}
If $\sqrt n\rho_N^2=o_\Pp(1)$ and $\sqrt n\J(q_z,e_N)=o_\Pp(1)$, the treatment limit is $N(0,I_z^{-1})$ at root-$n$ scale, with consistent sandwich variance. Numerical inference requires $\xi_n=o_\Pp(n^{-1/2})$. For $r_{n,N}=o_\Pp(1)$, the arm-specific survival predictions satisfy the bounds of Theorem~\ref{m:generalpredictiontheorem}.
\end{theorem}
Prediction needs only $\rho_N\to0$. The treatment estimate instead carries the bias $-I_z^{-1}\J(q_z,e_N)$, which is linear in the remaining error and is exactly the term ignored when a fitted historical score is treated as known. A rate $\rho_N=o_\Pp(n^{-1/4})$ controls the quadratic remainder but not this term, so accurate prediction does not by itself support valid treatment intervals.

\subsection*{Sharp local sensitivity to historical error}\label{m:sensitivitysection}
The linear term is nonzero because survival selects the risk sets. For a binary covariate $X$, no correction, independent censoring, and $g_0(X)=\kappa X$, the odds ratio between $A$ and $X$ in the weighted risk set at time $t$ is $\exp\{-\Lambda_{T,0}(t)(e^{\theta_0}-1)(e^{\kappa}-1)\}$ (Appendix D.7), which departs from one as the treatment effect, the prognosis, and the cumulative hazard grow. With $p_1=p_A$, $p_0=1-p_A$, and $w_a(t,x)=\Pr(U\ge t\mid A=a,X=x)e^{\theta_0a+g_0(x)}$, define
\begin{equation*}
\omega_z(x)=\int\sum_{a=0}^1p_aw_a(t,x)
 \{q_z(a,x)-Q_t(q_z)\}\,d\Lambda_{T,0}(t).
\end{equation*}
Then $\J(q_z,e)=\E(\omega_z e)$ for every bounded function $e$ of $X$, with $\E\omega_z=0$ and $\E(z\omega_z)=0$.

\begin{proposition}[Sharp calibrated sensitivity]\label{m:sensitivity}
Fix a target law satisfying R1 through R3, and let $\theta^\dagger(e)$ be the population treatment coefficient with offset $f_0+e$. For uniformly bounded errors with $\E e=\E(ze)=0$, $\|e\|_2\le r$, and roots in a common curvature neighborhood, as $r\to0$,
\begin{align*}
\theta^\dagger(e)-\theta_0=-I_z^{-1}\E(\omega_z e)+O(\|e\|_2^2),\qquad
\sup_e|\theta^\dagger(e)-\theta_0|=r\|\omega_z\|_2/I_z+O(r^2).
\end{align*}
When $\|\omega_z\|_2>0$, the first-order radius is attained by $e=\pm r\omega_z/\|\omega_z\|_2$. Otherwise it is $O(r^2)$.
\end{proposition}
Only the component of the remaining error along $\omega_z$ matters at first order, so errors of the same prediction norm can produce very different treatment biases (Figure~\ref{m:mechanism}). Drift outside the correction span enters $e_N$ and persists even with unlimited historical data.

\subsection*{Projected inference and the finite-dictionary trade-off}\label{m:projectionsection}
Let $a_n=\max_k\{|\widetilde\theta_k-\theta_0|+\inf_c\|\widetilde g_k-g_0-c\|_2\}$. For a fixed bounded dictionary containing $z$ modulo constants with $\Sigma_B=\J(B,B)$ positive definite, define $h_B=\J(A,B)\Sigma_B^{-1}B$, $q_B=A-h_B$, $I_B=\J(q_B,q_B)$, and $b_{B,n}=\max_k\inf_c\|\widehat h_k-h_B-c\|_2$.

\begin{theorem}[Finite-dictionary treatment expansion]\label{m:finitetheorem}
Under conditions R1 through R4 and $a_n,b_{B,n}=o_\Pp(1)$, the projected estimator \eqref{m:projected} satisfies
\begin{align*}
\widehat\theta^{\rm score}-\theta_0
=I_B^{-1}\{n^{-1}\textstyle\sum_i\phi_{q_B}(O_i)-\J(q_B,e_N)\}
+O_\Pp\{a_n^2+a_nb_{B,n}+(a_n+b_{B,n})n^{-1/2}+n^{-1}\}.
\end{align*}
If $\sqrt n(a_n^2+a_nb_{B,n})=o_\Pp(1)$ and $\sqrt n\J(q_B,e_N)=o_\Pp(1)$, its root-$n$ limit is $N(0,I_B^{-1})$ and $n\widehat{\operatorname{se}}_{\rm score}^2\to_\Pp I_B^{-1}$, provided numerical one-step errors are $o_\Pp(n^{-1/2})$.
\end{theorem}
Because $\J(q_B,z)=0$, the calibration error of each fold drops out at first order. With $d_B=\{\J(h_*-h_B,h_*-h_B)\}^{1/2}$ and a constant $C$ that depends only on the bounds in R1 to R3, projection geometry gives
\begin{equation}\label{m:tradeoff}
I_B-I_X=d_B^2,\quad I_X\le I_B\le I_z,\quad
\J(q_B,e_N)=\J(h_*-h_B,e_N),\quad
|\J(q_B,e_N)|\le C d_B\rho_N.
\end{equation}
In particular, the first-order bias term vanishes when the remaining error lies in the span of $B$, although a fixed error can leave a second-order bias (Proposition~\ref{m:fixed}). Enlarging nested dictionaries reduces $d_B$ but increases the limiting variance $I_B^{-1}$, and $B=z$ recovers the direct fit. A positive limiting ridge or active clipping adds residual terms (Proposition C.19).

\begin{theorem}[Unrestricted projected score limit]\label{m:fulltheorem}
Under conditions R1 through R4, let $b_n=\max_k\inf_c\|\widehat h_k-h_*-c\|_2$. If $a_n,b_n=o_\Pp(1)$ and $\sqrt n(a_n^2+a_nb_n)=o_\Pp(1)$, then, for numerical one-step errors $o_\Pp(n^{-1/2})$, the dictionary may grow with $n$ and
\[
\sqrt n(\widehat\theta^{\rm score}-\theta_0)
=I_X^{-1}n^{-1/2}\sum_i\phi_{A-h_*}(O_i)+o_\Pp(1)
\Rightarrow N(0,I_X^{-1}),\qquad n\widehat{\operatorname{se}}_{\rm score}^2\to_\Pp I_X^{-1}.
\]
\end{theorem}
The variance $I_X^{-1}$ is that of the target-only partially linear Cox model \citep{huang1999,zhong2022}, so history does not reduce the asymptotic variance of the projected estimator, and by \eqref{m:tradeoff} the direct fit is more precise whenever its bias is negligible. History instead helps through the nuisance rate. If $\rho_N=O_\Pp(r_N)$ and $b_n=O_\Pp(s_n)$, training-fold calibration gives $a_n=O_\Pp(r_N+n^{-1/2})$, and it suffices that $r_N,s_n\to0$ and $\sqrt n(r_N^2+r_Ns_n)\to0$ (Corollary C.9), whereas a target-only nuisance needs its own nonparametric rate. We therefore treat the direct fit as the default treatment analysis and the projected score as a robustness analysis. Theorems~\ref{m:finitetheorem} and~\ref{m:fulltheorem} are local, because they require $a_n\to0$. Drift outside the correction that does not shrink with the sample sizes leaves the training-fold fits at a misspecified population root, and the next result describes both estimators in that case.

\begin{proposition}[Fixed remaining error]\label{m:fixed}
Assume R1, R2, and R4 with $\lambda_n\to0$, and let the training-fold calibration fits converge to a population calibration root $\eta^\dagger=\theta^\dagger A+g^\dagger$, which may differ from $\eta_0$. Let $\J_\dagger$ be $\J$ with $Q_t$ computed under the risk-set weights $e^{\eta^\dagger}$, let $q_B^\dagger$ and $q_z^\dagger$ be the residuals of $A$ after $\J_\dagger$-projection onto $B$ and onto $z$, with $\J_\dagger(q_B^\dagger,A)>0$ and $\J_\dagger(q_z^\dagger,A)>0$, and let $\delta=\inf_c\|\eta^\dagger-\eta_0-c\|_2$. Then $\widehat\theta^{\rm score}\to_\Pp\theta_B^\dagger$, and
\[
\theta_B^\dagger-\theta_0=-\frac{\J_\dagger(q_B^\dagger,e_N)}{\J_\dagger(q_B^\dagger,A)}+O(\delta^2),\qquad
\theta^\dagger-\theta_0=-\frac{\J_\dagger(q_z^\dagger,e_N)}{\J_\dagger(q_z^\dagger,A)}+O(\delta^2).
\]
\end{proposition}
When $e_N$ lies in the span of $\{1,B\}$, the projected score has a bias of second order in $\delta$, although it is not consistent, whereas the direct fit keeps a first-order bias. In the strong-drift law of Section~\ref{m:biassection}, the population biases of the projected score and the direct fit are 0.004 and 0.064 (Table~S20). The training-fold fits also add a term of relative order $\delta$ to the variance, which the held-out SE omits and the full-procedure bootstrap includes (Proposition C.13).

An enlarged correction $z=B$ has the same first-order expansion as the projected score, but it estimates $M$ coefficients in the fit that also yields the treatment coefficient, which makes it unstable in small trials (Section~\ref{m:biassection}). Because Theorems~\ref{m:directtheorem} and~\ref{m:finitetheorem} hold on the same patients, the difference $\widehat\theta^{\rm score}-\widehat\theta$, standardized by its influence-based SE, checks whether the directional bias of the direct fit is negligible when $B$ absorbs $e_N$ (Corollary C.12). It does not test the treatment null, under which the difference degenerates.

\subsection*{Historical learning and attainable two-sample rates}\label{m:sourcesection}
Let $\mathcal C(\varepsilon,\mathcal F_N,\|\cdot\|_\infty)$ be the covering number of $\mathcal F_N$ in the supremum norm, $\mathcal E_N=\log\mathcal C(N^{-1},\mathcal F_N,\|\cdot\|_\infty)$, and suppose some $f_N^*\in\mathcal F_N$ has $\|f_N^*-f_0\|_\infty\le\varepsilon_N^{\rm app}$.

\begin{theorem}[Historical prediction bound]\label{m:source}
Assume conditionally independent historical censoring, continuous event times, $|f_0|\le B_f$, $0<\Lambda_{H,0}(\tau)<\infty$, and $\Pr(U^H\ge\tau\mid X^H)\ge c_H>0$. If $\widehat f$ maximizes \eqref{m:fitH} over $\mathcal F_N$ up to $\varepsilon_N^{\rm opt}$, then for $u\ge1$, with probability at least $1-3e^{-u}$,
\begin{equation*}
\inf_c\|\widehat f-f_0-c\|_{L_2(P_{H,X})}^2
\le C\{(\varepsilon_N^{\rm app})^2+\varepsilon_N^{\rm opt}+(\mathcal E_N+\log N+u)/N\},
\end{equation*}
where $P_{H,X}$ is the historical covariate law and $C$ depends on fixed risk, baseline, and observation bounds. A bounded ratio of target to historical covariate densities transfers this bound to $\rho_N$.
\end{theorem}
For bounded ReLU networks with $P_N$ bounded parameters, $\mathcal E_N$ grows as $P_N$ times a logarithmic factor, and an approximation bound $\varepsilon_N^{\rm app}\lesssim P_N^{-\beta_{\rm app}}$ with a sufficiently small optimization gap gives the rate $N^{-\beta_{\rm app}/(2\beta_{\rm app}+1)}$ up to logarithms (Corollary C.15), although for stochastic gradient training the gap $\varepsilon_N^{\rm opt}$ remains an assumption. A certified one-dimensional construction satisfies every remainder condition of Theorem~\ref{m:fulltheorem} when $N\asymp n$ (Corollary C.16).

\subsection*{Label-blind null testing}\label{m:nullsection}
Under $H_0^{\rm obs}$, let $R_g=\Delta-\int Ye^{g(X)}s_g^{-1}\,d\nu$ with $s_g(t)=\E\{Y(t)e^{g(X)}\}$ and $\nu=\E\cnt$, and let $g^\dagger=\widehat f+\gamma^{\dagger\top}z$, where $\gamma^\dagger$ is the population calibration coefficient with treatment fixed at zero.

\begin{theorem}[Observed-data null validity]\label{m:nulltheorem}
Assume $H_0^{\rm obs}$, $0<p_A<1$, positive event probability, $\Pr(U\ge\tau)\ge c_{\rm null}>0$, bounded working indices and $z$, a uniformly interior population calibration root with nonsingular Hessian, and $\E R_{g^\dagger}^2$ bounded away from zero. Then an interior fit with score residual $o_\Pp(n^{-1/2})$ satisfies
\[
\widehat\theta=\{np_A(1-p_A)\nu(\tau)\}^{-1}
\sum_i(A_i-p_A)R_{g^\dagger}(O_i)+o_\Pp(n^{-1/2}),
\qquad \widehat\theta/\widehat{\operatorname{se}}_{\rm sw}\Rightarrow N(0,1).
\]
No correct prognosis, PH model, or prediction overlap is required. The test \eqref{m:random} satisfies $\Pr(p_{\rm CR}\le\alpha\mid\widehat f,\{X_i,U_i,\Delta_i\},m_A)\le\alpha$ for every $\alpha\in(0,1)$ in finite samples, where $m_A$ is the number treated.
\end{theorem}
The robust variance uses $\E R_g^2$, which need not equal the event probability used by the information-only variance, and both conclusions are null guarantees.

\section*{Numerical studies}\label{m:simsection}
\subsection*{Design and comparisons}
To evaluate survival prediction and treatment inference on the same generated trial patients, we draw covariates $X\sim\operatorname{Uniform}(-1,1)^4$, a historical log risk $c_ff_0(x)$ with $f_0(x)=\sin(\pi x_1)+0.6\cos(\pi x_2)+0.5x_3x_4$ and prognostic amplitude $c_f=1$ unless stated otherwise, and a target drift $0.7x_1-0.5x_2$. Historical and target cumulative baselines are $0.28t^{1.4}$ and $0.35t^{1.2}$, censoring is exponential with rates 0.08 and 0.18, $\tau=2$, and $p_A=0.5$. Alternative cells have $\theta_0=\log(0.7)$ and $(n,N)=(300,3000)$, $(300,300)$, $(300,1000)$, and $(600,3000)$, with 1000 replications each, and null cells at $(300,3000)$ and $(600,3000)$ have 2000 each.

CT uses a historical neural network with rectified linear unit (ReLU) activations, hidden widths 24 and 12, a linear skip connection, and validation stopping. CT-spline uses a historical Cox fit on 18 fixed spline and product coordinates. Both use the four-coordinate linear correction. Target-only comparators are linear, spline (CSPLINE, on the same 18 coordinates), ridge-penalized spline (CSPLINE-ridge), and neural Cox fits, and a cross-fitted target-only neural score (CN-S). As a purpose-built transfer comparator, we implement an adapted TransCox-type estimator \citep{li2023} that shrinks deviations from the historical spline coefficients by a cross-validated lasso but, unlike the original, does not borrow the baseline hazard (Appendix O.4). PT-F uses five folds, a dictionary $B_{18}$ of four linear, eight first-frequency trigonometric, and six product terms informed by the generating law, ridge $0.001\sqrt{300/n_{\rm tr}}$ for training size $n_{\rm tr}$, and clipping at 3. Prediction error averages over 1000 fresh covariate vectors, both arms, and ten times from 0.2 to 2. Risk MSE is the mean squared error (MSE) of the centered control-arm log risk, and survival MSE, multiplied by 1000, is that of the predicted survival probabilities. Appendices I to O give all specifications and results and record which experiments were prespecified.

\subsection*{Historical information and prediction error}
At $(300,3000)$, CT has risk MSE 0.120 versus 0.250 for CSPLINE (Table~\ref{m:tableprediction}), a paired difference of $-0.130$ (Monte Carlo SE 0.004). At $N=300$ the comparison reverses, 0.402 versus 0.253, so calibration does not compensate for a poorly learned historical risk. CT-spline predicts more accurately than CT in all six cells, with risk MSE 0.062 at $(300,3000)$, and has treatment root mean squared error (RMSE) 0.180 versus 0.183 and the same 94.4\% coverage.

Penalizing the target-only fit narrows the advantage of transfer. In an independent replication of the four alternative cells (Appendix O.4), CSPLINE-ridge, with a cross-validated penalty, reduces the CSPLINE risk MSE at $(300,3000)$ from 0.264 to 0.192. CT-spline remains more accurate than CSPLINE-ridge whenever $N\ge1000$, by 0.054 to 0.130 in paired risk MSE, and is close to it at $N=300$, whereas neural CT is more accurate only at $(300,3000)$, by 0.069 (Monte Carlo SE 0.003). A TransCox-type estimator improves on CSPLINE-ridge and on neural CT in all four cells, but CT-spline is more accurate whenever $N\ge1000$, by 0.017 to 0.041, whereas with $N=300$ the TransCox-type estimator is more accurate by 0.047, because its penalized deviations adapt to a poorly learned historical fit (Appendix O.4).

\subsection*{Treatment precision and interval coverage}
At $(300,3000)$, PT-F has RMSE 0.206, coverage 93.9\%, and mean interval length 0.762, compared with 0.183, 94.4\%, and 0.693 for CT, which is the precision cost in \eqref{m:tradeoff}. With $N=3000$, CT rejects 5.2\% of nulls at $n=300$ and 600, and PT-F 6.2\% and 6.0\%. In the strong-prognosis law with $c_f=1.5$, $n=300$, and $N=6000$, both Wald intervals slightly undercover, and the full projected bootstrap raises PT-F coverage from 93.80\% to 95.56\%, with mean length 0.852 instead of 0.759 (Table~\ref{m:bootstrap}, paired gain 1.82 percentage points). Under four misspecified null laws, CR rejects 4.50 to 5.75\%, robust Wald 5.70 to 6.25\%, and information-only Wald 18.00 to 18.85\%.

\subsection*{When directional bias matters}\label{m:biassection}
\begin{figure}[p]\centering
\includegraphics[width=\textwidth]{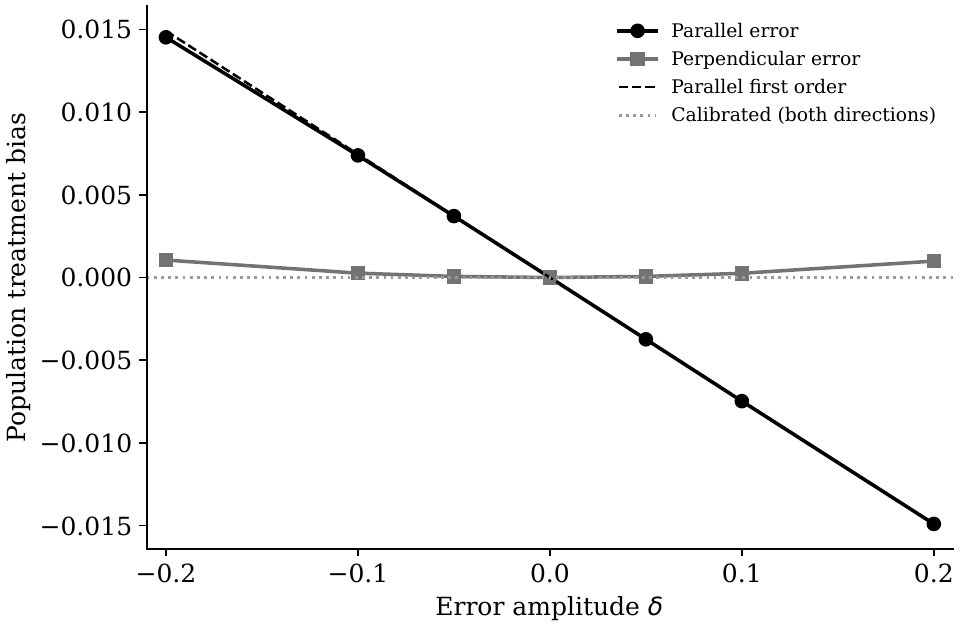}
\caption{Population treatment bias for fixed offsets $f_0+\delta v$ in the three-point covariate experiment, with error directions $v$ parallel and perpendicular to the representer $\omega_z$ without correction. Both directions have input MSE $\delta^2$. The parallel first-order curve is $-0.074542\delta$, and the perpendicular first-order term is zero. Calibration using the corresponding direction yields zero population bias.}\label{m:mechanism}
\end{figure}

With $X$ uniform on $\{-1,0,1\}$, $f_0(x)=0.8x$, and fixed errors of input MSE $\delta^2$ parallel and orthogonal to the representer $\omega_z$ without correction, the population treatment biases at $\delta=0.2$ are $-0.0149$ and $0.0010$, and the first-order prediction of Proposition~\ref{m:sensitivity} is $-0.0149$ (Figure~\ref{m:mechanism}). In the designs above, however, the CT bias is small relative to its SE, and the population score equations explain why. With hazard ratio 0.7 and drift $\sin(\pi x_2)+0.8x_1x_4$ outside the correction, the bias equals one SE only for target sizes above about 5000 (Table~S20).

\begin{figure}[p]\centering
\includegraphics[width=0.86\textwidth]{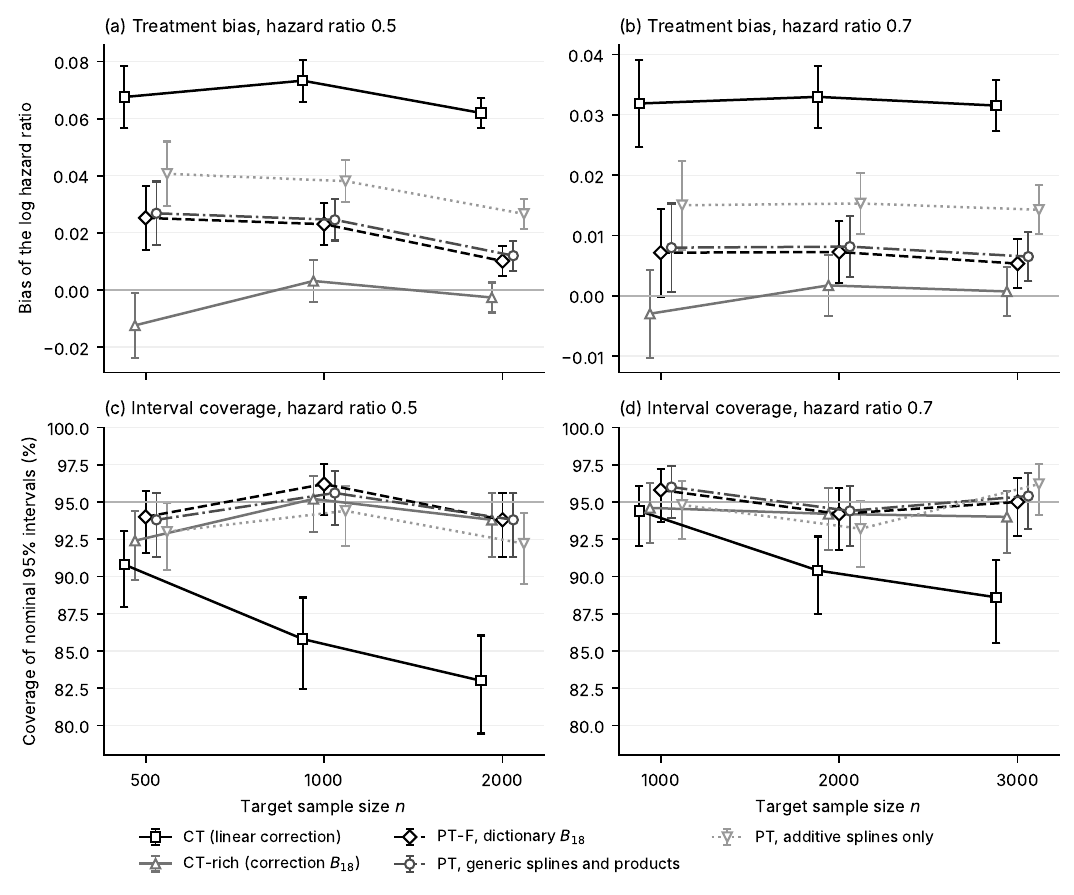}
\caption{Treatment bias (a, b) and coverage of nominal 95\% intervals (c, d) under drift outside the four-coordinate correction, with hazard ratio 0.5 (a, c) and 0.7 (b, d), $N=10000$, and 500 replications per target size.}\label{m:qbias}
\par\medskip\noindent Bars are 95\% Monte Carlo intervals, Wilson intervals for coverage. $B_{18}$ contains trigonometric terms of the generating law, and the generic dictionaries use additive natural cubic splines with or without all pairwise products.
\end{figure}

To show when the bias matters, we first evaluated a law chosen from this calculation, with a shared interaction outside the target spline span, drift outside the linear correction but inside $B_{18}$, $\theta_0=\log(0.5)$, $N=10000$, and about 63\% of target patients with events (Appendix O.3). The CT bias is 0.068, 0.073, and 0.062 at $n=500$, 1000, and 2000, close to the population value 0.064, so its coverage falls to 90.8\%, 85.8\%, and 83.0\% as its SE shrinks (Figure~\ref{m:qbias}a,c). PT-F covers 94.0\%, 96.2\%, and 93.8\%, with paired gains of 3.2, 10.4, and 10.8 percentage points (Monte Carlo SEs 1.3 to 1.6). CT-rich, which enlarges the correction to $B_{18}$, gives similar coverage, whereas the target-only CSPLINE and CN-S remain biased by 0.050 to 0.085.

To assess a less extreme law, we repeated this design with hazard ratio 0.7, as in OAK, and target sizes 1000, 2000, and 3000 (Appendix O.6). The CT bias is then 0.032 to 0.033, close to its population value 0.0335, and CT coverage falls to 94.4\%, 90.4\%, and 88.6\% (Figure~\ref{m:qbias}b,d), whereas PT-F covers 95.8\%, 94.2\%, and 95.0\%, with paired gains of 1.4, 3.8, and 6.4 percentage points (Monte Carlo SEs 1.1 to 1.4), a smaller RMSE at every size, and intervals at most 2.3\% longer. The standardized projected-minus-direct difference rejects in 39.0\%, 61.0\%, and 69.8\% of these trials.

Because $B_{18}$ contains trigonometric terms of the generating law, we repeated both designs with generic dictionaries (Appendix O.7). With additive natural cubic splines and all pairwise products, the projected score covers 93.8\% to 96.0\% under both laws, within 0.6 percentage points of PT-F (Figure~\ref{m:qbias}). Additive splines alone omit the product $x_1x_4$ of the drift and keep 43\% to 60\% of the CT bias, although they still cover 92.2\% to 96.2\%.

The two remedies differ in small trials (Appendix O.5). With $n=300$ and nested dictionaries of 18, 26, and 50 functions, CT-rich covers 94.4\%, 92.6\%, and 86.4\%, because more coefficients are estimated from about 190 events, whereas the projected score covers 95.2\% to 96.0\% with every dictionary. At $n=300$ and 600 the standardized difference rejects in 3.2\% to 5.8\% of trials without outside drift and in 16.2\% and 30.8\% with it.

\section*{External prognosis for survival prediction in OAK}\label{m:application}
\subsection*{Clinical question and analysis}
We examine retrospectively whether prognosis learned from POPLAR improves survival prediction in OAK when few target patients are available. Both trials compared atezolizumab with docetaxel in previously treated advanced non-small-cell lung cancer \citep{Fehrenbacher2016,Rittmeyer2017}. In the public clinical records accompanying \citet{Gandara2018}, the historical sample contains 143 POPLAR docetaxel recipients, and the target is the primary intention-to-treat population of OAK, with 850 patients. Censoring overall survival at 18 months gives 100 historical and 521 target deaths.

Using historical outcomes alone, repeated validation selects a bounded network with hyperbolic tangent (tanh) activations and a ReLU network for treatment analysis. Full calibration corrects all 11 shared clinical coordinates, and nested calibration selects among free score rescaling and three clinical correction spaces within target training samples. Comparators are clinical Cox, ridge clinical Cox with a penalty chosen by cross-validated partial likelihood \citep{verweij1993,verweij1994}, target spline and tuned neural Cox, classical recalibration \citep{vanhouwelingen2000} of a ridge-penalized POPLAR Cox score, and the adapted TransCox-type estimator of Section~\ref{m:simsection} with POPLAR clinical or ridge Cox coefficients as source. We assess prediction by the integrated Brier score (IBS) from 3 to 18 months in two designs, a full-size comparison with three repetitions of five-fold cross-validation and learning curves with four separate repetitions that reduce training samples to 100, 200, 400, or all patients, so full-size values differ slightly between the designs.

\subsection*{Prediction and treatment findings}
\begin{figure}[p]\centering
\includegraphics[width=\textwidth]{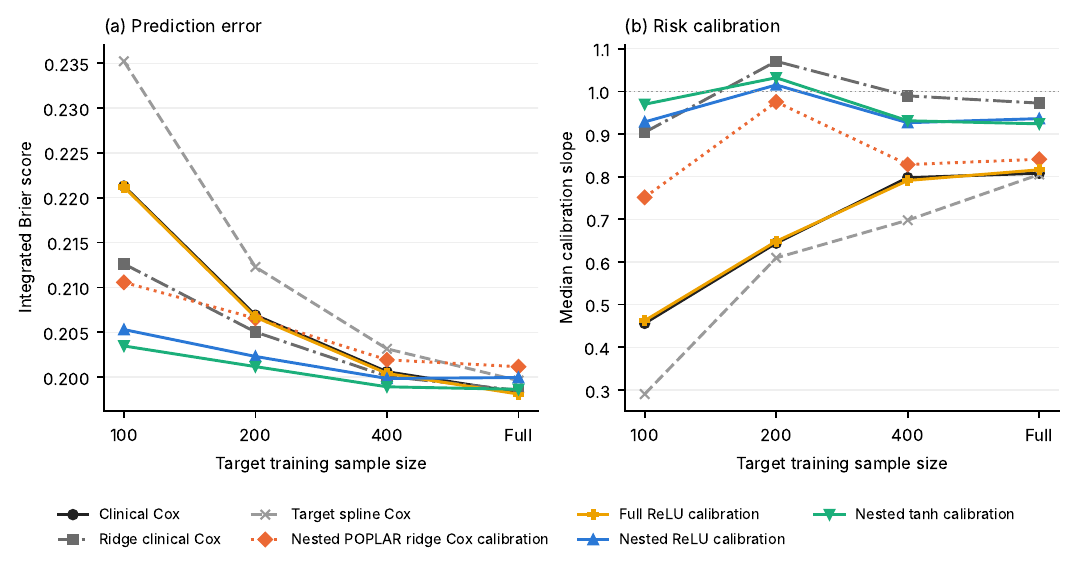}
\caption{OAK prediction across target training sizes.}\label{m:oakcurve}
\par\medskip\raggedright\noindent Each size uses 20 matched evaluations from four repetitions of five-fold validation. Full training samples contain 678 to 682 patients. Lower integrated Brier score is better. Calibration slopes are medians across evaluation folds, with one as the reference. Ridge clinical Cox chooses its penalty within each training sample. Nested calibration selects the correction within training samples, using either a final historical network score or a ridge-penalized POPLAR Cox score. Full ReLU calibration uses all 11 clinical corrections. The supplied historical scores remain fixed. The curves describe internal validation with overlapping samples and do not constitute independent replications or a test of superiority.
\end{figure}

In the full-size comparison, full ReLU calibration gives IBS 0.1969, compared with 0.1972 for clinical Cox and 0.1975 for ridge clinical Cox. In the learning curves with 100 training patients (Figure~\ref{m:oakcurve}), nested tanh calibration gives 0.2035, compared with 0.2213 for clinical Cox and 0.2126 for ridge clinical Cox, paired reductions of 8.1\% and 4.3\%, and median calibration slopes are 0.97, 0.46, and 0.90. The reduction relative to ridge clinical Cox is 1.9\% at 200 patients and at most 0.0012 in IBS at larger sizes, so the method helps mainly when the target sample is small.

Nested recalibration of the POPLAR ridge Cox score gives IBS 0.2106 at 100 patients, and the TransCox-type estimator gives 0.2139 with clinical and 0.2084 with ridge POPLAR coefficients. Nested tanh is lower than these by 3.4\%, 4.9\%, and 2.3\%, and its differences from the TransCox-type estimator are below 0.002 from 400 patients. A linear approximation of the tanh score, fitted to historical covariates alone, gives IBS 0.2036 at 100 patients. The gain therefore comes from better external clinical weights rather than from nonlinearity. In a second transfer, from the Rotterdam to the German Breast Cancer Study Group (GBSG) cohort with 552 historical patients, nested calibration and the TransCox-type estimator give IBS 0.1728 and 0.1713 at 100 patients, against 0.1816 for clinical Cox (Appendix Q).

Full ReLU calibration gives a working hazard ratio of 0.703 for atezolizumab relative to docetaxel, with nominal 95\% percentile interval (0.582, 0.834) from 999 target refits. As a robustness analysis, the projected score gives 0.723 (0.594, 0.885). The difference of 0.027 on the log scale has influence-based SE 0.018 and paired bootstrap standard deviation 0.036, which does not indicate directional bias in the direct fit, although with limited power. Bootstrap standard deviations of the log hazard ratio are 0.091 and 0.101, compared with 0.091 for clinical Cox, so prediction gains do not imply greater treatment precision. Given a global PH diagnostic with $p=0.010$ and unrecoverable allocation strata, these intervals are empirical working-model summaries (Appendix P).

\section*{Discussion}\label{m:discussion}
Calibration removes represented historical error exactly, but the remaining error has two roles. Its size determines prediction accuracy, and its direction within trial risk sets determines the first-order treatment bias. A transferred model can therefore predict well without valid treatment intervals, so prediction and precision gains should be reported separately.

The correction space governs how much is borrowed, and the accuracy of the historical learner, rather than its class, decides whether borrowing helps. In our simulations historical splines outperformed the neural learner and, with large historical samples, a coefficient-transfer estimator, whereas small historical samples produced negative transfer. In OAK nested calibration outperformed shrinkage toward POPLAR Cox coefficients at small target sizes, and in GBSG the two were similarly accurate. A linear approximation of the OAK score retained most of its gain, so a linear offset suffices when its weights are learned well.

For treatment inference, we recommend the direct fit as the primary analysis and the projected score as a robustness analysis. The direct fit is more precise when its bias is negligible, as in most of our designs and in OAK. Under drift outside the correction it undercovered at hazard ratios of 0.5 and 0.7, and the projected score restored coverage with law-informed and generic dictionaries. Under a fixed misspecification it keeps a second-order bias and is not consistent.

Our guarantees rest on a target Cox model, conditionally independent censoring, adequate follow-up, target covariate support within the historical support, comparable measurements, and, except in Proposition~\ref{m:fixed}, a vanishing remaining error. Trial inference concerns a constant conditional treatment coefficient under randomization. Estimated feature maps, inference after correction selection, time-varying effects, covariate-dependent allocation, joint efficiency using both cohorts, and historical sampling uncertainty remain open problems.

\section*{Data Availability}
The deidentified POPLAR and OAK data are available with \citet{Gandara2018}, and the Rotterdam and GBSG data with the \texttt{survival} package for R \citep{Royston2013}.

\section*{Supplementary Materials}
The Supplementary Materials, Appendices A to Q, follow the tables and figures of this article. Their equations, tables, figures, and algorithms are numbered S1, S2, and so on. A numerical archive and code for all results are available from the authors, and the Python package \texttt{ctcox} implementing the methods is available at \codeurl.

\begin{table}[p]\centering
\caption{Prediction and treatment estimation in the independent alternative cells.}\label{m:tableprediction}
\begin{tabular}{llrrrr}\toprule
$(n,N)$ & Method & Risk MSE & Surv. MSE & RMSE & Cover (\%)\\\midrule
300,3000 & CT & .120 & 5.654 & .183 & 94.4\\
 & CT-spline & .062 & 3.486 & .180 & 94.4\\
 & CSPLINE & .250 & 9.111 & .201 & 93.1\\
 & PT-F & NA & NA & .206 & 93.9\\
300,300 & CT & .402 & 17.988 & .180 & 94.7\\
 & CT-spline & .203 & 8.164 & .180 & 95.1\\
 & CSPLINE & .253 & 9.088 & .194 & 94.4\\
 & PT-F & NA & NA & .191 & 95.7\\
300,1000 & CT & .240 & 10.681 & .187 & 94.8\\
 & CT-spline & .085 & 4.302 & .181 & 94.3\\
 & CSPLINE & .248 & 9.096 & .200 & 93.5\\
 & PT-F & NA & NA & .201 & 94.3\\
600,3000 & CT & .099 & 4.351 & .126 & 94.4\\
 & CT-spline & .040 & 2.144 & .125 & 94.4\\
 & CSPLINE & .110 & 4.432 & .132 & 93.8\\
 & PT-F & NA & NA & .132 & 93.6\\\bottomrule
\end{tabular}
\par\medskip\noindent Each cell has 1000 replications and 100\% analysis availability. Risk MSE is the centered control-arm log-risk error. CT-spline uses a historical spline Cox fit with the same target calibration as CT. Survival MSE is multiplied by 1000. Treatment RMSE is on the log hazard ratio scale. Coverage is for nominal 95\% intervals. PT-F supplies treatment inference only. NA denotes a quantity that does not apply. Complete Monte Carlo uncertainty, other comparators, and null cells are in the Supplementary Materials.
\end{table}
\clearpage
\clearpage
\begin{table}[p]\centering
\caption{Projected bootstrap in the strong-prognosis setting, $n=300$ and $N=6000$.}\label{m:bootstrap}
\setlength{\tabcolsep}{8pt}
\begin{tabular}{llrrrr}\toprule
Law & Procedure & Available & Cover (\%) & Reject (\%) & Length\\\midrule
Null & CT Wald & 2000/2000 & 93.85 & 6.15 & .657\\
 & PT-F Wald & 2000/2000 & 93.35 & 6.65 & .727\\
 & PT-F bootstrap & 1991/2000 & 95.78 & 4.22 & .817\\
 & CR & 2000/2000 & NA & 5.25 & NA\\
Alternative & CT Wald & 1000/1000 & 93.70 & 51.50 & .682\\
 & PT-F Wald & 1000/1000 & 93.80 & 43.60 & .759\\
 & PT-F bootstrap & 992/1000 & 95.56 & 36.09 & .852\\
 & CR & 1000/1000 & NA & 49.50 & NA\\\bottomrule
\end{tabular}
\par\medskip\noindent NA denotes a quantity that does not apply. Rates condition on interval availability. Covering interval frequencies among all attempts are 95.35\% under the null and 94.80\% under the alternative for the projected bootstrap. The 95\% Monte Carlo intervals for its conditional coverage are [94.81,96.58] and [94.10,96.68], respectively, and the corresponding rejection intervals are [3.42,5.19] and [33.16,39.13]. The bootstrap shares the point estimate of PT-F and uses 999 full refits. Null rejection and alternative rejection must be read together, and these results do not establish size-controlled power for Wald procedures.
\end{table}
\clearpage
\clearpage

\clearpage
\section*{Supplementary Materials}

\noindent We give the assumptions, proofs, algorithms, and additional numerical results for the main article. Throughout the target analysis, we condition on the independent historical fit. We first give the prediction model for a general target cohort and then the treatment model for a randomized trial. We distinguish correction of historical prognosis from projection of the treatment score.
\paragraph{Relation to the main article.}
Appendices A and B expand the models, estimators, and algorithms of Sections 2 and 3 of the main article; Algorithms~\ref{alg:direct} and \ref{alg:score} give the computations summarized in Section 3.3 there. Appendix C restates every theoretical result with complete conditions, Appendix D discusses the conditions, and Appendices E to H give the proofs. Appendices I to O report the numerical designs and simulation results, Appendix P gives the application procedures and additional OAK results, and Appendix Q gives the GBSG prediction analysis. Complete records of exploratory and superseded follow-up experiments and secondary application tables are collected in a separate numerical archive, available from the authors, cited here as Archive Tables, Figures, and Sections. Equations and results in this supplement are numbered independently of the main article. The following table matches each result in the main article to its restatement and proof here.
{\footnotesize
\begin{longtable}{>{\raggedright\arraybackslash}p{6.5cm}>{\raggedright\arraybackslash}p{5.4cm}>{\raggedright\arraybackslash}p{2.4cm}}
\toprule
Main article & Supplement & Proof in Appendix\\
\midrule\endhead
\bottomrule\endfoot
Proposition 4.1 (calibration invariance) & Proposition~\ref{prop:invariance} & E.2\\
Theorem 4.2 (prediction without treatment) & Theorem~\ref{thm:generalprediction} & E.1\\
Corollary 4.3 (selection of the correction) & Corollary~\ref{cor:selection} & \ref{sec:selection}\\
Section 4.2 (existence of $h_*$, $\J$ seminorm bound) & Lemma~\ref{lem:geometry} & E.10\\
Theorem 4.4 (direct calibration and prediction) & Theorems~\ref{thm:plugin} and \ref{thm:prediction} & E.5, E.9\\
Proposition 4.5 (sharp sensitivity) & Proposition~\ref{prop:sensitivity} & E.11\\
Theorem 4.6 and display (7) (finite dictionary) & Theorem~\ref{thm:finite}, Proposition~\ref{prop:dictionarytradeoff} & E.13, E.12\\
Theorem 4.7 (unrestricted projection) & Theorem~\ref{thm:score}, Corollary~\ref{cor:rates} & E.6, E.7\\
Section 4.5 (projected minus direct contrast) & Corollary~\ref{cor:contrast} & \ref{sec:scoretheory}\\
Proposition 4.8 (fixed remaining error) & Proposition~\ref{prop:fixedmis} & \ref{sec:scoretheory}\\
Theorem 4.9 (historical prediction bound) & Theorem~\ref{thm:source}, Corollaries~\ref{cor:relu} and \ref{cor:feasible} & F.4, F.5, G.3\\
Theorem 4.10 (null validity) & Theorem~\ref{thm:null}, Proposition~\ref{prop:randomization} & H; \ref{sec:null}\\
Computations in Section 3.3 & Algorithms~\ref{alg:direct} and \ref{alg:score} & ---\\
\end{longtable}}\addtocounter{table}{-1}%

\setcounter{tocdepth}{1}
\tableofcontents
\clearpage

\section{Model and estimands}\label{sec:model}
\subsection{Notation}\label{sec:notation}
We use $H$ for the historical cohort and $T$ for the target cohort, which may be a randomized trial. Covariates take values in $\mathcal X\subseteq\mathbb R^d$, with dimension $d\ge1$ and generic value $x$. Historical covariates, event time, and censoring time are $X^H,T^H,C^H$. Their target counterparts are $X,T,C$. The subscript $T$ labels a cohort, whereas the random variable $T$ denotes an event time. For a fixed analysis horizon $\tau>0$ and indicator function $\ind$, define
\begin{align*}
U^H&=\min(T^H,C^H,\tau),&\Delta^H&=\ind\{T^H\le C^H,T^H\le\tau\},\\
U&=\min(T,C,\tau),&\Delta&=\ind\{T\le C,T\le\tau\}.
\end{align*}
For prediction without a treatment variable, the target record is $O=(U,\Delta,X)$. In the trial setting, historical patients receive background treatment. In the trial, $A=0$ denotes the comparator and $A=1$ the investigational treatment. We assume $A\sim\operatorname{Bernoulli}(p_A)$ independently of $X$, with $0<p_A<1$, and use $a\in\{0,1\}$ for an arm value. The independent samples contain $N$ historical records $O_j^H=(U_j^H,\Delta_j^H,X_j^H)$, $j=1,\ldots,N$, and $n$ target records $O_i=(U_i,\Delta_i,A_i,X_i)$, $i=1,\ldots,n$. Records within each cohort are independent and identically distributed, with laws $P_H$ and $P_T$. We write $P_{H,X},P_{T,X}$ for their covariate laws and $P_{T,A,X}$ for the target joint law of $(A,X)$.

Throughout, $\E$ and $\E_H$ denote target and historical expectation, and $\Pp$ denotes sampling probability. Vector norms are Euclidean. For functions, $\norm{\cdot}_2$ denotes the target $L_2$ norm unless another measure is stated, and $\norm{\cdot}_\infty$ the essential supremum. The symbol $\ll$ denotes absolute continuity. We use $O_\Pp,o_\Pp$, $\to_\Pp$, and $\Rightarrow$ for stochastic order and convergence. For positive sequences, $\lesssim$ denotes a bound up to a fixed constant, $\asymp$ bounds in both directions, and $\gg$ a diverging ratio. Unspecified constants $C$ are independent of sample size and may change between displays. Model, estimator, and projection notation is introduced where it is used below.

\subsection{Prediction in a general target cohort}\label{sec:generalmodel}
For a target cohort without a treatment variable, let $f_0$ be historical prognostic log risk, $z:\mathcal X\to\mathbb R^p$ a deterministic correction vector of fixed dimension, and $\gamma_0\in\mathbb R^p$ its target coefficient. Write $g_0=f_0+\gamma_0^\top z$. The prediction model is
\begin{equation}\label{eq:generalmodel}
\lambda_H(t\mid x)=\lambda_{H,0}(t)e^{f_0(x)},\qquad
\lambda_T(t\mid x)=\lambda_{T,0}(t)e^{g_0(x)}.
\end{equation}
The cohort baselines are unrelated. For $V\in\{H,T\}$, define $\Lambda_{V,0}(t)=\int_0^t\lambda_{V,0}(s)\,ds$. We estimate $S(t\mid x)=\exp\{-\Lambda_{T,0}(t)e^{g_0(x)}\}$. We assume independent observations within each independent cohort and conditionally independent censoring given its covariates. No treatment assignment assumption is imposed. Covariate distributions and censoring laws may differ between cohorts.

We exclude a correction intercept and use a fixed reference to identify the arbitrary log-risk constant. The positive observation and information conditions in Section~\ref{sec:generaltheory} identify the target correction and baseline once historical prognosis is identified on the target support. Specifically, equality of two target representations gives $b^\top z(X)$ equal to a constant. Positive correction information excludes a nonzero such $b$. For transfer of a historical prediction rate we use the support and density condition in \eqref{eq:overlap}. That bound is unnecessary if a target residual error rate is assumed directly.

Given an independently estimated bounded historical function $\widehat f$, put $\eta_\gamma(x)=\widehat f(x)+\gamma^\top z(x)$ and maximize
\begin{equation}\label{eq:generalfit}
\ell_n^{\rm pred}(\gamma)=\frac1n\sum_i\Delta_i
\left[\eta_\gamma(X_i)-\log\left\{\sum_{j:U_j\ge U_i}e^{\eta_\gamma(X_j)}\right\}\right].
\end{equation}
Let $\widehat\gamma$ be an interior maximizer and set $\widehat g=\widehat f+\widehat\gamma^\top z$. The prediction estimators are
\begin{align*}
\widehat\Lambda_{T,0}(t)&=\sum_{i:U_i\le t}\frac{\Delta_i}{\sum_{j:U_j\ge U_i}e^{\widehat g(X_j)}},\\
\widehat S(t\mid x)&=\exp\{-\widehat\Lambda_{T,0}(t)e^{\widehat g(x)}\}.
\end{align*}
The gradient and Hessian calculations in Section~\ref{sec:methods} apply with $W=z$, $\beta=\gamma$, and functions evaluated at $X$. Algorithm~\ref{alg:direct} therefore uses only the correction coordinates for this setting. It omits the treatment coordinate entirely. It does not fit a constant treatment column, which would create singular information. For $p=0$, the supplied risk is fixed and only the target baseline is estimated.

An invertible linear change of coordinates in a fixed correction space preserves its unpenalized fitted indices when coefficient constraints transform accordingly. More general estimated feature maps, imputation rules, and outcome-driven selection require stability or selection arguments for theoretical guarantees. During prediction validation, all estimated preprocessing and selection must use training data only. This prevents evaluation leakage but does not by itself prove the fixed feature assumptions.

\subsection{Randomized trial model and inferential targets}\label{sec:transfermodel}
For the randomized trial setting, we additionally seek an interpretable treatment coefficient and arm-specific survival predictions. We propose a calibrated transfer model with separate cohort baselines. Let $f_0:\mathcal X\to\mathbb R$ be historical prognostic log risk and $\lambda_{H,0}$ the historical baseline hazard. We model the historical conditional event hazard $\lambda_H(t\mid x)$ by the Cox proportional hazards (PH) model
\begin{equation}\label{eq:H}
\lambda_H(t\mid x)=\lambda_{H,0}(t)\exp\{f_0(x)\}.
\end{equation}
For the trial, let $z:\mathcal X\to\mathbb R^p$ be a prespecified deterministic correction vector of fixed dimension $p$, with no correction when $p=0$. Let $\theta_0\in\mathbb R$ and $\gamma_0\in\mathbb R^p$ be the treatment and correction coefficients, and $\lambda_{T,0}$ the target baseline hazard. We model the target conditional event hazard $\lambda_T(t\mid a,x)$ by the partially linear Cox model
\begin{equation}\label{eq:T}
\lambda_T(t\mid a,x)=\lambda_{T,0}(t)\exp\{\theta_0a+f_0(x)+\gamma_0^\top z(x)\}.
\end{equation}
We write $g_0=f_0+\gamma_0^\top z$ for target prognostic log risk. We transfer the nonlinear relative risk function and estimate its correction using current events. The linear treatment term and nonlinear prognostic function give the partially linear structure. In the baseline subscripts, $H$ or $T$ denotes cohort and $0$ denotes the baseline component. The baselines are unrestricted and unrelated. Neither a historical baseline nor a historical treatment effect is borrowed. The term $\gamma_0^\top z$ allows relative covariate effects to change between cohorts. Its adequacy is a substantive transport assumption. Enlarging $z$ permits more changes but requires more target information. Outcome driven or estimated correction coordinates need an additional selection or perturbation argument.

The OAK application uses prognosis learned from docetaxel recipients in POPLAR to adjust a comparison of atezolizumab with docetaxel. Separate baselines allow the cohorts to differ in overall survival, while calibration permits selected relative covariate effects to change. Comparable time origins, endpoints, and baseline measurements remain necessary for a useful transfer. Appendix O describes the implementation and distinguishes the empirical clinical summaries from guarantees under the model and allocation assumptions stated here.

For model-based inference, we assume $T^H\perp C^H\mid X^H$ and $T\perp C\mid(A,X)$, allowing different censoring distributions. Define the cumulative baseline hazards by
\[
\Lambda_{H,0}(t)=\int_0^t\lambda_{H,0}(s)\,ds,\qquad
\Lambda_{T,0}(t)=\int_0^t\lambda_{T,0}(s)\,ds.
\]
Our treatment estimand is $\theta_0$, with conditional hazard ratio $e^{\theta_0}$ under \eqref{eq:T}. We also estimate $g_0$ and target conditional survival,
\[
S_a(t\mid x)=\Pr(T>t\mid A=a,X=x)
=\exp[-\Lambda_{T,0}(t)e^{\theta_0a+g_0(x)}].
\]
The model excludes delayed effects and treatment and covariate interactions. Its conditional coefficient differs from a marginal hazard ratio or a restricted mean survival time contrast.

To fix log-risk levels, we choose a deterministic identifying reference $R\ll P_{H,X}$ and impose $\int f_0\,dR=0$ and $\int z\,dR=0$, with no correction intercept. Transport identification requires target support within historical support. For prediction rates, we use the stronger density bound, with fixed $C_{\rm ov}>0$,
\begin{equation}\label{eq:overlap}
P_{T,X}\ll P_{H,X},\qquad \frac{dP_{T,X}}{dP_{H,X}}\le C_{\rm ov}<\infty.
\end{equation}
The density bound holds almost surely under $P_{H,X}$. We use it for prediction and alternative effect guarantees, not for the robust null test.

\begin{proposition}[Identification on the observable interval]\label{prop:ident}
Let $0<\tau_{\rm id}\le\tau$ be an observable horizon. Assume conditionally independent censoring in both cohorts, $\Pr(U^H\ge\tau_{\rm id}\mid X^H)>0$ and $\Pr(U\ge\tau_{\rm id}\mid A,X)>0$ almost surely, and finite cumulative baseline hazards on $[0,\tau_{\rm id}]$, both strictly positive at $\tau_{\rm id}$. Suppose $0<p_A<1$, $P_{T,X}\ll P_{H,X}$, and the target covariance matrix $\operatorname{Var}_T\{z(X)\}$ is positive definite. Then $f_0$ on the historical support, $\theta_0$, $\gamma_0$, and both cumulative baselines on $[0,\tau_{\rm id}]$ are identified subject to the centering convention. Risk functions are identified up to their covariate law null sets and baseline hazard densities up to time null sets. No identification beyond the observable interval is asserted.
\end{proposition}
Historical cumulative hazard ratios identify differences in $f_0$. Randomization identifies both target arm hazards, and their ratio identifies $\theta_0$. The remaining target contrast identifies $\gamma_0$, because positive covariate variance excludes a nonzero correction combination equal to a constant. Centering fixes the otherwise arbitrary baseline multiplier.

\paragraph{A separate null testing target.}
We also study a treatment test that remains valid when the prognostic model is misspecified. For this analysis, we use the observed-data null
\begin{equation}\label{eq:null}
H_0^{\rm obs}:\quad A\perp(X,U,\Delta).
\end{equation}
Under randomization, equality of the arm-specific conditional event time distributions and a common conditional censoring law imply \eqref{eq:null} when censoring is conditionally independent. Thus the hypothesis represents no treatment effect with treatment-independent observation. It does not require PH or a correct historical risk model. Treatment dependent censoring that also changes the covariate composition of risk sets requires additional analysis. Enrollment randomization alone does not imply \eqref{eq:null}.

\section{Calibrated prediction and treatment analysis}\label{sec:methods}
\subsection{Historical learning and target calibration}
The historical fitting step is common to both settings. We give the target formulas with treatment below. Their prediction counterparts are specified in \eqref{eq:generalfit}.
Let $\mathcal F_N$ be a deterministic class of measurable historical log-risk functions bounded in absolute value by a fixed $B_f>0$. Any learner with bounded output can be used, and Section~\ref{sec:sourcerates} gives rates for bounded rectified linear unit (ReLU) networks. We assume a countable separable version of the class and a measurable fitted function. For a candidate $f\in\mathcal F_N$, define the normalized historical Cox partial log likelihood
\begin{equation}\label{eq:fitH}
\ell_H(f)=\frac1N\sum_{j=1}^N\Delta^H_j\left[f(X^H_j)-\log\left\{\sum_{k:U^H_k\ge U^H_j}e^{f(X^H_k)}\right\}\right].
\end{equation}
Let $\widehat f\in\mathcal F_N$ be the fitted historical function. Regularization or early stopping may be used computationally, but our theory retains the discrepancy from the unpenalized class optimum. For candidate coefficients, write $\beta=(\theta,\gamma^\top)^\top$ and $W(a,x)=(a,z(x)^\top)^\top$, with patient value $W_i=W(A_i,X_i)$. For a supplied offset $f$, define $\eta_{\beta,f}(a,x)=\theta a+f(x)+\gamma^\top z(x)$. With $f=\widehat f$, abbreviate this by $\eta_\beta$, with $\eta_{\beta,i}=\eta_\beta(A_i,X_i)$. We calibrate the transferred function by maximizing the target criterion
\begin{equation}\label{eq:fitT}
\ell_n(\beta,\widehat f)=\frac1n\sum_{i=1}^n\Delta_i\left[\eta_{\beta,i}-\log\left\{\sum_{j:U_j\ge U_i}e^{\eta_{\beta,j}}\right\}\right].
\end{equation}
Let $\widehat\beta=(\widehat\theta,\widehat\gamma^\top)^\top$ denote its fitted coefficients, and set $\widehat g=\widehat f+\widehat\gamma^\top z$ and $\widehat\eta(a,x)=\widehat\theta a+\widehat g(x)$. We also write $\ell_n(\theta,\gamma,f)$ or $\ell_n(\beta,f)$ for the same criterion with supplied offset $f$. We use each cohort's own risk sets. Combining them in one unstratified likelihood would impose a baseline relationship absent from our model.

For risk-set calculations, define $Y(t)=\ind(U\ge t)$ and $\cnt(t)=\ind(U\le t,\Delta=1)$, with patient versions $Y_i,\cnt_i$. Write $\nu(t)=\E \cnt(t)$ and $\nu_n(t)=n^{-1}\sum_i\cnt_i(t)$ for the population and empirical cumulative event measures. For a supplied index $\eta(a,x)$ and scalar or column vector function $v(a,x)$, put $\eta_i=\eta(A_i,X_i)$ and $v_i=v(A_i,X_i)$. We define weighted risk-set moments and the score by
\begin{align*}
s_{n,\eta}(t)&=n^{-1}\sum_iY_i(t)e^{\eta_i},\qquad
\bar v_{n,\eta}(t)=\frac{\sum_iY_i(t)e^{\eta_i}v_i}{\sum_iY_i(t)e^{\eta_i}},\\
\mathcal U_n(\eta,v)&=\frac1n\sum_i\int_0^\tau\{v_i-\bar v_{n,\eta}(t)\}\,d\cnt_i(t).
\end{align*}
Matrix valued averages are taken entry by entry. For scalar or column vector functions $u,v$, we define the empirical event-weighted covariance by
\begin{equation}\label{eq:empJ}
\widehat\J_n(\eta,u,v)=\int_0^\tau
\{\overline{uv^\top}_{n,\eta}(t)-\bar u_{n,\eta}(t)\bar v_{n,\eta}(t)^\top\}\,d\nu_n(t).
\end{equation}
A sample or fold subscript means that all averages, risk sets, and event measures use only that sample. Event integrals are over $[0,\tau]$ unless stated otherwise. The index argument is omitted only when specified by the fit. Ratios with an empty empirical risk set are set to zero. Such times contribute no events.

We obtain a finite-dimensional target optimization even when the historical learner is a deep network. Its gradient and negative Hessian are
\begin{equation}\label{eq:calibrationderivatives}
\nabla_\beta\ell_n=\mathcal U_n(\eta_\beta,W),\qquad
-\nabla_\beta^2\ell_n=\widehat\J_n(\eta_\beta,W,W).
\end{equation}
Thus the criterion is concave. An interior score root is a global maximum, with local uniqueness when the Hessian is positive definite. Newton updates or another concave optimizer can be used. Define the termination score residual by $\xi_n=\norm{\nabla_\beta\ell_n(\widehat\beta,\widehat f)}_2$. A nearly singular Hessian indicates weak event-weighted information and can amplify optimization and offset errors. A correction penalty changes the score equations and the calibration invariance proved below. Constant recentering of the historical output is harmless if the target baseline is rescaled accordingly.

We estimate the target cumulative baseline by the Breslow estimator and obtain conditional survival from
\begin{align}
\widehat\Lambda_{T,0}(t)&=\sum_{i:U_i\le t}\frac{\Delta_i}{\sum_{j:U_j\ge U_i}\exp\{\widehat\theta A_j+\widehat g(X_j)\}},\label{eq:breslow}\\
\widehat S_a(t\mid x)&=\exp[-\widehat\Lambda_{T,0}(t)\exp\{\widehat\theta a+\widehat g(x)\}].\label{eq:survival}
\end{align}
Thus survival calibration uses current events even when the historical risk function is well estimated. No historical baseline is transported.

\subsection{Empirical sandwich variance}
We estimate the influence vector of $\mathcal U_n(\eta,v)$ by
\begin{align}
\widehat\varphi_i(\eta,v)
&=\int_0^\tau\{v_i-\bar v_{n,\eta}(t)\}\,d\cnt_i(t)-\mathcal U_n(\eta,v)\notag\\
&\quad-\int_0^\tau\frac{Y_i(t)e^{\eta_i}}{s_{n,\eta}(t)}\{v_i-\bar v_{n,\eta}(t)\}\,d\nu_n(t).\label{eq:empIF}
\end{align}
The second integral accounts for estimation of the risk-set mean. Treating event contributions as independent score terms would omit it. Let $\widehat I_W=\widehat\J_n(\widehat\eta,W,W)$ be the negative Hessian at the trial fit. Estimate the coefficient covariance by
\begin{equation}\label{eq:sandwich}
\widehat{\operatorname{Cov}}(\widehat\beta)=\frac1n\widehat I_W^{-1}\left\{\frac1n\sum_i\widehat\varphi_i(\widehat\eta,W)\widehat\varphi_i(\widehat\eta,W)^\top\right\}\widehat I_W^{-1}.
\end{equation}
Define $\widehat{\operatorname{se}}_{\rm sw}$ as the square root of its $(1,1)$ entry, corresponding to treatment. This is the empirical form of robust Cox covariance \citep{linwei1989} and gives the null test in Section~\ref{sec:null}. Under alternatives, sandwich estimation does not by itself remove bias from historical error. Valid inference for $\theta_0$ then requires the direct fit conditions or the projected score construction below.

\subsection{A cross-fitted projected score}\label{sec:projectedmethod}
Write $\eta_0(a,x)=\theta_0a+g_0(x)$ for the true target index and $\eta_{0i}=\eta_0(A_i,X_i)$. For any supplied index $\eta$, we write $s_\eta(t)=\E\{Y(t)e^{\eta(A,X)}\}$ for the population weighted risk-set mass. For a supplied offset $f$, define the population calibration criterion by
\[
\ell_P(\beta,f)=\E\int_0^\tau
\{\eta_{\beta,f}(A,X)-\log s_{\eta_{\beta,f}}(t)\}\,d\cnt(t).
\]
We also write $\ell_P(\theta,\gamma,f)$. The expectation uses a fresh target record, holding an externally fitted $f$ fixed. This criterion corresponds to \eqref{eq:fitT} after removing its parameter independent $\log n$ term. Under the target model, set $s_0=s_{\eta_0}$. We define the true weighted risk-set expectation $Q_t$ and covariance form $\J$ by
\begin{align}
Q_t(v)&=\E\{Y(t)e^{\eta_0(A,X)}v(A,X)\}/s_0(t),\notag\\
\J(u,v)&=\int_0^\tau s_0(t)\{Q_t(uv^\top)-Q_t(u)Q_t(v)^\top\}\,d\Lambda_{T,0}(t).\label{eq:J}
\end{align}
For scalar $u$, its $\J$ seminorm is $\{\J(u,u)\}^{1/2}$. Constants have zero seminorm. Let $h_*(X)$ be the projection of $A$ onto $L_2(P_{T,X})$ functions, modulo constants, in this seminorm. Lemma~\ref{lem:geometry} establishes existence, closedness, and a bounded centered representative under the conditions below. Define its treatment residual $q_*$ and information $I_X$ by
\begin{equation*}
q_*=A-h_*(X),\qquad \J(q_*,h)=0\quad(h\in L_2(P_{T,X})),\qquad I_X=\J(q_*,A)=\J(q_*,q_*)>0.
\end{equation*}
The projection is weighted by event risk sets. The enrollment propensity $p_A$ need not satisfy this orthogonality under a treatment alternative.

We propose cross-fitting to separate nuisance training from score evaluation. Partition the target indices into $K\ge2$ folds $\mathcal I_1,\ldots,\mathcal I_K$, independently of patient records, with fixed $K$. Let $m_k=|\mathcal I_k|$, $w_k=m_k/n$, and $\mathcal I_{-k}$ be the training complement. On $\mathcal I_{-k}$, calibrate with $\widehat f$ held fixed to obtain $\widetilde\theta_k,\widetilde\gamma_k$, and define $\widetilde g_k=\widehat f+\widetilde\gamma_k^\top z$ and $\widetilde\eta_k(a,x)=\widetilde\theta_k a+\widetilde g_k(x)$.

To estimate the projection, choose a deterministic reference $R_P\ll P_{T,X}$ with bounded density ratio, and a deterministic dictionary $B(X)\in\mathbb R^M$ with coordinates centered under $R_P$. This projection reference may differ from the historical identifying reference $R$. Approximate $h_*$ by $h(X)=\mu^\top B(X)$, where $\mu\in\mathbb R^M$. Let $I_M$ be the identity matrix and choose fold ridge coefficients $0\le\lambda_k\le\lambda_n$, where $\lambda_n$ is a deterministic nonnegative sequence. We estimate the coefficient from
\[
\widehat\Sigma_{B,k}=\widehat\J_{\rm tr,k}(\widetilde\eta_k,B,B),\qquad
\widehat c_{B,k}=\widehat\J_{\rm tr,k}(\widetilde\eta_k,B,A),\qquad
\widehat\mu_k=(\widehat\Sigma_{B,k}+\lambda_kI_M)^{-1}\widehat c_{B,k}.
\]
Here the subscript ${\rm tr},k$ means evaluation on $\mathcal I_{-k}$, and the inverse is required to exist. Basis approximation and ridge error are included in the projection error.

For a fixed bound $L_h>0$, define $\operatorname{clip}_{[-L_h,L_h]}(v)=\min\{L_h,\max(-L_h,v)\}$. We align the fitted representative under $R_P$ and set $\widehat h_k(x)=\operatorname{clip}_{[-L_h,L_h]}\{\widehat\mu_k^\top B(x)\}$. Full projection inference requires the clipping interval to contain the corresponding representative of $h_*$. Finite-dictionary inference requires it to preserve the dictionary projection analyzed in Theorem~\ref{thm:finite}. Constants cancel from scores, but clipping does not commute with adding a constant, so alignment precedes clipping. The uniform reference construction in Appendix~\ref{app:projection} verifies this convention under its stated density assumptions.

On evaluation fold $k$, let $\widehat q_k(a,x)=a-\widehat h_k(x)$ and let $\widehat\varphi_{i,k}$ denote \eqref{eq:empIF} computed entirely within $\mathcal I_k$. Define its score $\mathcal U_k$, derivative $D_k$, and empirical influence variance $V_k$ by
\begin{align}
\mathcal U_k&=\mathcal U_{m_k}(\widetilde\eta_k,\widehat q_k),\qquad D_k=\widehat\J_k(\widetilde\eta_k,\widehat q_k,A),\label{eq:Uk}\\
V_k&=m_k^{-1}\sum_{i\in\mathcal I_k}\widehat\varphi_{i,k}(\widetilde\eta_k,\widehat q_k)^2.\notag
\end{align}
The cross-fitted one-step treatment estimator and its standard error are
\begin{align}
\widehat\theta^{\rm score}&=\sum_{k=1}^K w_k\{\widetilde\theta_k+\mathcal U_k/D_k\},\label{eq:scoreest}\\
\widehat{\operatorname{se}}_{\rm score}^{\,2}&=\sum_{k=1}^K\frac{w_k^2V_k}{m_kD_k^2}.\label{eq:se}
\end{align}
All evaluation risk sets and influence functions use only that fold. Under consistent nuisance estimation, $V_k$ and the model information variance have the same limit. The empirical influence version is used in the simulation. A nonpositive or nearly zero $D_k$ is treated as a numerical failure. Cross-fitting supplies independent evaluation observations. The projection supplies orthogonality.

\subsection{Fitting and score algorithms}\label{sec:algorithms}
We give the computational steps for the two analyses. In Algorithm~\ref{alg:direct}, $r=0,1,\ldots$ indexes coefficient iterates $\beta^{(r)}$, and $\eta^{(r)}=\eta_{\beta^{(r)},\widehat f}$. The backtracking step $\upsilon_r\in(0,1]$ keeps the iterate feasible and increases the target likelihood. We prespecify a normalized score tolerance $\varepsilon_{{\rm cal},n}>0$ and an iteration limit. The theorem's numerical root requirement is met if $\varepsilon_{{\rm cal},n}=o(n^{-1/2})$ and the algorithm returns an interior fit. A singular Hessian, separation, or failure to reach this tolerance is reported as a fitting failure.

\begin{algorithm}[htbp]
\caption{Historical learning and direct target calibration}\label{alg:direct}
\begin{algorithmic}[1]
\Require Historical records $\{O_j^H\}_{j=1}^N$, target records $\{O_i\}_{i=1}^n$, bounded source class $\mathcal F_N$, correction $z$, feasible $\beta^{(0)}$, and $\varepsilon_{{\rm cal},n}$.
\Ensure The fitted risk, target baseline, applicable survival predictions, coefficient covariance, and numerical diagnostics, or a fitting failure. A treatment standard error is returned only for a trial.
\State Fit $\widehat f\in\mathcal F_N$ using the historical criterion \eqref{eq:fitH}. Record the class and any certified likelihood gap. Algorithm~\ref{alg:source} gives one certified choice.
\State For prediction without treatment, use $W=z$, $\beta=\gamma$, and \eqref{eq:generalfit}. If $W$ has no coordinates, return the target baseline and survival fit. Otherwise set $r=0$ and evaluate the applicable target likelihood.
\While{$\norm{\mathcal U_n(\eta^{(r)},W)}_2>\varepsilon_{{\rm cal},n}$}
 \State Compute $\mathcal U_n(\eta^{(r)},W)$ and $\widehat\J_n(\eta^{(r)},W,W)$ from \eqref{eq:calibrationderivatives}. Report failure if the Hessian is singular or the iteration limit is reached.
 \State Backtrack $\upsilon_r$ until the following feasible update increases $\ell_n(\cdot,\widehat f)$. Report failure if no step is accepted:
 \Statex \hspace{\algorithmicindent}$\displaystyle \beta^{(r+1)}=\beta^{(r)}+\upsilon_r\{\widehat\J_n(\eta^{(r)},W,W)\}^{-1}\mathcal U_n(\eta^{(r)},W).$
 \State Set $r\gets r+1$.
\EndWhile
\State Set $\widehat\beta=\beta^{(r)}$, verify interiority and nonsingular $\widehat I_W$, and record $\xi_n=\norm{\mathcal U_n(\widehat\eta,W)}_2$. Report failure if these checks fail.
\State Form $\widehat g$ and return the baseline and survival fit in Section~\ref{sec:generalmodel} for prediction, or \eqref{eq:breslow} and \eqref{eq:survival} for a trial. Compute coefficient influence and sandwich variance by \eqref{eq:empIF} and \eqref{eq:sandwich} with the applicable $W$.
\end{algorithmic}
\end{algorithm}

In Algorithm~\ref{alg:score}, every training and evaluation quantity uses its own fold risk sets and sample-size normalization. We hold the same independent historical fit fixed across folds. A prespecified threshold $\underline D_n>0$ guards the score derivative. Choosing $\underline D_n\to0$ makes this guard asymptotically inactive under positive limiting information. The training score tolerance also has order $o(n^{-1/2})$, since training sizes are proportional to $n$. A failed fold makes the specified analysis fail. It is not omitted from aggregation.

\begin{algorithm}[htbp]
\caption{Cross-fitted projected treatment analysis}\label{alg:score}
\begin{algorithmic}[1]
\Require $\widehat f$, target records, folds $\mathcal I_k$, correction $z$, calibration initialization $\beta^{(0)}$ and tolerance $\varepsilon_{{\rm cal},n}$, deterministic dictionary $B\in\mathbb R^M$ centered under $R_P$, ridge coefficients $\lambda_k$, clipping bound $L_h$, and threshold $\underline D_n$.
\Ensure $\widehat\theta^{\rm score}$ and $\widehat{\operatorname{se}}_{\rm score}$, or an analysis failure.
\For{$k=1,\ldots,K$}
 \State Apply the target calibration steps of Algorithm~\ref{alg:direct} to $\mathcal I_{-k}$ with $\widehat f$ fixed, obtaining $\widetilde\theta_k,\widetilde\gamma_k,\widetilde g_k,\widetilde\eta_k$. Report failure if that fit fails.
 \State Compute $\widehat\Sigma_{B,k}=\widehat\J_{\mathrm{tr},k}(\widetilde\eta_k,B,B)$ and $\widehat c_{B,k}=\widehat\J_{\mathrm{tr},k}(\widetilde\eta_k,B,A)$ using \eqref{eq:empJ} on $\mathcal I_{-k}$.
 \State Solve $\widehat\mu_k=(\widehat\Sigma_{B,k}+\lambda_k I_M)^{-1}\widehat c_{B,k}$. Report failure if the inverse does not exist.
 \State Set $\widehat h_k(x)=\operatorname{clip}_{[-L_h,L_h]}\{\widehat\mu_k^\top B(x)\}$ and $\widehat q_k(a,x)=a-\widehat h_k(x)$.
 \State On $\mathcal I_k$ only, compute $\mathcal U_k,D_k,V_k$ from \eqref{eq:Uk} and the fold influence vectors \eqref{eq:empIF}.
 \State Report failure if $D_k\le\underline D_n$. Otherwise retain $\widetilde\theta_k+\mathcal U_k/D_k$ and $V_k/(m_kD_k^2)$.
\EndFor
\State Aggregate all folds using \eqref{eq:scoreest} and \eqref{eq:se}. Report the dictionary, ridge, clipping, training score residuals, and evaluation derivatives.
\end{algorithmic}
\end{algorithm}
For the general cohort, Algorithm~\ref{alg:direct} uses $W=z$ and returns $\widehat S$. The remaining algorithm discussion concerns randomized trials. The two algorithms use the same transferred risk but support different inferential claims. Algorithm~\ref{alg:direct} supplies prediction and the robust null test. Algorithm~\ref{alg:score} supplies alternative effect inference under the full projection rates, or under the additional finite-dictionary bias condition. A ridge that stays positive or an estimated dictionary is not covered automatically by these guarantees.

\paragraph{Implementation and choice of analysis.}
We prespecify the horizon, references, correction terms, projection dictionary, ridge, clipping bound, and numerical thresholds before held-out evaluation. Algorithms~\ref{alg:direct} and \ref{alg:score} implement the proposed estimators with cohort-specific risk sets. The bounded spline learner in Algorithm~\ref{alg:source} additionally certifies a source likelihood gap. A penalty or validation stopping rule does not itself provide that certificate. The choice of projection determines the claim supported by the analysis: alternative effect inference requires the growing dictionary conditions in \eqref{eq:rates}, or the additional finite-dictionary functional in \eqref{eq:finiteexp}. The direct sandwich test instead uses Theorem~\ref{thm:null} under \eqref{eq:null}.
Our one-step estimate targets the treatment coefficient. Equations~\eqref{eq:breslow} and \eqref{eq:survival} describe prediction from the direct calibration fit, with its rate established in Theorem~\ref{thm:prediction}. Substituting the one-step treatment coefficient into that prediction fit defines a different estimator and requires a corresponding baseline refit and analysis. Neither the treatment interval nor the prediction rate theorem provides a confidence band for an individual survival curve. Event time continuity is used in the proportional hazards results. Materially tied or discretized event times require a specified likelihood and an accompanying justification.

\subsection{A calibrated randomization test}\label{sec:randomizationmethod}
When the goal is to test \eqref{eq:null}, we can calibrate the same Cox treatment score using the known allocation law. We first fit the correction with treatment constrained to zero, obtaining a label-blind coefficient $\widehat\gamma_{\rm null}$ and risk $\widehat g_{\rm null}=\widehat f+\widehat\gamma_{\rm null}^{\top}z$. Every fitting choice, including stopping and any numerical fallback, uses the historical sample and target $(X,U,\Delta)$ only. Put $m_A=\sum_iA_i$, $\widehat p_A=m_A/n$, and define
\[
\widehat R_{{\rm null},i}=\Delta_i-\int_0^\tau
\frac{Y_i(t)e^{\widehat g_{\rm null}(X_i)}}{s_{n,\widehat g_{\rm null}}(t)}\,d\nu_n(t),\qquad
T_{\rm rnd}(\mathbf a)=\left|\sum_i(a_i-\widehat p_A)\widehat R_{{\rm null},i}\right|.
\]
Here $\mathbf a=(a_1,\ldots,a_n)^{\top}$ is an assignment vector. The residuals sum to zero, so $T_{\rm rnd}$ is the absolute treatment score multiplied by $n$ at the null-constrained fit. Under the independent Bernoulli design, conditional on $m_A$ the observed assignment is uniform on $\mathcal A_{m_A}=\{\mathbf a\in\{0,1\}^n:\sum_ia_i=m_A\}$. Draw $B_{\rm rnd}\ge1$ independent assignments $\mathbf A^{(b)}$ uniformly from this set and compute
\begin{equation}\label{eq:randomizationp}
p_{\rm CR}=\frac{1+\sum_{b=1}^{B_{\rm rnd}}
\ind\{T_{\rm rnd}(\mathbf A^{(b)})\ge T_{\rm rnd}(\mathbf A)\}}{B_{\rm rnd}+1}.
\end{equation}
We reject when $p_{\rm CR}\le\alpha$ for a prespecified level $\alpha\in(0,1)$. If all patients receive the same assignment or all residuals vanish, set $p_{\rm CR}=1$. Including the observed statistic and ties gives a valid Monte Carlo randomization test \citep{hemerik2018}. Proposition~\ref{prop:randomization} states its conditional guarantee. No neural refitting is needed for the sampled assignments. The procedure tests the observed-data null and does not supply a confidence interval for a nonzero conditional coefficient. Blocked or unequal probability designs require their actual allocation law, rather than the uniform assignment distribution used here.

\begin{algorithm}[tbp]
\caption{Calibrated randomization score test}\label{alg:randomization}
\begin{algorithmic}[1]
\Require Historical fit $\widehat f$, target records, correction $z$, number $B_{\rm rnd}$ of sampled assignments, and level $\alpha$.
\Ensure Null test $p$-value $p_{\rm CR}$ and rejection decision.
\State Fit $\widehat\gamma_{\rm null}$ with $\theta=0$ using only target $(X,U,\Delta)$ and the supplied historical risk.
\State Form $\widehat g_{\rm null}$ and compute $\widehat R_{{\rm null},i}$ with target risk sets.
\State Set $m_A\gets\sum_iA_i$, $\widehat p_A\gets m_A/n$, and compute $T_{\rm rnd}(\mathbf A)$.
\State If $m_A\in\{0,n\}$ or the residuals vanish, return $p_{\rm CR}=1$ and no rejection.
\For{$b=1,\ldots,B_{\rm rnd}$}
 \State Independently sample $\mathbf A^{(b)}$ uniformly from $\mathcal A_{m_A}$ and compute $T_{\rm rnd}(\mathbf A^{(b)})$ using the fixed residuals.
\EndFor
\State Compute \eqref{eq:randomizationp}. Reject if $p_{\rm CR}\le\alpha$.
\end{algorithmic}
\end{algorithm}

\section{Theoretical properties}\label{sec:theory}
We characterize the historical error that remains after calibration. We then derive treatment and prediction guarantees and give sufficient conditions for the required nuisance rates. We consider joint limits of $n,N$ with possibly different growth rates. Target arguments condition on the independent historical fit. We develop these results together, then give a separate robustness result for null testing.

\subsection{Prediction without a treatment variable}\label{sec:generaltheory}
Under model~\eqref{eq:generalmodel}, use the following sufficient conditions.
\begin{enumerate}[label=P\arabic*.]
\item Conditionally independent censoring holds in each independent cohort, event times are continuous, and $0<\Lambda_{T,0}(\tau)<\infty$.
\item The true and fitted historical risks, $g_0$, the fixed correction coordinates, and candidate target indices are uniformly bounded. The finite-dimensional coefficient set is compact. Relevant optima and the aligned comparison coefficient defined below are uniformly interior. There is a fixed $c_{\rm obs}>0$ such that $\Pr(U\ge\tau\mid X)\ge c_{\rm obs}$.
\item For $p>0$, the population negative Hessian for correction fitting has a common positive lower bound in a neighborhood of the aligned comparison coefficient. The mathematical fit is an interior score root. A numerical fit has score residual $\xi_n=\|\nabla_\gamma\ell_n^{\rm pred}(\widehat\gamma)\|_2$.
\end{enumerate}
All fixed bounds are uniform over the fitted historical functions under consideration. Let $c_N,b_N$ be the target $L_2$ projection coefficients of $\widehat f-f_0$ on constants and $z$. Set
\[
e_N=\widehat f-f_0-c_N-b_N^\top z,\qquad
\rho_N=\|e_N\|_2,\qquad \gamma_N=\gamma_0-b_N.
\]
The residual has $\E e_N=0$ and $\E(ze_N)=0$. For $p>0$, P3 excludes a nonconstant coefficient vector representing a constant function, so the projection coefficients are well defined. For $p=0$, retain only the constant projection.

\begin{theorem}[General cohort survival prediction]\label{thm:generalprediction}
Under P1 through P3, let $r_{n,N}=n^{-1/2}+\rho_N+\xi_n=o_\Pp(1)$. For the fit in \eqref{eq:generalfit},
\begin{align*}
\|\widehat\gamma-\gamma_N\|_2+\|\widehat g-g_0-c_N\|_2
&=O_\Pp(r_{n,N}),\\
\sup_{t\le\tau}|e^{c_N}\widehat\Lambda_{T,0}(t)-\Lambda_{T,0}(t)|
&=O_\Pp(r_{n,N}),\\
\left\|\sup_{t\le\tau}|\widehat S(t\mid\cdot)-S(t\mid\cdot)|\right\|_2
&=O_\Pp(r_{n,N}).
\end{align*}
Replacing an offset $f$ by $f+c+b^\top z$ preserves the fitted survival curves under unpenalized fitting whenever the corresponding correction shifts preserve the feasible fits. For $p=0$, omit the coefficient term and P3 and set $\xi_n=0$.
\end{theorem}
Theorem~\ref{thm:generalprediction} is proved in Section~\ref{sec:generalproof}. It provides consistency under a vanishing residual risk error, without a randomized treatment or a treatment information matrix. The time supremum is inside the target covariate $L_2$ norm. No uniform covariate bound, confidence band, or finite-sample superiority claim follows.

Under the source assumptions of Theorem~\ref{thm:source} and overlap in \eqref{eq:overlap}, the source error bound also bounds $\rho_N$ because projection can only reduce target $L_2$ error. Thus the historical approximation, sampling, and optimization terms enter the prediction rate directly. Prediction requires these terms to vanish, but does not require the root-$n$ directional bias condition used for treatment inference.

If the true target log risk is $f_0+\gamma_0^\top z+\delta$ for a bounded drift $\delta(X)$, decompose $(\widehat f-f_0)-\delta$ instead of $\widehat f-f_0$. The same proof applies provided the resulting aligned coefficients satisfy P2 and P3 and the total residual tends to zero. A fixed drift outside the correction span need not satisfy that requirement. The procedure is therefore not protected against arbitrary transport failure. Corollary~\ref{cor:selection} extends the prediction rate to data-dependent selection among finitely many such corrections.

\subsection{Calibration, direct inference, and prediction in trials}\label{sec:directtheory}
We use the following sufficient conditions for inference under \eqref{eq:H} and \eqref{eq:T}. Appendix~\ref{app:conditions} explains their roles, gives concrete sufficient cases, and relates them to \citet{zhong2022} and \citet{gu2026}.
\begin{enumerate}[label=R\arabic*.]
\item Conditionally independent censoring holds in both independent samples. Event times are continuous, and $0<\Lambda_{T,0}(\tau)<\infty$.
\item Covariates $z$, true and fitted risk functions, and fitted indices are bounded uniformly. The finite-dimensional coefficient set is compact with the relevant optima uniformly interior. Target observation probabilities satisfy $\Pr(U\ge\tau\mid A,X)\ge c_{\rm obs}>0$ for a fixed observation bound $c_{\rm obs}$.
\item The calibration information matrix $I_W=\J(W,W)$ is positive definite. Population negative Hessians have a common positive lower bound on the local finite-dimensional paths. The fitted mathematical maximum is an interior score root. A numerical score residual $\xi_n$ contributes $\Op(\xi_n)$ to coefficient errors.
\item For the projected score, $K$ is fixed, the fold fractions $m_k/n$ are bounded away from zero, and fitted projections have a common fixed bound. Dictionary coordinates are centered under a fixed reference $R_P\ll P_{T,X}$ with bounded density ratio, which may differ from the historical identifying reference $R$, and the clipping interval contains the centered representative of the limiting projection ($h_*$ in Theorem~\ref{thm:score} and $h_B$ in Theorem~\ref{thm:finite}).
\end{enumerate}
We use conditions R1 through R3 for the direct fit and add R4 for projected score inference. Follow-up positivity and transport compatibility are substantive restrictions. The appendix states when they hold and where they can fail. We first establish the event-weighted geometry needed for both analyses.

\begin{lemma}[Projection geometry and existence]\label{lem:geometry}
Under R1 and R2, write $\norm{u}_{2,\rm ctr}=\inf_{c\in\mathbb R}\norm{u-c}_{L_2(P_{T,A,X})}$ for the distance to constants, and let $B_\eta>0$ be a bound with $|\eta_0|\le B_\eta$. For every $u\in L_2(P_{T,A,X})$,
\begin{equation}\label{eq:coercivity}
c_{\rm obs}\,e^{-B_\eta}\Lambda_{T,0}(\tau)\norm{u}_{2,\rm ctr}^2
\le \J(u,u)\le e^{B_\eta}\Lambda_{T,0}(\tau)\norm{u}_{2,\rm ctr}^2.
\end{equation}
The subspace of functions of $X$, modulo constants, is closed in this norm. Its projection $h_*$ exists and is unique modulo constants, has a bounded representative, and satisfies
\[
I_X\ge c_{\rm obs}\,e^{-B_\eta}\Lambda_{T,0}(\tau)p_A(1-p_A)>0.
\]
Consequently existence, boundedness of the population projection, and nonzero treatment information follow from the stated observation and risk bounds and randomization. R4 retains boundedness of the fitted projection.
\end{lemma}
The observation bound is strong: it excludes covariate strata with vanishing follow-up probability at the analysis horizon. This condition provides the norm equivalence used below. Without it, closedness and series coercivity would require a different argument.

Let $c_N\in\mathbb R$ and $b_N\in\mathbb R^p$ minimize the target $L_2$ error of $\widehat f-f_0$ over a constant and the correction span. Define the remaining function $e_N=\widehat f-f_0-c_N-b_N^\top z$ and its target norm $\rho_N$ by
\begin{equation}\label{eq:error}
\widehat f-f_0=c_N+b_N^\top z+e_N,\qquad
\rho_N=\norm{e_N}_{L_2(P_{T,X})}=\inf_{c\in\mathbb R,\,b\in\mathbb R^p}\norm{\widehat f-f_0-c-b^\top z}_{L_2(P_{T,X})}.
\end{equation}
The residual satisfies $\E e_N(X)=0$ and $\E\{z(X)e_N(X)\}=0$. This is an analytical decomposition, not a computable error estimate. The comparison coefficient is $\beta_N=(\theta_0,(\gamma_0-b_N)^\top)^\top$, assumed interior, and $c_N$ enters the baseline. The error that matters is therefore a residual after calibration, rather than the entire historical prediction error.

\begin{proposition}[Correction space invariance]\label{prop:invariance}
For a supplied offset $f$, a constant $c\in\mathbb R$, and a correction vector $b\in\mathbb R^p$, replacing $f$ by $f+c+b^\top z$ in unpenalized target fitting leaves the fitted treatment coefficient and survival curves unchanged, provided the feasible fitted indices are unchanged. The correction becomes $\gamma-b$, and the Breslow baseline is multiplied by $e^{-c}$.
\end{proposition}
The new index is the old index plus $c$, which cancels in every partial likelihood contribution and in \eqref{eq:survival}. Penalties on $\gamma$ or constraints that block this substitution can break the invariance. The result holds exactly in finite samples, regardless of the size of the historical error in the representable directions.

\begin{proposition}[Finite-sample offset stability]\label{prop:offsetstability}
On a fixed trial sample, let $\widehat\beta_f$ and $\widehat\beta_{f+e}$ be interior score roots with offsets $f$ and $f+e$, where $e:\mathcal X\to\mathbb R$ is an offset perturbation. For bounds $L>0$ and $\varepsilon\ge0$, suppose $\max_i\norm{W_i}_2\le L$, $\max_i|e(X_i)|\le\varepsilon$, and the negative Hessian for offset $f+e$ has minimum eigenvalue at least $\kappa>0$ along the coefficient segment joining the roots. Then
\begin{equation}\label{eq:offsetstability}
\norm{\widehat\beta_{f+e}-\widehat\beta_f}_2
\le\frac{2L\nu_n(\tau)}{\kappa}\varepsilon.
\end{equation}
\end{proposition}
Here $\nu_n(\tau)$ is the observed event fraction. If an offset difference is $c+b^\top z+e$, apply Proposition~\ref{prop:invariance} first, provided the coefficient shifts remain feasible. The bound then controls the treatment change using only $e$. It is a deterministic sensitivity bound, not a confidence interval, and can be loose when curvature is small. The asymptotic results below use a sharper target $L_2$ residual analysis.

We now show how the residual historical error enters direct treatment estimation.
Let $\Pi_zA=\J(z,A)^\top\J(z,z)^{-1}z$ be the event-weighted projection onto the correction span. Define its residual $q_z=A-\Pi_zA$ and treatment information $I_z=\J(q_z,q_z)>0$. Under the target model, define the compensated event process $\mathcal M(t)$ and its patient versions $\mathcal M_i(t)$, followed by the direct fit score influence $\phi_z=\phi_{q_z}$, as
\begin{align*}
\mathcal M(t)&=\cnt(t)-\int_0^tY(s)e^{\eta_0(A,X)}\,d\Lambda_{T,0}(s),\qquad \mathcal M_i(t)=\mathcal M(t,O_i),\\
\phi_z(O_i)&=\int_0^\tau\{q_z(A_i,X_i)-Q_t(q_z)\}\,d\mathcal M_i(t).
\end{align*}\noindent More generally, for a scalar or column vector direction $q$, write $\phi_q(O)=\int_0^\tau\{q-Q_t(q)\}\,d\mathcal M(t)$. In particular, $\phi_z=\phi_{q_z}$.

\begin{theorem}[Direct fit expansion]\label{thm:plugin}
Under conditions R1 through R3 and $\rho_N=o_\Pp(1)$, the direct calibrated fit satisfies
\begin{align}
\norm{\widehat\beta-\beta_N}_2&=\Op(n^{-1/2}+\rho_N),\label{eq:betarate}\\
\norm{\widehat g-g_0-c_N}_{L_2(P_{T,X})}&=\Op(n^{-1/2}+\rho_N),\label{eq:grate}\\
\widehat\theta-\theta_0&=I_z^{-1}\left\{\frac1n\sum_i\phi_z(O_i)-\J(q_z,e_N)\right\}
+\Op(n^{-1}+\rho_N^2+\rho_Nn^{-1/2}).\label{eq:thetaexp}
\end{align}
These displays concern exact score roots. A numerical root adds $\Op(\xi_n)$ to their error bounds. If $\sqrt n\rho_N^2=o_\Pp(1)$ and $\sqrt n\J(q_z,e_N)=o_\Pp(1)$, then $\sqrt n(\widehat\theta-\theta_0)\Rightarrow N(0,I_z^{-1})$, and both information and sandwich variance estimates are consistent. Numerical inference additionally requires $\xi_n=o_\Pp(n^{-1/2})$.
\end{theorem}
In \eqref{eq:thetaexp}, the linear term $\J(q_z,e_N)$ is the part omitted when a fitted historical score is treated as known. A prediction rate $\rho_N=o_\Pp(n^{-1/4})$ controls the quadratic remainder but need not control this term. The stronger rate $o_\Pp(n^{-1/2})$ suffices, although it is unnecessary when the particular functional is smaller.

We illustrate the resulting risk-set association in Appendix~\ref{sec:biasexamples}. Even with enrollment randomization, the bias functional can be nonzero under a treatment alternative. This is why we distinguish treatment inference from survival prediction.

Our next result shows that the same direct calibration fit consistently estimates target survival when the residual risk error vanishes. Because a log-risk constant is unidentified without a baseline normalization, the appropriate baseline comparison must account for the constant $c_N$ in \eqref{eq:error}.

\begin{theorem}[Calibrated survival prediction]\label{thm:prediction}
Under the conditions of Theorem~\ref{thm:plugin}, suppose $\rho_N=o_\Pp(1)$ and the numerical score residual satisfies $\xi_n=o_\Pp(1)$. Set $\xi_n=0$ for an exact root. Put $r_{n,N}=n^{-1/2}+\rho_N+\xi_n$. For the estimators in \eqref{eq:breslow} and \eqref{eq:survival},
\begin{align}
\sup_{0\le t\le\tau}|e^{c_N}\widehat\Lambda_{T,0}(t)-\Lambda_{T,0}(t)|
&=\Op(r_{n,N}),\label{eq:baselinerate}\\
\max_{a\in\{0,1\}}\left\|\sup_{0\le t\le\tau}|\widehat S_a(t\mid\cdot)-S_a(t\mid\cdot)|\right\|_{L_2(P_{T,X})}
&=\Op(r_{n,N}).\label{eq:survivalrate}
\end{align}
Here $S_a(t\mid x)=\exp[-\Lambda_{T,0}(t)e^{\theta_0a+g_0(x)}]$. The rate is uniform over the analysis horizon and averaged over target covariates. It is not a uniform in-$x$ rate or a confidence band result.
\end{theorem}
The unknown alignment constant cancels in the product defining $\widehat S_a$, so survival prediction itself needs no estimate of $c_N$. Consistency requires a vanishing residual risk error, whereas treatment effect intervals require the stronger first order and remainder conditions of Theorem~\ref{thm:plugin} or \ref{thm:score}. Accurate predictions alone therefore do not establish treatment coverage.

We quantify the remaining treatment bias through a covariate space representation of its first-order functional. Write $p_1=p_A$ and $p_0=1-p_A$ for the allocation probabilities. Define the arm-specific weighted follow-up function $w_a$ and the covariate representer $\omega_z$ by
\begin{align}
w_a(t,x)&=\Pr(U\ge t\mid A=a,X=x)e^{\theta_0a+g_0(x)},\notag\\
\omega_z(x)&=\int_0^\tau\sum_{a=0}^1p_a w_a(t,x)
  \{q_z(a,x)-Q_t(q_z)\}\,d\Lambda_{T,0}(t),\label{eq:sensitivityweight}
\end{align}
The function $\omega_z$ is bounded, $\E\omega_z(X)=0$, $\E\{z(X)\omega_z(X)\}=0$, and, for any bounded covariate error function $e:\mathcal X\to\mathbb R$,
\[
\J(q_z,e)=\E\{\omega_z(X)e(X)\}.
\]
Its norm measures sensitivity to prediction errors that remain after calibration.

\begin{proposition}[Sharp local calibrated sensitivity]\label{prop:sensitivity}
Fix a target law satisfying conditions R1 through R3. Let $\theta^\dagger(e)$ be the treatment coordinate of the interior root of $\nabla_\beta\ell_P(\beta,f_0+e)=0$ with the target law fixed. For a radius $r>0$, restrict to uniformly bounded residuals satisfying $\E e=0$, $\E(ze)=0$ and $\norm e_2\le r$. For sufficiently small $r$, with roots in the common curvature neighborhood,
\begin{align}
\theta^\dagger(e)-\theta_0
  &=-I_z^{-1}\E(\omega_z e)+O(\norm e_2^2),\label{eq:localsensitivity}\\
\sup_e|\theta^\dagger(e)-\theta_0|
  &=r\norm{\omega_z}_2/I_z+O(r^2).\label{eq:sharpbias}
\end{align}
The constants are uniform over this residual class. If $\norm{\omega_z}_2>0$, the directions $e=\pm r\omega_z/\norm{\omega_z}_2$ attain the first-order radius. If it is zero, the radius is $O(r^2)$.
\end{proposition}
This is a local sensitivity statement under the target model, not an identified correction for unknown transport failure. Errors with the same $L_2$ size can have different treatment bias because only their component along $\omega_z$ enters at first order. Under the correctly specified observed-data null, $Q_t(A)=p_A$, $\J(A,z)=0$, and $\omega_z=0$. The robust null theorem below is stronger in a different direction: it does not require the Cox model.

We give an equal prediction error illustration in Appendix~\ref{sec:biasexamples}. It shows why a residual error's direction matters for treatment inference, even when its prediction norm is small.

For a bounded additional target drift $\delta:\mathcal X\to\mathbb R$, if the actual target risk is $f_0+\gamma_0^\top z+\delta$ rather than the transfer model, the total fitted discrepancy is
\[
(\widehat f-f_0)-\delta+(\widetilde\gamma_k-\gamma_0)^\top z.
\]
The sign of the drift contribution is negative. A fixed $\delta$ outside the correction span can leave a nonvanishing risk error even with infinite historical data. The local sensitivity expansion applies only when this total discrepancy is small around the true target index. Orthogonalization does not establish consistency for a permanently incorrect target model.

\subsection{Projected treatment inference and dictionary approximation}\label{sec:scoretheory}
We propose the projected score to remove the linear prognostic error term. Our results separate full nuisance orthogonality from the approximation supplied by a finite dictionary.
Define the maximum training fold index error $a_n$ and projection error $b_n$ by
\[
a_n=\max_{1\le k\le K}\left\{|\widetilde\theta_k-\theta_0|+\inf_c\norm{\widetilde g_k-g_0-c}_{L_2(P_{T,X})}\right\},\quad
b_n=\max_{1\le k\le K}\inf_c\norm{\widehat h_k-h_*-c}_{L_2(P_{T,X})}.
\]
\begin{theorem}[Projected score limit]\label{thm:score}
Under conditions R1 through R4, suppose
\begin{equation}\label{eq:rates}
a_n=o_\Pp(1),\qquad b_n=o_\Pp(1),\qquad\sqrt n(a_n^2+a_nb_n)=o_\Pp(1).
\end{equation}
For numerical one-step errors $o_\Pp(n^{-1/2})$,
\begin{equation}\label{eq:asympscore}
\sqrt n(\widehat\theta^{\rm score}-\theta_0)
=I_X^{-1}n^{-1/2}\sum_i\int_0^\tau\{q_*(A_i,X_i)-Q_t(q_*)\}\,d\mathcal M_i(t)+o_\Pp(1)
\Rightarrow N(0,I_X^{-1}).
\end{equation}
The variance in \eqref{eq:se} satisfies $n\widehat{\operatorname{se}}_{\rm score}^{\,2}\to_\Pp I_X^{-1}$.
\end{theorem}
Orthogonality removes the first-order prognostic term and leaves a squared risk error and a product of risk and projection errors. The limit corresponds to the unrestricted target partially linear Cox model. It need not be efficient for the more restricted joint historical/current model: $I_z\ge I_X$, so a known historical function can yield a smaller direct fit variance.

\begin{corollary}[Two sample rate condition]\label{cor:rates}
Let $r_N$ and $s_n$ be positive deterministic rate sequences. Suppose $\rho_N=\Op(r_N)$, Theorem~\ref{thm:plugin} applies on the training folds, and $b_n=\Op(s_n)$. Then it suffices that $r_N,s_n\to0$ and $\sqrt n(r_N^2+r_Ns_n)\to0$. If $r_N=N^{-\alpha_{\rm src}}$ up to logarithms for a source exponent $\alpha_{\rm src}>0$, the squared error condition requires $N\gg n^{1/(4\alpha_{\rm src})}$ up to those logarithms.
\end{corollary}
Numerical calibration error also enters the training fold nuisance rate $a_n$. For a positive deterministic tolerance sequence $t_n$, if the maximum training score residual is $\Op(t_n)$, Theorem~\ref{thm:plugin} gives $a_n=\Op(r_N+n^{-1/2}+t_n)$ when $\rho_N=\Op(r_N)$. Replace $r_N$ by $r_N+t_n$ in Corollary~\ref{cor:rates} to retain its sufficient conditions. Requiring $t_n=o(n^{-1/2})$ preserves the exact root rates. The same tolerance for the full target fit suffices for its numerical error requirement in direct treatment and null inference. Curvature and interior root conditions remain necessary. The evaluation fold one-step calculation has its separate $o_\Pp(n^{-1/2})$ error requirement.

Projection approximation must be included in $s_n$. Appendix~\ref{app:projection} gives a coefficient bound that separates approximation, empirical moments, and ridge error. Let $B_M$ denote a centered deterministic dictionary of $M$ uniformly bounded functions with well-conditioned population Gram matrices, and let $h_M$ be the $\J$ projection of $A$ onto its span. Define $\zeta_M=\inf_{c\in\mathbb R}\norm{h_M-h_*-c}_2$ and use the ridge bound $\lambda_n$ from Section~\ref{sec:projectedmethod}. The coefficient argument yields
\begin{equation}\label{eq:seriesrate}
b_n=\Op\{\zeta_M+M(a_n+n^{-1/2})+\lambda_n\},
\end{equation}
We use this bound to give a sufficient rate construction. Let $s>0$ be an approximation smoothness index and $u>0$ a dictionary growth exponent. For example, if $a_n=\Op(n^{-\kappa_{\rm rate}})$ for $1/3<\kappa_{\rm rate}<1/2$, and $\zeta_M\lesssim M^{-s/d}$ with $s\ge d$, choosing $M\asymp n^u$ with $(1/2-\kappa_{\rm rate})d/s<u<2\kappa_{\rm rate}-1/2$ and $\lambda_n=O(n^{-1/2})$ satisfies \eqref{eq:rates}. Proposition~\ref{prop:fourier} verifies these Gram and approximation assumptions for a deterministic Fourier dictionary when the target density and risk-set weights satisfy its stated bounds. A fixed dictionary needs the separate expansion below. Convergence to the full projection is not automatic.

We next show what changes when the projection is restricted to a finite dictionary. Such a dictionary may leave a first-order error functional even when its coefficients are consistently estimated. Let $B(X)\in\mathbb R^M$ be a fixed deterministic bounded dictionary whose span, modulo constants, contains the correction coordinates $z$. Remove redundant coordinates and assume $\Sigma_B=\J(B,B)$ is positive definite. Put
\[
h_B=\J(A,B)\Sigma_B^{-1}B,\qquad q_B=A-h_B,\qquad
I_B=\J(q_B,A)=\J(q_B,q_B)>0.
\]
For the cross-fitted estimator in \eqref{eq:scoreest}, denote the estimate using $h_B$ by $\widehat\theta_B^{\rm score}$, and define
$b_{B,n}=\max_k\inf_c\norm{\widehat h_k-h_B-c}_2$.

\begin{theorem}[Finite-dictionary expansion]\label{thm:finite}
Under conditions R1 through R3 and the fixed fold and bounded fit conditions, suppose $a_n,b_{B,n}=o_\Pp(1)$. With exact one-step calculations and $\phi_B(O)=\phi_{q_B}(O)=\int_0^\tau\{q_B-Q_t(q_B)\}\,d\mathcal M(t)$,
\begin{align}
\widehat\theta_B^{\rm score}-\theta_0
&=I_B^{-1}\left\{n^{-1}\sum_i\phi_B(O_i)-\J(q_B,e_N)\right\}\notag\\
&\quad+\Op\{a_n^2+a_nb_{B,n}+(a_n+b_{B,n})n^{-1/2}+n^{-1}\}.
\label{eq:finiteexp}
\end{align}
If $\sqrt n(a_n^2+a_nb_{B,n})=o_\Pp(1)$ and $\sqrt n\J(q_B,e_N)=o_\Pp(1)$, then
\[
\sqrt n(\widehat\theta_B^{\rm score}-\theta_0)\Rightarrow N(0,I_B^{-1}),
\qquad n\widehat{\operatorname{se}}_{\rm score}^{\,2}\to_\Pp I_B^{-1}.
\]
Numerical one-step errors must be $o_\Pp(n^{-1/2})$ for this limit.
\end{theorem}
Because $\J(q_B,z)=0$, the population nuisance term on every fold equals $\J(q_B,e_N)$ regardless of its estimated correction coefficient. A finite dictionary can therefore remove error directions contained in its span, but it leaves this functional for other directions. Taking $B=z$ gives the same first-order expansion as the direct fit. The following identity quantifies the associated variance and bias trade-off.

\begin{proposition}[Projection approximation, information, and bias]\label{prop:dictionarytradeoff}
Under conditions R1 through R3 and the stated conditions on the fixed dictionary $B$, define its nonnegative approximation distance $d_B=\{\J(h_*-h_B,h_*-h_B)\}^{1/2}$. Then
\begin{align}
I_B-I_X&=d_B^2,\qquad I_X\le I_B\le I_z,\notag
\\
\J(q_B,e_N)&=\J(h_*-h_B,e_N),\notag\\
|\J(q_B,e_N)|&\le d_B\{\J(e_N,e_N)\}^{1/2}\le C d_B\rho_N.\label{eq:dictionarybias}
\end{align}
The constant $C$ depends only on the upper bound in \eqref{eq:coercivity}.
\end{proposition}
For the same target law, the leading variance difference is
$I_X^{-1}-I_B^{-1}=d_B^2/(I_XI_B)$. Thus a finite dictionary can have a smaller nominal variance while leaving a first-order bias bounded by $C d_B\rho_N/I_B$. If it contains $h_*$ modulo constants, both this bias functional and the variance difference vanish. Enlarging nested dictionaries reduces $d_B$, but the realized bias need not decrease monotonically because it also depends on the direction of $e_N$. These population identities do not estimate the unknown historical error or replace the remainder conditions in Theorem~\ref{thm:finite}.

For this deterministic fixed dictionary, the coefficient bound in Appendix~\ref{app:projection} gives $b_{B,n}=\Op(a_n+n^{-1/2}+\lambda_n)$, provided the clipping convention preserves $h_B$. A ridge that stays positive changes the population projection. For a fixed $\lambda>0$, define $\mu_\lambda=(\Sigma_B+\lambda I_M)^{-1}\J(B,A)$. Then
\[
\J(A-\mu_\lambda^\top B,B)=\lambda\mu_\lambda^\top.
\]
Hence a nonvanishing ridge does not give exact dictionary orthogonality. A dictionary formed from an estimated risk function likewise needs an additional convergence or perturbation argument. Neither a fixed ridge nor an estimated dictionary inherits the unrestricted score theorem solely from cross-fitting.

\paragraph{Projected minus direct contrast.}
Both expansions hold on the same records, which yields a diagnostic for the direct fit.
\begin{corollary}[Projected minus direct contrast]\label{cor:contrast}
Assume the conditions of Theorems~\ref{thm:plugin} and~\ref{thm:finite} for the same target sample and fixed dictionary $B$, with exact roots and one-step calculations, $\sqrt n\rho_N^2=o_\Pp(1)$, and $\sqrt n(a_n^2+a_nb_{B,n})=o_\Pp(1)$. Put $\psi=I_B^{-1}\phi_B-I_z^{-1}\phi_z$, $\sigma_\Delta^2=\E\psi^2$, and $\delta_N=I_z^{-1}\J(q_z,e_N)-I_B^{-1}\J(q_B,e_N)$. Then
\[
\widehat\theta_B^{\rm score}-\widehat\theta=\delta_N+\frac1n\sum_{i=1}^n\psi(O_i)+\op(n^{-1/2}).
\]
If $\sigma_\Delta^2>0$ and $\sqrt n\delta_N\to_\Pp\delta$, then
\[
\sqrt n(\widehat\theta_B^{\rm score}-\widehat\theta)\Rightarrow N(\delta,\sigma_\Delta^2).
\]
With $\widehat\psi_{{\rm CT},i}=n^{-1}\{\widehat\J_n(\widehat\eta,W,W)^{-1}\widehat\varphi_i(\widehat\eta,W)\}_1$ and $\widehat\psi_{{\rm PT},i}=\widehat\varphi_{i,k}(\widetilde\eta_k,\widehat q_k)/(nD_k)$ for $i\in\mathcal I_k$, the estimate $\widehat\sigma_\Delta^2=n\sum_i(\widehat\psi_{{\rm PT},i}-\widehat\psi_{{\rm CT},i})^2$ is consistent.
\end{corollary}
\begin{proof}
Subtracting \eqref{eq:thetaexp} from \eqref{eq:finiteexp} gives the expansion, because both linear terms are averages over the same independent records and both remainders are $\op(n^{-1/2})$ under the stated rates. The summands $\psi(O_i)$ are independent, mean zero and square integrable by martingale isometry and R1 to R3, so the central limit theorem gives the limit. For the variance, the proofs of Theorems~\ref{thm:plugin} and~\ref{thm:score} show that the fitted influence contributions converge to $\phi_W$ and to the fold-wise $\phi_{q_B}$ in empirical squared norm (Appendices~\ref{sec:pluginproof} and~\ref{sec:scoreproof}; the finite-dictionary case in Appendix~\ref{sec:finiteproof} uses the same argument), and $\widehat\J_n(\widehat\eta,W,W)\to_\Pp I_W$ and $D_k\to_\Pp I_B$. Hence $n\sum_i(\widehat\psi_{{\rm PT},i}-\psi_{{\rm PT},i})^2$ and $n\sum_i(\widehat\psi_{{\rm CT},i}-\psi_{{\rm CT},i})^2$ are $\op(1)$, with $\psi_{{\rm PT},i}=\phi_B(O_i)/(nI_B)$ and $\psi_{{\rm CT},i}=\phi_z(O_i)/(nI_z)$, where the treatment row of $I_W^{-1}\phi_W$ equals $\phi_z/I_z$ by the partitioned inverse. The triangle inequality and the law of large numbers give $\widehat\sigma_\Delta^2\to_\Pp\sigma_\Delta^2$.
\end{proof}
The contrast compares the two first-order biases. When $B$ contains the remaining error modulo constants, or more generally when $\J(q_B,e_N)=o(n^{-1/2})$, it tests whether the direct fit's bias $-I_z^{-1}\J(q_z,e_N)$ is negligible. It does not test the treatment null. Under $\theta_0=0$ the risk-set association vanishes, so $q_B=q_z=A-p_A$ and $\sigma_\Delta^2=0$, and $\sigma_\Delta^2$ is of order $\theta_0^2$ near the null, where the normal approximation can be poor. The influence estimate also ignores the variability of the training-fold nuisance fits, which is of smaller order but need not be small relative to $\sigma_\Delta$ in finite samples; a paired bootstrap that refits both procedures on the same resamples is a useful check (Appendix~\ref{app:clinicaluncertainty}).

\paragraph{A fixed remaining error.}
Theorems~\ref{thm:plugin}, \ref{thm:score}, and~\ref{thm:finite} are local: they require $\rho_N\to0$, so that the training-fold calibration fits approach $\eta_0$ and $a_n\to0$. Drift outside $\operatorname{span}\{1,z\}$ that does not shrink with the sample sizes violates this, because the calibration fits then converge to a misspecified population root. The next result describes both estimators in that case. For a bounded index $\eta$ and bounded functions $u,v$, write $\bar u_\eta(t)=\E\{Y(t)e^{\eta}u\}/\E\{Y(t)e^{\eta}\}$ and
\[
\J_\eta(u,v)=\int_0^\tau\{\overline{uv^\top}_\eta(t)-\bar u_\eta(t)\bar v_\eta(t)^\top\}\,d\nu(t),\qquad \nu(t)=\E\cnt(t),
\]
so that $\J_{\eta_0}=\J$. The form $\J_\eta$ weights risk sets by the working index but integrates against the true event law, as the empirical forms do.
\begin{proposition}[Projected and direct limits under a fixed remaining error]\label{prop:fixedmis}
Assume R1 and R2, the fixed fold and bounded fit conditions of Theorem~\ref{thm:finite}, $\lambda_n\to0$, and that the clipping bound contains the limiting projection. Suppose the training-complement calibration fits converge in probability, in the supremum norm, to a bounded population calibration root $\eta^\dagger=\theta^\dagger A+g^\dagger$ with $g^\dagger=\widehat f+\gamma^{\dagger\top}z$, that is, $\mathcal U_P(\eta^\dagger,W)=0$. Write $\J_\dagger=\J_{\eta^\dagger}$, and suppose $\J_\dagger(B,B)$ is positive definite. Define
\[
q_B^\dagger=A-\J_\dagger(A,B)\J_\dagger(B,B)^{-1}B,\qquad q_z^\dagger=A-\J_\dagger(A,z)\J_\dagger(z,z)^{-1}z,
\]
assume $D^\dagger=\J_\dagger(q_B^\dagger,A)$ and $I_z^\dagger=\J_\dagger(q_z^\dagger,A)$ are positive, and let $\delta=\inf_c\norm{\eta^\dagger-\eta_0-c}_2$. Then $\widehat\theta_B^{\rm score}\to_\Pp\theta_B^\dagger=\theta^\dagger+\mathcal U_P(\eta^\dagger,q_B^\dagger)/D^\dagger$, and
\begin{equation}\label{eq:fixedmis}
\theta_B^\dagger-\theta_0=-\frac{\J_\dagger(q_B^\dagger,e_N)}{D^\dagger}+O(\delta^2),\qquad
\theta^\dagger-\theta_0=-\frac{\J_\dagger(q_z^\dagger,e_N)}{I_z^\dagger}+O(\delta^2),
\end{equation}
where the constants depend only on the bounds in R1 and R2 and on lower bounds for $D^\dagger$ and $I_z^\dagger$. If $e_N$ lies in $\operatorname{span}\{1,B\}$, then $\theta_B^\dagger-\theta_0=O(\delta^2)$. Moreover, the derivatives of the population one-step map $(\widetilde\theta,\widetilde\gamma,\mu)\mapsto\widetilde\theta+\mathcal U_P(\widetilde\eta,A-\mu^\top B)/\J_{\widetilde\eta}(A-\mu^\top B,A)$ at the limit $(\theta^\dagger,\gamma^\dagger,\mu^\dagger)$ are $O(\delta)$.
\end{proposition}
\begin{proof}
For bounded $q$, uniform convergence of the empirical risk-set ratios (Appendix E.3), the convergence of the training-fold fits, and $\lambda_n\to0$ give $\widehat\mu_k\to_\Pp\mu^\dagger=\J_\dagger(B,B)^{-1}\J_\dagger(B,A)$, $\mathcal U_k\to_\Pp\mathcal U_P(\eta^\dagger,q_B^\dagger)$, and $D_k\to_\Pp D^\dagger$ on every fold, so $\widehat\theta_B^{\rm score}\to_\Pp\theta_B^\dagger$. Let $\Delta=\eta^\dagger-\eta_0$, with the constant chosen to attain $\delta$, and $\eta_s=\eta^\dagger-s\Delta$, so $\eta_1=\eta_0$ up to a constant, which does not change risk-set ratios. By the derivative formulas in the proof of Lemma~\ref{lem:pop},
\[
Q_t(q)-\bar q_{\eta^\dagger}(t)=-\operatorname{Cov}_{\eta^\dagger,t}(q,\Delta)+r_t(q),\qquad \sup_t|r_t(q)|\le C\norm q_\infty\delta^2,
\]
where $\operatorname{Cov}_{\eta^\dagger,t}$ is the covariance under the risk-set weights $Y(t)e^{\eta^\dagger}$. Integrating against $d\nu=s_0\,d\Lambda_{T,0}$ gives
\[
\mathcal U_P(\eta^\dagger,q)=-\J_\dagger(q,\Delta)+R(q),\qquad |R(q)|\le C\norm q_\infty\delta^2.
\]
Decompose $\Delta=(\theta^\dagger-\theta_0)A+e_N+(b_N+\gamma^\dagger-\gamma_0)^\top z+c$ using \eqref{eq:error}. Because $z$ lies in $\operatorname{span}\{1,B\}$, $\J_\dagger(q_B^\dagger,z)=0$, and constants have zero covariance, so
\[
\mathcal U_P(\eta^\dagger,q_B^\dagger)=-(\theta^\dagger-\theta_0)D^\dagger-\J_\dagger(q_B^\dagger,e_N)+R(q_B^\dagger).
\]
Dividing by $D^\dagger$ and adding $\theta^\dagger$ gives the first identity in \eqref{eq:fixedmis}. For the direct fit, $\mathcal U_P(\eta^\dagger,W)=0$ and linearity in $q$ give $\mathcal U_P(\eta^\dagger,q_z^\dagger)=0$, and the same decomposition with $\J_\dagger(q_z^\dagger,z)=0$ gives the second identity. If $e_N\in\operatorname{span}\{1,B\}$, then $\J_\dagger(q_B^\dagger,e_N)=0$. For the last statement, differentiate the one-step map. The derivative in $\widetilde\theta$ is $1-\J_\dagger(q_B^\dagger,A)/D^\dagger+\mathcal U_P(\eta^\dagger,q_B^\dagger)\,O(1)$, the derivative in $\widetilde\gamma$ is $-\J_\dagger(q_B^\dagger,z)/D^\dagger+\mathcal U_P(\eta^\dagger,q_B^\dagger)\,O(1)$, and the derivative in $\mu$ is $-\mathcal U_P(\eta^\dagger,B)/D^\dagger+\mathcal U_P(\eta^\dagger,q_B^\dagger)\,O(1)$. The first two leading terms vanish, and $\mathcal U_P(\eta^\dagger,q)=O(\norm q_\infty\delta)$ for bounded $q$ by the display above.
\end{proof}
The projected estimator is therefore not consistent for $\theta_0$ under a fixed misspecification. When the remaining error lies in its dictionary, its limiting bias is of second order in the size $\delta$ of the misspecified working index, whereas the direct fit retains the first-order term, which reduces to $-I_z^{-1}\J(q_z,e_N)$ as $\delta\to0$. The population calculation in Appendix~\ref{sec:qpop} evaluates both limits exactly. Because the derivatives of the one-step map are $O(\delta)$ rather than zero, the training-fold estimation error contributes $O_\Pp(\delta n^{-1/2})$ to $\widehat\theta_B^{\rm score}-\theta_B^\dagger$, which the held-out standard error \eqref{eq:se} omits. The full-procedure bootstrap of Proposition~\ref{prop:fullscorebootstrap} targets $\theta_B^\dagger$ and includes this term.

\subsection{Source learning and attainable two-sample rates}\label{sec:sourcerates}
We now bound historical prediction error in order to check the rates required for treatment inference. We separate approximation error, sampling error, and optimization error. Each term must be controlled for the chosen learner.
\begin{theorem}[Source prediction bound]\label{thm:source}
Suppose historical conditionally independent censoring holds, event times are continuous, $\norm{f_0}_\infty\le B_f$, $0<\Lambda_{H,0}(\tau)<\infty$, and $\Pr(U^H\ge\tau\mid X^H)\ge c_H>0$ for a fixed historical observation bound $c_H$. Let $\mathcal F_N$ be deterministic and separable, consisting of measurable functions uniformly bounded by $B_f$, where $\mathcal C(\varepsilon,\mathcal F_N,\norm{\cdot}_\infty)$ is the minimum number of supremum norm balls of radius $\varepsilon>0$ covering the class. Define its entropy at radius $N^{-1}$ by $\mathcal E_N=\log\mathcal C(N^{-1},\mathcal F_N,\norm{\cdot}_\infty)$. Suppose a fixed $f_N^*\in\mathcal F_N$ has approximation error $\norm{f_N^*-f_0}_\infty\le\varepsilon_N^{\rm app}$ for a deterministic tolerance $\varepsilon_N^{\rm app}\ge0$. Let $\varepsilon_N^{\rm opt}\ge0$ denote a possibly random optimization gap and assume
\[
\ell_H(\widehat f)\ge\sup_{f\in\mathcal F_N}\ell_H(f)-\varepsilon_N^{\rm opt},\qquad\varepsilon_N^{\rm opt}\ge0.
\]
For a tail parameter $u\ge1$, with probability at least $1-3e^{-u}$,
\begin{equation}\label{eq:sourcebound}
\inf_c\norm{\widehat f-f_0-c}_{L_2(P_{H,X})}^2
\le C\{(\varepsilon_N^{\rm app})^2+\varepsilon_N^{\rm opt}+(\mathcal E_N+\log N+u)/N\}.
\end{equation}
The constant depends on the fixed risk, baseline, and observation bounds, not on $N$ or network parameter count.
\end{theorem}
The bound controls censored data likelihood error without treating dependent risk set terms as independent losses. It separates approximation, sampling, and optimization. Under \eqref{eq:overlap}, its centered historical norm controls the target norm up to $C_{\rm ov}^{1/2}$, and projection onto the correction space can only improve the comparison.

\begin{corollary}[Bounded ReLU classes]\label{cor:relu}
For ReLU networks on $[-1,1]^d$ with output clipped to $[-B_f,B_f]$, $L_N$ affine layers, maximum width $w_N\ge d$, $P_N$ free parameters in $[-K_N,K_N]$, and $K_N\ge1$, let $D_N=(w_N+1)(K_N+1)$ and $C_N=L_ND_N^{L_N}$. For an additional bounded linear skip, include its parameters in $P_N$ and use $C_N=L_ND_N^{L_N}+d+1$. Then
\[
\mathcal E_N\le P_N\log\{2+2NK_NC_N\}.
\]
Overlap implies
\begin{equation}\label{eq:relurate}
\rho_N=\Op\left[\varepsilon_N^{\rm app}+\sqrt{\varepsilon_N^{\rm opt}}
+\sqrt{\{P_N\log(2+2NK_NC_N)+\log N\}/N}\right].
\end{equation}
For an approximation exponent $\beta_{\rm app}>0$ and entropy exponent $c_{\rm ent}\ge0$, an approximation bound $\varepsilon_N^{\rm app}\lesssim P_N^{-\beta_{\rm app}}$, entropy of order $P_N\log^{c_{\rm ent}}N$, and an optimization gap no larger than the squared statistical error give rate $N^{-\beta_{\rm app}/(2\beta_{\rm app}+1)}$ up to logarithms after balancing class size.
\end{corollary}
The approximation exponent depends on the chosen function class. The optimization gap is part of the condition, rather than a guarantee for finite step Adam. These quantities determine whether the source sample is large enough for the treatment inference rates.

\begin{corollary}[One explicit feasible two-sample regime]\label{cor:feasible}
Suppose $d=1$, $X\in[0,1]$, the target covariate density is bounded above and away from zero, and overlap holds. Assume $f_0$ is twice continuously differentiable, $\norm{f_0}_\infty<B_f$, and the functions $w_a(t,\cdot)$ in \eqref{eq:sensitivityweight} have uniformly bounded periodic derivatives through order two. Assume the historical observation, baseline, and boundedness conditions of Theorem~\ref{thm:source}, conditions R1 through R3, and the fixed fold bounded fit conditions.

Let $J_N$ be the number of equally spaced historical spline intervals, with knots $j/J_N$ for $j=0,\ldots,J_N$. Use the linear interpolation $f_c$ of knot values $c=(c_0,\ldots,c_{J_N})^\top\in[-B_f,B_f]^{J_N+1}$, where $J_N\asymp(N/\log N)^{1/5}$. This is a one hidden layer ReLU class with fixed knots, tied output coefficients, and a linear skip. If its normalized likelihood gap is
$\varepsilon_N^{\rm opt}=O_\Pp\{(\log N/N)^{4/5}\}$, then
\[
\rho_N=O_\Pp\{(\log N/N)^{2/5}\}.
\]
For $N\asymp n$, use the deterministic Fourier dictionary of Proposition~\ref{prop:fourier} with dimension $M_n\asymp n^{1/5}$ and ridge $\lambda_n=O(n^{-1/2})$. Align its representative using the uniform covariate reference and clip to a fixed interval containing that representative of $h_*$. Then
\[
a_n=O_\Pp(n^{-2/5}\log^{2/5}n),\qquad
b_n=O_\Pp(n^{-1/5}\log^{2/5}n),
\]
and Theorem~\ref{thm:score} applies.
\end{corollary}
The historical spline likelihood $\ell_H(c)=\ell_H(f_c)$ is concave in its knot values. Let $\partial_j\ell_H$ denote its derivative with respect to $c_j$ and $\nabla\ell_H$ its coefficient gradient. For a fitted feasible vector $\widehat c$, its likelihood gap is bounded by the computable concavity certificate
\begin{equation}\label{eq:gapcertificate}
\sup_{c\in[-B_f,B_f]^{J_N+1}}\ell_H(c)-\ell_H(\widehat c)
\le \sum_{j=0}^{J_N}\{B_f|\partial_j\ell_H(\widehat c)|
 -\widehat c_j\partial_j\ell_H(\widehat c)\}.
\end{equation}
The spline criterion has gradient Lipschitz constant at most one in Euclidean coefficient norm. We therefore use projected gradient ascent with unit steps, as given in Algorithm~\ref{alg:source}, and stop when the certificate is at most $(\log N/N)^{4/5}$. The appendix proves that this positive tolerance is attained in finite iterations in exact arithmetic. Our construction provides an attainable smooth one-dimensional regime. It does not establish optimization or minimax rates for general deep networks.

\paragraph{Two feasible regimes and a limitation of the sufficient bound.}
Let $q>0$ be the source sample-size growth exponent. Write $N\asymp n^q$, $r_N=N^{-\alpha_{\rm src}}$ up to logarithms, and $\kappa_{\rm rate}=q\alpha_{\rm src}<1/2$. With $\zeta_M\lesssim M^{-s/d}$ and $M\asymp n^u$, the conservative projection argument requires
\[
\frac{(1/2-\kappa_{\rm rate})d}{s}<u<2\kappa_{\rm rate}-1/2,
\quad\text{equivalently}\quad
\kappa_{\rm rate}>\frac{s+d}{4s+2d}.
\]
The generic squared error condition only requires $\kappa_{\rm rate}>1/4$. These are different thresholds. The following are algebraic rate examples, not numerical experiments.
\begin{center}
\begin{tabular}{ccccc}
\toprule
$\alpha_{\rm src}$ & $q$ & $s/d$ & admissible $u$ & example/status\\
\midrule
$2/5$ & $1$ & $2$ & $(1/20,3/10)$ & $u=1/5$\\
$2/5$ & $9/10$ & $1$ & $(7/50,11/50)$ & $u=1/5$\\
$2/5$ & $7/10$ & $1$ & $(11/50,3/50)$ & empty interval\\
\bottomrule
\end{tabular}
\end{center}
The first regime is proved in Corollary~\ref{cor:feasible}. For an integer $s\ge1$, $C^s$ denotes functions with continuous derivatives through order $s$. The second uses the same $C^2$ historical spline construction, $N\asymp n^{9/10}$, and only periodic $C^1$ target risk-set weights, which give $\zeta_M\lesssim M^{-1}$ for $d=1$. It yields $a_n=\Op(n^{-9/25}\log^{2/5}n)$ and $b_n=\Op(n^{-4/25}\log^{2/5}n)$. Thus $\sqrt n a_n^2$ and $\sqrt n a_nb_n$ have polynomial exponents $-11/50$ and $-1/50$, respectively. In the third regime the source squared error condition holds, but no dimension exponent satisfies this conservative construction. This is failure of a sufficient bound, not a proof that valid inference is impossible.

\subsection{Robust null testing and inferential scope}\label{sec:null}
We finally show that direct calibration has a robust null testing justification that permits arbitrary historical error and target event-time relationships. For a bounded working function $g:\mathcal X\to\mathbb R$, define its weighted risk-set mass $s_g$ and observed record residual $R_g$ by
\[
s_g(t)=\E\{Y(t)e^{g(X)}\},\qquad
R_g(O)=\Delta-\int_0^\tau\frac{Y(t)e^{g(X)}}{s_g(t)}\,d\nu(t).
\]
For the realized historical fit, let $\gamma^\dagger$ maximize $\ell_P(0,\gamma,\widehat f)$ and define $g^\dagger=\widehat f+\gamma^{\dagger\top}z$. This is a working risk function, not a requirement that the transfer model be correct.

\begin{theorem}[Randomized null validity]\label{thm:null}
Assume \eqref{eq:null}, $0<p_A<1$, $\nu(\tau)>0$, $\Pr(U\ge\tau)\ge c_{\rm null}>0$ for a fixed bound $c_{\rm null}$, and uniformly bounded indices and $z$. Conditional on the independent historical fit, suppose the finite-dimensional population calibration optimum is uniformly interior, its Hessian is uniformly nonsingular, and $\E\{R_{g^\dagger}^2\}$ is bounded away from zero. For an interior target score root, or a fit with score residual $o_\Pp(n^{-1/2})$, the population treatment coefficient is zero, and
\begin{equation}\label{eq:nullexp}
\widehat\theta=\frac1{np_A(1-p_A)\nu(\tau)}\sum_i(A_i-p_A)R_{g^\dagger}(O_i)+o_\Pp(n^{-1/2}).
\end{equation}
With the treatment standard error from \eqref{eq:sandwich}, $\widehat\theta/\widehat{\operatorname{se}}_{\rm sw}\Rightarrow N(0,1)$. The result also holds for a sequence of historical fits under the stated uniform bounds. It does not require $\widehat f$ to be consistent, a correct correction model, PH, or historical/target overlap for prediction.
\end{theorem}
At the null, every risk-set mean of $A$ at a working index $g(X)$ is $p_A$. Its covariance with every correction covariate is zero. The treatment influence function therefore reduces to $(A-p_A)R_g$, and the sandwich estimates its actual variance. An information-only standard error instead uses $p_A(1-p_A)\nu(\tau)$ and need not equal $p_A(1-p_A)\E(R_g^2)$. Theorem~\ref{thm:null} supports a treatment test under \eqref{eq:null}. It does not imply consistent conditional effect estimation under an alternative, nor a variance gain over an unadjusted test.

\begin{proposition}[Finite-sample randomized null control]\label{prop:randomization}
Assume \eqref{eq:null}, independent Bernoulli allocation with common $0<p_A<1$, and an independent historical sample. Let $\widehat g_{\rm null}$ be any finite fitted risk obtained without target treatment labels, including a prespecified label-blind numerical fallback. Conditional on the historical sample, the target records $(X_i,U_i,\Delta_i)_{i=1}^n$, and $m_A$, the test in \eqref{eq:randomizationp} satisfies
\[
\Pr\{p_{\rm CR}\le\alpha\mid \text{historical sample},(X_i,U_i,\Delta_i)_{i=1}^n,m_A\}\le\alpha.
\]
This conclusion requires neither PH nor an accurate historical risk, a population interior optimum, or a large-sample normal approximation.
\end{proposition}
\begin{proof}
Under the conditioning, \eqref{eq:null} and common Bernoulli allocation make the observed assignment uniform on $\mathcal A_{m_A}$. The fitted risk and residual vector are fixed because their construction excludes treatment labels. The observed assignment and the $B_{\rm rnd}$ sampled assignments are therefore independent draws from the same conditional law, so their test statistics are exchangeable. With ties counted against rejection, the observed statistic's upper tail rank divided by $B_{\rm rnd}+1$ is super-uniform. This proves the display. The declared degenerate cases never reject. Averaging the conditional bound also gives unconditional control. We apply the randomization rank argument of \citet{hemerik2018} to the calibrated score.
\end{proof}
The finite-sample test complements the asymptotic sandwich analysis. It retains the same treatment-independent observation requirement and provides no alternative effect interval or automatic power gain. Its guarantee also explains why an outcome-trained nuisance using target treatment labels cannot simply be held fixed during reassignment.

\subsection{Residual orthogonality for ridge and clipped projections}\label{app:residualprojection}
The fixed dictionary theorem describes convergence to the unregularized, unclipped population projection. We now derive the expansion when the fitted projection converges to a different function. Let $h$ be a fixed bounded covariate function, $q=A-h(X)$, $D_q=\J(q,A)>0$, and $V_q=\J(q,q)>0$. Define $b_{q,n}=\max_k\inf_c\|\widehat h_k-h-c\|_2$ and $d_{g,k}=\widetilde g_k-g_0-c_k$, where any constant alignment may be used. Assume conditions R1 through R3, bounded fits, fixed positive fold fractions, and $a_n,b_{q,n}=o_\Pp(1)$.
\begin{proposition}[Expansion with residual nuisance derivative]\label{prop:residualprojection}
For the estimator in \eqref{eq:scoreest}, under the preceding conditions and exact one-step calculations,
\begin{align*}
\widehat\theta^{\rm score}-\theta_0
&=D_q^{-1}\left\{n^{-1}\sum_i\phi_q(O_i)
-\sum_k w_k\J(q,d_{g,k})\right\}\\
&\quad+\Op\{a_n^2+a_nb_{q,n}+(a_n+b_{q,n})n^{-1/2}+n^{-1}\},
\end{align*}
where $\phi_q=\int\{q-Q_t(q)\}\,d\mathcal M$. Moreover,
$n\widehat{\operatorname{se}}_{\rm score}^2\to_\Pp V_q/D_q^2$.
If the displayed remainder is $o_\Pp(n^{-1/2})$ and
$\sqrt n\sum_k w_k\J(q,d_{g,k})=o_\Pp(1)$, the standardized estimator converges to $N(0,1)$. Without this last condition, the held-out influence variance alone does not generally describe the estimator's full first-order distribution.
\end{proposition}
\begin{proof}
Apply the independent fold expansion in Lemma~\ref{lem:emp} with index error $(\widetilde\theta_k-\theta_0)A+d_{g,k}$ and replace the fitted score direction by $q$. The population Taylor expansion gives
\[
\mathcal U_k=m_k^{-1}\sum_{i\in\mathcal I_k}\phi_q(O_i)
-D_q(\widetilde\theta_k-\theta_0)-\J(q,d_{g,k})
+\Op\{a_n^2+a_nb_{q,n}+(a_n+b_{q,n})n^{-1/2}+n^{-1}\}.
\]
The derivative obeys $D_k=D_q+\Op(a_n+b_{q,n}+n^{-1/2})$. Inverting it and adding the pilot cancels its treatment term. Weighted aggregation over fixed $K$ yields the expansion. Overlap of training complements does not make their nuisance bias terms independent, and no such independence is used. Consistency of the empirical influence variance and the martingale covariance identity give $V_k\to_\Pp V_q$. Since $\sum_k w_k^2/m_k=1/n$, the variance limit follows. The i.i.d. central limit theorem and negligible nuisance term give the final conclusion.
\end{proof}

For a transferred pilot, constants cancel and
\[
\J(q,d_{g,k})=\J(q,e_N)+
\J(q,z)(\widetilde\gamma_k-\gamma_0+b_N).
\]
Thus residual dependence on the target calibration coefficients is absent only when the chosen score is orthogonal to $z$, or when that product is negligible. For a fixed ridge, write $h_\lambda=\mu_\lambda^\top B$ and $z=C_zB$ modulo constants, where $C_z$ is a $p\times M$ coefficient matrix. Then
$\J(A-h_\lambda,z)=\lambda\mu_\lambda^\top C_z^\top$.
With clipped limiting function $h_c=\operatorname{clip}_{[-L_h,L_h]}h_\lambda$, an additional term $\J(h_\lambda-h_c,z)$ remains. Also $D_q$ need not equal $V_q$. Thus ridge and clipping can change both the residual nuisance derivative and the limiting variance. They apply to a deterministic limiting dictionary. Learned dictionaries require convergence to such a limit and control of the resulting perturbation. The full procedure bootstrap below includes training complement variation under its separate regularity conditions, but still requires negligible working target bias for coverage of $\theta_0$.
\subsection{Prediction after selection among prespecified corrections}\label{sec:selection}
Let $z^{(1)},\ldots,z^{(L)}$ be a fixed finite set of correction vectors, each of fixed dimension and chosen independently of target outcomes. A coordinate equal to the independent historical fit $\widehat f$, as in free score rescaling, is allowed because the target analysis conditions on $\widehat f$. For candidate $l$, let $\widehat S^{(l)}$ be the calibrated survival estimator from \eqref{eq:generalfit} with $z=z^{(l)}$, with residual norm $\rho_N^{(l)}$ computed from the drift form of the decomposition in Section~\ref{sec:generaltheory} and score residual $\xi_n^{(l)}$.
\begin{corollary}[Selection among prespecified corrections]\label{cor:selection}
Suppose that, for every $l$, the conditions of Theorem~\ref{thm:generalprediction} hold for $z^{(l)}$ with $r_{n,N}^{(l)}=n^{-1/2}+\rho_N^{(l)}+\xi_n^{(l)}=o_\Pp(1)$. Let $\widehat l\in\{1,\ldots,L\}$ be any random index, possibly depending on target outcomes. Then
\[
\Big\|\sup_{t\le\tau}|\widehat S^{(\widehat l)}(t\mid\cdot)-S(t\mid\cdot)|\Big\|_2
=O_\Pp\Big(\max_{1\le l\le L}r_{n,N}^{(l)}\Big).
\]
\end{corollary}
\begin{proof}
Write $E_l=\|\sup_{t\le\tau}|\widehat S^{(l)}(t\mid\cdot)-S(t\mid\cdot)|\|_2$ and $r=\max_l r_{n,N}^{(l)}$, and fix $\epsilon>0$. By Theorem~\ref{thm:generalprediction}, for each $l$ there are $C_l<\infty$ and $n_l$ such that $\Pr\{E_l>C_lr_{n,N}^{(l)}\}\le\epsilon/L$ for $n\ge n_l$. Put $C=\max_lC_l$. Since $C\ge C_l$ and $r\ge r_{n,N}^{(l)}$ for every $l$, the event $E_{\widehat l}>Cr$ implies $E_l>C_lr_{n,N}^{(l)}$ for $l=\widehat l$. Hence, for $n\ge\max_ln_l$,
\[
\Pr(E_{\widehat l}>Cr)\le\Pr\Big[\bigcup_{l=1}^L\{E_l>C_lr_{n,N}^{(l)}\}\Big]\le\epsilon.
\]
Because $\epsilon$ is arbitrary, $E_{\widehat l}=O_\Pp(r)$.
\end{proof}
The argument uses no property of the selection rule, so it covers selection by inner cross-validation within target training data, as in Appendix~\ref{app:clinicalvalidation}. It requires a vanishing residual for every candidate. A candidate whose span misses part of the target drift is not covered, and selection then helps only if it avoids that candidate. The corollary does not show that selection improves on a fixed candidate; that question is empirical. For treatment inference, the first-order bias $\J(q_z,e_N)$ in \eqref{eq:thetaexp} depends on the selected correction, so post-selection intervals would require consistent selection or a separate argument. We therefore fix the correction for treatment analysis.

\section{Interpretation and sufficient cases for the assumptions}\label{app:conditions}
We explain the conditions in Section~\ref{sec:theory} and give settings in which they hold. Conditionally independent censoring identifies the event hazards. Bounded risks control likelihood ratios, curvature identifies calibration coefficients, and nuisance convergence rates control the remainder terms. These sufficient conditions require substantive consideration in an application, especially the assumptions on follow-up and transport.

\subsection{Prediction conditions P1 through P3}
The general cohort conditions concern a fixed correction space and contain no treatment requirement. Positive follow-up and bounded risk imply positive population risk-set denominators. If $\operatorname{Var}_T\{z(X)\}$ is positive definite, these bounds and positive event information also give correction curvature. For a coefficient direction $v$, the weighted covariance lower bound used below is proportional to $v^\top\operatorname{Var}_T(z)v$. The randomized treatment variance used for R3 is absent. The prediction proof is consequently a reduced parameter argument, not an application of a trial theorem with treatment set to zero.

The support bound transfers an error rate from historical to target covariates, but is separate from estimation of the target baseline. Neither a later recruitment period nor a different hospital guarantees adequate support. Training-based imputation and standardization require the same attention to the feature map as in Section~\ref{sec:generalmodel}. A validation design alone does not establish conditionally independent censoring or the transport model.

\subsection{Relation to partially linear Cox and transfer theory}
Our conditions serve some of the same purposes as those in \citet{zhong2022}. Their assumptions A2 and B1 control the covariate design and identification. B2 and B3 control observation and event information. Their architecture restrictions balance approximation and estimation. We instead use explicit event-weighted coercivity, bounded indices, and positive follow-up probability to support our projection and calibration arguments. Their continuous joint density assumption cannot be applied verbatim to a randomized binary treatment. We obtain treatment identification from randomization and derive correction identification below. Their target-only deep partially linear Cox results also do not establish rates for a transported, independently fitted offset. Our source bound and two-sample remainder conditions supply that additional argument.

\citet{gu2026} provide a different transfer approach based on source survival predictions and a target penalty. Conditions 1 to 3 in their Appendix A.1 address interior parameters, identification, and follow-up. Condition 6 assumes a unique interior population optimum. These conditions motivate comparable checks here, but their prediction penalty, transformation model, and optimization analysis differ from our offset calibration and projected score. We therefore use that paper for context rather than importing its assumptions or conclusions. In particular, source prediction accuracy alone does not establish our alternative effect inference: the residual functional in \eqref{eq:thetaexp} or \eqref{eq:finiteexp} must also be controlled.

\subsection{Observation and boundedness: R1 to R2}
Conditionally independent censoring in R1 is standard for interpreting a Cox event hazard, but it is an untestable restriction on the latent event and censoring times. We allow censoring to depend on measured covariates and, under the transfer model, on treatment. The finite, positive cumulative baseline excludes a horizon with no event information and an asymptotic regime in which that information vanishes. Continuous event times avoid a separate approximation for ties. Routinely rounded event dates may require such an extension.

The uniform observation bound in R2 is stronger than positive average follow-up. We give a sufficient case to make its content explicit. Write
\[
G_a(t\mid x)=\Pr(C\ge t\mid A=a,X=x),\qquad
G_H(t\mid x)=\Pr(C^H\ge t\mid X^H=x)
\]
for conditional censoring survival. Suppose both exceed a fixed $c_C>0$ at $\tau$, and the true indices satisfy $|\eta_0|\le B_\eta$ and $|f_0|\le B_f$. Conditionally independent censoring and the transfer model imply
\begin{align*}
\Pr(U\ge\tau\mid A=a,X=x)
&=G_a(\tau\mid x)\exp\{-\Lambda_{T,0}(\tau)e^{\eta_0(a,x)}\}\\
&\ge c_C\exp\{-\Lambda_{T,0}(\tau)e^{B_\eta}\}>0,\\
\Pr(U^H\ge\tau\mid X^H=x)
&\ge c_C\exp\{-\Lambda_{H,0}(\tau)e^{B_f}\}>0.
\end{align*}
This case permits covariate-dependent censoring and unequal cohort follow-up. A fixed horizon before follow-up becomes sparse can satisfy the bound. It fails if some strata are almost surely censored earlier, or if their survival through the horizon tends to zero. A small lower bound also produces poor finite-sample constants despite an asymptotically valid theorem.

Bounded correction coordinates and indices prevent extreme risk weights. Clipping the fitted source function enforces its computational bound, but does not establish bounded true risk, transport compatibility, or a small optimization gap. Compact coefficient sets can contain a fixed finite truth in their interior. The aligned coefficient $\beta_N$ and relevant population optima must also remain interior. A large absorbed correction shift can violate that requirement even when prediction errors cancel algebraically. We do not require every raw covariate to be continuous or to have a joint density.

\subsection{Identification, curvature, and numerical roots: R3}
We give a design condition under which the information requirement is transparent. For any $v\in\mathbb R^{p+1}$, Lemma~\ref{lem:geometry} yields
\[
v^\top I_Wv=\J(v^\top W,v^\top W)
\ge c_{\rm obs}e^{-B_\eta}\Lambda_{T,0}(\tau)
v^\top\operatorname{Var}_T(W)v.
\]
Here $\operatorname{Var}_T$ is variance under the target law. Randomization makes $\operatorname{Var}_T(W)$ block diagonal, with blocks $p_A(1-p_A)$ and $\operatorname{Var}_T\{z(X)\}$. Thus linearly independent nonconstant correction coordinates with positive definite covariance imply positive $I_W$. Duplicated coordinates or a correction intercept violate this sufficient condition. Treatment information for the unrestricted projection is positive by the same lemma, without assuming a continuous treatment distribution.

For population curvature, define $I_P(\beta,f)=-\nabla_\beta^2\ell_P(\beta,f)$. We enlarge $B_\eta$ to bound all candidate indices on the neighborhood in question. At each time their unnormalized risk weight is at least $c_{\rm obs}e^{-B_\eta}$, while its normalizing factor is at most $e^{B_\eta}$. The weighted variance identity as an infimum over constants then gives
\[
v^\top I_P(\beta,f)v
\ge c_{\rm obs}e^{-2B_\eta}\nu(\tau)
v^\top\operatorname{Var}_T(W)v.
\]
Consequently a uniform candidate bound and a fixed design covariance lower bound suffice for local curvature. These inequalities apply to a prespecified correction space of fixed dimension with no redundant coordinates. They do not guarantee a finite sample interior root: few events, separation, or a restrictive coefficient boundary can prevent one. Algorithm~\ref{alg:direct} reports these failures rather than treating a penalized or boundary fit as the root analyzed here. Its reported score residual verifies the numerical part of R3. Local population curvature and consistency are still statistical requirements.

\subsection{Projection representatives and dictionaries: R4}
The projection is identified only modulo constants, so clipping requires a common representative. Our deterministic $R_P$ makes this explicit. To see why the bounded density ratio condition is useful, take $v$ with $\int v\,dR_P=0$ and let $b$ minimize $\norm{v-b}_{L_2(P_{T,X})}$. Then
\[
|b|=\left|\int(v-b)\,dR_P\right|
\le\left\|\frac{dR_P}{dP_{T,X}}\right\|_{L_2(P_{T,X})}
\norm{v-b}_{L_2(P_{T,X})}.
\]
The centered unquotiented error is therefore bounded by a fixed multiple of the quotient error. Clipping to an interval containing the centered true projection cannot increase this error. A uniform reference on a compact domain with target density bounded away from zero is one sufficient choice, verified in Proposition~\ref{prop:fourier}. Estimated centering measures require their own error analysis.

Fixed $K$ and fold sizes proportional to $n$ keep training and evaluation rates comparable. Independence of fold assignment from patient records is easily enforced by design. A deterministic dictionary avoids an additional learning step, but its Gram and approximation conditions remain substantive. The Fourier construction proves both under explicit density and smoothness conditions. Its periodic boundary condition is specific to that construction, rather than to the general treatment theorem. A finite dictionary containing $z$ supports Theorem~\ref{thm:finite}. It need not approximate every nuisance direction. A fixed positive ridge can also leave a first-order residual even when the dictionary is rich. Neither choice establishes \eqref{eq:rates} merely by increasing sample size.

\subsection{Transport, source rates, and the separate null}
The common historical prognostic function with drift confined to $\operatorname{span}(z)$ in \eqref{eq:T} is a modeling assumption, not a technical consequence of randomization. Calibration absorbs only the directions stated in Proposition~\ref{prop:invariance}. The overlap bound \eqref{eq:overlap} gives, for any square-integrable function $v$ of $X$,
\[
\norm{v}_{L_2(P_{T,X})}^2
\le C_{\rm ov}\norm{v}_{L_2(P_{H,X})}^2.
\]
This elementary change of measure bound transports the source error rate. It cannot identify prognosis on target covariate regions absent from the historical cohort. Even with overlap, drift outside the correction space leaves the bias characterized by Proposition~\ref{prop:sensitivity}. A large historical sample does not remove it. The proposition is a local population sensitivity calculation, not an observed-data estimate of transport error.

The entropy, approximation, and likelihood gap assumptions in Theorem~\ref{thm:source} specify three separate sources of learning error. Bounded networks satisfy the stated finite cover bound, but an architecture does not certify approximation of an arbitrary true function or success of a nonconvex optimizer. Corollary~\ref{cor:feasible} and Algorithm~\ref{alg:source} show that the assumptions and treatment remainders can hold together for a smooth one-dimensional spline construction. The rate regimes following that corollary are sufficient algebraic choices, not guarantees for every neural learner or every clinical covariate distribution.

Finally, the observed-data null \eqref{eq:null} is different from the transfer model. Randomization, equal conditional event distributions across arms, and a common conditionally independent censoring law imply it, even without PH. Randomization alone does not eliminate treatment-dependent observation. Theorem~\ref{thm:null} still requires a stable interior working fit, bounded indices, positive follow-up, and nondegenerate variance. Its sandwich argument is consistent with the distinction between working model information and actual score variance emphasized by \citet{linwei1989}. It supplies a null test with arbitrary historical prediction error, not an alternative effect confidence interval or a survival prediction guarantee.

\subsection{Risk-set association and directional bias}\label{sec:biasexamples}
We give two analytical examples to clarify the restrictions behind alternative effect inference. Neither is a numerical experiment.

\paragraph{Binary risk-set association.}
Risk-set association explains the linear term. Consider independent binary $A,X$ at enrollment, a prognostic coefficient $b\in\mathbb R$ with $g_0(X)=bX$, and censoring independent of both. Define $p_t(a,x)=Q_t(\ind\{A=a,X=x\})$ as the weighted risk-set probability of each binary cell. Its odds ratio is
\begin{equation*}
\frac{p_t(1,1)p_t(0,0)}{p_t(1,0)p_t(0,1)}
=\exp\{-\Lambda_{T,0}(t)(e^{\theta_0}-1)(e^b-1)\}.
\end{equation*}
It differs from one when both coefficients are nonzero. Hence randomization need not make $\J(A,X)$ vanish. If the corresponding error is included in $z$, it is instead absorbed by Proposition~\ref{prop:invariance}.

\paragraph{A finite support illustration.}
Take $X\in\{-1,0,1\}$ with equal probabilities, $g_0(X)=bX$ for a scalar $b\in\mathbb R$, no nonadministrative censoring, no correction terms, and $\theta_0b\ne0$. The weighted risk-set treatment odds conditional on $X=x$ are
\[
\frac{p_A}{1-p_A}e^{\theta_0}
\exp\{-\Lambda_{T,0}(t)(e^{\theta_0}-1)e^{bx}\}.
\]
They are strictly monotone in $x$ whenever $\Lambda_{T,0}(t)>0$, so $\operatorname{Cov}_{Q_t}(A,X)$ has a fixed nonzero sign on those risk sets. Consequently $\J(A,X)\ne0$. Denote the uncalibrated representer by $\omega_{\varnothing}$, using \eqref{eq:sensitivityweight} with $q_z=A$. It is nonzero. Write $v_1=\omega_{\varnothing}(-1)$, $v_2=\omega_{\varnothing}(0)$, and $v_3=\omega_{\varnothing}(1)$. Define the covariate function $u$ by
\[
u(-1)=v_2-v_3,\qquad u(0)=v_3-v_1,\qquad u(1)=v_1-v_2.
\]
Then $\E u(X)=0$ and $\E\{u(X)\omega_{\varnothing}(X)\}=0$. The function is nonzero because $\omega_{\varnothing}$ is nonzero and mean zero. Define $e_\parallel=r\omega_{\varnothing}/\norm{\omega_{\varnothing}}_{L_2(P_{T,X})}$ and $e_\perp=ru/\norm u_{L_2(P_{T,X})}$. Both therefore have target prediction norm exactly $r$. The parallel error has a nonzero linear treatment bias, whereas the perpendicular error has bias $O(r^2)$, by Proposition~\ref{prop:sensitivity}. Including either error direction in the correction span absorbs it exactly. Including it only in the projection dictionary removes its first-order functional but still leaves the quadratic remainder. At $\theta_0=0$ both first-order effects vanish. All quantities in this example are bounded, and small $r$ gives feasible interior paths.

\section{Technical lemmas, prediction, and treatment inference}\label{app:target}
The arguments below condition on the historical fit, which is independent of target observations. In evaluation fold arguments they also condition on the corresponding target training observations. No entropy bound for the historical network class is used in a held-out target sample. Such a bound is still needed to prove a historical neural prediction rate.

For a target sample, use the completed, right-continuous filtration generated by baseline $(A_i,X_i)$, the independent historical fit, and the event and censoring counting processes through time $t$. Conditionally independent censoring and iid subjects give the event compensator $\int_0^tY_i(s)e^{\eta_{0i}}\,d\Lambda_{T,0}(s)$. The left-continuous at-risk indicator is predictable. For an evaluation fold, include that fold's independent training observations in the initial sigma field. The analogous historical filtration includes historical covariates and observed histories, with fixed candidate functions. The selected historical fit is handled afterward by the covering argument.

\subsection{Proof of Theorem~\ref{thm:generalprediction}}\label{sec:generalproof}
This proof concerns $O=(U,\Delta,X)$ under \eqref{eq:generalmodel}. Condition on the historical fit, which is independent of the target sample. Let $Y(t)=\ind(U\ge t)$, $\cnt(t)=\ind(U\le t,\Delta=1)$, $\nu(t)=\E \cnt(t)$, and $\nu_n(t)=n^{-1}\sum_i\cnt_i(t)$. For a bounded covariate index $v$, write
\[
s_v(t)=\E\{Y(t)e^{v(X)}\},\qquad
s_{n,v}(t)=n^{-1}\sum_iY_i(t)e^{v(X_i)},\qquad
\bar z_v(t)=\frac{\E\{Y(t)e^{v(X)}z(X)\}}{s_v(t)}.
\]
Let $\bar z_{n,v}$ denote the corresponding empirical ratio. P2 bounds $s_v$ away from zero uniformly over the horizon and the bounded candidate indices. For fixed bounded multipliers, the empirical averages indexed by $\ind(U\ge t)$ have uniform error $O_\Pp(n^{-1/2})$. Conditional on $\widehat f$, the family $e^{\widehat f+\gamma^\top z}$ is Lipschitz in the fixed-dimensional coefficient $\gamma$ on a compact set, with a bounded envelope. Its products with threshold indicators therefore have a finite uniform entropy integral. The same empirical rate holds uniformly over that coefficient set, also for first and second derivatives. This argument allows discrete or continuous covariates and uses no treatment variable.

First suppose $p>0$. At the comparison coefficient $\gamma_N$, remove the constant $c_N$ from the index to obtain $g_0+e_N$. The population score at any $v$ is
\[
\mathcal U_P(v,z)=\int_0^\tau\{\bar z_{g_0}(t)-\bar z_v(t)\}\,d\nu(t).
\]
It vanishes at $g_0$. Bounded exponential derivatives and the denominator lower bound give
\[
\sup_{t\le\tau}\|\bar z_{g_0+e_N}(t)-\bar z_{g_0}(t)\|_2
\le C\E|e_N(X)|\le C\rho_N.
\]
Because $\nu(\tau)\le1$, the comparison score is $O_\Pp(\rho_N)$. The empirical score differs from it by $O_\Pp(n^{-1/2})$. To justify the event integration step, bounded weighted tail functions and their ratios have uniformly bounded total variation when denominators are positive. Integration by parts against $\nu_n-\nu$ then bounds the event measure error by $O_\Pp(n^{-1/2})$.

By P3 and uniform Hessian convergence, the empirical negative Hessian has a positive lower bound on a fixed neighborhood of $\gamma_N$ with probability tending to one. The score at its center tends to zero. On the boundary of a sufficiently small ball, the radial derivative points inward. Concavity places an interior maximizer inside that ball and excludes a distinct maximum outside it. A numerical root with vanishing residual has the same localization. The mean value score equation then gives
\[
\|\widehat\gamma-\gamma_N\|_2
\le C\{\|\nabla_\gamma\ell_n^{\rm pred}(\gamma_N)\|_2+\xi_n\}
=O_\Pp(n^{-1/2}+\rho_N+\xi_n).
\]
Since $\widehat g-g_0-c_N=e_N+(\widehat\gamma-\gamma_N)^\top z$ and $z$ is bounded, the first conclusion follows. For $p=0$, there is no score equation and the risk error equals $e_N$ directly.

The projection coefficients $b_N,c_N$ are bounded. Indeed, for $p>0$ the fixed covariance of $z$ is nonsingular, as P3 would otherwise have a zero curvature direction. The projection normal equations then express $b_N$ through bounded covariances of $z$ and $\widehat f-f_0$, and $c_N$ through their means. For $p=0$, $c_N$ is the mean of a bounded historical error. Thus $v^{\rm a}=\widehat g-c_N$ is uniformly bounded and has $\|v^{\rm a}-g_0\|_2=O_\Pp(r_{n,N})$.

The empirical bound above and the Lipschitz exponential imply
\[
\sup_{t\le\tau}|s_{n,v^{\rm a}}(t)-s_{g_0}(t)|
\le O_\Pp(n^{-1/2})+C\E|v^{\rm a}(X)-g_0(X)|
=O_\Pp(r_{n,N}).
\]
The expectation evaluates a fresh target covariate conditional on the fitted function. The target model and independent censoring give $d\nu=s_{g_0}\,d\Lambda_{T,0}$. Hence
\begin{align*}
e^{c_N}\widehat\Lambda_{T,0}(t)-\Lambda_{T,0}(t)
&=\int_0^t\left\{s_{n,v^{\rm a}}(u)^{-1}-s_{g_0}(u)^{-1}\right\}\,d\nu_n(u)\\
&\quad+\int_0^t s_{g_0}(u)^{-1}\,d(\nu_n-\nu)(u).
\end{align*}
The first integral has uniform order $O_\Pp(r_{n,N})$ because $\nu_n(\tau)\le1$ and both denominators stay positive with probability tending to one. The second has uniform order $O_\Pp(n^{-1/2})$, since $1/s_{g_0}$ is bounded and nondecreasing, and the empirical event subdistribution has a uniform root-$n$ bound. This proves the baseline conclusion.

Write $\widehat\Lambda^{\rm a}=e^{c_N}\widehat\Lambda_{T,0}$. The prediction product is unchanged by alignment. On events whose probabilities tend to one, the baselines are bounded and nonnegative. Lipschitz bounds for the exponentials give
\[
\sup_{t\le\tau}|\widehat S(t\mid x)-S(t\mid x)|
\le C\left\{\sup_{t\le\tau}|\widehat\Lambda^{\rm a}(t)-\Lambda_{T,0}(t)|
+|v^{\rm a}(x)-g_0(x)|\right\}.
\]
Take the target covariate $L_2$ norm to obtain the survival bound. Finally, the replacement $f\mapsto f+c+b^\top z$, $\gamma\mapsto\gamma-b$ shifts every fitted index by $c$. The partial likelihood is unchanged, the Breslow estimator is multiplied by $e^{-c}$, and the survival product is unchanged. This proves the invariance statement under the stated feasibility condition. No treatment coordinate, randomization, or treatment information argument has been used.\hfill$\square$

\subsection{Identification and exact calibration invariance}
For Proposition~\ref{prop:ident}, conditionally independent censoring and positive observation probability identify the event hazard over the observed interval. Each positive cumulative baseline provides times with nonzero hazard. Equality of two historical representations then forces the difference in their log-risk functions to be constant across the identifiable support. Centering fixes that constant. In the target trial, the conditional arm ratio identifies $e^{\theta_0}$. Once $f_0$ is identified, the difference of two remaining target representations is $b^\top z(X)$ equal to a constant. Its variance is $b^\top\operatorname{Var}_T(z)b$, so positive definiteness gives $b=0$. The cumulative baselines follow from these representations on $[0,\tau_{\rm id}]$. No extrapolation past this interval is identified.

For Proposition~\ref{prop:invariance}, replacing $f$ with $f+c+b^\top z$ and $\gamma$ with $\gamma-b$ adds only $c$ to each target risk index. It cancels between numerator and denominator of the partial likelihood. Breslow increments acquire the factor $e^{-c}$, which cancels $e^c$ in individual risk, preserving every fitted survival curve. Feasibility and the absence of a correction penalty are essential.

For Proposition~\ref{prop:offsetstability}, hold the coefficient at $\widehat\beta_f$ and interpolate offsets as $f+se$, $0\le s\le1$. Differentiating the weighted risk-set mean of $W$ with respect to $s$ gives its risk-set covariance with $e$, whose Euclidean norm is at most $2L\varepsilon$. Integration over $s$ and the empirical event measure therefore gives
\[
\norm{\mathcal U_n(\eta_{\widehat\beta_f,f+e},W)
      -\mathcal U_n(\eta_{\widehat\beta_f,f},W)}_2
\le2L\nu_n(\tau)\varepsilon,
\]
where $\eta_{\beta,f}$ denotes the index at coefficient $\beta$ and offset $f$. The second score is zero. The first equals the integrated negative Hessian for offset $f+e$ times $\widehat\beta_{f+e}-\widehat\beta_f$, because the score at the latter root is zero. This symmetric matrix has minimum eigenvalue at least $\kappa$. Multiplying by its inverse proves \eqref{eq:offsetstability}. The correction space extension follows from the exact feasible reparameterization in Proposition~\ref{prop:invariance}. No stochastic or proportional hazards assumption is used in this fixed sample bound.

\subsection{Risk-set means and their empirical expansion}
For a bounded index $\eta(A,X)$ and bounded score direction $q(A,X)$, define
\[
 s_\eta(t)=\E\{Y(t)e^\eta\},\qquad
 \bar q_\eta(t)=\E\{Y(t)e^\eta q\}/s_\eta(t),\qquad
 d\nu(t)=\E\{d\cnt(t)\}=s_0(t)\,d\Lambda_{T,0}(t).
\]
For a sample of size $m$, write $s_{m,\eta}$ and $\bar q_{m,\eta}$ for the corresponding empirical ratios. Set
\begin{align*}
 \mathcal U_m(\eta,q)&=\frac1m\sum_{i=1}^m\int_0^\tau
       \{q_i-\bar q_{m,\eta}(t)\}\,d\cnt_i(t),\\
 \mathcal U_P(\eta,q)&=\int_0^\tau s_0(t)\{Q_t(q)-\bar q_\eta(t)\}\,d\Lambda_{T,0}(t).
\end{align*}

\begin{lemma}[An evaluation sample expansion]\label{lem:emp}
Under R1 and R2, let bounded $\eta,q$ be fixed independently of the evaluation sample. Uniform bounds are fixed and do not depend on $m$. Then
\begin{equation*}
 \mathcal U_m(\eta,q)=\mathcal U_P(\eta,q)+\frac1m\sum_{i=1}^m\varphi_{\eta,q}(O_i)+\Op(m^{-1}),
\end{equation*}
where the mean-zero influence function is
\begin{align}
 \varphi_{\eta,q}(O)
 &=\int_0^\tau\{q-\bar q_\eta(t)\}\,d\cnt(t)-\mathcal U_P(\eta,q)\notag\\
 &\quad-\int_0^\tau\frac{Y(t)e^\eta}{s_\eta(t)}\{q-\bar q_\eta(t)\}\,d\nu(t).\label{eq:IF}
\end{align}
Moreover, for a bounded $q_0$ and $u=\norm{\eta-\eta_0}_{L_2(P_{T,A,X})}$, $v=\norm{q-q_0}_{L_2(P_{T,A,X})}$,
\[
 \norm{\varphi_{\eta,q}-\varphi_{\eta_0,q_0}}_{L_2(P_T)}\le C(u+v).
\]
Consequently, conditionally on independently trained functions,
\begin{align}
 &\mathcal U_m(\eta,q)-\mathcal U_P(\eta,q)-\mathcal U_m(\eta_0,q_0)\notag\\
 &\hspace{12mm}=\Op\{(u+v)m^{-1/2}+m^{-1}\},\label{eq:stochdiff}
\end{align}
since $\mathcal U_P(\eta_0,q_0)=0$. At the true index,
\[
 \varphi_{\eta_0,q_0}(O)=\int_0^\tau\{q_0-Q_t(q_0)\}\,d\mathcal M(t).
\]
\end{lemma}
\begin{proof}
The survival lower bound and bounded indices imply $\inf_{t\le\tau}s_\eta(t)\ge c_1>0$. For each fixed independently trained multiplier, empirical averages indexed by $\ind(U\ge t)$ have a uniform $m^{-1/2}$ bound. This follows directly from the nested threshold sets and their bounded multipliers, or from the usual threshold empirical process bound. Thus $\inf_ts_{m,\eta}(t)\ge c_1/2$ with probability tending to one.

Let $a_m(t)=m^{-1}\sum_iY_i(t)e^{\eta_i}q_i$ and $a(t)=\E\{Y(t)e^\eta q\}$. Expanding $a_m/s_m$ about $a/s$ gives
\[
 \bar q_{m,\eta}(t)-\bar q_\eta(t)
 =\frac1m\sum_i\frac{Y_i(t)e^{\eta_i}}{s_\eta(t)}
       \{q_i-\bar q_\eta(t)\}+R_m(t),
\]
with $\sup_t|R_m(t)|=\Op(m^{-1})$, because the exact ratio remainder is a product of two empirical deviations divided by denominators bounded away from zero. Insert this into $\mathcal U_m$. To make the order two remainder explicit, put
\[
L_t(o)=\frac{Y_o(t)e^{\eta_o}}{s_\eta(t)}\{q_o-\bar q_\eta(t)\},\qquad
K(o,o')=\int_0^\tau L_t(o)\{d\cnt_{o'}(t)-d\nu(t)\}.
\]
Replacing the empirical event measure in the linear ratio term by $\nu$ leaves $m^{-2}\sum_{i,j}K(O_i,O_j)$. The kernel is bounded and has mean zero conditional on either argument, because $\E L_t=0$ and $\E(d\cnt-d\nu)=0$. For the off diagonal sum, pairs with disjoint indices or exactly one common index have zero covariance. Only $O(m^2)$ pairs remain. Its second moment is therefore $O(m^{-2})$. The $m$ diagonal terms are bounded deterministically after normalization by $m^2$, so contribute $O(m^{-1})$. The ratio remainder integrates against an event measure of mass at most one. These two errors are $\Op(m^{-1})$. The remaining linear terms are precisely \eqref{eq:IF}. Its expectation is zero because the weighted risk set residual has mean zero at every $t$.

For the Lipschitz bound, use $|e^a-e^b|\le C|a-b|$ on bounded intervals, $s_\eta\ge c_1$, and Cauchy--Schwarz. These imply
\[
 \sup_t|\bar q_\eta(t)-\bar q_{\eta_0,q_0}(t)|\le C(u+v),
 \qquad \sup_t|s_\eta(t)-s_0(t)|\le Cu.
\]
Here the notation $\bar q_{\eta_0,q_0}$ specifies both its index and residual function. Each term in \eqref{eq:IF} then has $L_2(P_T)$ difference at most $C(u+v)$: an event integral contains at most one event, and the compensator integrals have finite total baseline hazard and bounded weights. Conditional variance of the average influence function difference is at most $C(u+v)^2/m$. This proves \eqref{eq:stochdiff}. Finally substitute $d\nu=s_0d\Lambda_{T,0}$ at $\eta_0$ in \eqref{eq:IF}. The compensator subtraction gives the stated martingale integral.
\end{proof}

The same ratio argument applies to empirical event-weighted covariances $\widehat\J_m(u,v)$ with fixed bounded multipliers. At the true index they converge to $\J(u,v)$. Their population limits at a perturbed index differ by at most $C\norm{\eta-\eta_0}$ for bounded multipliers. For a multiplier tending to zero in $L_2$, the linear empirical fluctuation has order its $L_2$ norm times $m^{-1/2}$. The order two remainder above can be bounded by $O_\Pp(m^{-1})$. This weaker remainder is sufficient for the inference expansions below.

\subsection{Population score derivatives}
\begin{lemma}[First order and quadratic terms]\label{lem:pop}
For bounded $q$ and a bounded perturbation $v(A,X)$, with $\eta_0+v$ in the bounded index class,
\begin{equation}\label{eq:poptaylor}
 \mathcal U_P(\eta_0+v,q)=-\J(q,v)+R(v,q),\qquad |R(v,q)|\le C\norm v^2.
\end{equation}
If $q=q_*+d$ with bounded $q_*,q,d$, then
\[
 \mathcal U_P(\theta A+g,q)= -I_X(\theta-\theta_0)
       +O\left[(|\theta-\theta_0|+\norm{g-g_0})^2
       +\norm d(|\theta-\theta_0|+\norm{g-g_0})\right].
\]
Constants may be removed from $g-g_0$ or $d$ in these statements.
\end{lemma}
\begin{proof}
Along $\eta_s=\eta_0+sv$, differentiation of the risk-set ratio gives
\[
 \frac{d}{ds}\bar q_{\eta_s}(t)=\operatorname{Cov}_{Q_{t,s}}(q,v),\qquad
 \frac{d^2}{ds^2}\bar q_{\eta_s}(t)
 =Q_{t,s}\left[\{q-Q_{t,s}(q)\}\{v-Q_{t,s}(v)\}^2\right].
\]
The second derivative is bounded in absolute value by $C\E\{v(A,X)^2\}$, uniformly in $t,s$, using bounded likelihood ratios and denominators. The event intensity measure has finite mass. Taylor's formula inside $\mathcal U_P$ proves \eqref{eq:poptaylor}. In the second display take $v=(\theta-\theta_0)A+(g-g_0)(X)$. Projection gives $\J(q_*,g-g_0)=0$ and $\J(q_*,A)=I_X$. The remaining linear term satisfies
\[
 |\J(d,v)|\le \J(d,d)^{1/2}\J(v,v)^{1/2}\le C\norm d\norm v.
\]
Constants cancel from risk-set means and covariances, proving the last statement.
\end{proof}

\subsection{Proof of Theorem~\ref{thm:plugin}}\label{sec:pluginproof}
By Proposition~\ref{prop:invariance}, work with offset $f_0+e_N$ and comparison coefficient $\beta_0=(\theta_0,\gamma_0^\top)^\top$. Transform back by subtracting $b_N$ from the correction afterward. Let $S_n(\beta,e_N)$ be the vector score in the $W$ direction and let $\delta=\widehat\beta-\beta_N=(\delta_\theta,\delta_\gamma^\top)^\top$ denote the transformed coefficient error. Lemmas~\ref{lem:emp} and \ref{lem:pop}, component by component, give
\begin{equation}\label{eq:scorebeta}
 S_n(\beta_0,e_N)=\frac1n\sum_{i=1}^n\phi_W(O_i)-\J(W,e_N)
       +\Op(n^{-1}+\rho_N^2+\rho_Nn^{-1/2}),
\end{equation}
where
\[
 \phi_W(O)=\int_0^\tau\{W-Q_t(W)\}\,d\mathcal M(t).
\]
The population score perturbation is $O_\Pp(\rho_N)$ by Cauchy--Schwarz. Finite dimensional local convexity and the score law of large numbers place the interior maximum within a neighborhood of the true index with probability tending to one. More explicitly, in a fixed small ball the negative population criterion increases by at least a constant times $\norm\delta^2$ away from its local minimum, while the perturbation of its gradient is $O_\Pp(\rho_N)$ and its empirical gradient error is $O_\Pp(n^{-1/2})$. Positive curvature therefore yields
\[
 \norm\delta=\Op(n^{-1/2}+\rho_N).
\]
Concavity of the finite-dimensional partial likelihood excludes a different interior maximum outside this ball. This proves \eqref{eq:betarate}.

The integrated empirical negative Hessian between the score root and $\beta_0$ is
\[
 \bar I_n=I_W+\Op(n^{-1/2}+\rho_N+\norm\delta).
\]
To see this, use the event-weighted covariance ratio expansion for the independent offset at $\beta_0$, and a deterministic Lipschitz bound in the finite-dimensional parameter displacement. All coordinates of $W$ and the risk index are bounded. The score equation is $0=S_n(\beta_0,e_N)-\bar I_n\delta$. Its inverse expansion and \eqref{eq:scorebeta} give
\[
 \delta=I_W^{-1}\left\{\frac1n\sum_i\phi_W(O_i)-\J(W,e_N)\right\}
       +\Op(n^{-1}+\rho_N^2+\rho_Nn^{-1/2}).
\]
The first row of this inverse operation is the Schur complement against the $z$ coordinates. Thus its first coordinate is exactly \eqref{eq:thetaexp}, with $q_z=A-\J(A,z)\J(z,z)^{-1}z$ and $I_z=\J(q_z,q_z)$. The fitted risk error is $e_N+z^\top\delta_\gamma$ up to its constant, and bounded $z$ proves \eqref{eq:grate}.

Martingale isometry gives $\E\phi_z=0$ and $\E\phi_z^2=I_z$. Bounded covariates, finite integrated intensity, and independent subjects imply the ordinary iid central limit theorem for $n^{-1/2}\sum_i\phi_z(O_i)$. The two additional conditions in the theorem make the historical functional and quadratic remainder negligible on the $\sqrt n$ scale. The observed target information converges to $I_W$, so its treatment variance converges to $I_z^{-1}/n$. Finally, a numerical score residual $r_n$ adds $\bar I_n^{-1}r_n=\Op(\norm{r_n})$ to the coefficient expansion. This establishes the stated numerical qualification.\hfill$\square$

For sandwich consistency, align the fitted offset as in \eqref{eq:error}. Conditional on the historical function, the risk ratio and event measure bounds are uniform over the compact finite coefficient set. At a fixed comparison coefficient, the population influence vector differs from $\phi_W$ by $O_\Pp(\rho_N)$ in $L_2$, by Lemma~\ref{lem:emp}. Its empirical squared average has the same limit by boundedness and the conditional law of large numbers. Replacing that coefficient with the estimated root changes the influence average by $o_\Pp(1)$, using the coefficient rate, bounded derivatives, and uniformly positive denominators. The empirical influence construction in \eqref{eq:empIF} likewise differs by $o_\Pp(1)$ in empirical squared norm. To justify its event integral replacement, bounded weighted tail moments have bounded total variation, as do their ratios when the denominator is bounded below. Integration by parts against $\nu_n-\nu$ therefore controls these integrals uniformly over subjects and the finite coefficient class. Uniform convergence of the empirical risk ratios controls the remaining integrand differences. Thus
\[
\frac1n\sum_i\widehat\varphi_i(\widehat\eta,W)\widehat\varphi_i(\widehat\eta,W)^\top
\ \longrightarrow_\Pp\ \E(\phi_W\phi_W^\top)=I_W.
\]
The last equality is martingale isometry. Together with $\widehat I_W\to_\Pp I_W$, this proves that $n$ times the sandwich covariance converges to $I_W^{-1}$ and its treatment entry to $I_z^{-1}$. This argument also shows why treating the historical learner as fixed for empirical variance estimation does not, by itself, remove the functional bias in \eqref{eq:thetaexp}.

\subsection{Proof of Theorem~\ref{thm:score}}\label{sec:scoreproof}
Remove constants from the fitted nuisance differences. For fold $k$, write $s_k=\widetilde\theta_k-\theta_0$, $v_k=s_kA+(\widetilde g_k-g_0)(X)$, and $d_k=\widehat q_k-q_*$. Conditional on its independent training observations, $\norm{v_k}\le C a_n$ and $\norm{d_k}\le C b_n$. By Lemma~\ref{lem:pop},
\[
 \mathcal U_P(\widetilde\eta_k,\widehat q_k)
   =-I_Xs_k+\Op(a_n^2+a_nb_n).
\]
Lemma~\ref{lem:emp} supplies
\begin{align*}
 \mathcal U_k&=-I_Xs_k+\frac1{m_k}\sum_{i\in\mathcal I_k}\phi_{q_*}(O_i)\\
 &\quad+\Op\{a_n^2+a_nb_n+(a_n+b_n)m_k^{-1/2}+m_k^{-1}\},
\end{align*}
An empirical covariance expansion and the population Lipschitz bounds imply
\[
 D_k=I_X+\Op(a_n+b_n+m_k^{-1/2}),\qquad V_k=I_X+o_\Pp(1).
\]
For the empirical influence variance, the risk ratio expansion also gives convergence of each fitted empirical influence function to its population counterpart in empirical squared norm. Bounded independently trained multipliers control the empirical remainder, and Lemma~\ref{lem:emp} controls the nuisance perturbation. Since the true influence variance is $I_X$, this proves the same limit for the $V_k$ used in \eqref{eq:se}.

There are finitely many folds, so these conditional orders hold jointly by a union bound. No independence between fitted functions from different folds is assumed.

In the one-step update, the term $s_k$ cancels against $-I_Xs_k/D_k$. The leftover denominator error times $s_k$ is bounded by the same $a_n^2+a_nb_n$ and vanishing stochastic orders. Condition \eqref{eq:rates} therefore yields
\[
 \widetilde\theta_k+\mathcal U_k/D_k-\theta_0
 =I_X^{-1}\frac1{m_k}\sum_{i\in\mathcal I_k}\phi_{q_*}(O_i)+o_\Pp(n^{-1/2}).
\]
Multiplying by $w_k=m_k/n$ and summing includes each subject once, proving the expansion in \eqref{eq:asympscore}. The leading summands are iid mean zero with variance $I_X$ by martingale isometry. Their central limit theorem applies even though the estimated nuisance functions across folds are dependent, because those functions have disappeared from the leading term. Equation~\eqref{eq:se} satisfies
\[
 n\sum_k\frac{w_k^2V_k}{m_kD_k^2}
 =\sum_k\frac{m_k}{n}\frac{V_k}{D_k^2}\ \longrightarrow_\Pp I_X^{-1}.
\]
Slutsky's theorem proves the interval statement.\hfill$\square$

\subsection{Proof of Corollary~\ref{cor:rates} and the overlap comparison}
The direct fit rate on training samples proportional to $n$ gives $a_n=\Op(r_N+n^{-1/2})$. Substituting this and $b_n=\Op(s_n)$ into \eqref{eq:rates} leaves $\sqrt n(r_N^2+r_Ns_n)$, along with terms of order $r_N+s_n+n^{-1/2}$, which tend to zero. For $r_N=N^{-\alpha_{\rm src}}$, $\sqrt n r_N^2\to0$ is equivalent to $N/n^{1/(4\alpha_{\rm src})}\to\infty$. Logarithmic factors must also be dominated when present. Under \eqref{eq:overlap}, for any measurable error $e$,
\[
 \norm e_{L_2(P_{T,X})}^2
 =\int e^2\frac{dP_{T,X}}{dP_{H,X}}\,dP_{H,X}
 \le C_{\rm ov}\norm e_{L_2(P_{H,X})}^2.
\]
Taking an infimum over representable constant and correction errors only improves the target bound.\hfill$\square$

\subsection{Why the score variance need not be the smallest possible variance}
The projection space spanned by $z$ is contained in the closed space of all functions of $X$. Orthogonal projection onto the larger space leaves a residual with no larger squared $\J$ seminorm. Hence $I_X\le I_z$. Known $f_0$ and a correctly restricted correction model can thus give treatment variance $I_z^{-1}$ below the unrestricted target model variance $I_X^{-1}$. This is consistent with both inference theorems. It precludes an unsupported claim that the projected score is efficient for the restricted joint historical/target model.

\subsection{Proof of Theorem~\ref{thm:prediction}}
Write $\delta_g=\widehat g-g_0-c_N$ and $\widehat\eta^{\rm a}=\widehat\theta A+\widehat g-c_N$. Theorem~\ref{thm:plugin}, including its numerical root qualification, gives
\[
|\widehat\theta-\theta_0|+\norm{\delta_g}_{L_2(P_{T,X})}=\Op(r_{n,N}),\qquad
\norm{\widehat\eta^{\rm a}-\eta_0}_{L_2(P_{T,A,X})}=\Op(r_{n,N}).
\]
The constants $c_N$ and coefficients $b_N$ are uniformly bounded: the $L_2$ projection normal equations express $b_N$ through the inverse of the fixed positive definite $\operatorname{Var}_T(z)$ and the covariance of bounded $z$ with the bounded historical error. The intercept then follows from the means. Positive definiteness here follows also from R3 and Lemma~\ref{lem:geometry}. Thus the aligned indices remain uniformly bounded.

Conditional on the independent historical fit, the target index varies over a fixed-dimensional compact coefficient set. The weighted threshold empirical process bound, uniform in that coefficient set as in Appendix~\ref{app:projection}, yields
\[
\sup_{t\le\tau}|s_{n,\widehat\eta^{\rm a}}(t)-s_{\widehat\eta^{\rm a}}(t)|=\Op(n^{-1/2}).
\]
The bounded exponential is Lipschitz, so
\[
\sup_{t\le\tau}|s_{\widehat\eta^{\rm a}}(t)-s_0(t)|
\le C\E|\widehat\eta^{\rm a}-\eta_0|=\Op(r_{n,N}).
\]
The expectation denotes evaluation at a fresh covariate and assignment, conditional on the fitted functions. Hence the empirical aligned risk mean differs uniformly from $s_0$ by $\Op(r_{n,N})$. R2 implies $\inf_ts_0(t)>0$, and with probability tending to one the empirical denominator has the same property.

Let $\nu(t)=\E \cnt(t)$. The aligned Breslow estimator is exactly
\[
e^{c_N}\widehat\Lambda_{T,0}(t)=\int_0^t\frac{d\nu_n(s)}{s_{n,\widehat\eta^{\rm a}}(s)}.
\]
Under the target model, $d\nu=s_0\,d\Lambda_{T,0}$. Therefore its difference from $\Lambda_{T,0}(t)$ is
\[
\int_0^t\left\{\frac1{s_{n,\widehat\eta^{\rm a}}(s)}-\frac1{s_0(s)}\right\}d\nu_n(s)
+\int_0^t\frac1{s_0(s)}\,d\{\nu_n-\nu\}(s).
\]
The first term has uniform order $\Op(r_{n,N})$, since $\nu_n(\tau)\le1$ and the denominators are bounded below. For the second term, $s_0$ is nonincreasing and positive, so $1/s_0$ has bounded total variation. Integration by parts bounds its supremum by $C\sup_t|\nu_n(t)-\nu(t)|=\Op(n^{-1/2})$, using the empirical subdistribution of observed failures. This proves \eqref{eq:baselinerate}.

Put $\widehat\Lambda^{\rm a}=e^{c_N}\widehat\Lambda_{T,0}$. Both baselines are nonnegative and bounded in probability on the horizon. On bounded risk intervals, $|e^u-e^v|\le C|u-v|$, and for nonnegative $u,v$, $|e^{-u}-e^{-v}|\le|u-v|$. The prediction product is invariant to alignment. Consequently, on events with probability tending to one, for both assignments,
\[
\sup_{t\le\tau}|\widehat S_a(t\mid x)-S_a(t\mid x)|
\le C\left\{\sup_{t\le\tau}|\widehat\Lambda^{\rm a}(t)-\Lambda_{T,0}(t)|
+|\widehat\theta-\theta_0|+|\delta_g(x)|\right\}.
\]
Taking the target $L_2$ norm proves \eqref{eq:survivalrate}. No pointwise covariate convergence or survival confidence band is inferred.\hfill$\square$

\subsection{Proof of Lemma~\ref{lem:geometry}}
For $P=P_{T,A,X}$, set $w(t,A,X)=\Pr(U\ge t\mid A,X)e^{\eta_0}$. Then
\[
s_0(t)\operatorname{Var}_{Q_t}(u)=\inf_b\E\{w(t,A,X)(u-b)^2\}.
\]
Uniform observation positivity gives $c_{\rm obs}e^{-B_\eta}\le w\le e^{B_\eta}$ for every $t\le\tau$. Bound the infimum below by $c_{\rm obs}e^{-B_\eta}\inf_b\E(u-b)^2$, and above by evaluating it at $b=\E u$. Integration proves \eqref{eq:coercivity}. Therefore $L_2(P)$ modulo constants is a complete Hilbert space under $\J$. The $X$-measurable subspace is closed: its centered $L_2$ version is the range of the continuous conditional expectation projection. Norm equivalence preserves closedness. The Hilbert projection theorem yields $h_*$ and its uniqueness modulo constants.

Choose $\E_X h_*=0$. Projection contraction and the lower norm bound give a fixed bound on $\norm{h_*}_2$ and hence on $\norm{q_*}_2$. To prove boundedness rather than assume it, write
\[
b(t,x)=\sum_{a=0}^1p_a w_a(t,x),\quad
r(x)=\int_0^\tau b(t,x)\,d\Lambda_{T,0}(t),\quad
d(x)=\int_0^\tau p_A w_1(t,x)\,d\Lambda_{T,0}(t).
\]
Here $r(x)\ge c_{\rm obs}e^{-B_\eta}\Lambda_{T,0}(\tau)>0$. Let $\ell(t)=Q_t(q_*)$. Bounded weights and the positive denominator give $\sup_t|\ell(t)|\le C\norm{q_*}_2$. The normal equation $\J(q_*,v(X))=0$ for all $v\in L_2(P_{T,X})$ implies
\begin{equation}\label{eq:projectionformula}
h_*(x)=\frac{d(x)-\int_0^\tau b(t,x)\ell(t)\,d\Lambda_{T,0}(t)}{r(x)}
\quad\text{for }P_{T,X}\text{-almost every }x.
\end{equation}
Indeed, conditioning the covariance integrand on $X$ gives the inner product of $v(X)$ with $d-rh_*-\int b\ell\,d\Lambda_{T,0}$. This function is in $L_2$, so its vanishing inner products imply equality almost everywhere. All terms in the numerator are bounded, and $r$ has a positive lower bound. The resulting representative is bounded. Finally, randomization gives
\[
\inf_b\E(A-h_*-b)^2=p_A(1-p_A)+\operatorname{Var}\{h_*(X)\}\ge p_A(1-p_A).
\]
Apply \eqref{eq:coercivity} to $q_*$ to obtain the information lower bound.\hfill$\square$

\subsection{Proof of Proposition~\ref{prop:sensitivity}}
Fubini's theorem applied to \eqref{eq:J} gives $\J(q_z,e)=\E(\omega_z e)$. Taking $e=1$ and each coordinate of $z$ gives the two orthogonality relations. Bounded $q_z$, weights, and total baseline hazard give bounded $\omega_z$.

Work at the fixed target law and write $\delta(e)$ for the full population coefficient displacement under offset $f_0+e$. For bounded $W$, Lemma~\ref{lem:pop} gives the initial gradient $-\J(W,e)+O(\norm e_2^2)$. Positive curvature and the same local concavity argument as in the direct fit proof place its unique root at $\norm{\delta(e)}\le C\norm e_2$. The population Hessian differs from $I_W$ by $O(\norm e_2+\norm{\delta(e)})$, uniformly on these paths: bounded index ratio differentiation and Cauchy--Schwarz bound each entry. Thus
\[
\delta(e)=-I_W^{-1}\J(W,e)+O(\norm e_2^2).
\]
The Schur complement gives \eqref{eq:localsensitivity}. Cauchy--Schwarz bounds its linear part by $r\norm{\omega_z}_2/I_z$. If $\omega_z\ne0$, both signed directions $\pm r\omega_z/\norm{\omega_z}_2$ satisfy the residual orthogonality constraints and, for small $r$, the fixed bounded error restriction. Substitution supplies the matching lower bound. If $\omega_z=0$, only the uniform quadratic bound remains. This proves \eqref{eq:sharpbias}. At the correctly specified observed null, the weights do not depend on treatment, so $q_z=A$ and the conditional sum in \eqref{eq:sensitivityweight} is zero.\hfill$\square$

\subsection{Proof of Proposition~\ref{prop:dictionarytradeoff}}
Because $h_*-h_B$ is a function of $X$, full projection orthogonality gives $\J(q_*,h_*-h_B)=0$. The decomposition $q_B=q_*+(h_*-h_B)$ therefore yields $I_B=I_X+d_B^2$. Projection onto the dictionary containing $z$ leaves no more squared residual than projection onto $z$, proving $I_B\le I_z$. Since $e_N$ is a function of $X$, $\J(q_*,e_N)=0$ and hence $\J(q_B,e_N)=\J(h_*-h_B,e_N)$. Cauchy--Schwarz in the quotient Hilbert space gives the first bound in \eqref{eq:dictionarybias}. The upper inequality in \eqref{eq:coercivity} and $\inf_c\norm{e_N-c}_2\le\rho_N$ give the second. Subtracting the inverse informations gives the stated variance difference. For nested dictionaries, projection onto the larger space reduces the residual norm, and thus $d_B^2=I_B-I_X$ cannot increase. This norm ordering imposes no ordering on its inner product with a particular error.\hfill$\square$

\subsection{Proof of Theorem~\ref{thm:finite}}\label{sec:finiteproof}
Projection onto $B$ gives $\J(q_B,B)=0$. By Lemma~\ref{lem:geometry} and randomization, $I_B>0$. On fold $k$, let $s_k=\widetilde\theta_k-\theta_0$ and $v_k=\widetilde g_k-g_0$, with constants removed. Lemmas~\ref{lem:emp} and \ref{lem:pop} imply
\begin{align*}
\mathcal U_k&=-I_Bs_k-\J(q_B,v_k)+\frac1{m_k}\sum_{i\in\mathcal I_k}\phi_B(O_i)\\
&\quad+\Op\{a_n^2+a_nb_{B,n}+(a_n+b_{B,n})m_k^{-1/2}+m_k^{-1}\},\\
D_k&=I_B+\Op(a_n+b_{B,n}+m_k^{-1/2}).
\end{align*}
The historical fit is common to the folds, and \eqref{eq:error} gives
\[
v_k=c_N+e_N+(b_N+\widetilde\gamma_k-\gamma_0)^\top z.
\]
Consequently $\J(q_B,v_k)=\J(q_B,e_N)$, since constants and $z$ are orthogonal to $q_B$. This functional is also $O_\Pp(a_n)$ by the $L_2$ bound on $v_k$. Expanding the denominator in the one-step update introduces only $O_\Pp\{a_n(a_n+b_{B,n})+(a_n+b_{B,n})m_k^{-1/2}+m_k^{-1}\}$, including its product with this functional. The $s_k$ term cancels. A finite fold union bound and $w_k=m_k/n$ yield \eqref{eq:finiteexp}. Martingale isometry gives $\E\phi_B^2=I_B$. The empirical influence variance argument in the unrestricted score proof gives $V_k\to_\Pp I_B$, hence the stated standard error limit. The ordinary iid central limit theorem and the two negligible remainder conditions finish the proof.\hfill$\square$

\section{Proof of the source Cox prediction bound}\label{app:source}
This argument is independent of the target inference calculations. In this appendix all subjects are historical, sample size is $N$, and we suppress the historical superscript, so $X_j,U_j,\Delta_j$ denote $X^H_j,U^H_j,\Delta^H_j$. Write $Y_j(t)$, $\cnt_j(t)$ and $\mathcal M_j(t)$ for their at risk, event, and true martingale processes. Define
\[
 S_f(t)=\sum_{j=1}^NY_j(t)e^{f(X_j)},\qquad
 p_{f,j}(t)=Y_j(t)e^{f(X_j)}/S_f(t).
\]
When the risk set is empty, all subsequent event integrals are zero and the corresponding expressions are set to zero. At observed events the risk set contains the failing patient, so its denominator is positive. All functions in this argument are fixed candidates before observing the sample. Selection of the fitted function is handled by a finite cover.

\subsection{A likelihood ratio concentration inequality}
For $e=f-f_0$, the log partial likelihood ratio up to time $t$ is
\begin{equation*}
 L_f(t)=\sum_j\int_0^t\log\{p_{f,j}(s)/p_{0,j}(s)\}\,d\cnt_j(s).
\end{equation*}
At a jump for subject $j$, both probabilities are positive. The full ratio at $\tau$ is $N\{\ell_H(f)-\ell_H(f_0)\}$. Introduce the risk set affinity and its compensator,
\[
 a_f(t)=\sum_j\sqrt{p_{0,j}(t)p_{f,j}(t)},\qquad
 A_f(t)=\int_0^tS_0(s)\{1-a_f(s)\}\,d\Lambda_{H,0}(s).
\]

\begin{lemma}[Cox ratio martingale]\label{lem:LR}
The process $Z_f(t)=\exp\{L_f(t)/2+A_f(t)\}$ is a nonnegative local martingale starting at one and hence has expectation at most one. For a fixed finite set of $M$ candidate functions and $u>0$,
\[
 \Pr\left\{\max_f[A_f(\tau)+L_f(\tau)/2]>\log M+u\right\}\le e^{-u}.
\]
\end{lemma}
\begin{proof}
The event intensity for subject $j$ is $Y_j e^{f_0(X_j)}d\Lambda_{H,0}$. The jump multiplier for $\exp(L_f/2)$ is $\sqrt{p_{f,j}/p_{0,j}}$. Its conditional drift divided by its value is
\[
 \sum_jY_je^{f_0(X_j)}\{\sqrt{p_{f,j}/p_{0,j}}-1\}\,d\Lambda_{H,0}
 =S_0(a_f-1)\,d\Lambda_{H,0}.
\]
Multiplication by $\exp(A_f)$ cancels this drift. Conditionally independent censoring makes these intensities valid for the observed counting processes. All jump multipliers are bounded for bounded $f,f_0$. Standard stopping, or direct integrability on the finite interval, gives the nonnegative supermartingale inequality. Markov's inequality applied to $Z_f(\tau)$ followed by a union bound proves the display. No independence of event risk sets is asserted.
\end{proof}

\subsection{Affinity controls a covariate risk variance}
Because $|e|\le2B_f$,
\[
 \log\{p_{f,j}/p_{0,j}\}=e(X_j)-\log\left\{\sum_kp_{0,k}e^{e(X_k)}\right\}
\]
is bounded by $4B_f$ in absolute value. On this bounded interval, $(e^{v/2}-1)^2\ge c_Bv^2$. Therefore
\begin{align}
 2(1-a_f)
 &=\sum_jp_{0,j}\{\sqrt{p_{f,j}/p_{0,j}}-1\}^2\notag\\
 &\ge c_B\operatorname{Var}_{p_0}\{e(X)\}.\label{eq:affinity}
\end{align}
Define the predictable empirical information
\[
 J_N(e)=\frac1N\int_0^\tau S_0(t)\operatorname{Var}_{p_0(t)}\{e(X)\}\,d\Lambda_{H,0}(t).
\]
It follows that $A_f(\tau)\ge c_BNJ_N(e)$. Moreover,
\begin{align*}
 J_N(e)
 &=\int_0^\tau\inf_c\frac1N\sum_jY_j(t)e^{f_0(X_j)}\{e(X_j)-c\}^2\,d\Lambda_{H,0}(t)\\
 &\ge e^{-B_f}\Lambda_{H,0}(\tau)\inf_c\frac1N\sum_jY_j(\tau)\{e(X_j)-c\}^2.
\end{align*}
The infimum inside each risk set can vary with $t$. Replacing it by the common lower bound on the last line is valid for every $t$. This step uses only monotonicity of at-risk indicators.

Let $\mathcal G_N$ be a supremum norm $N^{-1}$ cover of $\mathcal F_N$ with $M_N$ centers. Clip any external centers to $[-B_f,B_f]$. Since all covered functions are bounded by $B_f$, clipping preserves the covering radius. The centers remain fixed, measurable candidates, and the ratio martingale argument applies whether or not they belong to $\mathcal F_N$. For each center $f$ and each constant $c\in[-2B_f,2B_f]$, the random variable
\[
 Y(\tau)\{f(X)-f_0(X)-c\}^2
\]
is nonnegative and bounded by a fixed constant. Its variance is at most a fixed constant times its expectation. Bernstein's inequality, the inequality $\sqrt{ab}\le a/2+b/2$ with appropriately rescaled constants, and a union bound over all centers and an $N^{-1}$ grid of constants give, with probability at least $1-e^{-u}$,
\begin{align}
 &\inf_c\frac1N\sum_jY_j(\tau)\{f(X_j)-f_0(X_j)-c\}^2\notag\\
 &\quad\ge\frac12\inf_c\E\left[Y(\tau)\{f(X)-f_0(X)-c\}^2\right]
       -C\frac{\log M_N+\log N+u}{N}\notag\\
 &\quad\ge\frac{c_H}{2}\operatorname{Var}_{H,X}\{f(X)-f_0(X)\}
       -C\frac{\log M_N+\log N+u}{N},\label{eq:varlower}
\end{align}
simultaneously for all cover centers. The minimizing constant is a weighted mean of a bounded error and hence lies in the stated interval. The squared error is Lipschitz in $c$ on this interval, so the constant grid adds only $C/N$. The last inequality uses $\Pr(U\ge\tau\mid X)\ge c_H$. Combining \eqref{eq:affinity} to \eqref{eq:varlower} controls population centered risk error by $A_f/N$ plus an entropy term.

\subsection{The likelihood of an approximating candidate}
Take the fixed approximator $f_N^*$ in Theorem~\ref{thm:source}, and put $e_*=f_N^*-f_0$. Its event log multiplier
\[
 v_j(t)=e_*(X_j)-\log\{\textstyle\sum_kp_{0,k}(t)e^{e_*(X_k)}\}
\]
has absolute value at most $2\varepsilon_N^{\rm app}$. Its log ratio decomposes as
\[
 L_*(\tau)=\sum_j\int_0^\tau v_j(t)\,d\mathcal M_j(t)
       -\int_0^\tau S_0(t)\operatorname{KL}\{p_0(t),p_*(t)\}\,d\Lambda_{H,0}(t).
\]
Bounded log multipliers imply the KL term is at most $CN(\varepsilon_N^{\rm app})^2$. The martingale has jumps at most $2\varepsilon_N^{\rm app}$ in absolute value and predictable variance at most $CN(\varepsilon_N^{\rm app})^2$, since $S_0(t)\le Ne^B_f$ and the integrated baseline is finite. The exponential martingale inequality for bounded jumps gives, with probability at least $1-e^{-u}$,
\begin{equation}\label{eq:approxLR}
 L_*(\tau)\ge-C\{N(\varepsilon_N^{\rm app})^2+\varepsilon_N^{\rm app}\sqrt{Nu}+\varepsilon_N^{\rm app}u\}
 \ge-C\{N(\varepsilon_N^{\rm app})^2+u\}.
\end{equation}
The last inequality uses Young's inequality and the fixed bound on $\varepsilon_N^{\rm app}$. For completeness, the bounded jump inequality follows by applying the exponential compensator bound
\[
 e^{tv}-1-tv\le\frac{t^2v^2}{2(1-tb/3)},\qquad |v|\le b,\quad 0<t<3/b,
\]
to the counting process martingale, then optimizing $t$. It yields the usual deviation $\sqrt{2Vu}+bu/3$ when its predictable variance is at most $V$. Thus \eqref{eq:approxLR} does not require independent partial likelihood terms.

\subsection{Proof of Theorem~\ref{thm:source}}
Take the three high probability events from Lemma~\ref{lem:LR}, \eqref{eq:varlower}, and \eqref{eq:approxLR}. Their joint probability is at least $1-3e^{-u}$. Choose a cover center $f_c$ within $N^{-1}$ of $\widehat f$. Each partial likelihood difference is Lipschitz with constant two in the supremum norm: the direct risk difference and the log risk sum difference are each bounded by that norm. There are at most $N$ events. Hence
\[
 L_{f_c}(\tau)\ge L_{\widehat f}(\tau)-2
 \ge L_*(\tau)-N\varepsilon_N^{\rm opt}-2.
\]
The martingale cover inequality gives $A_{f_c}\le\log M_N+u-L_{f_c}/2$. Combining this with \eqref{eq:varlower} and \eqref{eq:approxLR} yields
\[
 \operatorname{Var}_{H,X}\{f_c-f_0\}
 \le C\left\{(\varepsilon_N^{\rm app})^2+\varepsilon_N^{\rm opt}+\frac{\mathcal E_N+\log N+u}{N}\right\}.
\]
A supremum norm difference of $N^{-1}$ adds at most a constant times $N^{-2}$ to the squared centered $L_2$ comparison after the elementary inequality $(a+b)^2\le2a^2+2b^2$. Since $\inf_c\norm{e-c}_{L_2}^2=\operatorname{Var}(e)$, this proves \eqref{eq:sourcebound}. The inequality holds on this event for a random optimization gap as well: only its realized upper bound in the fitted likelihood comparison was used.\hfill$\square$

\subsection{Proof of Corollary~\ref{cor:relu}}
For parameters in $[-K,K]$, a network's layer-$l$ activation is bounded in supremum norm by $D^l$, where $D=(w+1)(K+1)$. If two parameter vectors differ by at most $\delta$ in every coordinate, induction using the one Lipschitz ReLU map gives a layer-$l$ output difference at most $lD^l\delta$. The final clipping map is also one Lipschitz. An additional linear skip has parameter Lipschitz constant at most $d+1$ on the covariate box. Thus use $C=LD^L$ without a skip and $C=LD^L+d+1$ with one, counting all free parameters. A parameter grid of radius $\varepsilon/C$ gives a function cover of radius $\varepsilon$. A box with $P$ coordinates has such a grid of cardinality at most
\[
 (2+2KC/\varepsilon)^P.
\]
Set $\varepsilon=N^{-1}$ and apply Theorem~\ref{thm:source}. The overlap inequality proved earlier transports the centered source norm to the target norm, and minimizing over the larger constant plus correction space can only improve the bound. Taking square roots gives \eqref{eq:relurate}. Balancing $P^{-2\beta_{\rm app}}$ and $P/N$ gives $P\asymp N^{1/(2\beta_{\rm app}+1)}$ up to logarithmic factors, with root rate $N^{-\beta_{\rm app}/(2\beta_{\rm app}+1)}$. This last statement presupposes the approximation construction and sufficiently small optimization gap. It does not prove them.\hfill$\square$

\section{Projection estimation and growing dictionaries}\label{app:projection}
Let $B(X)$ be a deterministic dictionary centered under a fixed reference, $\Sigma_B=\J(B,B)$, $c_B=\J(B,A)$, and $\mu_B=\Sigma_B^{-1}c_B$. Centering selects a representative of the constant equivalence class. Define $\zeta_B=\inf_c\norm{\mu_B^\top B-h_*-c}_{L_2(P_{T,X})}$.

\begin{proposition}[Projection coefficient bound]\label{prop:basis}
Suppose the population matrices satisfy
\[
\lambda_{\min}(\Sigma_B)\ge\kappa_B>0,\qquad \lambda_{\max}\{\E(BB^\top)\}\le C_B.
\]
With $\widehat\mu=(\widehat\Sigma_B+\lambda I_M)^{-1}\widehat c_B$ and $\lambda\ge0$, on the event $\norm{\widehat\Sigma_B-\Sigma_B}_{\rm op}+\lambda\le\kappa_B/2$,
\begin{align*}
\inf_c\norm{\widehat\mu^\top B-h_*-c}
&\le\zeta_B+\frac{2\sqrt{C_B}}{\kappa_B}
\left[\norm{\widehat c_B-c_B}_2
+\{\norm{\widehat\Sigma_B-\Sigma_B}_{\rm op}+\lambda\}\norm{\mu_B}_2\right].
\end{align*}
For a fixed bounded dictionary and the historical offset finite-dimensional training fit, its moment errors are $\Op(a_n+n^{-1/2})$. If $h_*$ belongs to its span, $b_n=\Op(a_n+n^{-1/2}+\lambda_n)$.
\end{proposition}

\subsection{Proof of Proposition~\ref{prop:basis}}
Subtract the population normal equation from its sample ridge version. The exact identity is
\[
 \widehat\mu-\mu_B=(\widehat\Sigma_B+\lambda I)^{-1}
 \{\widehat c_B-c_B-(\widehat\Sigma_B-\Sigma_B+\lambda I)\mu_B\}.
\]
The stated spectral event bounds the inverse norm by $2/\kappa_B$. The $L_2$ norm of $(\widehat\mu-\mu_B)^\top B$ is at most $\sqrt{C_B}\norm{\widehat\mu-\mu_B}_2$. The triangle inequality with the projection approximation error proves the deterministic bound.

For a fixed bounded basis, condition on the historical function. The target fitted index varies over a fixed-dimensional compact coefficient set, with bounded derivatives in those coefficients. Uniform empirical risk ratio and event moment bounds have order $n^{-1/2}$. Their covering numbers are those of a finite-dimensional smooth coefficient class and threshold time sets. Population covariance moments are Lipschitz in the $L_2$ risk index error by the same ratio bounds as in Lemma~\ref{lem:pop}, contributing $\Op(a_n)$. If $h_*$ belongs to the fixed basis, $\zeta_B=0$, giving the displayed rate. This assertion concerns a fixed prespecified basis. A basis constructed from fitted functions requires an additional bound for its own perturbation, and a growing basis requires dimension dependent stochastic control.\hfill$\square$

\begin{proposition}[A growing dictionary bound]\label{prop:growing}
Let $B_M$ be a deterministic centered dictionary of $M$ functions, each bounded by a common constant. Suppose $\lambda_{\min}\{\J(B_M,B_M)\}\ge\kappa>0$ and $\lambda_{\max}\{\E(B_MB_M^\top)\}\le C$ uniformly in $M$. Let $\zeta_M$ denote its approximation error. For the historical offset finite-dimensional training fit, if $M(a_n+n^{-1/2})+\lambda_n=o_\Pp(1)$,
\[
\inf_c\norm{\widehat\mu^\top B_M-h_*-c}
=\Op\{\zeta_M+M(a_n+n^{-1/2})+\lambda_n\}.
\]
The same order applies to a clipped fit if the unclipped and true projections are centered under a common reference $R_P\ll P_{T,X}$ with bounded density ratio, and the clipping interval contains that true representative. Bounds for data-dependent dictionaries require control of their additional perturbation.
\end{proposition}
\begin{proof}
Condition on the historical function. The fitted training index varies over a fixed-dimensional compact coefficient set. The bounded coordinate risk ratio argument of Appendix~\ref{app:target}, uniformly over that coefficient set, gives a second moment bound $C/n$ for each empirical covariance entry at its own population index. Its constant does not depend on $M$: each multiplier $B_jB_l$, $B_jA$, and each derivative with respect to the finite target coefficients is bounded uniformly. The threshold time class and the fixed-dimensional parameter class have uniform bounded empirical process integrals. Denominators are uniformly positive. The ratio remainder has the same or smaller second moment order. The exceptional small denominator event is controlled by the terminal at risk average.

For completeness, the uniform second moment bound can be obtained without taking a union bound over the dictionary. For each pair of coordinates separately, the threshold multipliers are
\[
Y(t)e^{\widehat f(X)+\beta^\top W}v(A,X),\qquad
v\in\{1,B_j,B_l,B_jB_l,A,B_jA\}.
\]
Conditional on the historical fit, their envelopes and parameter Lipschitz constants are uniform in $j,l,M$. Under any probability measure, cover the finite-dimensional $\beta$ set by an $\varepsilon$ grid and the nested threshold sets by an $L_2$ grid of order $\varepsilon^{-2}$. Their $L_2$ covering numbers are bounded by $(C/\varepsilon)^{p+3}$. Symmetrization and dyadic chaining give a mean supremum bound $C n^{-1/2}$, since $\sum_{r\ge0}2^{-r}\sqrt{r+1}<\infty$. Bounded difference concentration around this mean gives the corresponding second moment bound $C/n$.

On the terminal at risk event, covariance ratios are uniformly Lipschitz in these moments. The probability of its complement is exponentially small by the Bernoulli concentration bound for $n^{-1}\sum_iY_i(\tau)$. Covariance entries remain bounded there whenever defined, and are set to zero for an empty risk set. To integrate empirical covariance entries against events, first replace the empirical covariance by its population value, using the supremum bound and event mass at most one. The remaining empirical average is of the form $n^{-1}\sum_i\Delta_i C_\beta(U_i)$. Its family is bounded and Lipschitz in the fixed-dimensional $\beta$, so the same chaining argument gives second moment $C/n$. This proves the entrywise bounds uniformly over the fitted training coefficient.

Summing the entrywise second moment bounds gives
\[
\norm{\widehat\Sigma_M-\Sigma_M(\widetilde\eta)}_{\rm F}=\Op(M/\sqrt n),\qquad
\norm{\widehat c_M-c_M(\widetilde\eta)}_2=\Op(\sqrt{M/n}).
\]
The population risk set moment perturbation is at most $C a_n$ for each entry by bounded indices and the $L_2$ ratio Lipschitz bound. Thus
\[
\norm{\widehat\Sigma_M-\Sigma_M}_{\rm op}=\Op\{M(a_n+n^{-1/2})\},\quad
\norm{\widehat c_M-c_M}_2=\Op\{\sqrt M(a_n+n^{-1/2})\}.
\]
The bound is conservative. It uses Frobenius rather than a sharper matrix concentration inequality. Covariance moments are invariant to coordinatewise constant shifts. If the centering reference is changed to a training sample reference, the fitted representative must be aligned with the true one before clipping. Proposition~\ref{prop:fourier} instead uses a fixed, known uniform reference and proves the needed equivalence between centered and quotient errors.

Population projection implies $\mu_M^\top\Sigma_M\mu_M=\J(\mu_M^\top B_M,\mu_M^\top B_M)\le\J(A,A)$. Uniform coercivity therefore gives $\norm{\mu_M}_2\le C$. The spectral event in Proposition~\ref{prop:basis} has probability tending to one. Apply that proposition, use the upper $L_2$ Gram bound, and add $\zeta_M$. For the clipping claim, write $v$ for the difference of the two $R_P$-centered representatives and let $b$ minimize its target $L_2$ distance to constants. Since $\int v\,dR_P=0$, Cauchy--Schwarz and the bounded reference density give $|b|\le C\norm{v-b}_2$ and hence $\norm v_2\le C\inf_b\norm{v-b}_2$. Clipping is one Lipschitz relative to a true representative inside its interval. Applying this inequality before clipping proves the last statement. A common centering convention without a continuous reference functional would not suffice.
\end{proof}

To verify the sufficient example in \eqref{eq:seriesrate}, set $M\asymp n^u$, $a_n=\Op(n^{-\kappa_{\rm rate}})$, $\zeta_M\lesssim M^{-s/d}$, and $\lambda_n=O(n^{-1/2})$. The four terms in $\sqrt n a_nb_n$ have orders
\[
n^{1/2-\kappa_{\rm rate}-us/d},\quad n^{1/2-2\kappa_{\rm rate}+u},\quad n^{u-\kappa_{\rm rate}},\quad n^{-\kappa_{\rm rate}}.
\]
They vanish for $(1/2-\kappa_{\rm rate})d/s<u<2\kappa_{\rm rate}-1/2$, since $\kappa_{\rm rate}<1/2$ also gives $u<\kappa_{\rm rate}$. The interval is nonempty when $\kappa_{\rm rate}>1/3$ and $s\ge d$. The squared error term vanishes since $\kappa_{\rm rate}>1/4$. These are sufficient source/projection conditions, not minimax statements. The interval is nonempty more generally when $\kappa_{\rm rate}>(s+d)/(4s+2d)$. These exponents describe a sufficient region for this conservative moment bound, rather than an optimal sample size threshold. The next construction verifies the trigonometric dictionary assumptions.

\subsection{A deterministic dictionary with verified geometry}
The following construction verifies both the Gram condition and the approximation rate rather than assuming them independently. Periodicity is a sufficient boundary condition for this particular Fourier construction. It is not a condition for the general inference theorems.

\begin{proposition}[Fourier projection construction]\label{prop:fourier}
Suppose $X\in[0,1]^d$ has target density $p_X$ with $0<c_X\le p_X\le C_X$. Assume R1 to R2. Use the real sine and cosine Fourier functions with nonzero frequencies $k\in\mathbb Z^d$, $\norm k_\infty\le J$, choosing one representative from each pair $k,-k$, and the usual $\sqrt2$ normalization. Denote this $M\asymp J^d$ dictionary by $B_M$. Its reference mean is zero under the uniform measure. Its coordinates are bounded uniformly, and
\[
\lambda_{\min}\{\J(B_M,B_M)\}\ge\kappa>0,
\qquad\lambda_{\max}\{\E(B_MB_M^\top)\}\le C
\]
uniformly in $M$. If $w_a(t,\cdot)$ are periodic $C^s$ functions for an integer $s\ge1$, with uniformly bounded derivatives through order $s$, then the uniform reference centered representative of $h_*$ is in $H^s([0,1]^d)$ and
\[
\zeta_M=\inf_b\norm{\mu_M^\top B_M-h_*-b}_{L_2(P_{T,X})}
\le C M^{-s/d}.
\]
Clipping after uniform reference centering to an interval containing this representative preserves the rate in Proposition~\ref{prop:growing}.
\end{proposition}
\begin{proof}
The dictionary is orthonormal under the uniform measure and has zero uniform mean. For $u=\mu^\top B_M$, $\int u^2=\norm\mu_2^2$ and
\[
\inf_b\E_X(u-b)^2\ge c_X\inf_b\int(u-b)^2
 =c_X\norm\mu_2^2,\qquad
\E_X u^2\le C_X\norm\mu_2^2.
\]
Lemma~\ref{lem:geometry} gives the lower $\J$ Gram bound. The upper $L_2$ Gram bound follows from the second inequality. Constants in the dictionary would invalidate coercivity and are excluded.

In \eqref{eq:projectionformula}, the function $\ell(t)$ is independent of $x$ and bounded. Differentiation under the finite baseline measure is justified by the uniform derivative bounds. Thus $d,r$, and the integral in the numerator have bounded derivatives through order $s$, and $r$ has a positive lower bound. The quotient is periodic $C^s$, hence lies in the periodic Sobolev space $H^s$. Its Fourier truncation $T_Jh_*$ satisfies, by Parseval's identity,
\[
\norm{h_*-T_Jh_*}_{L_2(dx)}^2
\le C J^{-2s}\sum_{k\in\mathbb Z^d}(1+\norm k_2^2)^s|\widehat h_*(k)|^2
\le C J^{-2s}.
\]
Select the uniform centered representative, so its zero frequency coefficient is zero. The population $\J$ projection $h_M$ has no larger $\J$ error than $T_Jh_*$. The upper and lower norm comparisons and the density upper bound yield
\[
\inf_b\norm{h_M-h_*-b}_{L_2(P_{T,X})}
\le C\norm{h_M-h_*}_{\J}
\le C\norm{T_Jh_*-h_*}_{L_2(P_{T,X})}\le C J^{-s}.
\]
Since $M\asymp J^d$, this is the claimed approximation rate. Uniform reference centering is compatible with the quotient norm: for a function $v$ of zero uniform mean, if $b$ minimizes $\norm{v-b}_{L_2(P_{T,X})}$, then
$|b|=|\int(v-b)\,dx|\le c_X^{-1/2}\norm{v-b}_{L_2(P_{T,X})}$.
It follows that the centered, unquotiented $L_2$ error is bounded by a constant times the quotient error. Clipping is one Lipschitz relative to the centered true function inside the clipping interval. The coefficient perturbation bound and this comparison therefore preserve the stated rate.\end{proof}

This construction does not require appending $z$ to the Fourier dictionary. Appending covariates already approximated by the growing dictionary can destroy its uniform Gram lower bound. The unrestricted projection and the finite dictionary containing $z$ are distinct constructions with distinct guarantees.

\subsection{Proof of Corollary~\ref{cor:feasible}}
Let $t_j=j/J_N$, $j=0,\ldots,J_N$, and let $H_j$ be the piecewise linear interpolation hat functions. They are nonnegative and sum to one. Every $f_c=\sum_j c_jH_j$ with $|c_j|\le B_f$ is bounded by $B_f$. Interpolating $f_0$ gives supremum error at most $\norm{f_0''}_\infty/(8J_N^2)$. Moreover,
\[
\norm{f_c-f_{c'}}_\infty\le\norm{c-c'}_\infty,
\qquad \mathcal E_N\le(J_N+1)\log(2+2B_fN).
\]
Each interpolant is ReLU representable: with slopes $s_j=J_N(c_{j+1}-c_j)$,
\[
f_c(x)=c_0+s_0x+\sum_{j=1}^{J_N-1}(s_j-s_{j-1})(x-t_j)_+.
\]
Fixed knots and tied coefficients define the stated subclass. Its entropy is controlled directly by its knot values, not by treating all network parameters as free. Theorem~\ref{thm:source}, the gap condition, and overlap give
\[
\rho_N=\Op\{J_N^{-2}+\sqrt{J_N\log N/N}+\sqrt{\varepsilon_N^{\rm opt}}\}
 =\Op\{(\log N/N)^{2/5}\}.
\]
The spline partial likelihood is a linear term minus log sums of exponentials of linear functions of $c$, hence is concave. Its tangent at $\widehat c$ dominates its value everywhere in the box. Maximizing that tangent coordinate by coordinate proves \eqref{eq:gapcertificate}. A constrained global optimum exists because the box is compact and the criterion is continuous. Constant shift nonuniqueness of its maximizer does not affect its likelihood gap or centered prediction bound.

The hat vector $H(x)$ satisfies $\norm{H(x)}_2\le1$, since its coordinates are nonnegative and sum to one. The negative likelihood Hessian is an event-weighted average of risk-set covariance matrices of $H$. Its operator norm is at most one, as $N^{-1}\sum_j\Delta_j\le1$. Thus the gradient has Lipschitz constant at most one. If $c^+$ is the projected gradient step of size one and $d=c^+-c$, the projection variational inequality gives $\nabla\ell_H(c)^\top d\ge\norm d_2^2$. Smoothness then gives
\[
\ell_H(c^+)-\ell_H(c)\ge\tfrac12\norm d_2^2.
\]
The bounded above criterion implies $d\to0$. Compactness and continuity imply that every subsequential limit is a fixed point of the projected gradient map, hence satisfies the box first order optimality condition and is a global maximum. The certificate in \eqref{eq:gapcertificate} is continuous and is zero at every such maximum. It therefore tends to zero along the whole sequence. Every positive stopping tolerance is reached in finite iterations in exact arithmetic. This proves attainability of the specified gap, without imposing a rate on a generic deep network optimizer.

For $N\asymp n$, the training fold direct fit rate gives $a_n=\Op(n^{-2/5}\log^{2/5}n)$. Proposition~\ref{prop:fourier} with $s=2,d=1$ gives $\zeta_M\le CM^{-2}$. Apply Proposition~\ref{prop:growing} with $M\asymp n^{1/5}$ and $\lambda_n=O(n^{-1/2})$ to obtain
\[
b_n=\Op\{n^{-2/5}+n^{-1/5}\log^{2/5}n+n^{-3/10}+n^{-1/2}\}
 =\Op(n^{-1/5}\log^{2/5}n).
\]
The spectral perturbation tends to zero. The two inference remainders satisfy
\[
\sqrt n a_n^2=\Op(n^{-3/10}\log^{4/5}n),\qquad
\sqrt n a_nb_n=\Op(n^{-1/10}\log^{4/5}n),
\]
both tending to zero. The boundedness and centering requirements for clipping are supplied by Lemma~\ref{lem:geometry} and Proposition~\ref{prop:fourier}. These assumptions are compatible, for example with smooth bounded periodic prognostic and correction functions and smooth, bounded conditional censoring hazards over the fixed horizon. No numerical experiment is used in this construction.\hfill$\square$

\subsection{A source solver with a likelihood-gap certificate}\label{sec:sourcesolver}
We implement the spline construction of Corollary~\ref{cor:feasible} without introducing an optimization assumption for a generic network. Let $c^{(r)}$ be the knot vector at iteration $r=0,1,\ldots$. Define the nonnegative certificate
\[
\operatorname{gap}_{\rm cert}(c)
=\sum_{j=0}^{J_N}\{B_f|\partial_j\ell_H(c)|-c_j\partial_j\ell_H(c)\}.
\]
It is the right side of \eqref{eq:gapcertificate}, so it bounds the gap to the global maximum over the coefficient box. The clipping operator in Algorithm~\ref{alg:source} acts on every coordinate of this box. The derivative uses historical risk sets throughout.

\begin{algorithm}[htbp]
\caption{Certified bounded spline historical fit}\label{alg:source}
\begin{algorithmic}[1]
\Require Historical records, bound $B_f$, spline resolution $J_N$, and initial knot vector $c^{(0)}\in[-B_f,B_f]^{J_N+1}$, with $N>1$.
\Ensure $\widehat f=f_{\widehat c}$ and a certificate at most $(\log N/N)^{4/5}$.
\State Set $r=0$ and evaluate $\ell_H(c^{(0)})$, $\nabla\ell_H(c^{(0)})$, and $\operatorname{gap}_{\rm cert}(c^{(0)})$.
\While{$\operatorname{gap}_{\rm cert}(c^{(r)})>(\log N/N)^{4/5}$}
 \State Set $c^{(r+1)}=\operatorname{clip}_{[-B_f,B_f]^{J_N+1}}\{c^{(r)}+\nabla\ell_H(c^{(r)})\}$.
 \State Set $r\gets r+1$ and recompute the historical gradient and certificate.
\EndWhile
\State Return $\widehat c=c^{(r)}$, $\widehat f=f_{\widehat c}$, and $\operatorname{gap}_{\rm cert}(\widehat c)$.
\end{algorithmic}
\end{algorithm}

The preceding proof establishes termination at this positive tolerance in exact arithmetic. In floating point computation, we report the attained certificate and whether the requested tolerance was reached, including any stagnation or iteration limit. The certificate, rather than the number of iterations, is what verifies the optimization term in the source prediction bound. This algorithm establishes one feasible learner for our theory. It does not replace the flexible source class allowed in Algorithm~\ref{alg:direct}.

\section{Proof of randomized null validity}
The argument uses the observed-data law under \eqref{eq:null} and does not assume a true Cox model. It is conditional on a historical function fixed independently of all current patients. The same constants apply over historical functions within the stated uniform bounds.

Write $\eta_{\theta,\gamma}=\theta A+\widehat f(X)+\gamma^\top z(X)$, $s_\eta(t)=\E\{Y(t)e^\eta\}$, and $\overline v_\eta(t)=\E\{Y(t)e^\eta v\}/s_\eta(t)$. The population score is
\[
\mathcal U_P(\eta,v)=\E\left\{\int v\,d\cnt\right\}-\int\overline v_\eta(t)\,d\nu(t),\qquad\nu(t)=\E\{\cnt(t)\}.
\]
The omitted $\log n$ term in the normalized partial likelihood is parameter independent. Its population negative Hessian is
\[
I_P(\eta)=\int\left\{\overline{WW^\top}_\eta(t)-\overline W_\eta(t)\overline W_\eta(t)^\top\right\}\,d\nu(t),\quad W=(A,z^\top)^\top.
\]

\subsection{Population treatment coefficient and Hessian}
Under \eqref{eq:null}, $A$ is independent of the complete observed record and hence of every at risk function of $(X,U,\Delta)$. At $\theta=0$,
\[
\overline A_g(t)=p_A,\qquad
\overline{Az}_g(t)=p_A\overline z_g(t),\qquad
\E\left\{\int A\,d\cnt\right\}=p_A\nu(\tau).
\]
Thus $\mathcal U_P(g,A)=0$ for every $\gamma$. At the population calibration optimum $\gamma^\dagger$, the correction score is also zero. Concavity and the nonsingular Hessian identify the full working optimum as $(0,\gamma^\dagger)$, without requiring $g^\dagger$ to describe the event hazard. The negative Hessian has block form
\[
I_P(g^\dagger)=
\begin{pmatrix}
D&0\\0&H_{zz}
\end{pmatrix},\qquad D=p_A(1-p_A)\nu(\tau)>0.
\]
The zero cross block explains why estimating the finite correction has no first-order effect on treatment at this null.

\subsection{Influence expansion without proportional hazards}
For bounded fixed $\eta,v$, the empirical risk ratio expansion used in Lemma~\ref{lem:emp} is valid with arbitrary event measure $\nu$. Its influence function is
\begin{align*}
\varphi_{\eta,v}(O)
&=\int\{v-\overline v_\eta(t)\}\,d\cnt(t)-\mathcal U_P(\eta,v)\\
&\quad-\int\frac{Y(t)e^\eta}{s_\eta(t)}\{v-\overline v_\eta(t)\}\,d\nu(t).
\end{align*}
Expanding the empirical risk ratio leaves a product of empirical deviations and a bounded degenerate V statistic. Its remainder is $\Op(n^{-1})$. These claims do not use a martingale representation for the true hazard. The PH model is needed only for the final martingale simplification in Appendix~\ref{app:target}, which is not used here.

With $\eta=g^\dagger$ and $v=A$, the treatment influence function becomes
\[
\varphi_{g^\dagger,A}(O)=(A-p_A)\left[\Delta-\int\frac{Y(t)e^{g^\dagger(X)}}{s_{g^\dagger}(t)}\,d\nu(t)\right]
=(A-p_A)R_{g^\dagger}(O).
\]
It has mean zero and variance $p_A(1-p_A)\E(R_{g^\dagger}^2)$. Under the present observed-data independence, its covariance with the correction influence vector is also zero: that vector is a function of $(X,U,\Delta)$ alone, and $\E(A-p_A)=0$. The block diagonal Hessian already isolates the treatment coordinate in the inverse Hessian expansion.

For the finite-dimensional target coefficient class, bounded derivatives and compact parameters make the empirical score and Hessian bounds uniform in a neighborhood of $(0,\gamma^\dagger)$. Positive curvature and an interior optimum give a conditional coefficient error $\Op(n^{-1/2})$. Taylor expansion of the empirical score root then gives
\[
\widehat\beta-\beta^\dagger
=I_P(g^\dagger)^{-1}\frac1n\sum_i\varphi_{g^\dagger,W}(O_i)+o_\Pp(n^{-1/2}).
\]
Taking its treatment coordinate proves \eqref{eq:nullexp}. A numerical score error $o_\Pp(n^{-1/2})$ leaves the result unchanged.

\subsection{Studentization and a random historical fit}
The empirical ratios in \eqref{eq:empIF}, evaluated at the fitted finite-dimensional coefficient, converge uniformly to their population versions. Their squared influence averages converge to the population influence covariance because the functions are bounded and the coefficient displacement is $\Op(n^{-1/2})$. The empirical Hessian converges to $I_P(g^\dagger)$. Consequently the treatment entry of \eqref{eq:sandwich} estimates
\[
\frac{p_A(1-p_A)\E(R_{g^\dagger}^2)}{nD^2}.
\]
Conditional on the historical fit, the leading treatment influences are iid and bounded. The lower variance bound yields a uniform Lindeberg condition and a conditional central limit theorem after scaling by that possibly historical fit dependent variance. Slutsky's theorem gives the standard normal studentized limit. Uniform conditional convergence and dominated convergence then give the same unconditional limit even when the historical fit does not converge to a fixed function. No historical rate assertion is used.

This result does not use $\E(R_g^2)=\nu(\tau)$, an identity that generally fails for a misspecified prognostic index. Information only studentization would replace the actual variance with that expression. Nor does the proof identify an alternative treatment coefficient: when treatment affects survival, the factorization $\overline A_g(t)=p_A$ fails, and historical/correction misspecification may alter the working coefficient.\hfill$\square$

\section{Numerical designs and method definitions}\label{sec:sim}
\paragraph{Inventory of numerical experiments.}
The following table lists every simulation experiment reported in this supplement, in the order in which it appears. A prespecified experiment had its design, methods, and replication count fixed before any of its data were generated. A follow-up was designed after results of earlier experiments were examined, but its design was also fixed before its own data were generated. An exploratory experiment was chosen after inspecting earlier results and is reported descriptively. Pilot runs used separate seeds and are not reported: 60 replications before the nine-setting experiment, 20 per cell before the C-cell and certified-source experiments, and one timing replication before each experiment in Appendix~\ref{app:newsims}. Complete tables for the exploratory grid, its follow-ups and the tuning study, and secondary application records, are in a numerical archive, available from the authors, and are cited as Archive Tables and Figures.
{\footnotesize
\begin{longtable}{>{\raggedright\arraybackslash}p{2.9cm}>{\raggedright\arraybackslash}p{1.6cm}>{\raggedright\arraybackslash}p{3.9cm}>{\raggedright\arraybackslash}p{3.9cm}>{\raggedright\arraybackslash}p{1.9cm}}
\toprule
Experiment & Location & Purpose & Law and sizes & Status\\
\midrule\endhead
\bottomrule\endfoot
Nine settings (six alternatives, three nulls) & \ref{sec:sim}, \ref{app:implementation}, \ref{app:extension} & Transport scenarios, misspecification, and a non-PH null & Base law and variants; $n=300$, $N=3000$ or 300; 500 replications per setting & Prespecified\\
C01--C06 & \ref{sec:sim}, \ref{app:extension}; main Table 1 & Prediction and inference by $(n,N)$ & Base law; four alternatives with 1000 and two nulls with 2000 replications & Prespecified\\
CT-spline on C01--C06 & \ref{app:historicalspline} & Historical spline learner with the same target analysis & Records of the C cells & Follow-up\\
Certified source & \ref{sec:sim}, \ref{app:extension} & Attainable regime of Corollary~\ref{cor:feasible} & $d=1$, $N=n\in\{300,1000,3000\}$; 500 alternative and 500 null replications per size & Prespecified\\
Fixed-score stress & \ref{app:implementation} & Null studentization with incorrect fixed scores & Null linear-drift law; $n\in\{300,1000\}$; 1000 replications each & Prespecified\\
Adjusted log-rank & \ref{app:logrank} & External-score-adjusted log-rank test & Records of the nine-setting experiment & Prespecified\\
Fixed offset & \ref{app:mechanismexperiment}; main Figure 1 & Directional bias mechanism of Proposition~\ref{prop:sensitivity} & $X\in\{-1,0,1\}$; population paths and $n\in\{300,1000\}$ with 1000 replications per cell & Prespecified\\
R01--R04 & \ref{app:randomizationstress} & Randomization test under misspecified fixed scores & Complex PH and non-PH laws, $n\in\{300,600\}$; 2000 replications each & Prespecified\\
E01--E08 & \ref{sec:focused} & Larger historical samples at amplitudes 1 and 1.5 & $N\in\{6000,12000\}$, $n\in\{300,600\}$; 100 replications per cell & Exploratory\\
H01--H04 & \ref{app:confirmation} & Confirmation of the elevated E08 null rejection & Amplitude 1.5, $N=12000$, $n\in\{300,600\}$; 2000 null and 1000 alternative replications & Follow-up\\
B01--B02 & \ref{app:boundaryconfirmation} & Confirmation of E02 & Amplitude 1.5, $n=300$, $N=6000$; 2000 null and 1000 alternative replications & Follow-up\\
Tuning grid and V01--V02 & \ref{app:tuningvalidation} & Fold count and ridge sensitivity, then held-out validation & As B01--B02; 500 null and 500 alternative development datasets reused, then 2000 and 1000 new datasets & Follow-up\\
N01--N02, S01--S02, spline bootstrap & \ref{app:remainingflags} & Tuned target-only neural score and patient bootstraps for CT and CSPLINE & C01/C02 and strong-prognosis laws; 1000 alternative and 2000 null replications each & Follow-up\\
F01--F02 & \ref{app:fullbootstrap}; main Table 2 & Full-procedure bootstrap for the projected score & Amplitude 1.5, $n=300$, $N=6000$; 2000 null and 1000 alternative replications & Follow-up\\
Directional bias, Q500--Q2000 & \ref{sec:qbias}; main Figure 2 & When directional bias matters, and which remedies work & Prognosis outside the target spline span and drift outside the correction span; $N=10000$, $n\in\{500,1000,2000\}$; 500 replications per size & Follow-up\\
Penalized comparators, P1--P4 & \ref{sec:qpred} & Penalized target-only and penalized-transfer prediction & C-cell alternative laws with new seeds; 500 replications per cell & Follow-up\\
Dictionary size, D1--D4 & \ref{sec:qdict} & Enlarged correction versus projected score; PT$-$CT contrast & Directional bias law with and without outside drift; $N=10000$, $n\in\{300,600\}$; 500 replications per cell & Follow-up\\
Moderate drift, M1--M3 & \ref{sec:qmod}; main Figure 2 & Treatment inference at hazard ratio 0.7 and phase III sizes & Directional bias law with $\theta_0=\log(0.7)$; $N=10000$, $n\in\{1000,2000,3000\}$; 500 replications per size & Follow-up\\
Coefficient transfer & \ref{sec:qpred} & Adapted TransCox-type comparator for prediction & Records of P1--P4; 500 replications per cell & Follow-up\\
Generic dictionaries, G & \ref{sec:qgen}; main Figure 2 & Projected score with spline dictionaries chosen without the generating law & Laws of Q500--Q2000 and M1--M3; 500 replications per size & Follow-up\\
\end{longtable}}\addtocounter{table}{-1}

\subsection{Design}
We use MSE for mean squared error, RMSE for root mean squared error, SD for standard deviation, and SE for standard error. In tables, NA denotes an undefined quantity or one that does not apply.
The primary neural simulation comprises nine settings with 500 independent replications each and $n=300$ current trial patients. Historical size is $N=3000$, except in the small source setting with $N=300$. Unless specified otherwise, four covariates are independent $\operatorname{Uniform}(-1,1)$ and
\[
f_0(x)=\sin(\pi x_1)+0.6\cos(\pi x_2)+0.5x_3x_4.
\]
Current drift is zero, linear $0.7x_1-0.5x_2$, or complex $\sin(\pi x_2)+0.8x_1x_4$. The calibration basis contains the four linear covariates, so complex drift violates \eqref{eq:T}. Under linear truth, $f_0(x)=0.7x_1-0.5x_2+0.4x_3+0.2x_4$. A covariate shift setting draws target covariates as independent $2\operatorname{Beta}(2,1)-1$, giving a target/source density ratio at most 16 while preserving support.

Historical cumulative baseline hazard is $0.28t^{1.4}$ and target cumulative baseline hazard is $0.35t^{1.2}$. Independent exponential censoring has rates 0.08 and 0.18, respectively, and both samples have administrative censoring at $\tau=2$. Treatment is independent Bernoulli$(0.5)$. Six alternative settings use $\theta_0=\log(0.7)$: shared prognosis, linear drift, complex drift, small source, linear truth, and covariate shift. Three null settings use $\theta_0=0$: linear drift, complex drift, and non-PH. The latter changes the target cumulative hazard to
\[
\Lambda_T(t\mid a,x)=0.35t^{1.2\exp(0.45x_1)}\exp\{f_0(x)+0.7x_1-0.5x_2\},
\]
independently of treatment. It satisfies the observed data null but not the target Cox model.

The historical network has ReLU widths 24 and 12, a linear skip, output clipping to $[-4,4]$, and parameter clipping to $[-5,5]$. Full batch Adam uses learning rate 0.006 and weight penalty $10^{-4}$. A 20\% validation split chooses a step count from a run of at most 500 steps, followed by refitting on all fitting observations for that count. The current neural comparator uses the same architecture and stopping rule. These settings were fixed before the expanded run.

Calibrated transfer is compared with frozen transfer without $z$, a current linear Cox fit, a current neural fit, and oracle transfer using the true historical function. The oracle is a reference, not an implementable method. Projected transfer uses five folds, the four linear covariates and the first three powers of $\widetilde g/2$ as a small projection dictionary, ridge 0.002, and projection clipping to $[-2,2]$. This finite dictionary approximates the population projection. Its numerical coverage assesses that approximation. The growing dictionary result in \eqref{eq:seriesrate} is an asymptotic construction rather than a claim about this fixed dictionary.

Finite dimensional working Cox fits use sandwich variance, and projected transfer uses \eqref{eq:se}. The current neural comparator profiles treatment after training with its fitted neural function held fixed. Its information based interval is diagnostic because the same target observations train the nuisance function. It is not a reproduction of the efficient interval in \citet{zhong2022}. Information only intervals for other methods are retained to examine the variance issue at the null.

For hypothesis testing, we also compare an unadjusted treatment-only Cox Wald test and the covariate-adjusted log-rank analysis of \citet{ye2024}, using the same externally trained neural score as its additional covariate, as in the external score strategy of \citet{zhou2026}. The comparison shares the historical learner but not their martingale pseudo outcome training procedure. The analysis formula is given in Appendix~\ref{app:logrank}. Rejection rates compare the same null hypothesis. The unadjusted and marginal adjusted log-rank coefficients are not compared to $\theta_0$ as if they were conditional effect estimators.

We report treatment bias, RMSE, empirical SD, reported SE, and interval coverage. Centered control log risk MSE. And conditional survival probability MSE at ten times between 0.2 and 2 on 1000 fresh target distribution covariates. There is no static log risk target in the non-PH setting, so its prediction assessment uses survival MSE alone. Under model misspecification, coefficient errors are descriptive discrepancies from the generating coefficient, rather than evidence that the working estimator consistently targets it. Paired Monte Carlo comparisons account for the common generated datasets. Failure counts and all settings are retained. An earlier exploratory simulation informed the documented implementation changes. The expanded simulation uses independent seeds and is not combined with that pilot.

We extend the original experiment in three directions. First, we fit a stronger target-only comparator on the same 4500 patient datasets and evaluation covariates. This model, denoted CSPLINE, uses three natural cubic spline coordinates for each of the four variables, with internal knots $-1/3,1/3$ and boundary knots $-1,1$, together with the six pairwise products. Its 18 prognostic coordinates are fixed before observing outcomes. Deterministic centering and scaling under the uniform covariate reference leave this function space unchanged. We use the same numerical coefficient bounds and sandwich variance as for the other finite-dimensional working Cox fits.

Second, an independent six cell experiment uses the nonlinear historical risk and linear target drift specified above. Four alternative cells, each with 1000 replications and $\theta_0=\log(0.7)$, have $(n,N)=(300,3000),(300,300),(300,1000),(600,3000)$. The two matching null cells have $(n,N)=(300,3000),(600,3000)$ and 2000 replications each. These are cells C01 to C06 in this supplement, with C02 and C06 the null cells. Together they contain 8000 new patient datasets. Historical and current neural learners retain the original architecture, penalty, clipping and validation stopping rule.

We write CT for direct calibrated transfer, FT for frozen transfer, and PT-L for the original projected transfer implementation. The additional projected variant PT-F uses the deterministic dictionary $B_{18}$ of four linear coordinates, eight first-frequency sine and cosine terms, and six pairwise products. Its ridge is $0.001\sqrt{300/n_{\rm tr}}$ and its projection is clipped to $[-3,3]$. This dictionary is centered under the uniform reference and includes terms informed by the simulation structure. In C01 and C02 we additionally use $B_{26}$, which appends eight second-frequency trigonometric terms, and evaluate ridge multipliers $0.1,1,10$ for both dictionaries. All six specifications use identical patient records, historical fits and training complement calibration fits. Across the six, PT coverage ranges from 93.4\% to 94.1\% in C01 and from 93.8\% to 94.3\% in C02, null rejection from 5.7\% to 6.2\%, and alternative RMSE from 0.202 to 0.213 (Archive Table~24); this grid is a sensitivity analysis and was not used to choose PT-F.

A current-only score comparator, CN-S, trains the neural pilot on each training complement, freezes it, refits treatment and the four correction coefficients, and uses the same $B_{18}$ projection and held-out score as PT-F. CN-S is evaluated in C01, C02, C05 and C06. This comparison uses the same evaluation fold variance formula. It is a separately specified procedure rather than an implementation of the efficient estimator of \citet{zhong2022}. The full sample current neural fit, CNN, contributes point estimates and predictions only. Five folds are assigned independently of patient records. Projected methods do not supply survival predictions. Each direct prediction method uses its own treatment estimate and target Breslow baseline.

Prediction uses 1000 independent covariate vectors per replication and ten equally spaced times from 0.2 to 2.0. We evaluate each arm separately and then average equally across arms. Pairing with the original experiment uses its exact original random arm evaluation stream. For interval comparisons we report coverage, interval length and the ratio of mean reported SE to empirical SD. Treatment only and externally adjusted log-rank analyses are compared through rejection probability, without treating their coefficients as conditional treatment estimators. Paired differences use common successful replications and their Monte Carlo uncertainty. The appendix gives every failure and the all-attempt probability of returning an interval that covers the truth.

Third, a separate one-dimensional experiment implements the certified source learner and Fourier score construction in Corollary~\ref{cor:feasible}. We use $X\sim\operatorname{Uniform}(0,1)$, $f_0(x)=0.6\sin(2\pi x)+0.2\cos(2\pi x)$, no calibration covariates, and the preceding baseline, censoring and treatment laws. For each $N=n\in\{300,1000,3000\}$, we run 500 alternative and 500 null replications. The source uses $J_N=\lceil(N/\log N)^{1/5}\rceil$ linear interpolation intervals and knot values in $[-2,2]$. Unit projected gradient steps stop when the certificate in \eqref{eq:gapcertificate} is at most $\min\{(\log N/N)^{4/5},10^{-7}\}$. On each training complement, the Fourier dictionary has $J_F=\max\{1,\lfloor n_{\rm tr}^{1/5}/2\rfloor\}$ frequencies and ridge $n_{\rm tr}^{-1/2}$. The direct certified fit and its projected score are compared with the known-risk oracle and a target-only Cox model containing four fixed sine/cosine terms at frequencies one and two. The latter is a structure-informed reference containing the true risk function. Numerical certificates, clipping bounds and score tolerances are detailed in Appendix~\ref{app:extension}.

Table~\ref{tab:newfullprediction} compares prediction against both neural and spline current-only fits. At $(n,N)=(300,3000)$, CT log risk MSE is 0.120, compared with 0.250 for CSPLINE and 0.451 for CNN. The paired CT minus CSPLINE difference is -0.130 (Monte Carlo SE 0.004. 95\% Monte Carlo interval [-0.138, -0.123]). The survival MSE difference, multiplied by 1000, is -3.46 (0.09). When the historical sample is reduced to $N=300$, the corresponding risk errors are 0.402 and 0.253. CSPLINE is more accurate in this small source cell, so transfer provides no prediction advantage against that comparator. The original linear drift experiment gives a paired CT minus CSPLINE risk MSE difference of -0.129 (0.005). All nine original settings, including unfavorable and misspecified settings, remain in Appendix~\ref{app:extension}.

\section{Implementation and complete numerical results}\label{app:implementation}
\subsection{Likelihood and network fitting}
All observations are sorted by ascending observed time. For $w_i=\exp(r_i-\max_jr_j)$, the risk denominators are reverse cumulative sums $d_i=\sum_{j\ge i}w_j$. The normalized negative partial likelihood gradient with respect to $r_i$ is
\[
\frac1n\left\{w_i\sum_{j\le i}\Delta_j/d_j-\Delta_i\right\}.
\]
Event times are continuous in the simulation. Administrative censoring ties occur only at the final time, after all events, and so do not require an event tie approximation. Risk set covariance matrices are computed by reverse cumulative first and second weighted moments.

The finite-dimensional fit uses L-BFGS-B with coefficient bounds $[-5,5]$, maximum 250 iterations, objective tolerance $10^{-13}$, and gradient tolerance $10^{-8}$. A numerical ridge of $10^{-8}$ is included in the objective. Information and sandwich covariance use the unpenalized risk-set Hessian. The fixed ridge and finite stopping tolerances are numerical approximations to the unpenalized mathematical estimator. They are not vanishing sequences and do not themselves establish its asymptotic root conditions. Fits with a gradient norm above $2\times10^{-5}$ or a coefficient within 0.001 of its bound are recorded as failures. Asymptotic results concern roots, or sequences of approximate fits with the stated vanishing score residual.

The historical network has two hidden ReLU layers of widths 24 and 12, and a four variable linear skip. Its raw output is clipped to $[-4,4]$. Every parameter is clipped to $[-5,5]$ after each Adam update. Adam uses decay constants 0.9 and 0.999, learning rate 0.006, and numerical denominator constant $10^{-8}$. We penalize the squared weight and skip norms with coefficient $10^{-4}$. Biases and the finite treatment coefficient are not included in that penalty. Validation likelihood is checked every ten updates. The lowest validation loss selects the refit update count. The current neural comparator is trained with its treatment coefficient outside the clipped prognostic function and is then profiled with respect to treatment, holding the learned function fixed.

Output and parameter bounds give a bounded parameter class for the entropy specialization. It does not prove that the optimizer attains a small likelihood gap. A linear skip can be covered by an additional finite parameter grid of the same entropy order. Likewise, the finite projection dictionary is not asserted to exhaust the space of functions of $X$. The growing dictionary bound in Appendix~\ref{app:projection} concerns deterministic dictionaries, while the numerical dictionary includes powers of a fitted function. Its extra perturbation and approximation error remain part of $b_n$.

\subsection{Covariance computation}
For a vector design $W_i$, the empirical influence vector in \eqref{eq:empIF}, before centering, is computed as
\[
\Delta_i(W_i-\bar W_i)-w_i\left\{W_i\sum_{j\le i}\Delta_j/d_j-\sum_{j\le i}(\Delta_j/d_j)\bar W_j\right\}.
\]
Its sample mean is removed, which subtracts $U_n$. This includes both terms of the risk ratio influence function. Sandwich covariance is then formed from the empirical second moment and the unpenalized Hessian. The same routine evaluates the scalar residual $q$ on an independent score fold.

\subsection{Replication and Monte Carlo calculations}
The expanded run contains nine scenarios and 500 replications each. The base seed is 3141592659, with an increment of 100000 for scenario index and an increment of one for replication. Separate deterministic increments govern source initialization, target initialization, and fold randomization. No expanded run outcome is used to choose fitting parameters. The earlier 60 replication pilot informed the change to five folds and influence based variances, but is not combined with this independent expanded run.

For a mean metric $\bar Z$, its Monte Carlo SE is the replication SD divided by $\sqrt R$. For RMSE $\sqrt{\overline{E^2}}$, we use the delta method SE from $E^2/(2\sqrt{\overline{E^2}})$. Paired RMSE comparisons subtract those influence values within a common replication. For a rejection proportion $\widehat p$, the SE is $\sqrt{\widehat p(1-\widehat p)/R}$. Rejection and coverage calculations use the two-sided standard normal critical value 1.9599639845.

\begingroup\footnotesize\setlength{\tabcolsep}{4pt}
\begin{longtable}{@{}lrrrrrrr@{}}
\caption{Complete estimation and prediction results. SE is the mean reported standard error, coverage is in percent, and survival MSE is multiplied by 1000. Intervals under misspecification are descriptive. The current neural SE and coverage and projected transfer survival entries are omitted. NA denotes an undefined static log risk target.}\label{tab:full}\\
\toprule Method & Bias & RMSE & SD & SE & Coverage & Risk MSE & Survival\\\midrule\endfirsthead
\toprule Method & Bias & RMSE & SD & SE & Coverage & Risk MSE & Survival\\\midrule\endhead
\bottomrule\endfoot
\multicolumn{8}{l}{\textit{Shared prognosis}}\\
Calibrated transfer & -0.002 & 0.178 & 0.178 & 0.180 & 94.2 & 0.116 & 6.371\\
Projected transfer & 0.012 & 0.179 & 0.179 & 0.187 & 96.2 & 0.116 & NA\\
Frozen transfer & -0.001 & 0.174 & 0.175 & 0.178 & 95.2 & 0.083 & 4.778\\
Target linear & 0.036 & 0.183 & 0.180 & 0.179 & 94.0 & 0.441 & 24.454\\
Target neural & 0.021 & 0.187 & 0.186 & NA & NA & 0.434 & 21.924\\
Oracle transfer & -0.005 & 0.173 & 0.173 & 0.178 & 95.2 & 0.036 & 2.768\\
\addlinespace
\multicolumn{8}{l}{\textit{Linear drift}}\\
Calibrated transfer & 0.002 & 0.182 & 0.182 & 0.177 & 94.4 & 0.120 & 5.679\\
Projected transfer & 0.018 & 0.187 & 0.187 & 0.186 & 96.0 & 0.120 & NA\\
Frozen transfer & 0.044 & 0.172 & 0.167 & 0.167 & 94.6 & 0.317 & 14.548\\
Target linear & 0.036 & 0.183 & 0.180 & 0.178 & 94.4 & 0.455 & 22.130\\
Target neural & 0.015 & 0.185 & 0.185 & NA & NA & 0.445 & 18.107\\
Oracle transfer & -0.006 & 0.179 & 0.180 & 0.174 & 94.6 & 0.037 & 2.514\\
\addlinespace
\multicolumn{8}{l}{\textit{Complex drift}}\\
Calibrated transfer & 0.016 & 0.185 & 0.184 & 0.177 & 93.8 & 0.392 & 17.737\\
Projected transfer & 0.029 & 0.190 & 0.189 & 0.185 & 94.6 & 0.392 & NA\\
Frozen transfer & 0.041 & 0.197 & 0.192 & 0.183 & 93.2 & 0.653 & 33.196\\
Target linear & 0.056 & 0.189 & 0.181 & 0.179 & 93.2 & 0.737 & 37.219\\
Target neural & 0.016 & 0.190 & 0.190 & NA & NA & 0.610 & 27.336\\
Oracle transfer & 0.014 & 0.184 & 0.184 & 0.176 & 93.8 & 0.310 & 15.047\\
\addlinespace
\multicolumn{8}{l}{\textit{Small source}}\\
Calibrated transfer & 0.025 & 0.193 & 0.192 & 0.182 & 93.8 & 0.403 & 18.198\\
Projected transfer & 0.054 & 0.192 & 0.184 & 0.193 & 95.0 & 0.403 & NA\\
Frozen transfer & 0.058 & 0.188 & 0.179 & 0.174 & 94.4 & 0.651 & 29.676\\
Target linear & 0.047 & 0.188 & 0.182 & 0.178 & 94.2 & 0.457 & 22.287\\
Target neural & 0.023 & 0.188 & 0.187 & NA & NA & 0.434 & 17.867\\
Oracle transfer & -0.003 & 0.182 & 0.182 & 0.174 & 94.0 & 0.037 & 2.586\\
\addlinespace
\multicolumn{8}{l}{\textit{Linear truth}}\\
Calibrated transfer & 0.010 & 0.192 & 0.192 & 0.176 & 93.8 & 0.052 & 3.149\\
Projected transfer & 0.021 & 0.198 & 0.197 & 0.184 & 94.4 & 0.052 & NA\\
Frozen transfer & 0.062 & 0.187 & 0.177 & 0.164 & 92.6 & 0.267 & 11.591\\
Target linear & 0.009 & 0.191 & 0.191 & 0.176 & 94.4 & 0.036 & 2.554\\
Target neural & 0.024 & 0.194 & 0.192 & NA & NA & 0.197 & 8.339\\
Oracle transfer & 0.009 & 0.191 & 0.191 & 0.176 & 94.4 & 0.036 & 2.554\\
\addlinespace
\multicolumn{8}{l}{\textit{Covariate shift}}\\
Calibrated transfer & 0.010 & 0.158 & 0.158 & 0.158 & 95.8 & 0.093 & 5.765\\
Projected transfer & 0.027 & 0.160 & 0.158 & 0.165 & 95.8 & 0.093 & NA\\
Frozen transfer & 0.042 & 0.157 & 0.151 & 0.151 & 93.2 & 0.226 & 11.769\\
Target linear & 0.050 & 0.171 & 0.163 & 0.160 & 92.8 & 0.395 & 22.923\\
Target neural & 0.024 & 0.166 & 0.164 & NA & NA & 0.328 & 16.529\\
Oracle transfer & 0.006 & 0.152 & 0.152 & 0.156 & 95.2 & 0.032 & 2.570\\
\addlinespace
\multicolumn{8}{l}{\textit{Null, linear drift}}\\
Calibrated transfer & -0.007 & 0.160 & 0.160 & 0.168 & 96.2 & 0.115 & 5.755\\
Projected transfer & -0.005 & 0.168 & 0.168 & 0.176 & 96.4 & 0.115 & NA\\
Frozen transfer & -0.006 & 0.150 & 0.150 & 0.159 & 96.0 & 0.313 & 15.523\\
Target linear & -0.011 & 0.162 & 0.162 & 0.170 & 96.2 & 0.454 & 22.755\\
Target neural & -0.012 & 0.172 & 0.171 & NA & NA & 0.428 & 18.752\\
Oracle transfer & -0.005 & 0.156 & 0.156 & 0.166 & 95.4 & 0.034 & 2.425\\
\addlinespace
\multicolumn{8}{l}{\textit{Null, complex drift}}\\
Calibrated transfer & -0.006 & 0.174 & 0.174 & 0.170 & 94.2 & 0.387 & 19.059\\
Projected transfer & -0.006 & 0.174 & 0.174 & 0.176 & 94.0 & 0.387 & NA\\
Frozen transfer & -0.007 & 0.175 & 0.175 & 0.177 & 95.4 & 0.651 & 35.844\\
Target linear & -0.008 & 0.175 & 0.175 & 0.171 & 94.4 & 0.736 & 39.445\\
Target neural & -0.004 & 0.181 & 0.181 & NA & NA & 0.596 & 29.023\\
Oracle transfer & -0.004 & 0.169 & 0.170 & 0.168 & 94.8 & 0.307 & 16.071\\
\addlinespace
\multicolumn{8}{l}{\textit{Null, non-PH}}\\
Calibrated transfer & -0.007 & 0.164 & 0.164 & 0.163 & 95.8 & NA & 5.878\\
Projected transfer & -0.005 & 0.164 & 0.164 & 0.172 & 95.4 & NA & NA\\
Frozen transfer & -0.006 & 0.151 & 0.151 & 0.155 & 96.0 & NA & 17.554\\
Target linear & -0.010 & 0.170 & 0.170 & 0.164 & 95.0 & NA & 21.456\\
Target neural & -0.010 & 0.174 & 0.174 & NA & NA & NA & 16.548\\
Oracle transfer & -0.010 & 0.160 & 0.159 & 0.160 & 95.4 & NA & 2.776\\
\addlinespace
\end{longtable}
\endgroup

\subsection{Fixed score variance stress experiment}
To isolate null studentization, a separate experiment uses the null linear drift data generating law with $n=300$ and $n=1000$, and 1000 replications for each sample size. Three fixed external scores are specified before that experiment: zero, $-2\cos(\pi X_2)$, and $2\cos(\pi X_2)$. They are calibrated with the same four linear covariates. No historical sample or neural fitting is simulated here. The scores represent independently fixed external prognostic functions. This experiment evaluates the conditional null result, rather than the performance of a historical learner.

Both nonzero scores are deliberately incorrect for the target risk relationship. All results appear in Table~\ref{tab:stress}. Information only testing has substantial size distortion. Sandwich testing substantially reduces that distortion, with rejection still somewhat above 5\% in the more difficult $n=300$ cases. At $n=1000$ its rejection is closer to nominal. The experiment illustrates finite sample approximation as well as the difference between the variance formulas. It does not claim exact finite sample calibration.

\begin{table}[t]\centering\footnotesize
\caption{Fixed score null stress experiment, 1000 replications per sample size. Rejection rates are percentages (Monte Carlo SE). Scores are used as fixed offsets with four linear target corrections.}\label{tab:stress}
\begin{tabular}{@{}rlrrrr@{}}\toprule
$n$ & External score & Information & Sandwich & Empirical SD & Mean sandwich SE\\\midrule
300 & Zero & 5.5 (0.7) & 5.6 (0.7) & 0.176 & 0.170\\
300 & Negative cosine & 20.9 (1.3) & 6.7 (0.8) & 0.266 & 0.247\\
300 & Positive cosine & 16.9 (1.2) & 7.1 (0.8) & 0.248 & 0.227\\
1000 & Zero & 6.3 (0.8) & 5.7 (0.7) & 0.092 & 0.093\\
1000 & Negative cosine & 19.1 (1.2) & 5.9 (0.7) & 0.139 & 0.135\\
1000 & Positive cosine & 18.4 (1.2) & 5.8 (0.7) & 0.134 & 0.129\\
\bottomrule\end{tabular}\end{table}

\section{External score-adjusted log-rank comparison}\label{app:logrank}
We implement the simple randomization analysis of \citet{ye2024}, using one external covariate $F_i=\widehat f(X_i)$ as in the external score strategy of \citet{zhou2026}. The source function is the same neural Cox risk used in calibrated transfer. We do not reproduce their separate martingale pseudo outcome training algorithm or compare a marginal HR estimate to the conditional coefficient.

Let $\bar Y=n^{-1}\sum_iY_i$, $\bar\cnt=n^{-1}\sum_i\cnt_i$, and $\mu_n(t)=\sum_iA_iY_i(t)/\sum_iY_i(t)$. The unadjusted score and variance at zero are
\[
\widehat U_L=\frac1n\sum_i\int\{A_i-\mu_n(t)\}\,d\cnt_i(t),\qquad
\widehat\sigma_L^2=\frac1n\sum_i\int\mu_n(t)\{1-\mu_n(t)\}\,d\cnt_i(t).
\]
For subjects in each assigned arm, define the arm specific pseudo outcomes
\begin{align*}
\widehat O_{i1}&=\int\{1-\mu_n(t)\}\left\{d\cnt_i(t)-Y_i(t)d\bar\cnt(t)/\bar Y(t)\right\},\quad A_i=1,\\
\widehat O_{i0}&=\int\mu_n(t)\left\{d\cnt_i(t)-Y_i(t)d\bar\cnt(t)/\bar Y(t)\right\},\quad A_i=0.
\end{align*}
Let $\widehat\beta_1$ and $\widehat\beta_0$ be the ordinary least squares slopes from regressing these outcomes on $F$ separately within arms, with intercepts. With $\bar F=n^{-1}\sum_iF_i$, $\widehat p_A=n^{-1}\sum_iA_i$, and $\widehat\Sigma_F=n^{-1}\sum_i(F_i-\bar F)^2$, compute
\begin{align*}
\widehat U_{CL}&=\widehat U_L-\frac1n\sum_i\left\{A_i(F_i-\bar F)\widehat\beta_1-(1-A_i)(F_i-\bar F)\widehat\beta_0\right\},\\
\widehat\sigma_{CL}^2&=\widehat\sigma_L^2-\widehat p_A(1-\widehat p_A)(\widehat\beta_1+\widehat\beta_0)^2\widehat\Sigma_F.
\end{align*}
The statistic is $\sqrt n\widehat U_{CL}/\widehat\sigma_{CL}$. A nonpositive estimated variance is recorded as a failure. All 4500 test comparison replications are retained.

The variance reduction analysis and its assumptions belong to the cited adjusted log-rank methodology. Our comparison uses its published analysis formula at a common null, with a common source learner. Under alternatives it measures rejection probability, rather than the accuracy of a conditional hazard ratio estimate. This distinction avoids attributing changes of estimand to estimation error.

\section{Fixed offset directional bias experiment}\label{app:mechanismexperiment}
This experiment evaluates Proposition~\ref{prop:sensitivity} and the invariance in Proposition~\ref{prop:invariance}, independently of the historical network. Here $X$ is uniform on $\{-1,0,1\}$, $f_0(x)=0.8x$, $\theta_0=\log(0.7)$, $p_A=1/2$, $\Lambda_{T,0}(t)=0.35t^{1.2}$, and only administrative censoring occurs at $\tau=2$. Neither a historical sample size $N$ nor a source fitting error rate is assigned to these fixed offsets. The error norm concerns the input function before calibration, not the fitted calibrated prediction error.

\subsection{Population integration and fixed directions}
We integrate over the six $(A,X)$ cells and cumulative baseline time $u=\Lambda_{T,0}(t)\in[0,0.35\,2^{1.2}]$. The conditional survival weight is $\exp\{-u\exp(\theta_0a+0.8x)\}$. Gauss Legendre quadrature with 128 nodes supplies the working risk-set means, and the population score is integrated against the true event intensity. Scalar roots are solved independently of the first-order prediction. Two-dimensional calibrated roots are solved jointly. 

The uncalibrated information is $\J(A,A)=0.125708$ and $\norm{\omega_{\varnothing}}_2=0.0093705$, giving parallel bias derivative $-0.074542$. At $X=(-1,0,1)$ the fixed unit directions are, to six decimals,
\[
v_\parallel=(-0.787184,-0.623884,1.411067),\qquad
v_\perp=(-1.174880,1.269161,-0.094281).
\]
The directions are evaluated at full precision. Both directions have mean zero and squared norm one. The perpendicular inner product with the representer is below $10^{-12}$. The seven signed amplitudes and their solved roots appear in Table~\ref{tab:mechanismpopulation}. Replacing $\theta_0$ by zero makes the representer numerically zero and gives zero population treatment roots for both directions at every amplitude. This is a population null identity, not an additional Monte Carlo null study.

\subsection{Finite sample fitting and uncertainty}
For each $n\in\{300,1000\}$ and $\delta\in\{0,0.1,0.2\}$ we generate 1000 independent patient datasets. The two directions and both methods use common records within each size to amplitude pair. Counting the zero error cell once gives ten directional cells based on 6000 datasets. There are 24,000 fitting records, of which 2000 are duplicate zero error uncalibrated fits. These duplicates are not treated as independent observations.

The normalized unpenalized partial likelihood is solved by damped Newton iteration with at most 60 iterations, normalized score tolerance $10^{-10}$, and coefficient magnitude restricted below 5. Singular Hessians, invalid variances, unresolved scores, and bound or line search failures are retained. All 24,000 attempts succeeded. 

Tables~\ref{tab:mechanism300} and~\ref{tab:mechanism1000} report all cells. Information and sandwich intervals use critical value 1.9599639845. Means and bias use the replication SD divided by $\sqrt{1000}$. RMSE Monte Carlo SEs and paired RMSE contrasts use the delta method described in Appendix~\ref{app:implementation}. At $n=1000$, $\delta=0.2$ in the parallel direction, the paired calibrated minus uncalibrated RMSE difference is $-0.00568$ (Monte Carlo SE 0.00065). In the perpendicular direction at $n=300$, $\delta=0.1$, that difference is $0.00133$ (0.00042), illustrating a finite sample cost when calibration removes no first-order bias.

\begin{table}[tbp]
\centering\footnotesize
\caption{Population fixed offset paths. Both directions have input MSE $\delta^2$. The calibrated treatment bias is below $2\times10^{-16}$ in both directions at every displayed amplitude. The perpendicular first-order term is zero.}\label{tab:mechanismpopulation}
\begin{tabular}{@{}rrrrr@{}}\toprule
$\delta$ & Input MSE & Parallel bias & Parallel first-order & Perpendicular bias\\\midrule
-0.20 & 0.0400 & 0.014521 & 0.014908 & 0.001054\\
-0.10 & 0.0100 & 0.007381 & 0.007454 & 0.000262\\
-0.05 & 0.0025 & 0.003712 & 0.003727 & 0.000065\\
0.00 & 0.0000 & -0.000000 & -0.000000 & -0.000000\\
0.05 & 0.0025 & -0.003736 & -0.003727 & 0.000064\\
0.10 & 0.0100 & -0.007477 & -0.007454 & 0.000254\\
0.20 & 0.0400 & -0.014899 & -0.014908 & 0.000991\\
\bottomrule\end{tabular}\end{table}

\begin{table}[tbp]
\centering\footnotesize
\setlength{\tabcolsep}{3pt}
\caption{Fixed offset Monte Carlo results at $n=300$, with 1000 replications per cell. Bias parentheses give Monte Carlo SE. Coverage is a percentage. P and O denote parallel and perpendicular directions. U and C denote uncalibrated and known direction calibrated fits. The zero error uncalibrated rows share one estimate in each replication.}\label{tab:mechanism300}
\begin{tabular}{@{}rllrrrrrr@{}}\toprule
$\delta$ & Dir. & Fit & Bias (MC SE) & RMSE & SD & SW SE & I cov. & SW cov.\\\midrule
0.0 & P & C & 0.0003 (0.0051) & 0.162 & 0.162 & 0.164 & 95.9 & 95.8\\
0.0 & P & U & 0.0010 (0.0051) & 0.162 & 0.162 & 0.163 & 96.0 & 96.1\\
0.0 & O & C & -0.0000 (0.0051) & 0.162 & 0.162 & 0.163 & 95.7 & 96.0\\
0.0 & O & U & 0.0010 (0.0051) & 0.162 & 0.162 & 0.163 & 96.0 & 96.1\\
0.1 & P & C & -0.0074 (0.0053) & 0.167 & 0.167 & 0.164 & 94.4 & 94.4\\
0.1 & P & U & -0.0131 (0.0054) & 0.172 & 0.171 & 0.167 & 93.3 & 94.3\\
0.1 & O & C & -0.0064 (0.0053) & 0.167 & 0.167 & 0.163 & 94.3 & 94.1\\
0.1 & O & U & -0.0056 (0.0052) & 0.166 & 0.166 & 0.163 & 94.2 & 94.0\\
0.2 & P & C & -0.0006 (0.0051) & 0.163 & 0.163 & 0.164 & 95.3 & 95.2\\
0.2 & P & U & -0.0159 (0.0054) & 0.171 & 0.171 & 0.172 & 94.3 & 95.4\\
0.2 & O & C & -0.0004 (0.0052) & 0.163 & 0.163 & 0.163 & 95.2 & 95.1\\
0.2 & O & U & 0.0020 (0.0052) & 0.164 & 0.164 & 0.164 & 95.0 & 95.1\\
\bottomrule\end{tabular}
\par\smallskip\begin{minipage}{0.94\textwidth}\footnotesize
SW denotes sandwich and I information-only inference. The coverage Monte Carlo SE is $\{\widehat p(1-\widehat p)/1000\}^{1/2}$, approximately 0.7 percentage points near 95\%. All attempts succeeded. Wilson Monte Carlo intervals, mean information SE, interval lengths, SE/SD ratios, and paired contrasts are supplied in the machine-readable results.
\end{minipage}\end{table}

\begin{table}[tbp]
\centering\footnotesize
\setlength{\tabcolsep}{3pt}
\caption{Fixed offset Monte Carlo results at $n=1000$, with 1000 replications per cell. Bias parentheses give Monte Carlo SE. Coverage is a percentage. P and O denote parallel and perpendicular directions. U and C denote uncalibrated and known direction calibrated fits. The zero error uncalibrated rows share one estimate in each replication.}\label{tab:mechanism1000}
\begin{tabular}{@{}rllrrrrrr@{}}\toprule
$\delta$ & Dir. & Fit & Bias (MC SE) & RMSE & SD & SW SE & I cov. & SW cov.\\\midrule
0.0 & P & C & -0.0013 (0.0029) & 0.091 & 0.091 & 0.089 & 94.5 & 94.3\\
0.0 & P & U & -0.0010 (0.0029) & 0.090 & 0.090 & 0.089 & 94.2 & 94.1\\
0.0 & O & C & -0.0011 (0.0029) & 0.091 & 0.091 & 0.089 & 94.1 & 94.2\\
0.0 & O & U & -0.0010 (0.0029) & 0.090 & 0.090 & 0.089 & 94.2 & 94.1\\
0.1 & P & C & -0.0028 (0.0027) & 0.086 & 0.086 & 0.089 & 95.7 & 95.7\\
0.1 & P & U & -0.0100 (0.0028) & 0.089 & 0.088 & 0.091 & 95.6 & 95.7\\
0.1 & O & C & -0.0024 (0.0027) & 0.086 & 0.086 & 0.089 & 96.0 & 95.9\\
0.1 & O & U & -0.0021 (0.0027) & 0.086 & 0.086 & 0.089 & 95.7 & 95.8\\
0.2 & P & C & -0.0013 (0.0028) & 0.089 & 0.089 & 0.089 & 94.8 & 94.9\\
0.2 & P & U & -0.0157 (0.0030) & 0.095 & 0.093 & 0.094 & 93.0 & 94.3\\
0.2 & O & C & -0.0010 (0.0028) & 0.089 & 0.089 & 0.089 & 95.0 & 95.2\\
0.2 & O & U & -0.0001 (0.0028) & 0.090 & 0.090 & 0.090 & 94.8 & 94.7\\
\bottomrule\end{tabular}
\par\smallskip\begin{minipage}{0.94\textwidth}\footnotesize
SW denotes sandwich and I information-only inference. The coverage Monte Carlo SE is $\{\widehat p(1-\widehat p)/1000\}^{1/2}$, approximately 0.7 percentage points near 95\%. All attempts succeeded. Wilson Monte Carlo intervals, mean information SE, interval lengths, SE/SD ratios, and paired contrasts are supplied in the machine-readable results.
\end{minipage}\end{table}

\clearpage

\section{Additional implementation and complete comparisons}\label{app:extension}
\subsection{Fixed implementation choices}

The natural spline comparator uses the stated fixed knot sequence. Its three nonconstant coordinates per variable are centered and whitened under the known uniform reference before fitting. This is a deterministic change of coordinates. The six interaction columns are appended without outcome based selection. Finite dimensional fits use coefficient bounds $[-5,5]$, a numerical ridge of $10^{-8}$, and the unpenalized sandwich information. Scores exceeding $2\times10^{-5}$, coefficients within $0.001$ of a bound, nonpositive information, or invalid variance are retained as failures. One dimensional certified treatment fits instead solve the unpenalized scalar score, with residual at most $\min(1/n_{\rm tr},10^{-8})$ for a training pilot. The practical projected methods require each evaluation derivative to exceed $10^{-4}$. A failed fold fails the entire analysis.

For the constructive law, $|\eta_0|\le\sqrt{0.6^2+0.2^2}+|\log(0.7)|<1.36$. Hence
\[
\Pr(U\ge2\mid A,X)\ge
\exp\{-0.35\,2^{1.2}e^{1.36}-0.36\}>0.03.
\]
With $B=1.36$ and $c=0.03$, the centered representative in Lemma~\ref{lem:geometry} obeys
\[
\norm{h_*}_2\le\frac{e^B}{2\sqrt c},\qquad
\norm{q_*}_2\le\sqrt{\tfrac12+\frac{e^{2B}}{4c}},\qquad
\norm{h_*}_\infty\le
1+\frac{e^{2B}}c\sqrt{\tfrac12+\frac{e^{2B}}{4c}}<20000.
\]
The last inequality follows from \eqref{eq:projectionformula}, since $0\le d/r\le1$ and the remaining term averages $\ell(t)$ under nonnegative weights. We fix clipping at 20000 and use evaluation derivative threshold $0.001m^{-1/4}$. These constants are set from the generating law before evaluation. A validation loss or parameter gradient norm is never substituted for this certificate.

\subsection{Monte Carlo accounting}
The independent neural and certified source studies each exclude a separate 20 replication per cell pilot. Historical fits and training pilots are shared across the specified projection grid and log-rank comparisons.

Monte Carlo SEs for means and paired differences use the replication SD divided by the square root of the relevant replication count. RMSE differences use the paired-delta method influence values. Proportions additionally have Wilson Monte Carlo intervals with the tabulated follow-ups. Reported conditional coverage is accompanied by success fraction and all-attempt successful coverage. Comparison pairs report their common success fraction.

\subsection{Population nuisance diagnostics}
For replications 1 to 100 of C01, C03, C04 and C05, we integrate true survival and censoring weights on 10,000 independent covariate draws and 100 Gauss Legendre nodes in cumulative baseline time. The same law supplies the population $z$ and finite-$B_{18}$ Gram matrices and sensitivity weights. We remove the source error projection onto $(1,z)$, then evaluate $\rho_N$, $\J(q_z,e_N)$, $\J(q_B,e_N)$ and the implied first-order bias. Each fitted projection is compared with the population finite-$B_{18}$ projection, not the unrestricted $h_*$. 

\begin{table}[tbp]\centering\small
\caption{Population weighted diagnostics for 100 prespecified source fits per cell.}\label{tab:nuisance}
\setlength{\tabcolsep}{4pt}\begin{tabular}{lrrrrrrr}
\toprule
Cell & $\|e\|_2$ & $\rho_N$ & $\J_z$ & $\J_B$ & Bias$_z$ & Bias$_B$ & Proj. error \\
\midrule
C01 & 0.278 & 0.270 & -0.00030 & 0.00004 & 0.0027 & -0.0003 & 0.200 \\
C03 & 0.627 & 0.583 & -0.00357 & 0.00001 & 0.0316 & -0.0001 & 0.197 \\
C04 & 0.449 & 0.432 & -0.00154 & 0.00001 & 0.0137 & -0.0001 & 0.198 \\
C05 & 0.287 & 0.280 & -0.00041 & 0.00003 & 0.0036 & -0.0002 & 0.144 \\
\bottomrule\end{tabular}
\par\smallskip\begin{minipage}{\textwidth}\footnotesize Entries average over fitted source functions, conditional on one independent numerical integration grid. The first column of errors centers the source log risk. $\rho_N$ also removes its linear calibration component. Bias is the first-order quantity $-\J/I$. Projection error averages foldwise $L_2$ distance to the finite-$B_{18}$ population projection. It excludes dictionary approximation error to $h_*$.\end{minipage}\end{table}

The numerical target information is 0.11291 for the calibration residual and 0.11268 for the finite-$B_{18}$ residual. Doubling both resolutions changes source $L_2$ error by at most 0.0036, residual $\rho_N$ by 0.0039, and the finite-$B_{18}$ projection error estimate by 0.0011. The largest absolute changes in $\J_z$ and $\J_B$ are 3.2e-05 and 1e-05. These are observed integration differences, not certified population integration error bounds. In particular, small signed functionals can be sensitive to integration even when prediction error norms are stable. Per source values and the full refinement comparison are retained in the numerical results.

\subsection{Complete new results}

\begingroup\footnotesize\setlength{\tabcolsep}{3.3pt}
\renewcommand{\arraystretch}{0.88}
\begin{longtable}{lllrrrrrr}
\caption{Complete independent neural treatment results.}\label{tab:newfullinference}\\
\toprule
Cell & Method & Success & Bias & RMSE & Cover (\%) & All (\%) & SE/SD & Length \\
\midrule\endfirsthead
\multicolumn{9}{l}{\textit{\tablename~\thetable\ continued}}\\
\toprule
Cell & Method & Success & Bias & RMSE & Cover (\%) & All (\%) & SE/SD & Length \\
\midrule\endhead
\midrule\multicolumn{9}{r}{Continued on next page}\\\endfoot
\bottomrule\endlastfoot
C01 & CT & 1000/1000 & 0.003 & 0.183 & 94.4 (0.7) & 94.4 & 0.97 & 0.693 \\
C01 & FT & 1000/1000 & 0.046 & 0.173 & 94.0 (0.8) & 94.0 & 1.00 & 0.653 \\
C01 & CL & 1000/1000 & 0.041 & 0.187 & 93.9 (0.8) & 93.9 & 0.97 & 0.696 \\
C01 & CSPLINE & 1000/1000 & -0.019 & 0.201 & 93.1 (0.8) & 93.1 & 0.92 & 0.719 \\
C01 & CNN & 1000/1000 & 0.015 & 0.188 & NA & NA & NA & NA \\
C01 & Oracle & 1000/1000 & -0.001 & 0.178 & 94.5 (0.7) & 94.5 & 0.98 & 0.682 \\
C01 & PT-L & 1000/1000 & 0.023 & 0.189 & 95.1 (0.7) & 95.1 & 0.99 & 0.729 \\
C01 & PT-F & 1000/1000 & 0.014 & 0.206 & 93.9 (0.8) & 93.9 & 0.94 & 0.762 \\
C01 & CN-S & 1000/1000 & 0.019 & 0.229 & 92.2 (0.8) & 92.2 & 0.89 & 0.801 \\
C02 & CT & 2000/2000 & 0.000 & 0.175 & 94.8 (0.5) & 94.8 & 0.97 & 0.662 \\
C02 & FT & 2000/2000 & -0.002 & 0.160 & 94.8 (0.5) & 94.8 & 0.99 & 0.624 \\
C02 & CL & 2000/2000 & -0.004 & 0.174 & 94.7 (0.5) & 94.7 & 0.98 & 0.665 \\
C02 & CSPLINE & 2000/2000 & -0.004 & 0.190 & 93.3 (0.6) & 93.3 & 0.92 & 0.685 \\
C02 & CNN & 2000/2000 & -0.004 & 0.183 & NA & NA & NA & NA \\
C02 & Oracle & 2000/2000 & -0.001 & 0.171 & 94.7 (0.5) & 94.7 & 0.97 & 0.651 \\
C02 & PT-L & 2000/2000 & -0.001 & 0.177 & 95.0 (0.5) & 95.0 & 1.00 & 0.693 \\
C02 & PT-F & 2000/2000 & -0.004 & 0.192 & 93.8 (0.5) & 93.8 & 0.96 & 0.726 \\
C02 & CN-S & 2000/2000 & -0.007 & 0.209 & 93.2 (0.6) & 93.2 & 0.94 & 0.766 \\
C03 & CT & 1000/1000 & 0.019 & 0.180 & 94.7 (0.7) & 94.7 & 1.01 & 0.713 \\
C03 & FT & 1000/1000 & 0.057 & 0.180 & 94.6 (0.7) & 94.6 & 1.02 & 0.683 \\
C03 & CL & 1000/1000 & 0.034 & 0.180 & 94.6 (0.7) & 94.6 & 1.00 & 0.696 \\
C03 & CSPLINE & 1000/1000 & -0.032 & 0.194 & 94.4 (0.7) & 94.4 & 0.96 & 0.719 \\
C03 & CNN & 1000/1000 & 0.013 & 0.182 & NA & NA & NA & NA \\
C03 & Oracle & 1000/1000 & -0.013 & 0.171 & 95.9 (0.6) & 95.9 & 1.02 & 0.682 \\
C03 & PT-L & 1000/1000 & 0.032 & 0.190 & 95.5 (0.7) & 95.5 & 1.02 & 0.751 \\
C03 & PT-F & 1000/1000 & 0.005 & 0.191 & 95.7 (0.6) & 95.7 & 1.05 & 0.784 \\
C04 & CT & 1000/1000 & 0.008 & 0.187 & 94.8 (0.7) & 94.8 & 0.96 & 0.702 \\
C04 & FT & 1000/1000 & 0.052 & 0.175 & 93.8 (0.8) & 93.8 & 1.01 & 0.664 \\
C04 & CL & 1000/1000 & 0.042 & 0.188 & 93.4 (0.8) & 93.4 & 0.97 & 0.695 \\
C04 & CSPLINE & 1000/1000 & -0.018 & 0.200 & 93.5 (0.8) & 93.5 & 0.92 & 0.719 \\
C04 & CNN & 1000/1000 & 0.022 & 0.191 & NA & NA & NA & NA \\
C04 & Oracle & 1000/1000 & -0.005 & 0.181 & 94.6 (0.7) & 94.6 & 0.96 & 0.680 \\
C04 & PT-L & 1000/1000 & 0.025 & 0.185 & 95.2 (0.7) & 95.2 & 1.03 & 0.738 \\
C04 & PT-F & 1000/1000 & 0.010 & 0.201 & 94.3 (0.7) & 94.3 & 0.98 & 0.773 \\
C05 & CT & 1000/1000 & 0.009 & 0.126 & 94.4 (0.7) & 94.4 & 0.99 & 0.487 \\
C05 & FT & 1000/1000 & 0.051 & 0.127 & 92.4 (0.8) & 92.4 & 1.01 & 0.461 \\
C05 & CL & 1000/1000 & 0.049 & 0.135 & 92.3 (0.8) & 92.3 & 0.99 & 0.490 \\
C05 & CSPLINE & 1000/1000 & 0.001 & 0.132 & 93.8 (0.8) & 93.8 & 0.95 & 0.491 \\
C05 & CNN & 1000/1000 & 0.017 & 0.129 & NA & NA & NA & NA \\
C05 & Oracle & 1000/1000 & 0.007 & 0.125 & 94.4 (0.7) & 94.4 & 0.98 & 0.479 \\
C05 & PT-L & 1000/1000 & 0.018 & 0.125 & 94.5 (0.7) & 94.5 & 1.02 & 0.498 \\
C05 & PT-F & 1000/1000 & 0.010 & 0.132 & 93.6 (0.8) & 93.6 & 0.99 & 0.510 \\
C05 & CN-S & 1000/1000 & 0.018 & 0.142 & 93.4 (0.8) & 93.4 & 0.95 & 0.527 \\
C06 & CT & 2000/2000 & 0.007 & 0.122 & 94.8 (0.5) & 94.8 & 0.98 & 0.466 \\
C06 & FT & 2000/2000 & 0.006 & 0.114 & 94.5 (0.5) & 94.5 & 0.99 & 0.441 \\
C06 & CL & 2000/2000 & 0.005 & 0.120 & 95.0 (0.5) & 95.0 & 1.00 & 0.469 \\
C06 & CSPLINE & 2000/2000 & 0.007 & 0.122 & 94.0 (0.5) & 94.0 & 0.98 & 0.468 \\
C06 & CNN & 2000/2000 & 0.008 & 0.126 & NA & NA & NA & NA \\
C06 & Oracle & 2000/2000 & 0.007 & 0.118 & 94.8 (0.5) & 94.8 & 0.99 & 0.458 \\
C06 & PT-L & 2000/2000 & 0.007 & 0.122 & 94.5 (0.5) & 94.5 & 1.00 & 0.476 \\
C06 & PT-F & 2000/2000 & 0.008 & 0.126 & 94.0 (0.5) & 94.0 & 0.99 & 0.488 \\
C06 & CN-S & 2000/2000 & 0.007 & 0.129 & 94.0 (0.5) & 94.0 & 1.00 & 0.506 \\
\end{longtable}
\noindent C01: $(n,N,\theta_0)=(300,3000,\log 0.7)$. C02: $(300,3000,0)$. C03: $(300,300,\log 0.7)$. C04: $(300,1000,\log 0.7)$. C05: $(600,3000,\log 0.7)$. C06: $(600,3000,0)$. Cover is coverage among successful fits, with its Monte Carlo SE. All is the fraction of all attempts returning a covering interval. CNN has no inferential claim. Its interval entries are intentionally absent. All reported failures remain in the denominator of Success and All.
\endgroup

\begingroup\footnotesize\setlength{\tabcolsep}{3.3pt}
\begin{longtable}{llrrrr}
\caption{Complete independent neural prediction results.}\label{tab:newfullprediction}\\
\toprule
Cell & Method & Risk MSE & Survival $\times1000$ & Arm 0 $\times1000$ & Arm 1 $\times1000$ \\
\midrule\endfirsthead
\multicolumn{6}{l}{\textit{\tablename~\thetable\ continued}}\\
\toprule
Cell & Method & Risk MSE & Survival $\times1000$ & Arm 0 $\times1000$ & Arm 1 $\times1000$ \\
\midrule\endhead
\midrule\multicolumn{6}{r}{Continued on next page}\\\endfoot
\bottomrule\endlastfoot
C01 & CT & 0.120 (0.001) & 5.65 (0.05) & 5.96 & 5.35 \\
C01 & FT & 0.312 (0.001) & 14.34 (0.06) & 15.55 & 13.13 \\
C01 & CL & 0.458 (0.001) & 22.29 (0.07) & 23.15 & 21.44 \\
C01 & CSPLINE & 0.250 (0.004) & 9.11 (0.09) & 9.65 & 8.57 \\
C01 & CNN & 0.451 (0.006) & 18.46 (0.20) & 19.49 & 17.43 \\
C01 & Oracle & 0.039 (0.001) & 2.59 (0.05) & 2.68 & 2.51 \\
C02 & CT & 0.115 (0.001) & 5.80 (0.04) & 5.76 & 5.83 \\
C02 & FT & 0.313 (0.001) & 15.58 (0.05) & 15.56 & 15.61 \\
C02 & CL & 0.454 (0.001) & 22.90 (0.04) & 22.87 & 22.93 \\
C02 & CSPLINE & 0.226 (0.002) & 9.00 (0.06) & 8.99 & 9.01 \\
C02 & CNN & 0.436 (0.004) & 19.01 (0.15) & 19.00 & 19.02 \\
C02 & Oracle & 0.034 (0.001) & 2.48 (0.03) & 2.45 & 2.51 \\
C03 & CT & 0.402 (0.004) & 17.99 (0.14) & 18.96 & 17.02 \\
C03 & FT & 0.649 (0.006) & 29.86 (0.28) & 32.15 & 27.56 \\
C03 & CL & 0.456 (0.001) & 22.11 (0.06) & 22.94 & 21.29 \\
C03 & CSPLINE & 0.253 (0.004) & 9.09 (0.09) & 9.64 & 8.53 \\
C03 & CNN & 0.442 (0.006) & 18.08 (0.21) & 19.07 & 17.09 \\
C03 & Oracle & 0.035 (0.001) & 2.46 (0.04) & 2.52 & 2.39 \\
C04 & CT & 0.240 (0.002) & 10.68 (0.10) & 11.34 & 10.02 \\
C04 & FT & 0.428 (0.003) & 19.83 (0.13) & 21.49 & 18.18 \\
C04 & CL & 0.455 (0.001) & 22.10 (0.06) & 23.02 & 21.17 \\
C04 & CSPLINE & 0.248 (0.004) & 9.10 (0.09) & 9.68 & 8.51 \\
C04 & CNN & 0.451 (0.006) & 18.48 (0.20) & 19.57 & 17.40 \\
C04 & Oracle & 0.036 (0.001) & 2.51 (0.04) & 2.63 & 2.40 \\
C05 & CT & 0.099 (0.001) & 4.35 (0.04) & 4.63 & 4.07 \\
C05 & FT & 0.314 (0.001) & 13.90 (0.06) & 15.10 & 12.70 \\
C05 & CL & 0.439 (0.001) & 20.60 (0.04) & 21.43 & 19.78 \\
C05 & CSPLINE & 0.110 (0.001) & 4.43 (0.04) & 4.71 & 4.16 \\
C05 & CNN & 0.301 (0.003) & 11.86 (0.12) & 12.58 & 11.13 \\
C05 & Oracle & 0.018 (0.000) & 1.24 (0.02) & 1.29 & 1.19 \\
C06 & CT & 0.097 (0.001) & 4.52 (0.03) & 4.52 & 4.52 \\
C06 & FT & 0.313 (0.001) & 15.02 (0.04) & 15.01 & 15.02 \\
C06 & CL & 0.440 (0.001) & 21.43 (0.03) & 21.43 & 21.42 \\
C06 & CSPLINE & 0.102 (0.001) & 4.40 (0.03) & 4.39 & 4.41 \\
C06 & CNN & 0.293 (0.002) & 12.33 (0.09) & 12.31 & 12.34 \\
C06 & Oracle & 0.016 (0.000) & 1.19 (0.01) & 1.19 & 1.19 \\
\end{longtable}
\noindent Parentheses give Monte Carlo SEs. Survival MSE averages equally across the two arms and ten evaluation times on 1000 independent covariate vectors. Risk MSE centers the control arm log risk. The arm specific columns use the same covariates. Projected methods are treatment analyses and do not replace the direct fit prediction estimator.
\endgroup

\begingroup\footnotesize\setlength{\tabcolsep}{3.3pt}
\begin{longtable}{llrrrr}
\caption{Spline comparison on the original 4500 datasets.}\label{tab:legacyspline}\\
\toprule
Setting & Method & Risk MSE & Survival $\times1000$ & RMSE & Cover (\%) \\
\midrule\endfirsthead
\multicolumn{6}{l}{\textit{\tablename~\thetable\ continued}}\\
\toprule
Setting & Method & Risk MSE & Survival $\times1000$ & RMSE & Cover (\%) \\
\midrule\endhead
\midrule\multicolumn{6}{r}{Continued on next page}\\\endfoot
\bottomrule\endlastfoot
Shared & CT & 0.116 (0.002) & 6.37 (0.08) & 0.178 & 94.2 (1.0) \\
Shared & CSPLINE & 0.228 (0.004) & 10.33 (0.15) & 0.192 & 95.0 (1.0) \\
Shared & CNN & 0.434 (0.005) & 21.92 (0.21) & 0.187 & NA \\
Linear drift & CT & 0.120 (0.002) & 5.68 (0.08) & 0.182 & 94.4 (1.0) \\
Linear drift & CSPLINE & 0.249 (0.006) & 9.03 (0.13) & 0.194 & 95.4 (0.9) \\
Linear drift & CNN & 0.445 (0.008) & 18.11 (0.26) & 0.185 & NA \\
Complex drift & CT & 0.392 (0.002) & 17.74 (0.11) & 0.185 & 93.8 (1.1) \\
Complex drift & CSPLINE & 0.238 (0.005) & 9.44 (0.14) & 0.200 & 94.2 (1.0) \\
Complex drift & CNN & 0.610 (0.008) & 27.34 (0.31) & 0.190 & NA \\
Small source & CT & 0.403 (0.006) & 18.20 (0.21) & 0.193 & 93.8 (1.1) \\
Small source & CSPLINE & 0.245 (0.005) & 8.93 (0.12) & 0.207 & 92.2 (1.2) \\
Small source & CNN & 0.434 (0.008) & 17.87 (0.26) & 0.188 & NA \\
Linear truth & CT & 0.052 (0.001) & 3.15 (0.06) & 0.192 & 93.8 (1.1) \\
Linear truth & CSPLINE & 0.231 (0.006) & 8.97 (0.15) & 0.211 & 92.4 (1.2) \\
Linear truth & CNN & 0.197 (0.009) & 8.34 (0.26) & 0.194 & NA \\
Covariate shift & CT & 0.093 (0.002) & 5.76 (0.08) & 0.158 & 95.8 (0.9) \\
Covariate shift & CSPLINE & 0.245 (0.009) & 9.58 (0.14) & 0.167 & 94.4 (1.0) \\
Covariate shift & CNN & 0.328 (0.008) & 16.53 (0.28) & 0.166 & NA \\
Null, linear & CT & 0.115 (0.002) & 5.75 (0.07) & 0.160 & 96.2 (0.9) \\
Null, linear & CSPLINE & 0.237 (0.006) & 9.07 (0.14) & 0.171 & 94.6 (1.0) \\
Null, linear & CNN & 0.428 (0.007) & 18.75 (0.27) & 0.172 & NA \\
Null, complex & CT & 0.387 (0.002) & 19.06 (0.12) & 0.174 & 94.2 (1.0) \\
Null, complex & CSPLINE & 0.216 (0.004) & 9.44 (0.13) & 0.184 & 93.4 (1.1) \\
Null, complex & CNN & 0.596 (0.008) & 29.02 (0.32) & 0.181 & NA \\
Null, non-PH & CT & NA & 5.88 (0.07) & 0.164 & 95.8 (0.9) \\
Null, non-PH & CSPLINE & NA & 8.45 (0.12) & 0.176 & 95.8 (0.9) \\
Null, non-PH & CNN & NA & 16.55 (0.25) & 0.174 & NA \\
\end{longtable}
\noindent Each setting contains 500 replications. All spline fits succeeded. CSPLINE uses exactly the original random patient and evaluation streams, including the original random arm prediction evaluation. Its survival MSE is therefore paired with the original experiment, but not identical in definition to the new two arm evaluation. There is no static log risk target under non-PH. CNN intervals are not reported. Under misspecification, coefficient RMSE and coverage describe discrepancies from the generating coefficient, not a general consistency claim.
\endgroup

\begingroup\footnotesize\setlength{\tabcolsep}{3.3pt}
\begin{longtable}{rllrrrrrr}
\caption{Complete certified source experiment, with $N=n$.}\label{tab:constructivefull}\\
\toprule
$n$ & Law & Method & Bias & RMSE & Cover (\%) & SE/SD & Length & Reject (\%) \\
\midrule\endfirsthead
\multicolumn{9}{l}{\textit{\tablename~\thetable\ continued}}\\
\toprule
$n$ & Law & Method & Bias & RMSE & Cover (\%) & SE/SD & Length & Reject (\%) \\
\midrule\endhead
\midrule\multicolumn{9}{r}{Continued on next page}\\\endfoot
\bottomrule\endlastfoot
300 & Alt. & CT C & -0.009 & 0.173 & 96.0 (0.9) & 1.04 & 0.707 & 55.4 (2.2) \\
300 & Alt. & PT C & -0.004 & 0.174 & 96.2 (0.9) & 1.05 & 0.720 & 48.8 (2.2) \\
300 & Alt. & Oracle & -0.012 & 0.174 & 96.2 (0.9) & 1.03 & 0.703 & 55.4 (2.2) \\
300 & Alt. & Current Fourier & -0.014 & 0.178 & 95.4 (0.9) & 1.02 & 0.713 & 54.0 (2.2) \\
300 & Null & CT C & 0.015 & 0.170 & 95.8 (0.9) & 0.99 & 0.662 & 4.2 (0.9) \\
300 & Null & PT C & 0.016 & 0.174 & 95.4 (0.9) & 0.99 & 0.672 & 4.6 (0.9) \\
300 & Null & Oracle & 0.015 & 0.169 & 95.2 (1.0) & 0.99 & 0.657 & 4.8 (1.0) \\
300 & Null & Current Fourier & 0.016 & 0.172 & 95.2 (1.0) & 0.99 & 0.667 & 4.8 (1.0) \\
1000 & Alt. & CT C & -0.001 & 0.099 & 94.8 (1.0) & 0.99 & 0.385 & 96.0 (0.9) \\
1000 & Alt. & PT C & -0.001 & 0.100 & 94.0 (1.1) & 0.99 & 0.387 & 95.4 (0.9) \\
1000 & Alt. & Oracle & -0.002 & 0.099 & 95.0 (1.0) & 0.99 & 0.384 & 96.2 (0.9) \\
1000 & Alt. & Current Fourier & -0.003 & 0.099 & 95.2 (1.0) & 1.00 & 0.386 & 96.0 (0.9) \\
1000 & Null & CT C & 0.011 & 0.092 & 95.2 (1.0) & 1.01 & 0.361 & 4.8 (1.0) \\
1000 & Null & PT C & 0.010 & 0.093 & 94.8 (1.0) & 1.00 & 0.362 & 5.2 (1.0) \\
1000 & Null & Oracle & 0.011 & 0.092 & 95.0 (1.0) & 1.00 & 0.360 & 5.0 (1.0) \\
1000 & Null & Current Fourier & 0.011 & 0.093 & 95.2 (1.0) & 1.00 & 0.362 & 4.8 (1.0) \\
3000 & Alt. & CT C & 0.001 & 0.057 & 94.0 (1.1) & 0.99 & 0.222 & 100.0 (0.0) \\
3000 & Alt. & PT C & 0.000 & 0.057 & 94.4 (1.0) & 0.99 & 0.222 & 100.0 (0.0) \\
3000 & Alt. & Oracle & 0.000 & 0.057 & 93.8 (1.1) & 0.99 & 0.222 & 100.0 (0.0) \\
3000 & Alt. & Current Fourier & 0.000 & 0.057 & 93.8 (1.1) & 0.99 & 0.222 & 100.0 (0.0) \\
3000 & Null & CT C & 0.000 & 0.052 & 95.6 (0.9) & 1.02 & 0.208 & 4.4 (0.9) \\
3000 & Null & PT C & 0.000 & 0.052 & 95.8 (0.9) & 1.03 & 0.208 & 4.2 (0.9) \\
3000 & Null & Oracle & 0.000 & 0.052 & 95.6 (0.9) & 1.02 & 0.208 & 4.4 (0.9) \\
3000 & Null & Current Fourier & 0.000 & 0.052 & 96.0 (0.9) & 1.02 & 0.208 & 4.0 (0.9) \\
\end{longtable}
\noindent Each cell has 500 replications. CT C is direct certified transfer and PT C its cross-fitted Fourier score. All 12,000 method attempts succeeded. Current Fourier is a target-only four coordinate reference containing the true function. Parentheses give Monte Carlo SEs in percentage points. The known-risk oracle is not implementable.
\endgroup

\begin{table}[tbp]\centering\small
\caption{Attained source and score diagnostics in the certified construction.}\label{tab:constructivediagnostics}
\setlength{\tabcolsep}{4pt}\begin{tabular}{rlrrrrrrr}
\toprule
$n=N$ & Law & Knots & $M$ & $L_2$ error & Iter. & Gap $\times10^8$ & Min. deriv. & Clip (\%) \\
\midrule
300 & Alt. & 4 & 2 & 0.164 & 381 & 9.99971 & 0.0367 & 0.0 \\
300 & Null & 4 & 2 & 0.168 & 383 & 9.99791 & 0.0603 & 0.0 \\
1000 & Alt. & 4 & 2 & 0.114 & 377 & 9.99825 & 0.0774 & 0.0 \\
1000 & Null & 4 & 2 & 0.115 & 377 & 9.99983 & 0.0864 & 0.0 \\
3000 & Alt. & 5 & 4 & 0.072 & 472 & 9.99909 & 0.0870 & 0.0 \\
3000 & Null & 5 & 4 & 0.072 & 469 & 9.99967 & 0.0999 & 0.0 \\
\bottomrule\end{tabular}
\par\smallskip\begin{minipage}{\textwidth}\footnotesize Source $L_2$ error is centered and averaged over 500 historical fits per cell using 10,000 independent uniform covariates per fit. Iter. is the mean projected gradient iteration count. Gap is the largest computed certificate, multiplied by $10^8$. Min. deriv. and Clip are extrema across all 2500 evaluation folds in the cell. $M$ is the Fourier dictionary size. \end{minipage}\end{table}

\section{Additional comparisons and finite sample inference}\label{app:followups}
\subsection{Historical splines with the same target calibration}\label{app:historicalspline}
CT-spline fits a historical Cox model using twelve natural cubic spline columns (four knots at $-1,-1/3,1/3,1$ in each of four coordinates, with three centered columns per coordinate) and six pairwise products. The deterministic centering and whitening are identical to CSPLINE and do not use outcome information. Historical fitting uses separate historical risk sets, the same numerical ridge $10^{-8}$, coefficient bounds $[-5,5]$, score tolerance $2\times10^{-5}$, and 250 iteration cap as the existing spline implementation. The function is frozen for target fitting. The target estimates treatment and the four linear correction coefficients, with its own Breslow baseline. This is a comparator for the historical learner, with the same target correction structure as CT.

The full benchmark was specified after reviewing the earlier neural results, with no pilot or tuning on the benchmark outcomes. All six existing cells and every replication are retained. As in the other direct fits, source induced bias under an alternative is not removed by sandwich variance alone.

\begin{table}[tbp]\centering\footnotesize
\caption{Historical spline and neural borrowing on the identical C cell records. Survival MSE is multiplied by $10^3$. Coverage and rejection are percentages.}\label{tab:historicalspline}
\setlength{\tabcolsep}{3.5pt}\begin{tabular}{llrrrrr}\toprule
Cell & Method & Risk MSE & Survival & RMSE & Cover & Reject\\\midrule
C01 & CT-spline & 0.062 & 3.486 & 0.180 & 94.4 & 51.9 \\
C01 & CT & 0.120 & 5.654 & 0.183 & 94.4 & 51.2 \\
C01 & CSPLINE & 0.250 & 9.111 & 0.201 & 93.1 & 52.6 \\
\addlinespace
C02 & CT-spline & 0.057 & 3.439 & 0.171 & 95.3 & 4.7 \\
C02 & CT & 0.115 & 5.795 & 0.175 & 94.8 & 5.2 \\
C02 & CSPLINE & 0.226 & 8.999 & 0.190 & 93.3 & 6.7 \\
\addlinespace
C03 & CT-spline & 0.203 & 8.164 & 0.180 & 95.1 & 52.0 \\
C03 & CT & 0.402 & 17.988 & 0.180 & 94.7 & 43.2 \\
C03 & CSPLINE & 0.253 & 9.088 & 0.194 & 94.4 & 56.0 \\
\addlinespace
C04 & CT-spline & 0.085 & 4.302 & 0.181 & 94.3 & 52.9 \\
C04 & CT & 0.240 & 10.681 & 0.187 & 94.8 & 50.5 \\
C04 & CSPLINE & 0.248 & 9.096 & 0.200 & 93.5 & 52.6 \\
\addlinespace
C05 & CT-spline & 0.040 & 2.144 & 0.125 & 94.4 & 80.2 \\
C05 & CT & 0.099 & 4.351 & 0.126 & 94.4 & 80.7 \\
C05 & CSPLINE & 0.110 & 4.432 & 0.132 & 93.8 & 80.1 \\
\addlinespace
C06 & CT-spline & 0.039 & 2.148 & 0.119 & 94.2 & 5.8 \\
C06 & CT & 0.097 & 4.523 & 0.122 & 94.8 & 5.2 \\
C06 & CSPLINE & 0.102 & 4.401 & 0.122 & 94.0 & 5.9 \\
\addlinespace
\bottomrule\end{tabular}\end{table}
\begin{table}[tbp]\centering\footnotesize
\caption{Paired differences, historical spline minus neural CT. Entries are differences (Monte Carlo SE). Survival differences are multiplied by $10^3$.}\label{tab:historicalsplinepaired}
\setlength{\tabcolsep}{4pt}\begin{tabular}{lrrr}\toprule
Cell & Risk MSE & Survival MSE & Treatment RMSE\\\midrule
C01 & -0.0577 (0.0007) & -2.1675 (0.0288) & -0.0033 (0.0012) \\
C02 & -0.0579 (0.0005) & -2.3558 (0.0213) & -0.0035 (0.0008) \\
C03 & -0.1991 (0.0040) & -9.8242 (0.1373) & -0.0006 (0.0031) \\
C04 & -0.1550 (0.0019) & -6.3793 (0.0867) & -0.0054 (0.0020) \\
C05 & -0.0587 (0.0007) & -2.2073 (0.0278) & -0.0008 (0.0009) \\
C06 & -0.0585 (0.0005) & -2.3754 (0.0207) & -0.0029 (0.0006) \\
\bottomrule\end{tabular}
\par\smallskip\begin{minipage}{\textwidth}\footnotesize Treatment RMSE differences use the paired-delta method Monte Carlo SE. Negative entries favor CT-spline for that metric. Machine readable files also give comparisons with target-only splines, current-only neural fits and the oracle, and all-attempt failure accounting.\end{minipage}\end{table}

All 8000 historical spline attempts are retained. 0 fail the declared source or target checks. This comparison concerns the four-dimensional generating function in these cells. A spline space that works well here need not represent a different high-dimensional clinical prognostic function. Conversely, the present experiments do not show that a neural source is required for useful borrowing.
\subsection{Fixed score randomization stress under misspecification}\label{app:randomizationstress}
This separate mechanism experiment tests the observed data null with an incorrect fixed input score. It is not a comparison of historical learners. We use $X\sim\operatorname{Uniform}(-1,1)^4$, independent $A\sim\operatorname{Bernoulli}(1/2)$, exponential censoring with rate 0.18, and administrative horizon 2. The fixed offset is $f_{\rm fix}(x)=-2\cos(\pi x_2)$. Target calibration estimates four linear corrections. Under the complex PH law the prognostic risk is $f_0(x)+\sin(\pi x_2)+0.8x_1x_4$, with the original $f_0$ and cumulative baseline $0.35t^{1.2}$. Under the non-PH law the conditional survival is
\[
S(t\mid x)=\exp[-0.35t^{k(x)}\exp\{f_0(x)+0.7x_1-0.5x_2\}],\qquad k(x)=1.2\exp(0.45x_1).
\]
Both event laws are treatment independent. The offset plus linear corrections is misspecified in both laws, and the second has no static Cox log risk representation.

R01 and R02 use the complex PH law at $n=300$ and 600. R03 and R04 use the non-PH law at those sizes. Each has 2000 replications. No pilot or tuning is used. The fitting script and configuration were fixed before this block, which shares no records with the other experiments. We compare the information-only and sandwich Wald tests from the same direct fit with the calibrated randomization score from a label blind null fit. 

\begin{table}[tbp]\centering\footnotesize
\caption{Fixed score null rejection: percentage (Monte Carlo SE), followed by Wilson 95\% Monte Carlo interval.}\label{tab:randomizationstress}
\setlength{\tabcolsep}{4pt}\begin{tabular}{llrr}\toprule
Cell/law & Test & Rejection & MC interval\\\midrule
R01 & Sandwich & 5.70 (0.52) & [4.77, 6.80] \\
R01 & Information only & 18.00 (0.86) & [16.38, 19.74] \\
R01 & CR & 4.85 (0.48) & [3.99, 5.88] \\
\addlinespace
R02 & Sandwich & 6.10 (0.54) & [5.13, 7.24] \\
R02 & Information only & 18.85 (0.87) & [17.20, 20.62] \\
R02 & CR & 5.75 (0.52) & [4.81, 6.86] \\
\addlinespace
R03 & Sandwich & 6.25 (0.54) & [5.27, 7.40] \\
R03 & Information only & 18.35 (0.87) & [16.71, 20.11] \\
R03 & CR & 4.50 (0.46) & [3.68, 5.50] \\
\addlinespace
R04 & Sandwich & 5.70 (0.52) & [4.77, 6.80] \\
R04 & Information only & 18.75 (0.87) & [17.10, 20.52] \\
R04 & CR & 4.65 (0.47) & [3.81, 5.66] \\
\addlinespace
\bottomrule\end{tabular}
\par\smallskip\begin{minipage}{\textwidth}\footnotesize R01/R02 are complex PH at $n=300/600$. R03/R04 are non-PH at those sizes. Every cell has 2000 replications and a nominal 5\% test. A randomization rejection estimate can exceed 5\% through Monte Carlo variation even though its conditional population level is bounded by 5\%. Its interval is reported rather than claiming every estimate is below nominal.\end{minipage}\end{table}

All 24,000 method attempts return a result. 0 direct fits and 0 null constrained fits fail the declared optimizer checks. Under treatment dependent censoring that violates the observed data null, neither this simulation nor the randomization proposition supplies a validity guarantee.
\subsection{Strong-prognosis follow-up chain}\label{sec:focused}\label{app:confirmation}\label{app:boundaryconfirmation}\label{app:tuningvalidation}
Several follow-ups examine one law with a stronger shared prognosis. Historical risk is $c_ff_0$ and target risk $c_ff_0+0.7X_1-0.5X_2$, with the base baselines, censoring, network and four linear corrections, and prognostic amplitude $c_f=1.5$. An exploratory grid, E01 to E08, was chosen after inspecting earlier results and has 100 replications per cell. It crosses $c_f\in\{1,1.5\}$ with $(n,N)\in\{(300,6000),(300,12000),(600,12000)\}$ under $\theta_0=\log(0.7)$ and adds two $c_f=1.5$ nulls with $N=12000$. It raised two flags. In E08 ($n=600$, null) CT and CSPLINE rejected in 12\% of replications, PT-F in 9\%, and the known-risk oracle in 11\%. In E02 ($n=300$, $N=6000$, alternative) CT and PT-F covered 92\% and 90\%.

Each flag was re-examined in an independent follow-up whose law, methods and replication counts were fixed before its data were generated: H01 to H04 for E07 and E08 ($N=12000$), B01/B02 for E02 ($N=6000$), a bounded tuning study validated on V01/V02, a CT patient bootstrap on S01/S02 (Appendix~\ref{app:remainingflags}), and the full-procedure projected bootstrap F01/F02 (Appendix~\ref{app:fullbootstrap}). Table~\ref{tab:strongchain} collects the primary entries. The complete exploratory grid, every method in every cell, Wilson intervals, coefficient summaries, and the tuning grid are in Archive Tables~1 to~13. Historical fits are shared across the null and alternative cells of a replication index, so cells are dependent across a row but each has independent replications.

\begin{table}[tbp]\centering\footnotesize
\caption{Strong-prognosis follow-up chain ($c_f=1.5$ unless stated): null rejection / alternative coverage, in percent, for nominal 5\% tests and 95\% intervals.}\label{tab:strongchain}
\setlength{\tabcolsep}{4pt}\begin{tabular}{lrrrllll}\toprule
Block & $n$ & $N$ & Null/alt. & CT Wald & PT-F Wald & CR & Companion interval \\\midrule
E07/E04 & 300 & 12000 & 100/100 & 5.0 / 95.0 & 6.0 / 95.0 & NA & \\
E08/E06 & 600 & 12000 & 100/100 & 12.0 / 93.0 & 9.0 / 90.0 & NA & \\
E02 & 300 & 6000 & --/100 & -- / 92.0 & -- / 90.0 & NA & \\
\midrule
H01/H03 & 300 & 12000 & 2000/1000 & 5.75 / 93.9 & 6.05 / 93.4 & 5.05 & \\
H02/H04 & 600 & 12000 & 2000/1000 & 5.15 / 95.6 & 4.85 / 94.9 & 4.70 & \\
B01/B02 & 300 & 6000 & 2000/1000 & 5.45 / 93.40 & 5.75 / 93.40 & 4.45 & \\
V01/V02 & 300 & 6000 & 2000/1000 & 5.55 / 93.40 & 6.15 / 94.00 & 4.80 & CT boot.\ 5.10 / 94.40 \\
S01/S02 & 300 & 6000 & 2000/1000 & 6.70 / 93.60 & 7.10 / 94.40 & 5.75 & CT boot.\ 6.30 / 94.00 \\
F01/F02 & 300 & 6000 & 2000/1000 & 6.15 / 93.70 & 6.65 / 93.80 & 5.25 & PT-F boot.\ 4.22 / 95.56 \\
\addlinespace
Pooled B--F & 300 & 6000 & 8000/4000 & 5.96 / 93.53 & 6.41 / 93.90 & 5.06 & \\
\bottomrule\end{tabular}
\par\smallskip\begin{minipage}{\textwidth}\footnotesize E cells are exploratory and the others are follow-ups. Exploratory proportions have Monte Carlo SE about 2.2 percentage points near 5\% and 3 points near 90\%; follow-up null rejections about 0.5 and alternative coverages about 0.75. Pooled Wilson 95\% Monte Carlo intervals: CT 5.46 to 6.50 and 92.72 to 94.25, PT-F 5.90 to 6.97 and 93.12 to 94.60, CR 4.60 to 5.56. The pooled row combines complete batches of the unchanged procedures with different seeds; no batch is omitted. All follow-up method attempts succeeded except 17 unavailable PT-F bootstrap intervals in F01/F02. CR is a null test.\end{minipage}\end{table}

Neither flag recurred at its exploratory magnitude. In H02 CT rejects 5.15\% of 2000 nulls (Wilson interval 4.26 to 6.21) rather than 12\%, and in B02 CT and PT-F both cover 93.40\% rather than 92\% and 90\%. The follow-ups do show a small, consistent departure at $n=300$ under this law. Pooling the four independent $N=6000$ batches, the CT and PT-F Wald intervals cover 93.53\% and 93.90\% of 4000 alternatives and reject 5.96\% and 6.41\% of 8000 nulls, whereas CR rejects 5.06\%. The known-risk oracle departs less in the same batches, with coverage 93.8 to 94.8\% and null rejection 5.35 to 6.55\% (Archive Tables~7 and~10, Table~\ref{tab:fullbootstrapcomplete}), so part of the departure reflects finite-sample studentization at this target size rather than the historical learner.

\paragraph{Bounded tuning.} Because B02 undercovered, the first 500 null and 500 alternative B01/B02 datasets were reused as development data for fold counts $K\in\{2,3,5\}$ and ridge multipliers $c_\lambda\in\{0.3,1,3\}$, with a selection rule fixed before those outcomes were examined: screen on Wilson intervals for coverage and null rejection, then minimize alternative RMSE (Archive Table~8). The rule selected $(K,c_\lambda)=(2,3)$. On the 2000 null and 1000 alternative V01/V02 datasets, generated after the choice was frozen, the selected variant covered 92.80\% with RMSE 0.222, against 94.00\% and 0.202 for the prespecified PT-F (paired coverage difference $-1.20$ percentage points, 95\% Monte Carlo interval $-2.74$ to 0.34). The tuning attempt therefore did not improve coverage, and the prespecified procedure remains primary. A sandwich-scaled profile likelihood interval for CT, defined and justified in Archive Section~1.5, changed V02 coverage by $-0.10$ points ($-0.30$ to 0.10) and is not a remedy.

\subsection{Patient bootstrap companions and a tuned target-only neural score}\label{app:remainingflags}
\paragraph{A fixed patient bootstrap interval.}
We next examine direct fit uncertainty by resampling complete target patient records, following the subject resampling approach to censored data \citep{efron1981}. We hold $\widehat f$ fixed and refit both treatment and correction coefficients in every resample. Denote the number of resamples by $B_{\rm bt}$. For $b=1,\ldots,B_{\rm bt}$, draw independent indices $i_{b,r}^*$ uniformly from $\{1,\ldots,n\}$ for $r=1,\ldots,n$, and set $O_{b,r}^*=O_{i_{b,r}^*}$. Let $\widehat\theta_b^*$ and $\widehat{\operatorname{se}}_{{\rm sw},b}^*$ be the direct estimate and its sandwich standard error in that resample. Compute
\[
T_b^*=\frac{\widehat\theta_b^*-\widehat\theta}{\widehat{\operatorname{se}}_{{\rm sw},b}^*}.
\]
For $p\in(0,1)$, let $\widehat c_p^*$ be the empirical $p$ quantile of these studentized values. Our companion interval is
\[
\mathcal C_{\rm bt}=\left[
\widehat\theta-\widehat c_{1-\alpha/2}^*\widehat{\operatorname{se}}_{\rm sw},\,
\widehat\theta-\widehat c_{\alpha/2}^*\widehat{\operatorname{se}}_{\rm sw}
\right].
\]
Algorithm~\ref{alg:patientbootstrap} specifies the calculation. Covariates, treatment, time and event indicator are resampled together. We recompute all risk sets, allowing repeated event times with the Breslow convention. The repeated records are bootstrap multiplicities, not new original patients. The index remains bounded by the existing output and coefficient constraints. A single failed resample makes the interval fail. No failed resample is discarded.

\begin{algorithm}[tbp]
\caption{Patient bootstrap companion to direct calibration}\label{alg:patientbootstrap}
\begin{algorithmic}[1]
\Require Independent historical fit $\widehat f$, target records $O_1,\ldots,O_n$, correction $z$, direct fit $\widehat\beta$, sandwich SE $\widehat{\operatorname{se}}_{\rm sw}$, level $\alpha$ and resample count $B_{\rm bt}$.
\For{$b=1,\ldots,B_{\rm bt}$}
 \State Draw $i_{b,1}^*,\ldots,i_{b,n}^*$ uniformly with replacement and form $O_{b,r}^*=O_{i_{b,r}^*}$.
 \State Hold $\widehat f$ fixed. Refit target calibration, initialized at $\widehat\beta$, and recompute \eqref{eq:empIF} and \eqref{eq:sandwich} on these records.
 \State If the fit or sandwich variance fails the declared checks, report interval failure and retain its record. Otherwise compute $T_b^*$.
\EndFor
\State If every resample succeeds, compute $\widehat c_{\alpha/2}^*,\widehat c_{1-\alpha/2}^*$ and report $\mathcal C_{\rm bt}$.
\end{algorithmic}
\end{algorithm}

We use $B_{\rm bt}=999$, $\alpha=0.05$ and linear empirical quantile interpolation. These choices, the 2000/1000 validation counts and the failure rule are recorded before any bootstrap outcomes. Resamples within each dataset use an independent random stream. A development check on the first 100 null and 100 alternative B01/B02 indices is a diagnostic only and chose no setting. Validation reuses the complete V01/V02 targets and sources, so it adds method comparisons rather than more patients. It involves 2,997,000 bootstrap refits. Each resample has the original coefficient bounds, numerical ridge, score tolerance and information guard. Initialization at the full target fit changes computation, not the objective or inferential target.

We give a first-order score expansion to justify the bootstrap, following the argument for regular $M$ estimation \citep{chenghuang2010}. It requires the direct fit negligible bias conditions under an alternative. It does not follow from randomization alone. Conditional on $\widehat f$ and the target data, denote bootstrap probability by $\Pp^*$ and write $m_{b,i}^*$ for the number of times record $i$ appears in resample $b$. Under the bounded risk, positive follow-up and finite-dimensional interior root conditions used for the direct influence expansion, linearization of the resampled score gives
\[
\sqrt n(\widehat\beta_b^*-\widehat\beta)
=\widehat I_W^{-1}\frac1{\sqrt n}\sum_i(m_{b,i}^*-1)
\widehat\varphi_i(\widehat\eta,W)+o_{\Pp^*}(1)
\]
in probability over the original records. One way to verify the conditional remainder is to express the Cox score using empirical event integrals and risk-set averages as in \eqref{eq:calibrationderivatives}. Their denominators remain uniformly positive with probability tending to one. With the independent offset held fixed, the time-indexed indicators and bounded finite-dimensional coefficient paths have integrable uniform entropy. Their conditional bootstrap empirical process is tight, and Taylor expansion of the ratios yields the same centered influence derivative \eqref{eq:empIF}. Products of empirical perturbations are $o_{\Pp^*}(n^{-1/2})$. Local positive curvature then yields the displayed root expansion. Breslow multiplicities represent the resampled empirical measure in this argument. The same bounds give consistency of its resampled influence covariance.

The conditional multinomial central limit theorem therefore gives $T_b^*\Rightarrow N(0,1)$ in probability when the treatment variance stays positive and numerical score errors are $o_\Pp(n^{-1/2})$, also conditionally for the refits. For ideal interior roots, $B_{\rm bt}\to\infty$ makes the empirical quantiles converge to the corresponding normal quantiles. Algorithm~\ref{alg:patientbootstrap}, which fails if any refit fails, additionally requires $B_{\rm bt}\Pp^*(\text{a refit fails})\to0$ in probability along such an asymptotic sequence. A union bound then makes its failure guard inactive. Combined with Theorem~\ref{thm:plugin} and its negligible historical bias conditions, this gives asymptotic coverage $1-\alpha$ for $\theta_0$. The analogous working fit argument under Theorem~\ref{thm:null} gives asymptotic inclusion of zero under the observed data null. The fixed 999 resample implementation has additional Monte Carlo quantile variation and is assessed empirically here. This argument does not assert that 999 resamples guarantee exact coverage or certify the vanishing failure requirement. Holding the source fixed is justified only for these first-order statements: it does not repair a first-order historical error functional, a misspecified alternative coefficient, or arbitrary nonregular learning. We make no higher order accuracy claim.

\paragraph{Evaluation.} On the V01/V02 records, the CT bootstrap interval covers 94.40\% of alternatives (Wilson interval 92.80 to 95.66) against 93.40\% for the Wald interval from the same fits, a paired difference of 1.00 percentage point (95\% Monte Carlo interval 0.32 to 1.68). Its mean length is 0.701 instead of 0.682, and it rejects 5.10\% of nulls (4.22 to 6.15). None of the 2,997,000 bootstrap refits failed. Archive Tables~9 and~10 give every procedure in this evaluation, including the tuned PT-F variant and the profile interval.

\paragraph{Tuned target-only neural score and spline bootstrap.}
The current neural tuning study uses 200 development targets (100 alternatives and 100 nulls) at $n=300,N=3000$, under the C01/C02 laws. A five-fold split supplies 1000 overlapping training sets. On each training set, an inner 20\% validation split selects the update count and supplies held-out Cox loss. The grid combines widths $(24,12)$, $(12,6)$ and $(8,4)$ with penalties $10^{-4},10^{-3},10^{-2}$. Other fitting choices are unchanged. The configuration minimizing mean inner validation loss is frozen before independent evaluation on 1000 alternatives and 2000 nulls. Treatment truth and coverage are not selection criteria. The selected widths are $(24,12)$ with penalty $0.01$. Prespecified and tuned CN-S share evaluation patients and folds and use the deterministic $B_{18}$ projection. All one-step analyses succeed. Archive Table~11 gives the entire development grid and Table~\ref{tab:remainingneural} gives the matched evaluation.

The spline bootstrap applies Algorithm~\ref{alg:patientbootstrap} with zero historical offset and all 18 spline columns as target nuisance functions. It refits those coefficients and treatment on each of 999 complete patient resamples, using Breslow multiplicities and the complete sandwich SE. A failed refit fails the interval. The C01/C02 evaluation contains 1000 alternatives and 2000 nulls independent of the original C cell experiment. The same bootstrap is evaluated for CT at $n=300,N=6000$ and amplitude 1.5, again with 1000 alternatives and 2000 nulls. These follow-ups were motivated by earlier interval diagnostics and retain fixed methods and replication counts. Bootstrap intervals in Table~\ref{tab:remainingbootstrap} are distinct from the projected bootstrap below. No bootstrap corrects persistent bias of a misspecified alternative working model.

{\footnotesize
\begin{longtable}{llrrrrrr}
\caption{Current neural tuning: fixed independent validation for the one-step CN-S estimator.}\label{tab:remainingneural}\\
\toprule Cell & Procedure & Success & Bias & RMSE & Cover (\%) & Reject (\%) & Length\\
\midrule\endfirsthead
\multicolumn{8}{l}{\textit{\tablename~\thetable\ continued}}\\
\toprule Cell & Procedure & Success & Bias & RMSE & Cover (\%) & Reject (\%) & Length\\
\midrule\endhead
\midrule\multicolumn{8}{r}{Continued on next page}\\\endfoot
\bottomrule\endlastfoot
N01 & Prespecified CN-S & 1000/1000 & 0.018 & 0.226 & 91.80 & 39.30 & 0.802 \\
N01 & Tuned CN-S & 1000/1000 & 0.020 & 0.208 & 93.20 & 39.60 & 0.780 \\
N02 & Prespecified CN-S & 2000/2000 & 0.008 & 0.208 & 92.75 & 7.25 & 0.764 \\
N02 & Tuned CN-S & 2000/2000 & 0.006 & 0.195 & 94.15 & 5.85 & 0.743 \\
\end{longtable}
}

\begin{table}[tbp]\centering\small
\caption{Bootstrap companions on complete fixed validation batches.}\label{tab:remainingbootstrap}
\setlength{\tabcolsep}{3.5pt}\begin{tabular}{llrrr}\toprule
Cell & Procedure & Cover (\%) & Cover 95\% MC interval & Reject (\%)\\\midrule
C01 & Spline bootstrap & 95.00 & [93.47, 96.19] & 44.90 \\
C02 & Spline bootstrap & 95.60 & [94.61, 96.41] & 4.40 \\
S01 & CT bootstrap & 93.70 & [92.55, 94.68] & 6.30 \\
S02 & CT bootstrap & 94.00 & [92.35, 95.31] & 51.30 \\
\bottomrule\end{tabular}
\par\smallskip\begin{minipage}{\textwidth}\footnotesize C01/S02 are alternatives (1000 repetitions each). C02/S01 are nulls (2000 each). Each interval uses 999 full target refits. These four comparisons total 5,994,000 bootstrap refits on reused targets, not more independent target datasets.\end{minipage}\end{table}

The S01/S02 records also give CT Wald intervals, with alternative coverage 93.60\% and null rejection 6.70\%. Relative to them, the CT bootstrap changes alternative coverage by $+0.40$ percentage points (paired 95\% Monte Carlo interval $-0.15$ to 0.95) and null rejection by $-0.40$ ($-0.86$ to 0.06). Pooling S01/S02 with the independent V01/V02 evaluation, CT bootstrap alternative coverage is 94.20\% (Wilson interval 93.09 to 95.14, 2000 replications) and null rejection is 5.70\% (5.02 to 6.46, 4000 replications). The bootstrap therefore improves CT coverage slightly but leaves a small excess null rejection at this law.

\subsection{Full-procedure bootstrap for the projected score}\label{app:fullbootstrap}
We evaluate full-procedure resampling for projected intervals. The historical function, five-fold point estimator, deterministic $B_{18}$ dictionary, ridge and clipping remain unchanged. Here we resample complete patients and refit both target calibration and projection, so their dependence across overlapping training complements enters the bootstrap calculation. This is a new interval companion, not a new source learner or an unrestricted projection guarantee.

\paragraph{Resampling and interval.}
We condition on the independent $\widehat f$ and original partition $\mathcal I_1,\ldots,\mathcal I_K$. Retain $m_k$ and $w_k$ from Section~\ref{sec:projectedmethod}. For each resample $b$ and fold $k$, draw $m_k$ indices independently with replacement from $\mathcal I_k$, and denote the resulting multiset of complete records by $\mathcal I_{k,b}^*$. Its complement is the union of the other resampled multisets. All copies of an original patient stay in its original fold, so an evaluation patient's copies never enter its own nuisance training complement. We hold $\widehat f$ fixed, but repeat every target calibration and projection step of Algorithm~\ref{alg:score}, including training risk sets, ridge and clipping, on these multisets. Breslow multiplicities handle repeated observed event times.

Let $\widehat\theta_b^{{\rm score},*}$ and $\widehat{\operatorname{se}}_{{\rm score},b}^*$ be the complete resampled outputs of \eqref{eq:scoreest} and \eqref{eq:se}. Reusing $B_{\rm bt}$ for a bootstrap count, define
\[
T_{{\rm score},b}^*=
\frac{\widehat\theta_b^{{\rm score},*}-\widehat\theta^{\rm score}}
{\widehat{\operatorname{se}}_{{\rm score},b}^*}.
\]
Write $\widehat c_{{\rm score},p}^*$ for its empirical $p$ quantile. The companion interval is
\[
\mathcal C_{{\rm score},{\rm bt}}=
\left[\widehat\theta^{\rm score}-\widehat c_{{\rm score},1-\alpha/2}^*
\widehat{\operatorname{se}}_{\rm score},\,
\widehat\theta^{\rm score}-\widehat c_{{\rm score},\alpha/2}^*
\widehat{\operatorname{se}}_{\rm score}\right].
\]
The point estimate and original studentizer are unchanged. Interval shape and width can differ, and this cannot improve point estimation RMSE by construction. We fix $B_{\rm bt}=999$, $\alpha=0.05$ and linear quantile interpolation. Any failed original fold or resampled fold makes the companion interval unavailable. We retain failed resamples and distinguish conditional on success coverage from the fraction of all attempts that return a covering interval.

\begin{algorithm}[tbp]
\caption{Full procedure patient bootstrap for the projected score}\label{alg:fullscorebootstrap}
\begin{algorithmic}[1]
\Require Independent historical fit $\widehat f$, target records, original folds $\mathcal I_k$, the settings of Algorithm~\ref{alg:score}, level $\alpha$ and $B_{\rm bt}$.
\State Run Algorithm~\ref{alg:score}. Retain $\widehat\theta^{\rm score}$, $\widehat{\operatorname{se}}_{\rm score}$ and original training coefficients, or report failure.
\For{$b=1,\ldots,B_{\rm bt}$}
 \State Independently draw $m_k$ complete patients with replacement within each $\mathcal I_k$, forming $\mathcal I_{k,b}^*$. Preserve every original patient's fold membership.
 \State Hold $\widehat f$ fixed. Repeat Algorithm~\ref{alg:score} on these multisets, refitting target calibration and projection and rebuilding all risk sets.
 \State Record every fitting/derivative failure. If all folds succeed, compute $T_{{\rm score},b}^*$.
\EndFor
\State Report $\mathcal C_{{\rm score},{\rm bt}}$ only if every resample succeeds. Otherwise report interval failure. Retain all resampled records and estimates.
\end{algorithmic}
\end{algorithm}

\paragraph{Conditional finite-dimensional justification.}
We justify the bootstrap under explicit regularity conditions on the fitted target functional. The empirical process bootstrap and its delta method apply to a regular target functional \citep[Chapters 20 and 23]{vandervaart1998}. \citet{chenghuang2010} give related semiparametric $M$ estimation theory. Here the resampled nuisance is finite-dimensional because the historical learner is external and held fixed.

For a fixed supplied bounded function $f$, let $P_{n,k}$ be the empirical measure on $\mathcal I_k$. Denote by $\mathcal T_n(P_1,\ldots,P_K,f)$ the entire population version of the specified calibration, projection, and one-step calculation, retaining its deterministic ridge and clipping conventions. Define its working target by $\theta_{{\rm work},n}(f)=\mathcal T_n(P_T,\ldots,P_T,f)$. This definition need not equal $\theta_0$ under an incompatible offset or inadequate projection.

\begin{proposition}[Conditional finite-dimensional bootstrap]\label{prop:fullscorebootstrap}
Fix $K$ and a bounded deterministic finite dictionary. Suppose the fold fractions $w_k=m_k/n$ converge to positive constants, the supplied $f$ and covariate/correction functions are bounded, and follow-up risk-set denominators are bounded away from zero. Assume regular interior calibration roots, uniformly nonsingular calibration and regularized projection matrices, and positive limiting evaluation score derivatives and variance. Require zero probability at each clipping boundary. Suppose the time-indexed score classes have a square integrable Donsker and conditional bootstrap limit, the local derivatives of $\mathcal T_n$ converge with uniformly negligible quadratic/remainder terms, and numerical root errors are $o(n^{-1/2})$ both originally and conditionally in resamples. If the full functional limiting variance is positive, the conditional distribution of $T_{{\rm score},b}^*$ consistently estimates that of
\[
\{\widehat\theta^{\rm score}-\theta_{{\rm work},n}(f)\}/
\widehat{\operatorname{se}}_{\rm score}.
\]
With $B_{\rm bt}\to\infty$ and $B_{\rm bt}\Pp^*(\text{a complete refit fails})\to0$, the companion interval has asymptotic coverage $1-\alpha$ for this working target. Coverage for $\theta_0$ additionally requires $\sqrt n\{\theta_{{\rm work},n}(f)-\theta_0\}\to0$. For a random historical fit, these statements require the stated bounds and remainders uniformly on a set of historical fits with probability tending to one.
\end{proposition}
\begin{proof}
The implicit function derivative of each interior calibration root exists by positive local curvature. Matrix inversion differentiates the regularized projection. Clipping is differentiable in integrated scores when its boundaries have no atoms. Positive denominators and evaluation derivatives make the subsequent Cox ratios and one-step aggregation regular compositions. Let $\psi_{n,k}$ denote the centered influence derivative of the entire map with respect to its $k$th empirical measure. It includes all training complement contributions, not just the held-out influence in \eqref{eq:se}. The assumed bootstrap delta conditions give
\[
\widehat\theta^{\rm score}-\theta_{{\rm work},n}(f)
=\sum_{k=1}^K\frac1{m_k}\sum_{i\in\mathcal I_k}\psi_{n,k}(O_i)
+o_\Pp(n^{-1/2}).
\]
If $M_{b,i}^*$ is the multiplicity of original record $i$ in its own resampled fold, the corresponding conditional expansion is
\[
\widehat\theta_b^{{\rm score},*}-\widehat\theta^{\rm score}
=\sum_{k=1}^K\frac1{m_k}\sum_{i\in\mathcal I_k}
(M_{b,i}^*-1)\widehat\psi_{n,k}(O_i)+o_{\Pp^*}(n^{-1/2}),
\]
where the plug in derivatives consistently approximate $\psi_{n,k}$ in $L_2$. The empirical measures of the disjoint original folds are independent. Their independent multinomial resampling therefore reproduces the same Gaussian limit. Overlap among training complements is already included in each full map derivative. Continuity of \eqref{eq:se} gives $\sqrt n\widehat{\operatorname{se}}_{\rm score}\to s_f>0$ and the ratio of resampled to original studentizers converges conditionally to one. It is not necessary here to identify $s_f^2$ with the full functional variance: both original and bootstrap statistics use the same limiting scale. The continuous limiting distribution and convergent bootstrap quantiles give working target coverage. A union bound makes the declared failure guard inactive under the stated vanishing failure condition. Negligible working target bias then transfers coverage to $\theta_0$.
\end{proof}

For bounded, fixed dimensional Cox coefficient paths, threshold indicators in time and their finite products have the empirical process regularity above. Positive follow-up and curvature allow Taylor expansions of the risk-set ratios. These conditions are plausible for regular calibration fits, but separation, a vanishing score derivative, clipping atoms or an increasing unstable dictionary can invalidate them. The regularity and negligible bias conditions are not verified by a successful numerical run. This proposition neither proves a general neural bootstrap theorem nor certifies unrestricted projection rates. Fixed 999 resample quantiles add Monte Carlo error. Holding the source fixed does not remove historical transport bias or omitted target prognosis.

\paragraph{Independent finite sample evaluation.}
The evaluation uses $n=300,N=6000$, prognostic amplitude 1.5, 2000 null and 1000 alternative replications, with the historical network and generating law above. This follow-up was specified after earlier diagnostics and uses new source and target streams. A separate eight target pilot is excluded. The full-procedure bootstrap uses 999 resamples per target. Nine null and eight alternative intervals are unavailable because a resampled evaluation derivative fails $D_k>10^{-4}$. Calibration refits succeed. Failed draws are not discarded or replaced.

\begin{table}[tbp]\centering\footnotesize
\caption{Complete final projected bootstrap validation and references.}\label{tab:fullbootstrapcomplete}
\setlength{\tabcolsep}{2.6pt}\begin{tabular}{llrrrrrrrr}\toprule
Cell & Procedure & Success & Bias & RMSE & SE/SD & Cover (\%) & Reject (\%) & Length & All cover (\%) \\\midrule
F01 & PT-F Wald & 2000/2000 & 0.004 & 0.198 & 0.936 & 93.35 & 6.65 & 0.727 & 93.35 \\
F01 & PT-F full bootstrap & 1991/2000 & 0.004 & 0.198 & 0.936 & 95.78 & 4.22 & 0.817 & 95.35 \\
F01 & CT Wald & 2000/2000 & 0.001 & 0.176 & 0.951 & 93.85 & 6.15 & 0.657 & 93.85 \\
F01 & Oracle & 2000/2000 & 0.002 & 0.174 & 0.947 & 93.85 & 6.15 & 0.647 & 93.85 \\
F01 & CR & 2000/2000 & NA & NA & NA & NA & 5.25 & NA & NA \\
F02 & PT-F Wald & 1000/1000 & 0.007 & 0.203 & 0.956 & 93.80 & 43.60 & 0.759 & 93.80 \\
F02 & PT-F full bootstrap & 992/1000 & 0.007 & 0.203 & 0.956 & 95.56 & 36.09 & 0.852 & 94.80 \\
F02 & CT Wald & 1000/1000 & -0.003 & 0.181 & 0.960 & 93.70 & 51.50 & 0.682 & 93.70 \\
F02 & Oracle & 1000/1000 & -0.005 & 0.180 & 0.955 & 94.60 & 54.60 & 0.672 & 94.60 \\
F02 & CR & 1000/1000 & NA & NA & NA & NA & 49.50 & NA & NA \\
\bottomrule\end{tabular}
\par\smallskip\begin{minipage}{\textwidth}\footnotesize F01 is the null and F02 the alternative. PT-F and its bootstrap share the point estimate and original SE, so their bias, RMSE and SE/SD use all successful original point fits and are identical. Coverage, rejection and width condition on interval availability. All cover is the proportion of every attempted analysis that returns an interval covering the generating coefficient. Alternative rejection for an empirically miscalibrated procedure is not size controlled power.\end{minipage}\end{table}

Table~\ref{tab:fullbootstrapcomplete} shows every prespecified procedure, and Archive Table~12 gives Wilson Monte Carlo intervals and interval availability. The following paired differences use common successful records except for the explicitly identified all-attempt covering fraction. They are descriptive Monte Carlo comparisons, not simultaneous guarantees after method selection.
\begin{table}[tbp]\centering\small
\caption{Paired full bootstrap minus original PT-F changes.}\label{tab:fullbootstrappaired}
\begin{tabular}{llrrr}\toprule
Cell & Metric & Paired records & Difference & 95\% MC interval \\\midrule
F02 & Coverage (pp) & 992 & +1.815 & [+0.938, +2.691] \\
F01 & Null rejection (pp) & 1991 & -2.461 & [-3.156, -1.766] \\
F02 & Interval length & 992 & +0.093 & [+0.090, +0.096] \\
F02 & All attempt cover (pp) & 1000 & +1.000 & [-0.036, +2.036] \\
F02 & Alt. rejection (pp) & 992 & -7.661 & [-9.341, -5.982] \\
\bottomrule\end{tabular}\end{table}

\paragraph{Precision from all complete unchanged method batches.}
Pooling the four complete independent batches at this law (B01/B02, V01/V02, S01/S02 and F01/F02) gives 8000 null and 4000 alternative replications for the unchanged CT, PT-F and CR procedures (last row of Table~\ref{tab:strongchain}; Archive Table~13). Selected variants and development outcomes are not pooled, and the 100-replication exploration remains separate.

\section{When directional bias matters, and penalized prediction comparators}\label{app:newsims}
\subsection{Purpose and status}
This appendix reports six follow-up experiments and one population calculation. They address five questions that the earlier designs left open. First, in every earlier design the direct fit's directional bias was small relative to its standard error, so the projected score could only cost precision there. Appendix~\ref{sec:qpop} uses the population score equations to locate laws in which the bias is material, and Appendix~\ref{sec:qbias} evaluates one such law at three target sizes. Second, the earlier prediction comparisons used unpenalized target-only models. Appendix~\ref{sec:qpred} adds a ridge-penalized target spline Cox model and a penalized transfer rule under the C-cell laws. Third, the directional bias experiment shows that enlarging the correction to the dictionary can replace projection in large trials; Appendix~\ref{sec:qdict} asks whether it can in small ones, and evaluates the projected-minus-direct contrast of Corollary~\ref{cor:contrast}. Fourth, Appendix~\ref{sec:qmod} evaluates treatment inference under a less extreme law, with hazard ratio 0.7 and phase III target sizes, and Appendix~\ref{sec:qpred} also adds an adapted TransCox-type coefficient-transfer comparator. Fifth, Appendix~\ref{sec:qgen} replaces the law-informed dictionary $B_{18}$ by generic spline dictionaries.

All six experiments are follow-ups in the sense of Appendix~\ref{sec:sim}: they were designed after earlier results were examined, and their laws, methods, seeds, replication counts, and summaries were recorded in design files before their own data were generated. The directional bias experiment was first specified with 1000 replications per size and reduced to 500, with indices 0 to 499, for computational reasons before any replication result was examined. One timing replication for each experiment used a separate seed and is excluded. No replication is replaced or omitted. The code and design files are in the folder \texttt{simulations/directional\_bias\_and\_followups} of the supplementary code and records, available from the authors.

\subsection{Population bias under drift outside the correction span}\label{sec:qpop}
Let the historical log risk be
\[
f_0(x)=\sin(\pi x_1)+0.6\cos(\pi x_2)+0.5x_3x_4+\iota\sin(\pi x_1)\cos(\pi x_3),
\]
and let the target log risk be $g_0(x)=f_0(x)+0.7x_1-0.5x_2+\kappa\{\sin(\pi x_2)+0.8x_1x_4\}$, with $X\sim\operatorname{Uniform}(-1,1)^4$. The linear part of the drift lies in the four-coordinate correction space $z=x$. The part multiplied by $\kappa$ lies outside it but inside the projection dictionary $B_{18}$. The interaction multiplied by $\iota$ is shared by both cohorts. It lies outside the span of the 18 target-only spline coordinates used by CSPLINE and outside $B_{18}$.

With the historical risk learned without error, we solve the population Cox score equations for CT, the population one-step equation for PT-F with dictionary $B_{18}$, and the population equation for CSPLINE. Expectations over covariates use 3000 fixed uniform draws, and time integrals use 160 points on $[0,2]$. The target cumulative baseline is $s\,0.35t^{1.2}$ with event scale $s\in\{1,3\}$, censoring is exponential with rate 0.18, and $\tau=2$. Table~\ref{tab:qpopgrid} reports the bias and the target size $n_1$ at which the CT bias equals one CT standard error.

\begin{table}[tbp]\centering\footnotesize
\caption{Population treatment bias under unrepresented drift. Bias is the infinite-sample treatment coefficient minus $\theta_0$ with a correctly learned historical risk. $n_1$ is the target size at which the CT bias equals one CT standard error, $({\rm unit\ SD}/{\rm bias})^2$.}\label{tab:qpopgrid}
\setlength{\tabcolsep}{4pt}\begin{tabular}{rrrrrrrrr}\toprule
HR & Drift & Interaction & Event prob. & CT bias & PT-F bias & CSPLINE bias & Unit SD & $n_1$\\\midrule
0.7 & 0.5 & 0.0 & 0.45 & +0.007 & +0.000 & +0.002 & 2.99 & 165,963\\
0.7 & 0.5 & 0.0 & 0.67 & +0.010 & +0.000 & +0.002 & 2.45 & 65,582\\
0.7 & 0.5 & 1.0 & 0.45 & +0.007 & +0.000 & +0.023 & 3.00 & 173,870\\
0.7 & 0.5 & 1.0 & 0.67 & +0.009 & +0.000 & +0.032 & 2.46 & 70,522\\
0.7 & 1.0 & 0.0 & 0.45 & +0.029 & +0.002 & +0.001 & 2.99 & 10,617\\
0.7 & 1.0 & 0.0 & 0.67 & +0.035 & +0.003 & +0.002 & 2.45 & 4,958\\
0.7 & 1.0 & 1.0 & 0.45 & +0.027 & +0.002 & +0.024 & 2.99 & 11,903\\
0.7 & 1.0 & 1.0 & 0.66 & +0.033 & +0.002 & +0.033 & 2.46 & 5,412\\
0.5 & 0.5 & 0.0 & 0.42 & +0.013 & +0.000 & +0.003 & 3.16 & 55,233\\
0.5 & 0.5 & 0.0 & 0.64 & +0.018 & +0.000 & +0.003 & 2.55 & 20,000\\
0.5 & 0.5 & 1.0 & 0.42 & +0.013 & +0.000 & +0.043 & 3.16 & 56,125\\
0.5 & 0.5 & 1.0 & 0.64 & +0.017 & +0.000 & +0.060 & 2.55 & 21,541\\
0.5 & 1.0 & 0.0 & 0.42 & +0.054 & +0.004 & +0.003 & 3.15 & 3,355\\
0.5 & 1.0 & 0.0 & 0.64 & +0.066 & +0.005 & +0.003 & 2.55 & 1,485\\
0.5 & 1.0 & 1.0 & 0.42 & +0.051 & +0.003 & +0.043 & 3.15 & 3,749\\
0.5 & 1.0 & 1.0 & 0.63 & +0.064 & +0.004 & +0.062 & 2.55 & 1,616\\
\bottomrule\end{tabular}
\par\smallskip\begin{minipage}{\textwidth}\footnotesize Drift multiplies $\sin(\pi x_2)+0.8x_1x_4$, which lies outside the four linear correction coordinates but inside $B_{18}$. Interaction multiplies $\sin(\pi x_1)\cos(\pi x_3)$ in the shared historical and target prognosis, which lies outside the target-only spline span. Event probability is by $\tau=2$ under the target law with baseline $0.35t^{1.2}$ times the event scale (1 or 3; the higher value corresponds to probability above 0.6). Integrals use 3000 fixed covariate draws and 160 time points. The design of Appendix~\ref{sec:qbias} corresponds to HR 0.5, drift 1, interaction 1, and event scale 3.\end{minipage}\end{table}

Three features explain the earlier results. The CT bias grows with $\kappa$, with the treatment effect, and with the event probability, as the risk-set association in Proposition~\ref{prop:sensitivity} predicts. With hazard ratio 0.7 and $\kappa\le1$, as in the earlier complex drift designs, $n_1$ exceeds 4900, so experiments with $n\le600$ could not reveal the bias; the earlier complex drift setting had CT bias 0.016 with empirical SD 0.184. With hazard ratio 0.5, $\kappa=1$, and event probability near 0.64, $n_1$ falls to between 1485 and 1616. PT-F removes most of the CT bias in every configuration. Its remaining population bias, at most 0.005, is the second-order remainder in Proposition~\ref{prop:fixedmis}, whose first-order term vanishes because the drift lies in the span of $B_{18}$. These population values are the limits $\theta^\dagger$ and $\theta_B^\dagger$ of that proposition for a fixed drift. The CSPLINE bias depends on $\iota$ rather than $\kappa$, because a target-only fit has its own span limitation.

\subsection{Directional bias experiment}\label{sec:qbias}
\paragraph{Design.} We use $\iota=1$, $\kappa=1$, $\theta_0=\log(0.5)$, and event scale $s=3$, the last row of Table~\ref{tab:qpopgrid}. Each replication draws $N=10000$ historical patients with cumulative baseline $0.28t^{1.4}$ and exponential censoring rate 0.08. It then draws independent target trials with $n=500$, 1000, and 2000, treatment probability 0.5, target cumulative baseline $1.05t^{1.2}$, and censoring rate 0.18, all with $\tau=2$. One historical network is fitted per replication and shared by the three target sizes. It has the architecture, penalty, clipping, and validation stopping rule of the earlier neural simulations, with 20\% of historical patients reserved for stopping. Historical seeds are $20261013+1000r$, and target seeds add $17j$ for size index $j=1,2,3$.

\paragraph{Methods.} CT uses the historical network as offset and the four linear coordinates as correction. CT-rich uses the same offset with the 18 coordinates of $B_{18}$ as correction. It shows what is gained when the analyst can enlarge the correction to contain the drift. PT-F uses five folds, the CT pilot on each training complement, dictionary $B_{18}$, ridge $0.001\sqrt{300/n_{\rm tr}}$, clipping at 3, and the evaluation derivative guard $10^{-4}$, exactly as in the C cells. PT-S replaces $B_{18}$ by the 18 whitened spline coordinates. CN-S replaces the historical offset by a target-only neural pilot with the same architecture and the penalty $10^{-2}$ selected in Appendix~\ref{app:remainingflags}, trained on each training complement, followed by the CT-type refit and the $B_{18}$ projection. CSPLINE is the target-only Cox model with 18 whitened spline and product coordinates. The oracle uses the true target log risk as offset. All intervals are nominal 95\% Wald intervals using the sandwich SE for direct fits and the held-out influence SE for projected fits. Direct fits also report centered control log-risk MSE on 4000 fresh covariates.

{\footnotesize\setlength{\tabcolsep}{3.5pt}\begin{longtable}{rlrrrrrrrr}
\caption{Directional bias experiment: treatment estimation and nominal 95\% interval coverage, 500 replications per target size. Bias is $\widehat\theta-\theta_0$ with its Monte Carlo SE. The Wilson Monte Carlo interval accompanies coverage. Risk MSE is the centered control log-risk MSE of the direct prediction fits.}\label{tab:qresults}\\
\toprule $n$ & Method & Avail. & Bias (MC SE) & SE/SD & RMSE & Cover & MC interval & Length & Risk MSE\\\midrule\endfirsthead
\multicolumn{10}{l}{\textit{Table \thetable\ continued}}\\\toprule $n$ & Method & Avail. & Bias (MC SE) & SE/SD & RMSE & Cover & MC interval & Length & Risk MSE\\\midrule\endhead
\bottomrule\endfoot
500 & Oracle & 500/500 & +0.005 (0.005) & 0.97 & 0.116 & 94.4 & [92.0, 96.1] & 0.443 & NA\\
500 & CT & 500/500 & +0.068 (0.006) & 0.96 & 0.141 & 90.8 & [87.9, 93.0] & 0.469 & 0.348\\
500 & CT-rich & 500/500 & -0.012 (0.006) & 0.92 & 0.129 & 92.4 & [89.7, 94.4] & 0.465 & 0.125\\
500 & PT-F & 500/500 & +0.025 (0.006) & 0.99 & 0.129 & 94.0 & [91.6, 95.8] & 0.492 & NA\\
500 & PT-S & 500/500 & +0.027 (0.006) & 0.99 & 0.130 & 93.8 & [91.3, 95.6] & 0.493 & NA\\
500 & CN-S & 500/500 & +0.085 (0.006) & 0.96 & 0.151 & 88.2 & [85.1, 90.7] & 0.472 & NA\\
500 & CSPLINE & 500/500 & +0.050 (0.006) & 0.92 & 0.138 & 90.8 & [87.9, 93.0] & 0.463 & 0.369\\
\addlinespace
1000 & Oracle & 500/500 & +0.009 (0.004) & 1.02 & 0.079 & 95.2 & [93.0, 96.8] & 0.314 & NA\\
1000 & CT & 500/500 & +0.073 (0.004) & 1.00 & 0.112 & 85.8 & [82.5, 88.6] & 0.332 & 0.339\\
1000 & CT-rich & 500/500 & +0.003 (0.004) & 0.99 & 0.084 & 95.2 & [93.0, 96.8] & 0.325 & 0.080\\
1000 & PT-F & 500/500 & +0.023 (0.004) & 1.05 & 0.086 & 96.2 & [94.1, 97.6] & 0.341 & NA\\
1000 & PT-S & 500/500 & +0.025 (0.004) & 1.05 & 0.086 & 95.6 & [93.4, 97.1] & 0.341 & NA\\
1000 & CN-S & 500/500 & +0.077 (0.004) & 0.97 & 0.114 & 82.2 & [78.6, 85.3] & 0.318 & NA\\
1000 & CSPLINE & 500/500 & +0.062 (0.004) & 0.97 & 0.105 & 86.8 & [83.6, 89.5] & 0.323 & 0.328\\
\addlinespace
2000 & Oracle & 500/500 & -0.002 (0.003) & 1.00 & 0.056 & 95.0 & [92.7, 96.6] & 0.222 & NA\\
2000 & CT & 500/500 & +0.062 (0.003) & 1.01 & 0.086 & 83.0 & [79.5, 86.0] & 0.235 & 0.336\\
2000 & CT-rich & 500/500 & -0.003 (0.003) & 0.98 & 0.060 & 93.8 & [91.3, 95.6] & 0.229 & 0.062\\
2000 & PT-F & 500/500 & +0.010 (0.003) & 1.01 & 0.061 & 93.8 & [91.3, 95.6] & 0.238 & NA\\
2000 & PT-S & 500/500 & +0.012 (0.003) & 1.01 & 0.061 & 93.8 & [91.3, 95.6] & 0.238 & NA\\
2000 & CN-S & 500/500 & +0.063 (0.003) & 0.94 & 0.087 & 81.0 & [77.3, 84.2] & 0.220 & NA\\
2000 & CSPLINE & 500/500 & +0.055 (0.003) & 0.95 & 0.083 & 82.2 & [78.6, 85.3] & 0.228 & 0.311\\
\end{longtable}}

\begin{figure}[tbp]\centering
\includegraphics[width=\textwidth]{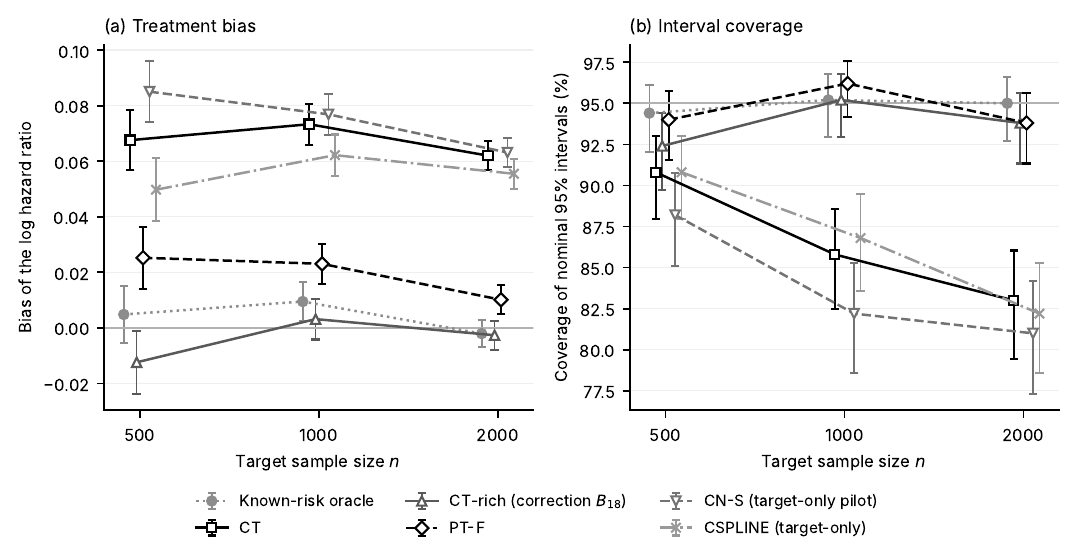}
\caption{Directional bias experiment. Panel (a) shows mean bias of the log hazard ratio with 95\% Monte Carlo intervals, and panel (b) shows coverage of nominal 95\% intervals with Wilson Monte Carlo intervals, at each target size. Each point uses 500 replications.}\label{fig:qbias}
\end{figure}

\paragraph{Results.} All 10,500 analyses succeed. The historical networks stop after a median of 230 steps, and their centered $L_2$ error on fresh covariates averages 0.223. Table~\ref{tab:qresults} and Figure~\ref{fig:qbias} give the results. The CT bias, 0.068, 0.073, and 0.062 at the three sizes, agrees with the population value 0.064 in Table~\ref{tab:qpopgrid} and does not shrink with $n$. Its SE/SD ratio stays between 0.96 and 1.01, so the undercoverage, 90.8\%, 85.8\%, and 83.0\%, is caused by bias rather than by the standard error. PT-F reduces the bias to 0.025, 0.023, and 0.010 and covers in 94.0\%, 96.2\%, and 93.8\% of replications. Its paired coverage gains over CT are 3.2, 10.4, and 10.8 percentage points, with Monte Carlo SEs 1.3, 1.5, and 1.6, and its paired squared error differences are $-0.0033$, $-0.0051$, and $-0.0037$, with Monte Carlo SEs 0.0010, 0.0005, and 0.0003. Its intervals are 4.9\%, 2.7\%, and 1.3\% longer than those of CT. The remaining PT-F bias at $n=500$ and 1000 reflects historical learning error and second-order terms, and it falls to 0.010 at $n=2000$. PT-S, with the whitened spline dictionary, behaves like PT-F, with coverage within 0.6 percentage points, because $\sin(\pi x_2)$ is well approximated by the spline coordinates and $x_1x_4$ is among them.

CT-rich, which uses $B_{18}$ as its calibration correction, removes the bias at $n=1000$ and 2000 and covers in 95.2\% and 93.8\%. At $n=500$ it has bias $-0.012$, SE/SD 0.92, and coverage 92.4\%, because its 18 correction coefficients are estimated from about 315 events. Its paired coverage differences from PT-F are within 1.6 percentage points, with Monte Carlo SEs near one, and its squared error matches that of PT-F, as the first-order equivalence of calibration and projection on the same span predicts. CT-rich also gives the best prediction, with risk MSE 0.125, 0.080, and 0.062 against 0.348, 0.339, and 0.336 for CT. An analyst who can anticipate the drift directions can therefore enlarge the correction when the trial is large. The projected score reaches the same treatment inference without changing the prediction model, and Appendix~\ref{sec:qdict} shows that it remains stable in smaller trials and with larger dictionaries, where the enlarged correction does not.

The target-only methods remain biased. CSPLINE has bias 0.050, 0.062, and 0.056, and CN-S has 0.085, 0.077, and 0.063, with coverage between 81.0\% and 90.8\%. Both miss the shared interaction $\sin(\pi x_1)\cos(\pi x_3)$, which lies outside the CSPLINE span and outside $B_{18}$. The CN-S network would need many more target events to learn it. The historical network learns most of it from 10000 patients, so the remaining PT-F pilot error is dominated by the drift, whose centered $L_2$ norm is about 0.76 and which lies in the span of $B_{18}$. This is the nuisance-rate role of history described after Theorem~\ref{thm:score}. The oracle, which uses the true target log risk, covers in 94.4\% to 95.2\%.

\subsection{Penalized comparators for prediction}\label{sec:qpred}
\paragraph{Design.} We regenerate the four alternative C-cell laws, with $(n,N)=(300,3000)$, $(300,300)$, $(300,1000)$, and $(600,3000)$ and $\theta_0=\log(0.7)$, using new seeds $31415+100000c+r$ for cell $c$ and replication $r$, and 500 replications per cell. CT, CT-spline, target linear Cox, and unpenalized CSPLINE are defined as in Appendices~\ref{sec:sim} and \ref{app:historicalspline}. CSPLINE-ridge uses the same 18 whitened spline coordinates with ridge penalty $\lambda\|\beta_{\rm spline}\|_2^2/2$ on the summed partial likelihood, leaves treatment unpenalized, and selects $\lambda\in\{0\}\cup\{n10^{-3+k/2}:k=0,\ldots,6\}$ by five-fold cross-validated partial likelihood \citep{verweij1993}. Penalized transfer adds the same ridge-penalized spline correction to the historical network offset, with the same tuning. Prediction metrics are those of the C cells, evaluated on 1000 fresh covariates per replication.

{\footnotesize\begin{longtable}{llrrrr}
\caption{Penalized comparators under the C-cell alternative laws with new seeds, 500 replications per cell. Entries are means (Monte Carlo SE). Differences are paired, method minus CSPLINE-ridge, so negative values favor the listed method. Survival MSE is multiplied by 1000.}\label{tab:qpred}\\
\toprule Cell & Method & Risk MSE & Survival MSE & Risk diff. & Survival diff.\\\midrule\endfirsthead
\multicolumn{6}{l}{\textit{Table \thetable\ continued}}\\\toprule Cell & Method & Risk MSE & Survival MSE & Risk diff. & Survival diff.\\\midrule\endhead
\bottomrule\endfoot
(300, 3000) & CT & 0.123 (0.002) & 5.91 (0.07) & -0.069 (0.003) & -2.90 (0.11)\\
(300, 3000) & CT-spline & 0.062 (0.001) & 3.54 (0.06) & -0.130 (0.003) & -5.27 (0.10)\\
(300, 3000) & Penalized transfer & 0.187 (0.002) & 8.59 (0.10) & -0.005 (0.002) & -0.22 (0.09)\\
(300, 3000) & CSPLINE-ridge & 0.192 (0.003) & 8.81 (0.12) & NA & NA\\
(300, 3000) & CSPLINE & 0.264 (0.006) & 9.52 (0.13) & +0.072 (0.005) & +0.71 (0.08)\\
(300, 3000) & Target linear & 0.456 (0.002) & 22.11 (0.09) & +0.264 (0.003) & +13.30 (0.11)\\
\addlinespace
(300, 300) & CT & 0.411 (0.006) & 18.42 (0.23) & +0.222 (0.006) & +9.79 (0.24)\\
(300, 300) & CT-spline & 0.206 (0.004) & 8.30 (0.12) & +0.017 (0.004) & -0.33 (0.14)\\
(300, 300) & Penalized transfer & 0.232 (0.006) & 10.41 (0.24) & +0.043 (0.006) & +1.78 (0.23)\\
(300, 300) & CSPLINE-ridge & 0.189 (0.003) & 8.63 (0.12) & NA & NA\\
(300, 300) & CSPLINE & 0.262 (0.006) & 9.27 (0.13) & +0.073 (0.005) & +0.64 (0.07)\\
(300, 300) & Target linear & 0.455 (0.002) & 22.10 (0.09) & +0.266 (0.003) & +13.47 (0.12)\\
\addlinespace
(300, 1000) & CT & 0.240 (0.003) & 10.76 (0.13) & +0.048 (0.004) & +1.97 (0.16)\\
(300, 1000) & CT-spline & 0.086 (0.002) & 4.36 (0.07) & -0.105 (0.003) & -4.42 (0.11)\\
(300, 1000) & Penalized transfer & 0.223 (0.003) & 10.31 (0.14) & +0.031 (0.003) & +1.52 (0.13)\\
(300, 1000) & CSPLINE-ridge & 0.192 (0.003) & 8.79 (0.12) & NA & NA\\
(300, 1000) & CSPLINE & 0.254 (0.005) & 9.30 (0.13) & +0.063 (0.004) & +0.51 (0.08)\\
(300, 1000) & Target linear & 0.454 (0.002) & 22.08 (0.09) & +0.262 (0.003) & +13.29 (0.12)\\
\addlinespace
(600, 3000) & CT & 0.101 (0.001) & 4.46 (0.05) & +0.006 (0.002) & +0.16 (0.06)\\
(600, 3000) & CT-spline & 0.040 (0.001) & 2.16 (0.03) & -0.054 (0.001) & -2.14 (0.05)\\
(600, 3000) & Penalized transfer & 0.135 (0.001) & 5.78 (0.06) & +0.040 (0.001) & +1.47 (0.05)\\
(600, 3000) & CSPLINE-ridge & 0.094 (0.001) & 4.31 (0.05) & NA & NA\\
(600, 3000) & CSPLINE & 0.115 (0.002) & 4.54 (0.06) & +0.020 (0.001) & +0.24 (0.02)\\
(600, 3000) & Target linear & 0.440 (0.001) & 20.75 (0.06) & +0.346 (0.002) & +16.45 (0.06)\\
\end{longtable}}

\paragraph{Results.} All 12,000 analyses succeed. Ridge penalization improves the target-only spline model in every cell, reducing its risk MSE by 0.020 to 0.073. Its penalty is never zero; the median selected $\lambda/n$ is 0.1 when $n=300$ and $N\ge1000$, and 0.032 otherwise. CT-spline is more accurate than CSPLINE-ridge in the three cells with $N\ge1000$, with paired risk MSE differences of $-0.130$, $-0.105$, and $-0.054$ and survival MSE differences of $-5.27$, $-4.42$, and $-2.14$. At $N=300$ it has risk MSE 0.017 higher and survival MSE 0.33 lower. Neural CT is more accurate than CSPLINE-ridge only at $(300,3000)$, with risk and survival differences $-0.069$ and $-2.90$. At $(300,1000)$ it is less accurate ($+0.048$ and $+1.97$), and at $(600,3000)$ the two are close ($+0.006$ and $+0.16$). The small historical cell reproduces the negative transfer of the C cells.

Penalized transfer shrinks the neural offset's spline correction, including its linear coordinates, toward zero. It therefore moves toward frozen transfer when the penalty is large, and the largest penalty is selected in 21\% of replications at $(300,3000)$. It improves on CT when the historical network is poor, at $N=300$ and 1000, but is less accurate than CT at $N=3000$ and improves on CSPLINE-ridge only at $(300,3000)$, by 0.005 in risk MSE. Treatment RMSE varies little across methods within a cell, ranging from 0.168 to 0.204 at $n=300$ and from 0.124 to 0.131 at $n=600$. These results temper the prediction comparison in Appendix~\ref{app:historicalspline}. A strong target-only model must be penalized, and transfer is worthwhile only when the historical learner estimates the shared prognosis better than such a model can.

\paragraph{Coefficient-transfer comparator.} To compare with a purpose-built survival transfer method, we implement an adapted TransCox-type estimator based on \citet{li2023} on exactly these records. It is not the authors' software, and its complete specification and its differences from the original method follow the results. The historical spline Cox fit supplies source coefficients $b_H$ on the 18 whitened spline coordinates, and the target coefficients are $b_H+\zeta$ with an $L_1$ penalty on the deviation $\zeta$, the treatment coefficient unpenalized, and a target Breslow baseline. The penalty is chosen from zero and twelve values between $\lambda_{\max}$ and $10^{-3}\lambda_{\max}$ by five-fold cross-validated partial likelihood with the folds above, where $\lambda_{\max}$ is the smallest penalty that sets every deviation to zero. The experiment regenerates the data of every replication from its seed without refitting the neural networks, and the recomputed CT-spline risk MSE equals the stored value in every replication. This comparator is a follow-up.

\begin{table}[tbp]\centering\footnotesize
\caption{Coefficient-transfer comparator on the penalized-comparator records: the adapted TransCox-type estimator specified in Appendix~\ref{sec:qpred} against the methods of Table~\ref{tab:qpred}. Entries are risk MSE, with paired differences TransCox-type minus each method (Monte Carlo SE).}\label{tab:qtrans}
\setlength{\tabcolsep}{4pt}\begin{tabular}{lrrrrr}\toprule
$(n,N)$ & TransCox-type & minus CT & minus CT-spline & minus CSPLINE-ridge & Survival MSE\\\midrule
$(300,3000)$ & 0.103 & $-0.020$ (0.003) & $+0.041$ (0.003) & $-0.089$ (0.002) & 5.23\\
$(300,300)$ & 0.159 & $-0.252$ (0.006) & $-0.047$ (0.004) & $-0.030$ (0.002) & 7.35\\
$(300,1000)$ & 0.114 & $-0.125$ (0.003) & $+0.028$ (0.003) & $-0.077$ (0.002) & 5.71\\
$(600,3000)$ & 0.058 & $-0.043$ (0.001) & $+0.017$ (0.001) & $-0.037$ (0.001) & 2.79\\
\bottomrule\end{tabular}
\par\smallskip\begin{minipage}{\textwidth}\footnotesize Each cell has 500 replications paired with Table~\ref{tab:qpred}. Survival MSE is multiplied by 1000. Negative differences favor the TransCox-type estimator. The penalty is chosen by five-fold cross-validated partial likelihood from 13 values including zero.\end{minipage}
\end{table}

The TransCox-type estimator has risk MSE 0.103, 0.159, 0.114, and 0.058 in the four cells (Table~\ref{tab:qtrans}). It improves on CSPLINE-ridge and on neural CT in every cell, by 0.030 to 0.089 and by 0.020 to 0.252. CT-spline, which calibrates the same historical spline score in the prespecified linear correction, is more accurate whenever $N\ge1000$, by 0.041, 0.028, and 0.017, because the drift lies in that correction and the remaining coefficients need not be re-estimated. With $N=300$ the TransCox-type estimator is more accurate by 0.047, because its penalized deviations adapt to the poorly learned historical fit. Survival MSE gives the same ordering. The selected penalty is never zero, the largest value is selected in at most 4\% of replications, and the median number of nonzero deviations is 7 to 11 of 18. Its treatment RMSE, 0.176 to 0.182 at $n=300$ and 0.124 at $n=600$, is in the range of the other methods, although inference after lasso selection is not covered by our results.

\paragraph{Specification of the adapted TransCox-type estimator.} The same estimator is used here, in OAK (Appendix~\ref{app:oakcomparators}), and in GBSG (Appendix~\ref{app:gbsg}), and the \texttt{AdaptedTransCox} class of the \texttt{ctcox} package (\codeurl) implements it.
\begin{enumerate}[leftmargin=*,itemsep=2pt]
\item \emph{Source coefficients.} A Cox model fitted to the historical sample alone on coordinates $x\in\mathbb R^r$ gives $b_H$. These are the 18 whitened spline coordinates in the simulations, the 11 clinical coordinates in OAK, and the seven clinical coordinates in GBSG, with unpenalized or ridge source fits as stated in each appendix.
\item \emph{Target model.} The target hazard is $\lambda_0(t)\exp\{\theta a+x^\top(b_H+\zeta)\}$, with offset $x^\top b_H$, deviation $\zeta\in\mathbb R^r$, and an unpenalized treatment coefficient $\theta$, which is omitted when there is no treatment variable.
\item \emph{Criterion.} The estimator minimizes $-n^{-1}\ell_n(\theta,\zeta)+\lambda\sum_j s_j|\zeta_j|$, where $\ell_n$ is the target log partial likelihood with Breslow ties. In OAK and GBSG, $s_j$ is the training standard deviation of coordinate $j$. In the simulations the coordinates are whitened and $s_j=1$, with the criterion on the summed scale, which only rescales $\lambda$.
\item \emph{Penalty grid and tuning.} With $\widetilde\theta$ the fit with $\zeta=0$, $\lambda_{\max}=\max_j|\partial_{\zeta_j}\ell_n(\widetilde\theta,0)|/(ns_j)$ is the smallest penalty that sets every deviation to zero. The grid is $\{\lambda_{\max}10^{-3k/11}:k=0,\ldots,11\}\cup\{0\}$, and $\lambda=0$ reproduces target-only Cox regression on $(a,x)$. The penalty maximizes the five-fold cross-validated partial likelihood $\sum_k\{\ell_n(\widehat\beta_{-k})-\ell_{n,-k}(\widehat\beta_{-k})\}$ of \citet{verweij1993}, with the folds stated in each appendix.
\item \emph{Algorithm.} Proximal Newton steps use the gradient and the information matrix at the current value, cyclic coordinate descent with soft thresholding on the quadratic model until coordinates change by less than $10^{-12}$ ($10^{-10}$ in the simulations), and step halving on the penalized criterion. Iterations stop when the step is below $10^{-9}$ ($10^{-8}$ in the simulations), with warm starts along the decreasing grid. In OAK and GBSG a fit is accepted when its Karush-Kuhn-Tucker residual is below $10^{-6}$.
\item \emph{Prediction.} The target cumulative baseline hazard is the Breslow estimate at the fitted linear predictor, and survival predictions are $\exp[-\widehat\Lambda_0(t)\exp\{\widehat\theta a+x^\top(b_H+\widehat\zeta)\}]$. The treatment coefficient is reported as a point estimate only, because inference after lasso selection is not covered here.
\end{enumerate}
The original TransCox method models the target baseline hazard as the source baseline plus a deviation with a second lasso penalty, estimates both deviations jointly from the target likelihood, and selects its two penalties by the Bayesian information criterion in the authors' software. Our adaptation does not borrow the baseline hazard, because no design assumes equal historical and target baselines and CT does not borrow them either. It uses one cross-validated penalty on the partial likelihood and its own solver. Differences between the two estimators may therefore come from the baseline borrowing, the tuning rule, or the optimization, and our results describe the adapted estimator only.

\subsection{Enlarged correction versus projected score in small trials}\label{sec:qdict}
\paragraph{Design.} In the directional bias experiment, an enlarged correction matched the projected score with $n\ge1000$. Cells D1 to D4 ask whether it remains an adequate substitute with fewer patients and larger dictionaries, and how the contrast of Corollary~\ref{cor:contrast} behaves. They reuse the historical law, replication seeds, and fitted networks of Appendix~\ref{sec:qbias} ($N=10000$) and draw new target trials with $n=300$ (D1, D3) and $n=600$ (D2, D4). D1 and D2 use the target law of Appendix~\ref{sec:qbias}, with drift $\sin(\pi x_2)+0.8x_1x_4$ outside the linear correction. D3 and D4 remove that drift ($\kappa=0$), so the correction contains the target change and only historical learning error remains. Target seeds are $20261013+1000r+300+c$ and fold seeds $77+1000r+c$ for cell $c$. We compare three nested dictionaries: $B_{18}$; $B_{26}$, which adds $\sin(2\pi x_j)$ and $\cos(2\pi x_j)$; and $B_{50}$, which adds the 24 products $\sin(\pi x_i)\cos(\pi x_j)$ ($i\ne j$), $\sin(\pi x_i)\sin(\pi x_j)$, and $\cos(\pi x_i)\cos(\pi x_j)$ ($i<j$), and so contains the shared interaction $\sin(\pi x_1)\cos(\pi x_3)$. For each dictionary $B_M$, CT-rich uses it as the calibration correction with sandwich intervals, and PT uses it as the projection dictionary with the PT-F settings, the CT pilot on each training complement shared across dictionaries. CT and the oracle are as before. For each PT, the contrast $\widehat\theta^{\rm score}-\widehat\theta$ with CT is standardized by its influence-based standard error. Each cell has 500 replications, the design file was written before the data were generated, and one timing replication is excluded.

\begingroup\footnotesize\setlength{\tabcolsep}{3.4pt}
\begin{longtable}{llrrrrlrr}
\caption{Enlarged correction versus projected score (D1 to D4). Bias, root mean squared error, and interval length are on the log hazard ratio scale; coverage is for nominal 95\% intervals, with Wilson 95\% Monte Carlo intervals; risk MSE is the centered control log-risk error on 4000 fresh covariates. Each cell has 500 replications.}\label{tab:qdict}\\
\toprule
Cell & Method & Bias (MC SE) & SE/SD & RMSE & Cover (\%) & MC interval & Length & Risk MSE\\
\midrule\endfirsthead
\multicolumn{9}{l}{\textit{\tablename~\thetable\ continued}}\\
\toprule
Cell & Method & Bias (MC SE) & SE/SD & RMSE & Cover (\%) & MC interval & Length & Risk MSE\\
\midrule\endhead
\bottomrule\endlastfoot
D1, $n=300$ & Oracle & $+0.004$ (0.006) & 1.08 & 0.136 & 96.0 & [93.9, 97.4] & 0.572 & NA\\
 & CT & $+0.062$ (0.007) & 1.00 & 0.166 & 93.6 & [91.1, 95.4] & 0.605 & 0.357\\
 & CT-rich, $B_{18}$ & $-0.029$ (0.007) & 0.99 & 0.160 & 94.4 & [92.0, 96.1] & 0.611 & 0.195\\
 & CT-rich, $B_{26}$ & $-0.048$ (0.007) & 0.96 & 0.174 & 92.6 & [90.0, 94.6] & 0.628 & 0.272\\
 & CT-rich, $B_{50}$ & $-0.118$ (0.009) & 0.91 & 0.231 & 86.4 & [83.1, 89.1] & 0.713 & 0.697\\
 & PT, $B_{18}$ & $+0.025$ (0.007) & 1.04 & 0.163 & 96.0 & [93.9, 97.4] & 0.658 & NA\\
 & PT, $B_{26}$ & $+0.026$ (0.008) & 1.03 & 0.170 & 96.0 & [93.9, 97.4] & 0.678 & NA\\
 & PT, $B_{50}$ & $+0.030$ (0.008) & 1.02 & 0.187 & 95.2 & [93.0, 96.8] & 0.739 & NA\\
\addlinespace
D2, $n=600$ & Oracle & $+0.001$ (0.005) & 0.98 & 0.105 & 93.6 & [91.1, 95.4] & 0.406 & NA\\
 & CT & $+0.060$ (0.005) & 0.96 & 0.129 & 91.2 & [88.4, 93.4] & 0.428 & 0.344\\
 & CT-rich, $B_{18}$ & $-0.013$ (0.005) & 0.95 & 0.114 & 93.2 & [90.6, 95.1] & 0.423 & 0.108\\
 & CT-rich, $B_{26}$ & $-0.020$ (0.005) & 0.94 & 0.117 & 93.4 & [90.9, 95.3] & 0.425 & 0.134\\
 & CT-rich, $B_{50}$ & $-0.053$ (0.006) & 0.89 & 0.138 & 89.8 & [86.8, 92.2] & 0.446 & 0.254\\
 & PT, $B_{18}$ & $+0.019$ (0.005) & 1.02 & 0.114 & 94.6 & [92.3, 96.3] & 0.447 & NA\\
 & PT, $B_{26}$ & $+0.021$ (0.005) & 1.03 & 0.114 & 94.8 & [92.5, 96.4] & 0.453 & NA\\
 & PT, $B_{50}$ & $+0.023$ (0.005) & 1.00 & 0.123 & 95.2 & [93.0, 96.8] & 0.474 & NA\\
\addlinespace
D3, $n=300$ & Oracle & $+0.008$ (0.006) & 1.03 & 0.141 & 94.8 & [92.5, 96.4] & 0.571 & NA\\
 & CT & $+0.004$ (0.007) & 1.01 & 0.148 & 94.2 & [91.8, 95.9] & 0.587 & 0.075\\
 & CT-rich, $B_{18}$ & $-0.026$ (0.007) & 0.95 & 0.166 & 93.6 & [91.1, 95.4] & 0.609 & 0.194\\
 & CT-rich, $B_{26}$ & $-0.039$ (0.008) & 0.91 & 0.179 & 92.2 & [89.5, 94.2] & 0.625 & 0.271\\
 & CT-rich, $B_{50}$ & $-0.112$ (0.009) & 0.87 & 0.237 & 87.0 & [83.8, 89.7] & 0.711 & 0.682\\
 & PT, $B_{18}$ & $+0.021$ (0.007) & 1.00 & 0.165 & 93.8 & [91.3, 95.6] & 0.641 & NA\\
 & PT, $B_{26}$ & $+0.021$ (0.008) & 0.97 & 0.174 & 93.6 & [91.1, 95.4] & 0.659 & NA\\
 & PT, $B_{50}$ & $+0.018$ (0.009) & 0.95 & 0.195 & 92.2 & [89.5, 94.2] & 0.720 & NA\\
\addlinespace
D4, $n=600$ & Oracle & $+0.003$ (0.005) & 1.01 & 0.102 & 95.8 & [93.7, 97.2] & 0.405 & NA\\
 & CT & $+0.004$ (0.005) & 0.99 & 0.107 & 94.8 & [92.5, 96.4] & 0.416 & 0.061\\
 & CT-rich, $B_{18}$ & $-0.012$ (0.005) & 0.95 & 0.114 & 94.4 & [92.0, 96.1] & 0.422 & 0.109\\
 & CT-rich, $B_{26}$ & $-0.020$ (0.005) & 0.93 & 0.118 & 93.8 & [91.3, 95.6] & 0.424 & 0.136\\
 & CT-rich, $B_{50}$ & $-0.053$ (0.006) & 0.90 & 0.136 & 90.8 & [87.9, 93.0] & 0.444 & 0.252\\
 & PT, $B_{18}$ & $+0.006$ (0.005) & 0.98 & 0.113 & 93.8 & [91.3, 95.6] & 0.434 & NA\\
 & PT, $B_{26}$ & $+0.006$ (0.005) & 0.98 & 0.115 & 94.4 & [92.0, 96.1] & 0.441 & NA\\
 & PT, $B_{50}$ & $+0.000$ (0.005) & 0.98 & 0.120 & 94.0 & [91.6, 95.8] & 0.461 & NA\\
\addlinespace
\end{longtable}
\endgroup

\begin{figure}[tbp]\centering
\includegraphics[width=\textwidth]{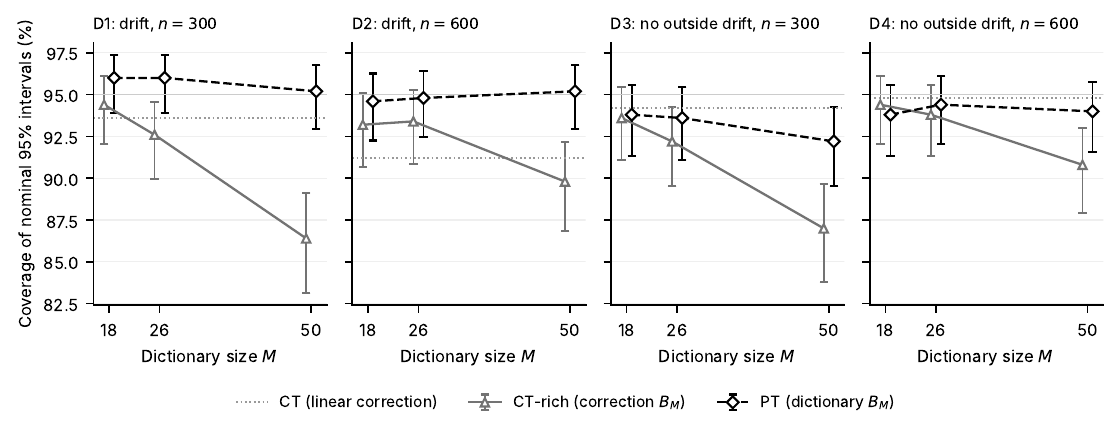}
\caption{Coverage of nominal 95\% intervals against dictionary size in D1 to D4, with Wilson 95\% Monte Carlo intervals. CT-rich uses $B_M$ as its calibration correction; PT uses it as its projection dictionary. The dotted line is CT with the linear correction. Each point uses 500 replications.}\label{fig:qdict}
\end{figure}

\paragraph{Results.} All 22,000 analyses and contrasts succeed. Target trials have about 191 events at $n=300$ and 383 at $n=600$. Table~\ref{tab:qdict} and Figure~\ref{fig:qdict} give the results.

Under the outside drift, CT has bias 0.062 at $n=300$ and 0.060 at $n=600$, close to the population value 0.064, and covers 93.6\% and 91.2\%; at these sizes its standard error is still large relative to the bias. Enlarging the correction removes this bias but introduces another. With $n=300$, CT-rich has bias $-0.029$, $-0.048$, and $-0.118$ for $B_{18}$, $B_{26}$, and $B_{50}$, its SE/SD falls from 0.99 to 0.91, and its coverage falls from 94.4\% to 92.6\% and 86.4\%. Its risk MSE, 0.195 with $B_{18}$, rises to 0.272 and then to 0.697, above the 0.357 of CT. With 51 coefficients and about 191 events, the partial likelihood estimate of the treatment coefficient is biased away from zero, as is familiar for Cox regression with few events per coefficient \citep{peduzzi1995}, and its sandwich standard error is too small. With $n=600$ the same pattern is milder: CT-rich covers 93.2\%, 93.4\%, and 89.8\%, with risk MSE 0.108, 0.134, and 0.254. The projected score is stable across dictionaries. With $n=300$ it has bias 0.025 to 0.030 and coverage 96.0\%, 96.0\%, and 95.2\%, and with $n=600$ bias 0.019 to 0.023 and coverage 94.6\%, 94.8\%, and 95.2\%. Its paired coverage gains over CT-rich are 1.6, 3.4, and 8.8 percentage points at $n=300$ (Monte Carlo SEs 0.8, 0.9, and 1.6), and over CT 2.4 and 3.4 points with $B_{18}$ at $n=300$ and 600 (SEs 1.1 and 1.2). Its RMSE grows with the dictionary, from 0.163 to 0.187 at $n=300$, so a larger dictionary costs precision without improving coverage here; with $B_{50}$ its squared error is 0.019 below that of CT-rich (SE 0.003).

Without the outside drift, CT is adequate: bias 0.004 and coverage 94.2\% and 94.8\%. CT-rich degrades with $M$ as before, with coverage 93.6\%, 92.2\%, and 87.0\% at $n=300$ and risk MSE up to 0.682, against 0.075 for CT. The projected score covers 93.8\%, 93.6\%, and 92.2\% at $n=300$ and 93.8\% to 94.4\% at $n=600$, with RMSE 0.165 to 0.195 against 0.148 for CT at $n=300$, the variance cost predicted by Proposition~\ref{prop:dictionarytradeoff}.

\begin{table}[tbp]\centering\footnotesize
\caption{Projected minus direct contrast in D1 to D4: rejection of equal first-order bias at the two-sided 5\% level, with Wilson 95\% Monte Carlo intervals, and the standardized contrast $z$. D1 and D2 have drift outside the correction; D3 and D4 do not.}\label{tab:qdictcontrast}
\setlength{\tabcolsep}{4pt}\begin{tabular}{llrrrrrr}\toprule
Cell & Dictionary & Reject (\%) & MC interval & Mean difference & SD difference & Mean SE & SD($z$)\\\midrule
D1, $n=300$ & $B_{18}$ & 16.2 & [13.2, 19.7] & $-0.037$ & 0.107 & 0.081 & 1.30\\
 & $B_{26}$ & 12.6 & [10.0, 15.8] & $-0.036$ & 0.110 & 0.091 & 1.19\\
 & $B_{50}$ & 9.0 & [6.8, 11.8] & $-0.032$ & 0.137 & 0.119 & 1.13\\
\addlinespace
D2, $n=600$ & $B_{18}$ & 30.8 & [26.9, 35.0] & $-0.041$ & 0.063 & 0.040 & 1.54\\
 & $B_{26}$ & 24.2 & [20.7, 28.1] & $-0.038$ & 0.064 & 0.045 & 1.42\\
 & $B_{50}$ & 16.2 & [13.2, 19.7] & $-0.037$ & 0.074 & 0.058 & 1.30\\
\addlinespace
D3, $n=300$ & $B_{18}$ & 5.8 & [4.1, 8.2] & $+0.017$ & 0.079 & 0.079 & 0.97\\
 & $B_{26}$ & 5.8 & [4.1, 8.2] & $+0.016$ & 0.089 & 0.089 & 1.00\\
 & $B_{50}$ & 7.6 & [5.6, 10.3] & $+0.014$ & 0.129 & 0.116 & 1.11\\
\addlinespace
D4, $n=600$ & $B_{18}$ & 3.2 & [2.0, 5.1] & $+0.002$ & 0.037 & 0.039 & 0.94\\
 & $B_{26}$ & 3.8 & [2.4, 5.9] & $+0.002$ & 0.042 & 0.044 & 0.95\\
 & $B_{50}$ & 5.6 & [3.9, 8.0] & $-0.004$ & 0.056 & 0.056 & 1.01\\
\addlinespace
\bottomrule\end{tabular}
\par\smallskip\begin{minipage}{\textwidth}\footnotesize The difference is $\widehat\theta^{\rm score}-\widehat\theta$ for PT with the stated dictionary and CT with the linear correction; its standard error is $\{\sum_i(\widehat\psi_{{\rm PT},i}-\widehat\psi_{{\rm CT},i})^2\}^{1/2}$ (Corollary~\ref{cor:contrast}). SD($z$) near one indicates a calibrated standard error. Every contrast was available.\end{minipage}
\end{table}

The projected-minus-direct contrast (Table~\ref{tab:qdictcontrast}) is approximately calibrated when the direct fit is unbiased: without the outside drift it rejects in 3.2\% to 7.6\% of trials, and the standardized contrast has SD 0.94 to 1.11. Under the drift it rejects in 16.2\% of trials with $B_{18}$ at $n=300$ and 30.8\% at $n=600$, with mean difference $-0.04$, and its power falls as the dictionary grows, because the projected estimate becomes more variable. The contrast thus detects a directional bias of 0.06 only partially at these sizes, consistent with the CT coverage loss being small there. Both its power and the CT coverage loss grow with $n$, because each depends on the ratio of a fixed bias to a standard error of order $n^{-1/2}$.

\subsection{Moderate-effect drift at phase III target sizes}\label{sec:qmod}
\paragraph{Design.} The directional bias experiment uses a hazard ratio of 0.5. Cells M1 to M3 evaluate the less extreme law with $\theta_0=\log(0.7)$, a hazard ratio typical of oncology trials and close to the OAK estimate, and target sizes $n=1000$, 2000, and 3000. The law is otherwise that of Appendix~\ref{sec:qbias}, including the drift $\sin(\pi x_2)+0.8x_1x_4$ outside the linear correction and the shared interaction, and it is the row of Table~\ref{tab:qpopgrid} with hazard ratio 0.7, drift and interaction 1, and event probability 0.66, where the population CT bias is 0.0335, the PT-F bias 0.0020, and the CT bias equals one SE at about 5400 patients. The historical law, replication seeds, and fitted networks are those of Appendix~\ref{sec:qbias}. Target seeds are $20261013+1000r+500+c$ and fold seeds $88+1000r+c$ for cell $c$. CT, CT-rich with correction $B_{18}$, PT-F, and the oracle are as before, and the PT-F minus CT contrast of Corollary~\ref{cor:contrast} is recorded. Each cell has 500 replications, the design file was written before the data were generated, and one timing replication is excluded.

\begin{table}[tbp]\centering\footnotesize
\caption{Moderate-effect drift at phase III target sizes (M1 to M3): hazard ratio 0.7, drift outside the linear correction, $N=10000$, 500 replications per size. Bias and RMSE are on the log hazard ratio scale; coverage is for nominal 95\% intervals with Wilson 95\% Monte Carlo intervals; risk MSE is the centered control log-risk error on 4000 fresh covariates.}\label{tab:qmod}
\setlength{\tabcolsep}{3.6pt}\begin{tabular}{llrrrrlrr}\toprule
$n$ & Method & Bias (MC SE) & SE/SD & RMSE & Cover (\%) & MC interval & Length & Risk MSE\\\midrule
1000 & Oracle & $-0.0004$ (0.0036) & 0.97 & 0.0798 & 94.8 & [92.5, 96.4] & 0.304 & NA\\
 & CT & $+0.0319$ (0.0037) & 0.99 & 0.0890 & 94.4 & [92.0, 96.1] & 0.322 & 0.339\\
 & CT-rich & $-0.0030$ (0.0037) & 0.96 & 0.0830 & 94.6 & [92.3, 96.3] & 0.312 & 0.079\\
 & PT-F & $+0.0071$ (0.0037) & 1.01 & 0.0837 & 95.8 & [93.7, 97.2] & 0.329 & NA\\
\addlinespace
2000 & Oracle & $+0.0019$ (0.0025) & 0.98 & 0.0559 & 94.8 & [92.5, 96.4] & 0.215 & NA\\
 & CT & $+0.0330$ (0.0026) & 0.99 & 0.0672 & 90.4 & [87.5, 92.7] & 0.228 & 0.335\\
 & CT-rich & $+0.0017$ (0.0026) & 0.97 & 0.0577 & 94.2 & [91.8, 95.9] & 0.220 & 0.061\\
 & PT-F & $+0.0072$ (0.0026) & 1.01 & 0.0585 & 94.2 & [91.8, 95.9] & 0.231 & NA\\
\addlinespace
3000 & Oracle & $+0.0004$ (0.0020) & 0.98 & 0.0456 & 93.8 & [91.3, 95.6] & 0.176 & NA\\
 & CT & $+0.0315$ (0.0022) & 0.97 & 0.0582 & 88.6 & [85.5, 91.1] & 0.186 & 0.335\\
 & CT-rich & $+0.0007$ (0.0021) & 0.98 & 0.0468 & 94.0 & [91.6, 95.8] & 0.180 & 0.055\\
 & PT-F & $+0.0053$ (0.0021) & 1.03 & 0.0468 & 95.0 & [92.7, 96.6] & 0.187 & NA\\
\addlinespace
\bottomrule\end{tabular}
\par\smallskip\begin{minipage}{\textwidth}\footnotesize Paired differences (Monte Carlo SE): $n=1000$: PT-F minus CT coverage $+1.4$ (1.1) percentage points, squared error $-0.00092$ (0.00035), contrast rejection 39.0\% [34.8, 43.3]; $n=2000$: PT-F minus CT coverage $+3.8$ (1.2) percentage points, squared error $-0.00110$ (0.00022), contrast rejection 61.0\% [56.7, 65.2]; $n=3000$: PT-F minus CT coverage $+6.4$ (1.4) percentage points, squared error $-0.00120$ (0.00015), contrast rejection 69.8\% [65.6, 73.7]. Every analysis succeeded.\end{minipage}
\end{table}

\paragraph{Results.} All 7500 analyses and contrasts succeed. Target trials have about 669, 1335, and 2005 events. Table~\ref{tab:qmod} gives the results. The CT bias is 0.032, 0.033, and 0.032, close to the population value 0.0335, and does not shrink with $n$, so CT coverage falls from 94.4\% at $n=1000$ to 90.4\% and 88.6\%. For a bias of 0.0335 and correct standard errors, the population calculation predicts 92.9\%, 90.7\%, and 88.4\%. PT-F has bias 0.005 to 0.007 and covers 95.8\%, 94.2\%, and 95.0\%. Its paired coverage gains over CT are 1.4, 3.8, and 6.4 percentage points (Monte Carlo SEs 1.1, 1.2, and 1.4), its squared error is smaller at every size, and its intervals are 2.3\%, 1.3\%, and 0.8\% longer. Its remaining bias exceeds the population value 0.0020, which is consistent with historical learning error that lies partly outside $B_{18}$ and enters the first-order term of Proposition~\ref{prop:fixedmis}. CT-rich covers 94.6\%, 94.2\%, and 94.0\%, as expected at these sizes, with risk MSE 0.055 to 0.079 against 0.34 for CT. The projected-minus-direct contrast rejects in 39.0\%, 61.0\%, and 69.8\% of trials (Wilson intervals 34.8 to 43.3, 56.7 to 65.2, and 65.6 to 73.7), so at phase III sizes it detects the bias that undermines the direct intervals. Its standardized values have SD 2.0 to 3.0 because the shift varies with the historical fit across replications. Under this plausible law, the direct fit is adequate for a trial of 1000 patients, but the projected score is preferable from about 2000 patients, where it improves both coverage and RMSE.

\subsection{Generic projection dictionaries}\label{sec:qgen}
\paragraph{Design.} The dictionary $B_{18}$ of PT-F contains $\sin(\pi x_j)$ and $\cos(\pi x_j)$, which are informed by the generating law. This experiment replaces it by generic dictionaries that an analyst could choose without that knowledge, under the hazard ratio 0.5 law of Appendix~\ref{sec:qbias} and the hazard ratio 0.7 law of Appendix~\ref{sec:qmod}, with their historical networks, target samples, and folds. $S_{12}$ has additive natural cubic splines with knots at $-1$, $-1/3$, $1/3$, and $1$ for each covariate, that is $x_j$ and two nonlinear terms. It approximates $\sin(\pi x_2)$ but does not contain the product $x_1x_4$ in the drift. $S_{18}$ adds the six pairwise products and is the PT-S dictionary of Appendix~\ref{sec:qbias}. $S_{22}$ uses knots at $-1$, $-1/2$, $0$, $1/2$, and $1$ with the six products. Each dictionary is centered and whitened under the uniform reference law with a fixed draw. The projected score is otherwise PT-F, with five folds, the CT pilot on each training complement, ridge $0.001\sqrt{300/n_{\rm tr}}$, clipping at 3, and the derivative guard $10^{-4}$, and the projected-minus-direct contrast is recorded for each dictionary. Each law and size has 500 replications. The design file was written before the data were generated, and one timing replication with a separate seed is excluded. CT is recomputed in every replication and PT-F with $B_{18}$ in replications 0 to 49 as checks against the stored records.

\begin{table}[tbp]\centering\footnotesize
\caption{Generic projection dictionaries under the hazard ratio 0.5 law of Appendix~\ref{sec:qbias} and the hazard ratio 0.7 law of Appendix~\ref{sec:qmod}, 500 replications per size. $S_{12}$ has additive natural cubic splines only, $S_{18}$ adds the six pairwise products, and $S_{22}$ uses five knots per covariate with the products. Coverage is for nominal 95\% intervals with Wilson 95\% Monte Carlo intervals. The last column is the rejection rate of the standardized projected-minus-direct contrast.}\label{tab:qgen}
\setlength{\tabcolsep}{3.4pt}\begin{tabular}{llrrrrlrr}\toprule
Law, $n$ & Method & Bias (MC SE) & SE/SD & RMSE & Cover (\%) & MC interval & Length & Contrast (\%)\\\midrule
HR 0.5, 500 & CT & $+0.0676$ (0.0056) & 0.96 & 0.1413 & 90.8 & [87.9, 93.0] & 0.469 & NA\\
 & PT-F, $B_{18}$ & $+0.0252$ (0.0057) & 0.99 & 0.1292 & 94.0 & [91.6, 95.8] & 0.492 & NA\\
 & PT, $S_{12}$ & $+0.0407$ (0.0058) & 0.96 & 0.1354 & 93.0 & [90.4, 94.9] & 0.486 & 23.6\\
 & PT, $S_{18}$ & $+0.0269$ (0.0057) & 0.99 & 0.1296 & 93.8 & [91.3, 95.6] & 0.493 & 27.2\\
 & PT, $S_{22}$ & $+0.0270$ (0.0058) & 0.99 & 0.1315 & 94.4 & [92.0, 96.1] & 0.498 & 24.4\\
\addlinespace
HR 0.5, 1000 & CT & $+0.0733$ (0.0038) & 1.00 & 0.1117 & 85.8 & [82.5, 88.6] & 0.332 & NA\\
 & PT-F, $B_{18}$ & $+0.0230$ (0.0037) & 1.05 & 0.0860 & 96.2 & [94.1, 97.6] & 0.341 & NA\\
 & PT, $S_{12}$ & $+0.0381$ (0.0037) & 1.03 & 0.0918 & 94.4 & [92.0, 96.1] & 0.339 & 46.4\\
 & PT, $S_{18}$ & $+0.0246$ (0.0037) & 1.05 & 0.0862 & 95.6 & [93.4, 97.1] & 0.341 & 52.6\\
 & PT, $S_{22}$ & $+0.0237$ (0.0037) & 1.05 & 0.0866 & 95.0 & [92.7, 96.6] & 0.343 & 50.8\\
\addlinespace
HR 0.5, 2000 & CT & $+0.0620$ (0.0027) & 1.01 & 0.0857 & 83.0 & [79.5, 86.0] & 0.235 & NA\\
 & PT-F, $B_{18}$ & $+0.0101$ (0.0027) & 1.01 & 0.0608 & 93.8 & [91.3, 95.6] & 0.238 & NA\\
 & PT, $S_{12}$ & $+0.0266$ (0.0027) & 1.01 & 0.0656 & 92.2 & [89.5, 94.2] & 0.238 & 70.0\\
 & PT, $S_{18}$ & $+0.0120$ (0.0027) & 1.01 & 0.0611 & 93.8 & [91.3, 95.6] & 0.238 & 79.4\\
 & PT, $S_{22}$ & $+0.0109$ (0.0027) & 1.01 & 0.0615 & 93.2 & [90.6, 95.1] & 0.239 & 79.4\\
\addlinespace
HR 0.7, 1000 & CT & $+0.0319$ (0.0037) & 0.99 & 0.0890 & 94.4 & [92.0, 96.1] & 0.322 & NA\\
 & PT-F, $B_{18}$ & $+0.0071$ (0.0037) & 1.01 & 0.0837 & 95.8 & [93.7, 97.2] & 0.329 & NA\\
 & PT, $S_{12}$ & $+0.0150$ (0.0038) & 0.99 & 0.0853 & 94.8 & [92.5, 96.4] & 0.327 & 37.2\\
 & PT, $S_{18}$ & $+0.0080$ (0.0037) & 1.01 & 0.0837 & 96.0 & [93.9, 97.4] & 0.329 & 37.4\\
 & PT, $S_{22}$ & $+0.0070$ (0.0037) & 1.01 & 0.0839 & 95.8 & [93.7, 97.2] & 0.331 & 35.6\\
\addlinespace
HR 0.7, 2000 & CT & $+0.0330$ (0.0026) & 0.99 & 0.0672 & 90.4 & [87.5, 92.7] & 0.228 & NA\\
 & PT-F, $B_{18}$ & $+0.0072$ (0.0026) & 1.01 & 0.0585 & 94.2 & [91.8, 95.9] & 0.231 & NA\\
 & PT, $S_{12}$ & $+0.0153$ (0.0026) & 1.01 & 0.0602 & 93.2 & [90.6, 95.1] & 0.230 & 56.8\\
 & PT, $S_{18}$ & $+0.0081$ (0.0026) & 1.02 & 0.0583 & 94.4 & [92.0, 96.1] & 0.231 & 59.8\\
 & PT, $S_{22}$ & $+0.0080$ (0.0026) & 1.02 & 0.0585 & 95.0 & [92.7, 96.6] & 0.231 & 57.4\\
\addlinespace
HR 0.7, 3000 & CT & $+0.0315$ (0.0022) & 0.97 & 0.0582 & 88.6 & [85.5, 91.1] & 0.186 & NA\\
 & PT-F, $B_{18}$ & $+0.0053$ (0.0021) & 1.03 & 0.0468 & 95.0 & [92.7, 96.6] & 0.187 & NA\\
 & PT, $S_{12}$ & $+0.0142$ (0.0021) & 1.03 & 0.0482 & 96.2 & [94.1, 97.6] & 0.187 & 62.4\\
 & PT, $S_{18}$ & $+0.0065$ (0.0021) & 1.03 & 0.0467 & 95.4 & [93.2, 96.9] & 0.187 & 67.2\\
 & PT, $S_{22}$ & $+0.0056$ (0.0021) & 1.03 & 0.0468 & 95.4 & [93.2, 96.9] & 0.187 & 66.8\\
\bottomrule\end{tabular}
\par\smallskip\begin{minipage}{\textwidth}\footnotesize Paired coverage differences in percentage points (Monte Carlo SE): HR 0.5, $n=500$: $S_{18}$ minus CT $+3.0$ (1.3), $S_{18}$ minus PT-F $-0.2$ (0.3), $S_{12}$ minus CT $+2.2$ (1.1); HR 0.5, $n=1000$: $S_{18}$ minus CT $+9.8$ (1.5), $S_{18}$ minus PT-F $-0.6$ (0.4), $S_{12}$ minus CT $+8.6$ (1.4); HR 0.5, $n=2000$: $S_{18}$ minus CT $+10.8$ (1.6), $S_{18}$ minus PT-F $+0.0$ (0.4), $S_{12}$ minus CT $+9.2$ (1.4); HR 0.7, $n=1000$: $S_{18}$ minus CT $+1.6$ (1.1), $S_{18}$ minus PT-F $+0.2$ (0.3), $S_{12}$ minus CT $+0.4$ (1.1); HR 0.7, $n=2000$: $S_{18}$ minus CT $+4.0$ (1.2), $S_{18}$ minus PT-F $+0.2$ (0.3), $S_{12}$ minus CT $+2.8$ (1.2); HR 0.7, $n=3000$: $S_{18}$ minus CT $+6.8$ (1.4), $S_{18}$ minus PT-F $+0.4$ (0.3), $S_{12}$ minus CT $+7.6$ (1.3). CT rows are recomputed and agree with the stored records of Appendices~\ref{sec:qbias} and~\ref{sec:qmod}, and PT-F rows are those stored records.\end{minipage}
\end{table}

\paragraph{Results.} All 9000 projected analyses and contrasts succeed. CT, recomputed in every replication, agrees with the stored estimates to $8\times10^{-10}$ under the hazard ratio 0.5 law and exactly under the hazard ratio 0.7 law, and PT-F with $B_{18}$, recomputed in replications 0 to 49, agrees to $10^{-10}$. The design file stated a tolerance of $10^{-10}$ for CT, and the larger difference under the first law is at the level of floating-point error in the historical network fits. Table~\ref{tab:qgen} and main Figure 2 give the results.

With $S_{18}$ or $S_{22}$, the projected score behaves like PT-F under both laws. Its bias is 0.011 to 0.027 at hazard ratio 0.5 and 0.006 to 0.008 at hazard ratio 0.7, against 0.010 to 0.025 and 0.005 to 0.007 for PT-F. Its coverage is 93.2\% to 96.0\%, its paired coverage differences from PT-F are within 1.2 percentage points, and its intervals are at most 1.1\% longer than those of PT-F. The spline coordinates approximate $\sin(\pi x_2)$ and the products contain $x_1x_4$, so the drift lies close to the span of these dictionaries. $S_{12}$ does not contain $x_1x_4$, and its bias, 0.027 to 0.041 at hazard ratio 0.5 and 0.014 to 0.015 at hazard ratio 0.7, is 43\% to 60\% of the CT bias. Its coverage, 92.2\% to 96.2\%, is above that of CT at every size, because this remaining bias is still small relative to the SE, but the bias does not shrink with $n$, as the first-order term $\J(q_B,e_N)$ of Theorem~\ref{thm:finite} predicts for error outside the dictionary. The standardized contrast rejects in 27.2\% to 79.4\% of trials with $S_{18}$ and in 23.6\% to 70.0\% with $S_{12}$, so it remains informative with generic dictionaries. A generic dictionary of additive splines and pairwise products is therefore a reasonable default when the analyst has no prior knowledge of the drift, and a dictionary that misses a component of the drift still reduces, but does not remove, the directional bias.

\section{Clinical analysis procedures and additional OAK results}\label{app:clinical}
\subsection{Purpose and scope}
We use the public clinical data in order to examine calibration with an external prognostic score, compare prediction across target training sizes, and describe the resulting treatment summaries. The source cohorts are independent of their target cohorts. The clinical analyses are retrospective. The horizons, correction comparisons, and diagnostic sensitivities were specified for this analysis after earlier inspection of the same cohorts. Nested validation separates fitting from evaluation within each comparison, but does not make this a prospective validation study.

The OAK analysis transfers prognosis from the docetaxel arm of POPLAR \citep{Fehrenbacher2016,Rittmeyer2017,Gandara2018}. Both trials measured overall survival from randomization. The target is the first 850 randomized OAK patients constituting the primary intention-to-treat population, rather than the subset with an available blood biomarker measurement. No post treatment response enters the prognostic model. We use all 143 POPLAR docetaxel patients and all 850 target patients. Censoring includes the end of observed follow-up and the administrative horizon of 18 months. There are 100 source deaths and 521 target deaths by that horizon. In OAK, 278 of 425 docetaxel patients and 243 of 425 atezolizumab patients die within the analysis window.

The public records do not contain the original immune cell component of the allocation strata or the allocation sequence. The available combined biomarker category cannot recover that component. We therefore do not attempt an unrestricted permutation test or claim exact inference under the original stratified block design. Our empirical Cox summaries describe the fitted working models. The diagnostics below also identify departures from the constant PH specification.

\subsection{Clinical coordinates and historical learning}\label{app:clinicalspec}
Table~\ref{tab:clinicalcoords} specifies the continuous transformations. OAK also uses indicators for Asian and other race, male sex, squamous histology, performance status one, two prior therapies, current smoking, and never smoking. Their reference categories are White race, female sex, nonsquamous histology, performance status zero, one prior therapy, and former smoking. The resulting vector has 11 coordinates. GBSG has seven coordinates, comprising the four continuous transformations in the table, postmenopausal status, tumor size greater than 20 and at most 50 mm, and tumor size greater than 50 mm. Tumor size at most 20 mm is the reference. For Rotterdam, the published middle size category is used as supplied. No additional grade variable is included because grading protocols differ.

\begin{table}[tbp]\centering\small
\caption{Continuous clinical transformations used before fitting.}\label{tab:clinicalcoords}
\begin{tabular}{llp{.48\textwidth}}\toprule
Cohort pair & Measurement & Transformation\\\midrule
POPLAR/OAK & Age in years & $(\mathrm{age}-60)/10$\\
 & Tumor burden in mm & $\log(1+\mathrm{burden})-\log(51)$\\
 & Metastatic site count & $\{\log(1+\mathrm{sites})-\log(4)\}/0.7$\\
Rotterdam/GBSG & Age in years & $(\mathrm{age}-55)/10$\\
 & Positive node count & $\log(1+\mathrm{nodes})-\log(4)$\\
 & Progesterone receptor & $\{\log(1+\mathrm{pgr})-\log(51)\}/2$\\
 & Estrogen receptor & $\{\log(1+\mathrm{er})-\log(51)\}/2$\\\bottomrule
\end{tabular}
\par\smallskip\noindent Receptor measurements use their supplied dataset scale. Every clinical coordinate is clipped to $[-4,4]$. No coordinate requires clipping in these cohorts. One missing OAK tumor burden value is replaced by the source median, 74 mm. The other shared coordinates are complete in the selected public records. Natural logarithms are used throughout.
\end{table}

For each continuous coordinate, the clinical cubic basis uses four fixed knots at $-2,-0.7,0.7,2$ with natural endpoint conditions. We use three coordinates spanning this cubic interpolation space modulo constants, represented relative to the clinical point zero, and retain the categorical indicators. Outside the outer knots the implementation continues the boundary polynomial segments on the bounded covariate domain. The resulting basis has 17 coordinates for OAK and 15 for GBSG and contains the linear clinical correction modulo constants. The knot locations and clinical transformations are not estimated from target outcomes.

Historical clinical and spline Cox fits maximize the unpenalized partial likelihood on the corresponding source coordinates. We retain a reference historical neural learner for sensitivity comparisons and for GBSG. It has two ReLU hidden layers of widths 8 and 4, a linear skip, and output clipped to $[-4,4]$. The weight penalty is 0.1 for POPLAR and 0.03 for Rotterdam. This reference rule was chosen using source data. The final OAK learners are defined below and are distinct from this reference rule. The neural score averages the log risk predictions from three fits with different initializations. For each fit, one of five source folds determines stopping by validation partial likelihood, evaluated every ten iterations up to 500 iterations. We then fit all source patients for the chosen iteration count. Adam uses learning rate 0.006, moment coefficients 0.9 and 0.999, and numerical constant $10^{-8}$. Parameters are clipped to $[-5,5]$ during training. These choices define a finite learning procedure. They do not certify the unpenalized likelihood gap required by Theorem~\ref{thm:source}.

The fixed target neural comparator uses widths 24 and 12, weight penalty $10^{-4}$, and the same stopping procedure within each target training sample. Its treatment coefficient is estimated jointly with prognosis. We retain this fixed comparator for GBSG and as an OAK sensitivity. The primary OAK neural comparator instead uses the nested tuning procedure below. Other target fits, apart from the penalized comparators in Appendix~\ref{app:oakcomparators}, are unpenalized and unconstrained Cox fits with a normalized score tolerance of $2\times10^{-5}$ and an information eigenvalue guard of $10^{-12}$. The finite tolerance is a numerical criterion, not an asymptotic rate statement. Every Cox calculation uses the complete Breslow risk set for tied events. The continuous event assumptions in the theory are not asserted for the rounded clinical times.

\subsection{Historical tuning and the OAK target neural comparator}\label{app:oaktuning}
We select the final OAK historical learners using POPLAR outcomes alone. We vary architecture, activation, initialization, output bound, learning rate, stopping, and penalties. The loss is negative Breslow partial likelihood divided by the number of fitting patients. For hidden weight matrices $W_j$, output weights $w_{\rm out}$, and linear skip weights $w_{\rm lin}$, the penalty is
\[
\frac{\lambda_{\rm hid}}{2}\sum_j\|W_j\|_F^2+
\frac{\lambda_{\rm out}}{2}\|w_{\rm out}\|_2^2+
\frac{\lambda_{\rm lin}}{2}\|w_{\rm lin}\|_2^2.
\]
The three $\lambda$ values are nonnegative training penalties. Biases are unpenalized and parameters remain in $[-5,5]$. Historical validation uses held-out partial likelihood. Its values can be compared across rules on the same folds, but not directly with likelihood values from different risk sets.

The search first evaluates 64 configurations with three historical folds. Twelve finalists, including at least three ReLU configurations, receive two repetitions of three-fold validation and three fixed initializations. A local search examines 32 further configurations and intensively evaluates four. In total, 96 configurations require 672 historical validation fits. Stopping uses an additional validation subset within each fitting sample, followed by refitting for the selected number of steps. Selection favors tanh overall and the ReLU rule shown in Table~\ref{tab:oaktuning}. Both supplied scores average three fixed starts fitted to all historical patients. We use the selected ReLU score for treatment analysis. Tanh is a bounded prediction alternative. The ReLU source rate in Theorem~\ref{thm:source} does not automatically apply to tanh, and finite penalization and stopping do not establish the assumed optimization gap for either fitted rule.

\begin{table}[tbp]\centering\small
\caption{Historical training choices for the final OAK scores.}\label{tab:oaktuning}
\begin{tabular}{lrr}
\toprule
Training choice & ReLU & Tanh \\
\midrule
Hidden widths & 16, 8 & 16, 8 \\
Hidden weight penalty & 0.0229269 & 0.0115814 \\
Output weight penalty & 0.00229269 & 0.00579070 \\
Linear skip penalty & 0.000101003 & 0.000805689 \\
Initial learning rate & 0.00713359 & 0.01622246 \\
Learning rate schedule & Cosine & Cosine \\
Initialization & He & Xavier \\
Initial hidden bias & 0.2 & 0.2 \\
Output bound & 4 & 4 \\
Ensemble members & 3 & 3 \\
Historical validation loss & 2.2135 & 2.2053 \\
Median validation stopping steps & 40 & 10 \\
\bottomrule
\end{tabular}

\par\smallskip\noindent Validation loss is mean held-out partial likelihood loss of the ensemble. Median steps summarizes historical validation fits and is not a convergence certificate. Predictions average all three starts without selecting a favorable member.
\end{table}

The tuned target neural comparator examines 12 configurations combining ReLU or tanh activation, hidden widths 8 and 4 or 16 and 8, and hidden and output penalties $10^{-4}$, $10^{-3}$, or $10^{-2}$. Its linear skip penalty is the larger of the hidden weight penalty and $10^{-3}$, initial learning rate is 0.004 with a cosine schedule, initial hidden bias is 0.1, and output bound is 4. Three inner target folds select by IBS, with stopping also confined to fitting samples. Each selected outer fit averages three final starts. Across the 15 outer evaluations this requires 540 inner fits and 45 final fits. The tuned comparator is evaluated in the full sample validation. We do not have a corresponding tuned target neural learning curve and do not infer its small-sample performance from the fixed clinical and spline comparisons.

Archive Table~14 describes variability across fixed starts. Each member is evaluated with its own target calibration and baseline. These are diagnostics, and the primary rules use the equal ensembles. Fixed start variability is distinct from uncertainty due to sampling historical patients.

\subsection{Penalized target, classical recalibration, and coefficient-transfer comparators}\label{app:oakcomparators}
Ridge clinical Cox fits treatment and the 11 clinical coordinates by minimizing the negative Breslow partial likelihood divided by the training size plus $(\lambda/2)\sum_j s_j^2\beta_j^2$, where $s_j$ is the training sample standard deviation of clinical coordinate $j$. The treatment coefficient is unpenalized. Within each training sample, five folds balanced by treatment and event status select $\lambda$ from $\{0\}\cup\{10^{-4+k/4}:k=0,\ldots,20\}$ by maximizing the cross-validated partial likelihood \citep{verweij1993,verweij1994}. The selected penalty is then refitted on the entire training sample with a single Breslow baseline. Its median selected value is 1.00 at 100 patients and 0.18 at full training size. The largest grid value is selected in 1 of 20 fits at 100 patients and never at larger sizes.

The POPLAR ridge Cox score applies the same rule to the 143 historical patients and selects $\lambda=0.316$. Free and nested recalibration then treat this linear score exactly as the network scores, with the same inner folds as the final nested network rules. We also report free and nested recalibration of the unpenalized POPLAR clinical Cox score at every learning-curve size. The reference nested clinical row in Table~\ref{tab:oakallprediction} used method-specific inner folds and is retained unchanged. Shrunk rules add a ridge-penalized correction of all 11 clinical coordinates to a freely scaled score and choose the penalty by the rule above. As the penalty grows they approach free rescaling, which explains why the largest grid value is often selected at 100 patients. These rules resemble shrinkage toward an external model, and the TransCox-type rule below follows transfer learning for the Cox model \citep{li2023} more closely.

The adapted TransCox-type rule of Appendix~\ref{sec:qpred} shrinks target coefficients toward historical ones, following \citet{li2023}. With source coefficients $b_H$ from the unpenalized POPLAR clinical Cox fit, or from the POPLAR ridge Cox fit above, it fits the linear predictor $\theta a+x^\top(b_H+\zeta)$ and minimizes the negative partial likelihood divided by the training size plus $\lambda\sum_j s_j|\zeta_j|$, with the treatment coefficient unpenalized. The penalty is chosen from $\{\lambda_{\max}10^{-3k/11}:k=0,\ldots,11\}\cup\{0\}$, where $\lambda_{\max}$ is the smallest value that gives $\zeta=0$, by five-fold cross-validated partial likelihood with the inner folds of ridge clinical Cox, and the fits are computed by proximal Newton steps along the warm-started path. As for every other rule, the baseline is the target Breslow estimate. With $\lambda=0$ the rule reproduces clinical Cox, and with $\zeta=0$ it keeps the POPLAR coefficients as a fixed offset.

Table~\ref{tab:oaktranscox} gives its results on the cached splits. At 100 patients it gives IBS 0.2139 with clinical and 0.2084 with ridge source coefficients. Both are below the 0.2213 of clinical Cox, and the ridge source is also below the 0.2126 of ridge clinical Cox, but nested tanh is lower still, by 0.0104 and 0.0049, with patient resampling ranges that exclude zero. The selected penalty keeps the POPLAR coefficients unchanged in 7 and 12 of the 20 fits at this size. At 200 patients nested tanh remains lower, and from 400 patients every difference is below 0.002 with a range that includes zero. The TransCox-type rule was added after the comparators above, and its design was recorded before any fit. All 190 fits complete, with Karush-Kuhn-Tucker residuals below $2\times10^{-13}$, and clinical Cox recomputed in the same loop reproduces the stored values exactly.

\begin{table}[tbp]\centering\footnotesize\setlength{\tabcolsep}{4pt}
\caption{OAK TransCox-type comparator. CV denotes the 15 evaluation comparison, and other sizes use 20 evaluations. Differences are nested tanh minus the TransCox-type estimator with 95\% patient resampling ranges, so negative values favor nested tanh.}\label{tab:oaktranscox}
\begin{tabular}{lrrrrll}\toprule
 & \multicolumn{2}{c}{TransCox-type IBS} & Nested & Ridge & \multicolumn{2}{c}{Nested tanh minus TransCox-type}\\\cmidrule(lr){2-3}\cmidrule(lr){6-7}
Size & Clinical source & Ridge source & tanh & clinical Cox & Clinical source & Ridge source\\\midrule
100 & 0.2139 & 0.2084 & 0.2035 & 0.2126 & -0.0104 (-0.0142, -0.0067) & -0.0049 (-0.0077, -0.0022) \\
200 & 0.2086 & 0.2051 & 0.2012 & 0.2050 & -0.0074 (-0.0106, -0.0044) & -0.0040 (-0.0064, -0.0014) \\
400 & 0.2001 & 0.2005 & 0.1989 & 0.2001 & -0.0011 (-0.0037, 0.0013) & -0.0016 (-0.0038, 0.0004) \\
Full & 0.1983 & 0.1988 & 0.1986 & 0.1985 & 0.0004 (-0.0018, 0.0029) & -0.0002 (-0.0021, 0.0019) \\
CV & 0.1975 & 0.1976 & 0.1983 & 0.1975 & 0.0009 (-0.0015, 0.0033) & 0.0007 (-0.0015, 0.0028) \\
\bottomrule\end{tabular}

\end{table}

These comparators were specified after the earlier OAK evaluation, because that evaluation contained no penalized target-only model and no classical recalibration in the learning curves. They reuse the cached training and evaluation patient sets. Before any new rule is summarized, the code reproduces the cached split-level IBS of clinical Cox and of full, free, and nested calibration with the final ReLU and tanh scores to within $4\times10^{-10}$. All 1235 new and reproduced fits complete.

\subsection{Fixed calibration and prediction selection}\label{app:clinicalvalidation}
For treatment summaries, the primary correction is the full linear clinical vector, fixed across target fits. Fixed sensitivities use no correction, the core clinical coordinates, the full linear vector, or the clinical cubic basis. The OAK core comprises tumor burden, metastatic site count, and performance status. The GBSG core comprises positive node count and progesterone receptor measurement. Frozen prognosis still estimates the target treatment coefficient and baseline hazard.

We also fit a free coefficient on the historical score. Writing $\gamma_{\rm sc}$ for its correction coefficient, the prognostic part is
\[
\widehat f(X)+\gamma_{\rm sc}\widehat f(X)
  =(1+\gamma_{\rm sc})\widehat f(X).
\]
An unrestricted $\gamma_{\rm sc}$ therefore gives free score rescaling. This source dependent coordinate is a conditional prediction benchmark and is not the deterministic clinical correction of the joint source and target theorem. We report it once rather than as two different methods.

Primary prediction validation uses three repetitions of five outer folds, balanced by treatment and event status. These folds are used only for prediction validation. They differ from the independent identifier folds used for projected treatment estimation. For each adaptive fit, three inner folds compare free rescaling, core, linear, and cubic corrections. Let $L_j$ be the mean inner fold IBS for candidate $j$, and let $s_j$ be its standard deviation across the three inner folds. If $j_*$ minimizes $L_j$ among candidates with three successful fits, we choose the candidate with the fewest coefficients among those satisfying
\[
L_j\le L_{j_*}+s_{j_*}/\sqrt{3}.
\]
Ties are resolved by the mean loss and then the fixed candidate name. This rule favors parsimony but is not a formal confidence bound because the training sets overlap. After selection, we refit the chosen correction and a single Breslow baseline on the entire outer training sample. No evaluation patient contributes to fitting, stopping, or correction selection.

For learning curves we use four repetitions of five outer folds. Within each outer training pool, seeded permutations of patient identifiers define nested samples of 100, 200, 400, and all available training patients. Every method uses the same patients at each size. The historical fit remains fixed. Full training sizes range from 678 to 682 in OAK and 547 to 550 in GBSG. These curves have 20 evaluations at every size and are distinct from the primary 15 fit comparison. Reusing patients across evaluations induces dependence. The curves describe performance within these cohorts and are not 20 independent clinical replications.

\subsection{Prediction measures and censoring}
Let $\widehat G_a(t)$ be the training sample Kaplan--Meier estimate of the censoring survival function for treatment group $a$. For an evaluation patient $i$, set $\widehat S_i(t)=\widehat S_{A_i}(t\mid X_i)$. We calculate the inverse probability of censoring weighted Brier contribution \citep{graf1999} using
\[
\frac{\Delta_i\ind(U_i\le t)\widehat S_i(t)^2}
 {\max\{\widehat G_{A_i}(U_i-),0.05\}}
 +\frac{\ind(U_i>t)\{1-\widehat S_i(t)\}^2}
 {\widehat G_{A_i}(t)}.
\]
At the administrative horizon, patients censored exactly at that horizon contribute as known survivors, and the censoring estimate is evaluated just before the administrative censoring jump. At other grid times the implementation also includes patients censored exactly at the evaluation time as survivors at that time. An evaluation is unavailable when the relevant training censoring survival at a grid time is below 0.05. The death time denominator is truncated at 0.05. Such truncation controls extreme weights but can introduce bias when it is active.

We average contributions over evaluation patients and integrate by the trapezoidal rule, dividing by the length of the integration window. OAK uses 31 equally spaced times from 3 to 18 months. GBSG uses 21 times from 182.625 to 1826.25 days. For these group specific censoring weights to estimate prediction error, sufficient independent censoring within the groups and follow-up support are needed. Conditional independence given the full covariate vector alone does not generally justify these simpler weights.

The C index uses pairs within an evaluation fold in which the earlier observed time is an event before the horizon. Tied risks receive half credit. We aggregate concordant and comparable pairs across folds. This is a descriptive concordance measure and does not correct for arbitrary covariate dependent censoring. Calibration slopes are Cox coefficients fitted to the entire predicted risk index in evaluation folds and summarized by their median. A separate absolute calibration gap is the censoring weighted observed event risk minus predicted event risk at the final horizon. The calibration figures compare average predicted survival for each treatment group with its Kaplan--Meier curve.

For descriptive uncertainty in the primary IBS differences, we first average each patient's validation loss across repetitions. We resample patients while keeping those predictions and the historical fit fixed. The resulting 95\% percentile ranges describe variation in the evaluated losses, conditional on the fitted models. They omit variation from fitting the target and historical models. We do not use them as tests of superiority or equivalence. Complete OAK differences are in Table~\ref{tab:oaklossdiff}, and GBSG differences are in Archive Table~17.

\subsection{Projection and conditional uncertainty}\label{app:clinicaluncertainty}
The primary projected fit uses five folds from a seeded permutation of canonical patient identifiers. The permutation is independent of treatment and outcomes. Calibration uses the full linear clinical correction on each training complement. The projection dictionary is the fixed clinical cubic basis. Its ridge is $0.001\sqrt{300/n_{\rm tr}}$, where $n_{\rm tr}$ is the training complement size. The projection function is clipped at 3 and every evaluation derivative must exceed $10^{-4}$. Projection clipping is inactive in the primary evaluation folds.

The implemented coordinates are represented relative to the clinical point zero. That convention does not establish the dominated reference measure assumption in the theorem. Risk set covariances are invariant to additive coordinate constants, and clipping is inactive in these fits, but population approximation and rate requirements remain assumptions. Ridge also leaves a nonzero correction orthogonality residual. Consequently, the variance from evaluation scores alone need not represent the variability of the complete fitted procedure in these finite samples. We report its scale as a diagnostic and use complete target refits for empirical intervals.

For each cohort and the reference historical learners, we use 999 target bootstrap samples for treatment-only Cox, target clinical Cox, target spline Cox, full calibration of each of the three historical learners, and projected neural estimation. These are seven reference procedures, including the historical clinical calibration that reproduces target clinical Cox. Direct fits resample patients and refit all target coefficients. Projected fits resample independently within the original inference folds. All copies of a sampled patient remain in that patient's original fold. We refit the calibration, risk sets, projection, and fold updates in every bootstrap sample. The source fit is held fixed. The application interval endpoints are the 0.025 and 0.975 quantiles of the bootstrap log coefficient, transformed to the hazard ratio scale. These are percentile intervals. They are not the studentized intervals in Appendix~\ref{app:fullbootstrap}, and the coverage results for that construction are not evidence of coverage here. All 13986 reference method fits completed. We additionally apply the same 999 resampling indices to full calibration and projected estimation with the final OAK ReLU score. All 1998 additional fits complete. These final intervals also condition on the supplied score and do not include historical tuning or historical sampling uncertainty. Failed resamples would not have been replaced, and any required failure would have made the corresponding interval unavailable.

Direct sandwich scales follow \citet{linwei1989}. Percentile refitting addresses empirical target fitting variation under its resampling model, but does not remove model bias or establish the missing trial assignment law. We use 100 separate historical patient resamples to examine sensitivity to the supplied prognosis. Each resample refits the fixed historical learning procedure with the target patients held fixed. For the reference learner, all 600 method fits completed across the two applications and three target procedures. For the final OAK ReLU rule, each resample refits all three initializations with their original seeds and step budget. Its stopping sample is formed from distinct resampled patients, so copies of one patient never enter both the fitting and stopping samples. Without duplicates this reproduces the supplied ReLU members exactly. All 300 final rule fits complete. When the stopping sample is instead formed from resampled rows, duplicated patients lengthen training and inflate the nonlinear component, and the full calibration hazard ratio ranges from 0.574 to 0.729. We regard that construction as unsuitable for a learner with internal stopping. These perturbations are reported separately and are not combined into a joint confidence interval.

\paragraph{Projected minus direct contrast.} Corollary~\ref{cor:contrast} gives an influence-based standard error for the difference between the projected and full calibration estimates on the same patients. With the final ReLU score the difference is $-0.3250-(-0.3522)=0.027$ on the log hazard ratio scale, with standard error 0.018 ($z=1.50$, two-sided $p=0.13$); the corresponding ratio of projected to direct hazard ratios is 1.028 (0.992, 1.065). The influence contributions of the two estimators have correlation 0.98, which is why the difference is estimated much more precisely than either coefficient. The tanh score gives 0.025 (0.018, $z=1.36$). Across the ten five-fold identifier partitions the ReLU difference ranges from $-0.010$ to 0.039 and $z$ from $-0.48$ to 1.79. Because the influence estimate holds the training-fold nuisance fits fixed, we also refit both procedures on 999 paired patient resamples drawn within the original inference folds (seed $880000+b$), all of which completed. The bootstrap SD of the difference is 0.036, twice the influence value, and the percentile interval is $(-0.032,0.108)$. Neither calculation indicates a directional bias of the direct fit, although the contrast has limited power at this size (Appendix~\ref{sec:qdict}), and the earlier OAK bootstrap draws cannot be used for this purpose because they resampled the two procedures independently. Records are in \texttt{final\_contrast.csv} and \texttt{final\_contrast\_boot.csv}.

Table~\ref{tab:clinicalbootstrapcompare} compares percentile intervals for the reference learners with studentized intervals calculated from the same complete target resamples. Studentization uses the direct sandwich scale or projected evaluation score scale in each resample and the corresponding original scale, as in the algorithms of Appendix~\ref{app:followups}. It changes interval endpoints without changing the fitted coefficient. Both constructions remain empirical working-model summaries. Agreement between them does not establish coverage, source learning rates, or validity under the original OAK allocation law.
\begin{table}[tbp]\centering\small
\caption{Percentile and studentized target bootstrap intervals for the reference learners.}\label{tab:clinicalbootstrapcompare}
\begin{tabular}{lrr}
\toprule Procedure & Percentile interval & Studentized interval \\ \midrule
\multicolumn{3}{l}{\textit{OAK}}\\
Treatment only & (0.606, 0.847) & (0.608, 0.848) \\
Target clinical Cox & (0.581, 0.831) & (0.592, 0.844) \\
Target spline Cox & (0.591, 0.849) & (0.601, 0.859) \\
Historical clinical calibration & (0.581, 0.831) & (0.592, 0.844) \\
Historical spline calibration & (0.572, 0.827) & (0.580, 0.833) \\
Full neural calibration & (0.581, 0.831) & (0.592, 0.844) \\
Projected neural score & (0.591, 0.881) & (0.596, 0.869) \\
\addlinespace
\multicolumn{3}{l}{\textit{GBSG}}\\
Treatment only & (0.514, 0.859) & (0.517, 0.861) \\
Target clinical Cox & (0.501, 0.853) & (0.496, 0.839) \\
Target spline Cox & (0.478, 0.824) & (0.488, 0.839) \\
Historical clinical calibration & (0.501, 0.853) & (0.496, 0.839) \\
Historical spline calibration & (0.485, 0.830) & (0.481, 0.822) \\
Full neural calibration & (0.499, 0.851) & (0.496, 0.841) \\
Projected neural score & (0.488, 0.871) & (0.492, 0.852) \\
\addlinespace
\bottomrule\end{tabular}

\par\smallskip\noindent All intervals are nominal 95\% and condition on the historical learner. The same 999 complete resamples are used for both constructions. GBSG coefficients are associations. These sensitivities do not replace the primary percentile intervals.
\end{table}

\subsection{Complete OAK prediction comparisons}\label{app:oakprediction}
Table~\ref{tab:oakallprediction} includes all 32 available rules. Full denotes calibration with all clinical coordinates, free denotes an estimated score coefficient, nested denotes correction selection within training samples, and frozen denotes a fixed prognostic offset with an estimated target treatment coefficient and baseline. Reference ReLU denotes the fixed 8 and 4 network. ReLU and tanh without that qualifier denote the final historically selected scores. Ridge clinical score denotes the POPLAR ridge Cox score, and shrunk denotes a freely scaled score with a ridge-penalized clinical correction (Appendix~\ref{app:oakcomparators}). These definitions apply throughout the OAK tables. Projected inference shares the direct prediction rule and has no separate prediction performance claim.

\begin{table}[tbp]\centering\small
\caption{All OAK prediction comparisons with three repetitions of five-fold validation.}\label{tab:oakallprediction}
\begin{tabular}{lrrrr}
\toprule
Prediction rule & IBS & C index & Slope & Fits \\
\midrule
Nested clinical score & 0.2033 & 0.634 & 0.91 & 15/15 \\
Nested reference ReLU & 0.1996 & 0.645 & 0.92 & 15/15 \\
Nested spline score & 0.2050 & 0.635 & 0.82 & 15/15 \\
Clinical Cox & 0.1972 & 0.655 & 0.93 & 15/15 \\
Fixed target neural & 0.2028 & 0.641 & 0.77 & 15/15 \\
Target spline Cox & 0.1984 & 0.650 & 0.86 & 15/15 \\
Full clinical score & 0.1972 & 0.655 & 0.93 & 15/15 \\
Full reference ReLU & 0.1972 & 0.655 & 0.93 & 15/15 \\
Full spline score & 0.2008 & 0.646 & 0.85 & 15/15 \\
Free clinical score & 0.2066 & 0.622 & 1.04 & 15/15 \\
Free reference ReLU & 0.2039 & 0.632 & 1.00 & 15/15 \\
Free spline score & 0.2081 & 0.618 & 1.03 & 15/15 \\
Frozen clinical score & 0.2142 & 0.621 & 0.55 & 15/15 \\
Frozen reference ReLU & 0.2037 & 0.632 & 0.92 & 15/15 \\
Frozen spline score & 0.2172 & 0.616 & 0.51 & 15/15 \\
Treatment only Cox & 0.2214 & 0.537 & 1.02 & 15/15 \\
Nested tanh & 0.1983 & 0.651 & 0.99 & 15/15 \\
Nested ReLU & 0.1993 & 0.646 & 0.94 & 15/15 \\
Tuned target neural & 0.1996 & 0.652 & 1.18 & 15/15 \\
Full tanh & 0.1972 & 0.654 & 0.94 & 15/15 \\
Full ReLU & 0.1969 & 0.655 & 0.93 & 15/15 \\
Free tanh & 0.1984 & 0.651 & 0.99 & 15/15 \\
Free ReLU & 0.1995 & 0.646 & 0.94 & 15/15 \\
Frozen tanh & 0.2015 & 0.648 & 1.49 & 15/15 \\
Frozen ReLU & 0.2046 & 0.640 & 1.65 & 15/15 \\
Linearized tanh & 0.1983 & 0.651 & 1.01 & 15/15 \\
\midrule
Ridge clinical Cox & 0.1975 & 0.654 & 1.14 & 15/15 \\
Free ridge clinical score & 0.2045 & 0.630 & 1.00 & 15/15 \\
Nested ridge clinical score & 0.2000 & 0.644 & 0.90 & 15/15 \\
Shrunk ridge clinical score & 0.1976 & 0.653 & 1.00 & 15/15 \\
Shrunk ReLU & 0.1974 & 0.654 & 0.97 & 15/15 \\
Shrunk tanh & 0.1974 & 0.653 & 1.00 & 15/15 \\
\bottomrule
\end{tabular}

\par\smallskip\noindent IBS integrates prediction loss from 3 to 18 months. C index aggregates comparable pairs within evaluation folds. Slope is the median evaluation fold calibration slope. Every rule completes all 15 fits. The linearized tanh rule is a subsequent diagnostic, not a candidate used to choose the primary learner. Rows below the rule were added with the penalized and classical recalibration comparators.
\end{table}

Full ReLU calibration gives IBS 0.1969 versus 0.1972 for clinical Cox and 0.1975 for ridge clinical Cox. The paired differences are approximately $-0.00025$ and $-0.0006$, with descriptive intervals spanning zero. Ridge clinical Cox has median calibration slope 1.14 in this comparison, so penalization does not improve on clinical Cox with about 680 training patients. Recalibration of the POPLAR ridge Cox score gives 0.2045 with a free coefficient and 0.2000 with nested correction. Tuning the target neural comparator reduces its IBS from 0.2028 to 0.1996, which remains above clinical Cox. Nested tanh gives 0.1983 and free tanh gives 0.1984. Thus, neither a favorable source validation loss nor neural flexibility establishes a material gain over the clinical model in full target training samples.

Table~\ref{tab:oaklossdiff} gives the paired contrasts for the final rules. The intervals resample each patient's average evaluation loss with prediction functions held fixed. They omit training, tuning, and historical sampling uncertainty and are not tests of superiority or equivalence. Fold loss dispersion also does not give an independent sampling standard error because training samples overlap. An additional decimal place preserves the sign of the difference smaller than 0.0001.
\begingroup\setstretch{1}\footnotesize\setlength{\tabcolsep}{3pt}
\begin{longtable}{l>{\raggedright\arraybackslash}p{2.4cm}>{\raggedright\arraybackslash}p{2.7cm}rrr}
\caption{OAK paired IBS differences with fitted predictions held fixed. CV denotes the 15 evaluation comparison. Other sizes use 20 evaluations. Negative differences favor the first rule.}\label{tab:oaklossdiff}\\
\toprule
Size & Rule & Reference & Difference & Relative \% & 95\% range \\
\midrule\endfirsthead
\multicolumn{6}{l}{\small Table \thetable{} continued}\\
\toprule
Size & Rule & Reference & Difference & Relative \% & 95\% range \\
\midrule\endhead
\midrule\endfoot
\bottomrule\endlastfoot
CV & Full ReLU & Clinical Cox & -0.0003 & -0.13 & (-0.0008, 0.0004) \\
CV & Full ReLU & Nested reference ReLU & -0.0027 & -1.34 & (-0.0054, 0.0002) \\
CV & Nested tanh & Clinical Cox & 0.0012 & 0.60 & (-0.0017, 0.0041) \\
CV & Nested tanh & Nested reference ReLU & -0.0012 & -0.62 & (-0.0033, 0.0009) \\
CV & Tuned target neural & Clinical Cox & 0.0024 & 1.21 & (0.0002, 0.0046) \\
CV & Tuned target neural & Nested reference ReLU & -0.00003 & -0.02 & (-0.0029, 0.0028) \\
CV & Tuned target neural & Fixed target neural & -0.0032 & -1.59 & (-0.0058, -0.0007) \\
CV & Full ReLU & Ridge clinical Cox & -0.0006 & -0.32 & (-0.0018, 0.0006) \\
CV & Nested tanh & Ridge clinical Cox & 0.0008 & 0.41 & (-0.0018, 0.0034) \\
CV & Nested tanh & Nested ridge clinical score & -0.0016 & -0.82 & (-0.0038, 0.0005) \\
100 & Nested ReLU & Clinical Cox & -0.0160 & -7.23 & (-0.0208, -0.0109) \\
100 & Nested ReLU & Nested reference ReLU & -0.0048 & -2.29 & (-0.0073, -0.0021) \\
100 & Nested tanh & Clinical Cox & -0.0178 & -8.06 & (-0.0226, -0.0131) \\
100 & Nested tanh & Nested reference ReLU & -0.0066 & -3.16 & (-0.0092, -0.0040) \\
100 & Nested ReLU & Ridge clinical Cox & -0.0073 & -3.44 & (-0.0116, -0.0028) \\
100 & Nested ReLU & Nested ridge clinical score & -0.0053 & -2.50 & (-0.0078, -0.0026) \\
100 & Nested tanh & Ridge clinical Cox & -0.0092 & -4.30 & (-0.0132, -0.0048) \\
100 & Nested tanh & Nested ridge clinical score & -0.0071 & -3.37 & (-0.0099, -0.0044) \\
200 & Nested ReLU & Clinical Cox & -0.0046 & -2.23 & (-0.0089, -0.0003) \\
200 & Nested ReLU & Nested reference ReLU & -0.0043 & -2.07 & (-0.0068, -0.0019) \\
200 & Nested tanh & Clinical Cox & -0.0057 & -2.77 & (-0.0098, -0.0017) \\
200 & Nested tanh & Nested reference ReLU & -0.0054 & -2.62 & (-0.0078, -0.0031) \\
200 & Nested ReLU & Ridge clinical Cox & -0.0027 & -1.33 & (-0.0065, 0.0009) \\
200 & Nested ReLU & Nested ridge clinical score & -0.0043 & -2.06 & (-0.0068, -0.0016) \\
200 & Nested tanh & Ridge clinical Cox & -0.0039 & -1.88 & (-0.0071, -0.0005) \\
200 & Nested tanh & Nested ridge clinical score & -0.0054 & -2.61 & (-0.0078, -0.0029) \\
400 & Nested ReLU & Clinical Cox & -0.0008 & -0.38 & (-0.0043, 0.0028) \\
400 & Nested ReLU & Nested reference ReLU & -0.0018 & -0.90 & (-0.0038, 0.0002) \\
400 & Nested tanh & Clinical Cox & -0.0017 & -0.84 & (-0.0048, 0.0014) \\
400 & Nested tanh & Nested reference ReLU & -0.0027 & -1.36 & (-0.0047, -0.0008) \\
400 & Nested ReLU & Ridge clinical Cox & -0.0002 & -0.12 & (-0.0033, 0.0030) \\
400 & Nested ReLU & Nested ridge clinical score & -0.0021 & -1.03 & (-0.0042, 0.0002) \\
400 & Nested tanh & Ridge clinical Cox & -0.0012 & -0.59 & (-0.0039, 0.0015) \\
400 & Nested tanh & Nested ridge clinical score & -0.0030 & -1.49 & (-0.0050, -0.0011) \\
Full & Nested ReLU & Clinical Cox & 0.0016 & 0.83 & (-0.0012, 0.0045) \\
Full & Nested ReLU & Nested reference ReLU & -0.0017 & -0.83 & (-0.0034, 0.0002) \\
Full & Nested tanh & Clinical Cox & 0.0003 & 0.17 & (-0.0026, 0.0032) \\
Full & Nested tanh & Nested reference ReLU & -0.0030 & -1.49 & (-0.0049, -0.0012) \\
Full & Nested ReLU & Ridge clinical Cox & 0.0014 & 0.72 & (-0.0012, 0.0040) \\
Full & Nested ReLU & Nested ridge clinical score & -0.0012 & -0.60 & (-0.0032, 0.0008) \\
Full & Nested tanh & Ridge clinical Cox & 0.0001 & 0.06 & (-0.0023, 0.0028) \\
Full & Nested tanh & Nested ridge clinical score & -0.0025 & -1.26 & (-0.0044, -0.0005) \\
\end{longtable}

\endgroup

Archive Table~15 retains every available training size and comparison. At 100 patients, nested tanh and ReLU give IBS 0.2035 and 0.2053, compared with 0.2213 for clinical Cox, 0.2126 for ridge clinical Cox, 0.2106 for nested recalibration of the POPLAR ridge Cox score, 0.2129 for nested recalibration of the unpenalized POPLAR clinical Cox score, and 0.2101 for nested reference ReLU. Ridge penalization therefore removes about half of the small-sample gap between clinical Cox and nested tanh, and raises the median calibration slope from 0.46 to 0.90, but nested tanh remains 4.3\% lower. Shrunk tanh gives 0.2037, which is no better than nested selection. The gain contracts with more target information. At full learning-curve training size, nested tanh gives 0.1986 versus 0.1983 for clinical Cox. The full size values differ from the 15 evaluation comparison because the learning curves use a separate 20 evaluation design. SD($\widehat\theta$) describes dispersion across overlapping training samples, not bias against a known truth. Median SE is the within training sample sandwich scale. No treatment SE is attached to nested prediction fits.

\subsection{What the historical score contributes}
We describe the empirical nonlinear component by projecting each supplied score on an intercept and the full clinical vector in OAK, then dividing residual variance by total score variance. The fractions are 2.51\% for the final ReLU score, 0.36\% for tanh, and 3.66\% for the historical spline score. Historical clinical Cox and the reference ReLU score belong to the clinical span to numerical precision. This diagnostic concerns the fitted score and does not estimate the complexity of true prognosis.

The reference ReLU network has a negligible hidden contribution and a nearly zero output weight norm, while the linear skip remains active. Output clipping is inactive and hidden activations are often positive. These observations support shrinkage of the nonlinear component as the explanation for its nearly linear score. They do not identify which individual training change accounts for the final learners' performance.

To assess whether the tanh gain requires its nonlinear component, we regress its supplied log risk on an intercept and historical clinical coordinates using historical patients alone. We apply this linear approximation in OAK and freely estimate its target coefficient and baseline. This is a diagnostic constructed after evaluation, not a replacement selected for favorable target performance. It gives IBS 0.1983 in ordinary validation and 0.2036, 0.2012, 0.1992, and 0.1986 at the four learning-curve sizes. The corresponding free tanh values are 0.1984 and 0.2037, 0.2012, 0.1994, and 0.1987. Their proximity suggests that external clinical weighting and restricted target calibration explain most of the observed benefit. It does not establish formal equivalence or isolate the effects of architecture, regularization, and stopping.

Table~\ref{tab:oakselection} gives the correction selection counts. Selection occurs inside each target training sample. The final ReLU and tanh rules retain free rescaling in 13 to 19 of 20 fits at every learning-curve size and in 13 and 14 of the 15 primary fits. The POPLAR ridge Cox score moves toward the core correction as training size grows.
\begin{table}[tbp]\centering\small
\caption{OAK correction selection counts.}\label{tab:oakselection}
\begin{tabular}{llrrrr}
\toprule
Training size & Learner & Scale & Core & Linear & Cubic \\
\midrule
Primary & Reference ReLU & 4 & 11 & 0 & 0 \\
Primary & Spline & 3 & 6 & 4 & 2 \\
Primary & Clinical & 6 & 7 & 2 & 0 \\
100 & Reference ReLU & 17 & 3 & 0 & 0 \\
100 & Spline & 19 & 1 & 0 & 0 \\
200 & Reference ReLU & 17 & 3 & 0 & 0 \\
200 & Spline & 16 & 1 & 3 & 0 \\
400 & Reference ReLU & 11 & 9 & 0 & 0 \\
400 & Spline & 10 & 8 & 0 & 2 \\
Full & Reference ReLU & 8 & 10 & 2 & 0 \\
Full & Spline & 3 & 13 & 2 & 2 \\
\midrule
Primary & ReLU & 13 & 2 & 0 & 0 \\
Primary & Tanh & 14 & 1 & 0 & 0 \\
Primary & Ridge clinical & 4 & 11 & 0 & 0 \\
100 & ReLU & 18 & 2 & 0 & 0 \\
100 & Tanh & 19 & 1 & 0 & 0 \\
100 & Ridge clinical & 17 & 3 & 0 & 0 \\
200 & ReLU & 18 & 2 & 0 & 0 \\
200 & Tanh & 19 & 1 & 0 & 0 \\
200 & Ridge clinical & 16 & 4 & 0 & 0 \\
400 & ReLU & 17 & 3 & 0 & 0 \\
400 & Tanh & 17 & 3 & 0 & 0 \\
400 & Ridge clinical & 8 & 12 & 0 & 0 \\
Full & ReLU & 13 & 6 & 1 & 0 \\
Full & Tanh & 17 & 3 & 0 & 0 \\
Full & Ridge clinical & 6 & 12 & 2 & 0 \\
\bottomrule
\end{tabular}

\par\smallskip\noindent Primary denotes the 15 fit validation. Other rows use the 20 evaluation learning curves. Scale denotes free rescaling, linear denotes all clinical coordinates, and cubic denotes the clinical cubic basis.
\end{table}

\subsection{Treatment summaries with the final ReLU score}
Table~\ref{tab:oaktreatment} reports full calibration and projected treatment estimates with the final ReLU score, together with clinical Cox and reference ReLU comparisons. Full calibration gives HR 0.703 (0.582, 0.834), and projection gives 0.723 (0.594, 0.885). Their bootstrap standard deviations are 0.091 and 0.101 on the log hazard ratio scale. Clinical Cox gives 0.091. These results do not show a material precision advantage from historical tuning. We cannot assess treatment bias against an unknown true effect.
\begin{table}[tbp]\centering\small
\caption{OAK treatment summaries through 18 months.}\label{tab:oaktreatment}
\begin{tabular}{lrrr}
\toprule
Treatment summary & HR & 95\% interval & Bootstrap SD \\
\midrule
Clinical Cox & 0.702 & (0.581, 0.831) & 0.091 \\
Full ReLU calibration & 0.703 & (0.582, 0.834) & 0.091 \\
Projected ReLU & 0.723 & (0.594, 0.885) & 0.101 \\
Full reference ReLU & 0.702 & (0.581, 0.831) & 0.091 \\
Projected reference ReLU & 0.720 & (0.591, 0.881) & 0.101 \\
\bottomrule
\end{tabular}

\par\smallskip\noindent HR compares atezolizumab with docetaxel. Intervals are nominal 95\% percentile intervals from 999 complete target refits conditional on each supplied historical score. Projected refits preserve the original five inference folds. Bootstrap SD is on the log hazard ratio scale. These empirical intervals do not reproduce the original stratified assignment law.
\end{table}

Table~\ref{tab:oakreluph} gives the PH diagnostics for full calibration with the final ReLU score. The global $p$ value is 0.010 and the treatment term gives 0.079. The treatment diagnostic does not establish a constant effect. The time-varying nuisance and restricted mean survival analyses below provide additional clinical context, but were performed with the reference specifications.
\begin{table}[tbp]\centering\small
\caption{OAK PH diagnostics for full calibration with the final ReLU score.}\label{tab:oakreluph}
\begin{tabular}{lrrr}
\toprule
Term & $\chi^2$ & df & $p$ value \\
\midrule
Treatment & 3.08 & 1 & 0.079 \\
Age & 2.41 & 1 & 0.121 \\
Tumor burden & 1.22 & 1 & 0.269 \\
Metastatic site count & 0.45 & 1 & 0.503 \\
Asian race & 6.59 & 1 & 0.010 \\
Other race & 0.06 & 1 & 0.815 \\
Male sex & 0.38 & 1 & 0.536 \\
Squamous histology & 0.36 & 1 & 0.546 \\
Performance status one & 7.97 & 1 & 0.005 \\
Two prior therapies & 3.96 & 1 & 0.047 \\
Current smoking & 0.95 & 1 & 0.331 \\
Never smoking & 0.64 & 1 & 0.424 \\
Global & 26.34 & 12 & 0.010 \\
\bottomrule
\end{tabular}

\par\smallskip\noindent Individual $p$ values are descriptive and are not adjusted for multiple comparisons.
\end{table}

\subsection{OAK treatment sensitivities}
The correction, dictionary, partition, horizon, and historical perturbation analyses below use both the reference learners and the final ReLU score, and the fixed corrections also use the final tanh score. The reference rows are unchanged from the earlier analysis.

Table~\ref{tab:oakcorrections} reports every fixed correction and free rescaling fit. With the final ReLU score, the hazard ratio ranges from 0.703 with the full linear correction to 0.714 without correction, and the free score coefficient is 2.01. The corresponding tanh range is 0.700 to 0.711. These intervals use the working sandwich scale and therefore differ from the primary percentile intervals. The no correction fits estimate the target treatment coefficient and baseline with a frozen prognosis. Full cubic correction absorbs the historical cubic score by the same calibration invariance argument that applies to a linear historical score.

\begin{table}[tbp]\centering\small
\caption{OAK fixed correction and free rescaling.}\label{tab:oakcorrections}
\begin{tabular}{lrrrr}
\toprule
Source & Correction & HR & Working Wald interval & Score coef. \\
\midrule
Clinical & None & 0.713 & (0.586, 0.867) & NA \\
Clinical & Core & 0.709 & (0.592, 0.849) & NA \\
Clinical & Linear & 0.702 & (0.588, 0.837) & NA \\
Clinical & Cubic & 0.713 & (0.598, 0.851) & NA \\
Clinical & Scale & 0.718 & (0.604, 0.854) & 0.525 \\
Spline & None & 0.708 & (0.581, 0.862) & NA \\
Spline & Core & 0.701 & (0.583, 0.842) & NA \\
Spline & Linear & 0.693 & (0.579, 0.829) & NA \\
Spline & Cubic & 0.713 & (0.598, 0.851) & NA \\
Spline & Scale & 0.715 & (0.601, 0.850) & 0.485 \\
Reference ReLU & None & 0.715 & (0.600, 0.853) & NA \\
Reference ReLU & Core & 0.707 & (0.592, 0.843) & NA \\
Reference ReLU & Linear & 0.702 & (0.588, 0.837) & NA \\
Reference ReLU & Cubic & 0.713 & (0.598, 0.851) & NA \\
Reference ReLU & Scale & 0.716 & (0.601, 0.852) & 0.910 \\
\midrule
ReLU & None & 0.714 & (0.603, 0.846) & NA \\
ReLU & Core & 0.705 & (0.591, 0.840) & NA \\
ReLU & Linear & 0.703 & (0.590, 0.839) & NA \\
ReLU & Cubic & 0.714 & (0.599, 0.852) & NA \\
ReLU & Scale & 0.713 & (0.599, 0.849) & 2.010 \\
Tanh & None & 0.709 & (0.599, 0.841) & NA \\
Tanh & Core & 0.702 & (0.589, 0.838) & NA \\
Tanh & Linear & 0.700 & (0.586, 0.834) & NA \\
Tanh & Cubic & 0.711 & (0.596, 0.848) & NA \\
Tanh & Scale & 0.707 & (0.594, 0.843) & 1.602 \\
\bottomrule
\end{tabular}

\par\smallskip\noindent Intervals are nominal 95\% sandwich Wald intervals conditional on the historical learner. The score coefficient applies only to free rescaling. NA denotes not applicable.
\end{table}

Table~\ref{tab:oakdictionary} compares ridge multipliers 0.0001, 0.001, and 0.01 with the fixed clinical dictionary and with the supplied neural score appended. The appended reference neural score is redundant to numerical precision in OAK. The final ReLU score has a nonzero component outside the clinical span and gives full observed rank 18 when appended. Across the six final ReLU fits, the projected hazard ratio ranges from 0.717 to 0.729 and the score scale from 0.092 to 0.093. The unregularized information rank is 17 for 18 columns in the augmented dictionary. A ridge permits numerical inversion but does not establish positive definite population information for that redundant space. The augmented dictionary is a source conditional diagnostic and is not used for the deterministic dictionary claim.

\begin{table}[tbp]\centering\small
\caption{OAK dictionary and ridge sensitivities for projected estimation.}\label{tab:oakdictionary}
\begin{tabular}{llrrr}
\toprule
Score & Dictionary & Ridge multiplier & HR & Score scale \\
\midrule
Reference ReLU & base & 0.0001 & 0.722 & 0.092 \\
Reference ReLU & base & 0.0010 & 0.720 & 0.092 \\
Reference ReLU & base & 0.0100 & 0.715 & 0.092 \\
Reference ReLU & base plus source & 0.0001 & 0.722 & 0.092 \\
Reference ReLU & base plus source & 0.0010 & 0.721 & 0.092 \\
Reference ReLU & base plus source & 0.0100 & 0.716 & 0.092 \\
\midrule
ReLU & base & 0.0001 & 0.725 & 0.092 \\
ReLU & base & 0.0010 & 0.723 & 0.092 \\
ReLU & base & 0.0100 & 0.717 & 0.092 \\
ReLU & base plus source & 0.0001 & 0.729 & 0.093 \\
ReLU & base plus source & 0.0010 & 0.725 & 0.092 \\
ReLU & base plus source & 0.0100 & 0.718 & 0.092 \\
\bottomrule
\end{tabular}

\par\smallskip\noindent The ridge equals the displayed multiplier times $\sqrt{300/n_{\rm tr}}$. Score scale denotes the evaluation score studentizer on the log hazard ratio scale, not the bootstrap standard deviation of the complete procedure. The same original five folds are used throughout.
\end{table}

Across ten identifier partitions for each fold count, the projected OAK hazard ratio with the reference learner ranges from \OakThreePartitionRange{} with three folds and \OakFivePartitionRange{} with five folds. With the final ReLU score the ranges are \OakThreePartitionRangeFinal{} and \OakFivePartitionRangeFinal{}. No partition is selected on the basis of its result. With the original overall survival follow-up in both cohorts, target clinical Cox gives HR \OakFullOsClin{}, full calibration with the reference learner 0.701, and five-fold projection 0.726. Refitting the final ReLU rule with the original follow-up gives \OakFullOsCtFinal{} for full calibration and \OakFullOsPtFinal{} for projection. These horizon sensitivities refit the historical learner with its original follow-up.

Table~\ref{tab:oaksourceperturb} describes the 100 historical perturbations for each learner. Full correction is relatively insensitive to the reference scores, consistent with their largely linear form. The final ReLU rule is more sensitive, because its refitted nonlinear component varies: the median fraction of score variance outside the clinical span is 2.4\%, and its 90th percentile is 4.5\%. Its full calibration and projected hazard ratios still remain within 0.684 to 0.707 and 0.702 to 0.726. The ranges are sensitivity summaries, not confidence intervals.
\begin{table}[tbp]\centering\small
\caption{OAK sensitivity to refitting the historical neural learners.}\label{tab:oaksourceperturb}
\begin{tabular}{lrrr}
\toprule
Target procedure & HR range & SD & Fits \\
\midrule
Full reference ReLU correction & 0.698 to 0.702 & 0.0013 & 100/100 \\
Free reference ReLU coefficient & 0.699 to 0.730 & 0.0092 & 100/100 \\
Projected reference ReLU score & 0.718 to 0.720 & 0.0008 & 100/100 \\
\midrule
Full ReLU correction & 0.684 to 0.707 & 0.0051 & 100/100 \\
Free ReLU coefficient & 0.700 to 0.735 & 0.0110 & 100/100 \\
Projected ReLU score & 0.702 to 0.726 & 0.0052 & 100/100 \\
\bottomrule
\end{tabular}

\par\smallskip\noindent The OAK patients are fixed. Each row uses all 100 historical patient resamples. Final ReLU refits use stopping samples of distinct patients. SD is on the log hazard ratio scale.
\end{table}

\subsection{OAK model diagnostics and clinical context}
\begin{figure}[tbp]\centering\includegraphics[width=.98\textwidth]{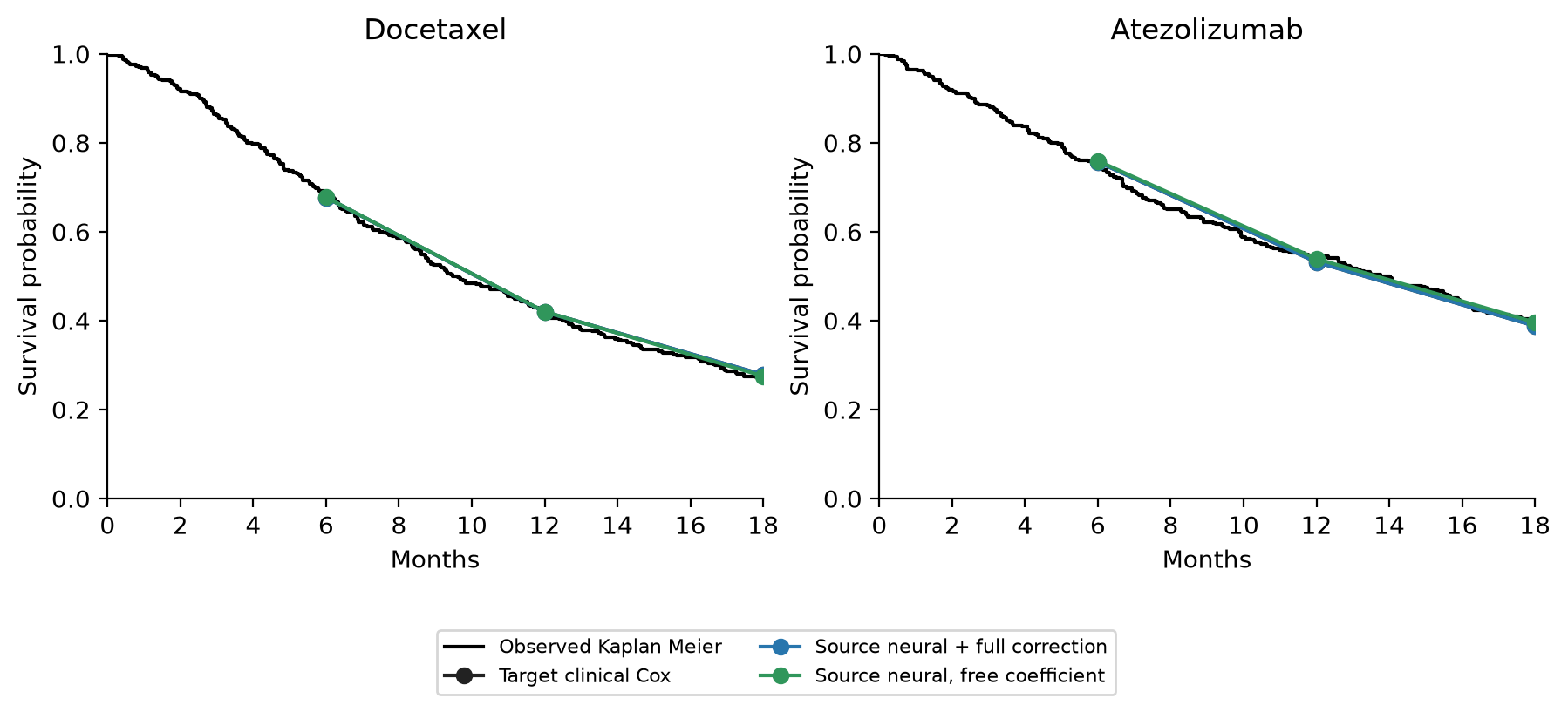}
\caption{OAK group average survival calibration for reference model fits. Curves describe absolute calibration within treatment arms. They are not simultaneous confidence bands.}\label{fig:oakcalibration}
\end{figure}

Archive Figure~1 describes measured case mix. The largest absolute standardized difference is 0.34 for the Asian race indicator. The standardized differences use the target minus source mean divided by the square root of the average within cohort variances. They do not establish overlap of the joint covariate laws. Figure~\ref{fig:oakcalibration} compares group average predicted survival with observed survival. Its predictions use the reference historical learners and repeated outer validation fits, so the target baseline is always estimated without the corresponding evaluation patients.

Table~\ref{tab:oakph} gives the Schoenfeld diagnostics \citep{grambsch1994} for all terms in target clinical Cox. The global $p$ value is \OakGlobalPh{}. The treatment term has $p=0.075$, which neither establishes a constant treatment effect nor excludes a clinically relevant departure. The targeted nuisance sensitivity adds performance status and Asian race multiplied by $\log(t/12)$, where $t$ is in months. All original clinical coordinates and a constant treatment coefficient remain in the fit. The resulting treatment HR is \OakTimeHr{}, using a sandwich Wald interval. The selected nuisance terms were informed by the earlier diagnostics, so this is a retrospective sensitivity. It is outside the constant nuisance theorem and does not demonstrate that all PH departures are resolved. Archive Figure~2 shows the corresponding residual diagnostics. Race is an adjustment descriptor and its fitted association is not interpreted as a biological effect.

\begin{table}[tbp]\centering\small
\caption{OAK PH diagnostics for target clinical Cox.}\label{tab:oakph}
\begin{tabular}{lrrr}
\toprule
Term & $\chi^2$ & df & $p$ \\
\midrule
Treatment & 3.17 & 1 & 0.075 \\
Age & 2.32 & 1 & 0.127 \\
Tumor burden & 1.32 & 1 & 0.251 \\
Metastatic sites & 0.47 & 1 & 0.493 \\
Asian race & 6.77 & 1 & 0.009 \\
Other race & 0.06 & 1 & 0.811 \\
Male sex & 0.40 & 1 & 0.528 \\
Squamous histology & 0.40 & 1 & 0.529 \\
Performance status & 7.95 & 1 & 0.005 \\
Two prior therapies & 3.92 & 1 & 0.048 \\
Current smoking & 0.91 & 1 & 0.340 \\
Never smoking & 0.61 & 1 & 0.433 \\
Global & 26.58 & 12 & 0.009 \\
\bottomrule
\end{tabular}

\par\smallskip\noindent Diagnostics use scaled Schoenfeld residuals. The individual $p$ values are descriptive and are not adjusted for multiple comparisons.
\end{table}

The restricted mean survival through 18 months is the area under the separate treatment arm Kaplan--Meier curves. Estimates are 10.39 months with docetaxel and 11.85 months with atezolizumab. The difference is \OakRmst{} months with percentile interval (\OakRmstLow{}, \OakRmstHigh{}) from 999 empirical patient resamples within treatment arms. This unadjusted absolute contrast is a separate estimand. It does not validate the conditional Cox coefficient or reproduce the full stratified assignment mechanism.

\section{Prediction from Rotterdam to GBSG}\label{app:gbsg}
\subsection{Cohorts, endpoint, and purpose}
We use the Rotterdam and German Breast Cancer Study Group (GBSG) data distributed with the R \texttt{survival} package. They have previously been used to develop and externally validate Cox prognostic models \citep{Royston2013}. Rotterdam contains 2982 primary breast cancer patients. Our historical sample contains the 552 patients with positive nodes, no hormonal therapy, and recorded chemotherapy. The public GBSG sample contains 686 patients with complete prognostic records from the original 720 patient comprehensive cohort. Of these, 440 have no hormonal therapy and 246 have hormonal therapy. The original study included both randomized participants and patients treated according to preference \citep{Schmoor2008}. The public dataset does not identify the randomized subgroup, so we treat hormonal therapy as an observed covariate and do not interpret its coefficient causally.

The endpoint is recurrence or death, with time measured on the supplied recurrence-free follow-up scale after primary breast cancer treatment. Both cohorts are administratively censored at 1826.25 days, or five years. For Rotterdam we use recurrence when recorded and otherwise death observed within recurrence follow-up. A death recorded after an earlier recurrence censoring time is not counted as an observed recurrence-free event. There are no excluded late deaths of this type in the restricted historical sample. By five years, there are 258 historical events and 285 target events. GBSG has 199 events among patients without hormonal therapy and 86 among those with hormonal therapy.

We use the seven clinical coordinates in Table~\ref{tab:clinicalcoords}. Grade is excluded because measurement protocols differ, and the public GBSG data do not provide the original chemotherapy duration assignment. The endpoint harmonization and restricted source population support comparison, but they cannot establish identical prognosis across cohorts. We retain measured case-mix differences in the analysis. We do not select a more favorable historical population using GBSG outcomes.

\subsection{Prediction results}
The analysis uses the reference historical learners, correction rules, and nested validation in Appendix~\ref{app:clinical}. The additional OAK tuning does not change the GBSG analysis. Prediction loss is integrated from six months to five years. Table~\ref{tab:gbsgprediction} gives the principal comparisons. Target clinical Cox gives IBS \GbsgClinIbs{}, target spline Cox \GbsgCurrentSplineIbs{}, and target neural Cox \GbsgTargetNeuralIbs{}. Full historical neural calibration gives \GbsgNeuralIbs{} and full historical spline calibration \GbsgSplineIbs{}. Nested spline calibration gives \GbsgAdaptiveSplineIbs{}, a small numerical difference from target spline Cox. Free neural rescaling gives \GbsgFreeIbs{}. These results do not establish a material prediction advantage with the available target sample.

Only \GbsgNonlinearPercent{}\% of the supplied neural score variance lies outside the linear clinical span on GBSG. For the historical spline score the corresponding fraction is 13.65\%. These descriptive fractions use the same residual variance calculation as OAK. Full neural calibration consequently resembles target clinical Cox. The learner labels should not be interpreted as evidence that this application requires a complex nonlinear prognosis.

\begin{figure}[tbp]\centering\includegraphics[width=\textwidth]{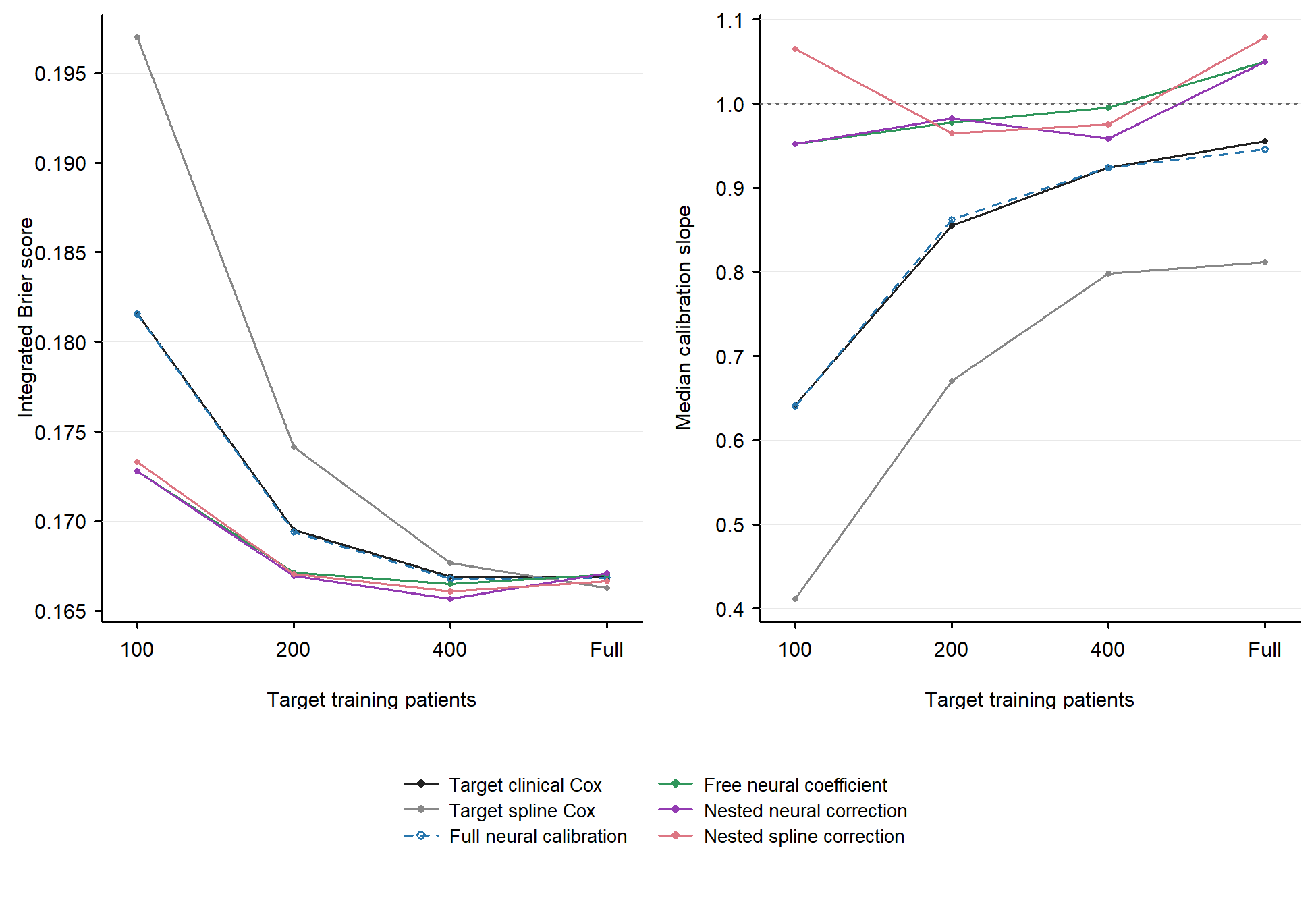}
\caption{GBSG learning curves with 20 evaluations at every size. All methods use the same training and evaluation patients. Full training sizes range from 547 to 550. Repeated samples overlap and the curves are descriptive.}\label{fig:gbsgcurve}
\end{figure}

Figure~\ref{fig:gbsgcurve} shows the learning curves. With 100 target training patients, nested neural correction gives IBS \GbsgAdaptiveSmallIbs{}, compared with \GbsgClinSmallIbs{} for target clinical Cox and \GbsgCurrentSplineSmallIbs{} for target spline Cox. Median calibration slopes are \GbsgAdaptiveSmallCal{} and \GbsgClinSmallCal{} for the first two fits. Free neural rescaling also gives IBS \GbsgFreeSmallIbs{}. The advantage diminishes at larger training sizes. This pattern agrees with OAK and supports studying restricted calibration when target information is limited. It does not justify a general neural superiority claim.

\begin{table}[tbp]\centering\small
\caption{GBSG primary prediction comparisons with 15 validation fits per method.}\label{tab:gbsgprediction}
\begin{tabular}{lrrr}
\toprule
Prediction fit & IBS & C index & Slope \\
\midrule
Target clinical Cox & 0.1658 & 0.690 & 0.94 \\
Target spline Cox & 0.1644 & 0.698 & 0.88 \\
Target neural Cox & 0.1694 & 0.678 & 0.92 \\
Frozen source neural & 0.1670 & 0.686 & 1.11 \\
Source spline + full correction & 0.1649 & 0.696 & 0.90 \\
Source neural + full correction & 0.1657 & 0.691 & 0.95 \\
Source neural, free coefficient & 0.1667 & 0.686 & 1.04 \\
Nested source neural correction & 0.1669 & 0.685 & 1.04 \\
Nested source spline correction & 0.1640 & 0.696 & 1.00 \\
\bottomrule
\end{tabular}

\par\smallskip\noindent IBS integrates Brier loss from six months to five years. C index aggregates comparable pairs within evaluation folds. Slope is the median evaluation fold Cox calibration slope. Hormonal treatment is an observed covariate, not a randomized intervention in this analysis.
\end{table}

The TransCox-type comparator of Appendix~\ref{app:oakcomparators} uses source coefficients from the stored Rotterdam clinical Cox fit or from a Rotterdam ridge Cox fit chosen by the same rule, which selects $\lambda=0.1$. It is applied to the 95 splits regenerated from the analysis code, which reproduce the stored IBS of clinical Cox and nested neural calibration to within $3\times10^{-9}$. Table~\ref{tab:gbsgtranscox} shows IBS 0.1713 and 0.1714 at 100 patients, below 0.1728 for nested neural calibration, and every difference from nested neural calibration is at most 0.0015 in absolute value, with a range that includes zero. With 552 historical patients and seven coordinates, shrinkage toward the historical Cox coefficients is therefore as accurate as nested calibration of the neural score. All 190 fits complete.

\begin{table}[tbp]\centering\footnotesize\setlength{\tabcolsep}{4pt}
\caption{GBSG TransCox-type comparator. CV denotes the 15 evaluation comparison, and other sizes use 20 evaluations. Differences are nested neural calibration minus the TransCox-type estimator with 95\% patient resampling ranges, so negative values favor nested neural calibration.}\label{tab:gbsgtranscox}
\begin{tabular}{lrrrrll}\toprule
 & \multicolumn{2}{c}{TransCox-type IBS} & Nested & Clinical & \multicolumn{2}{c}{Nested neural minus TransCox-type}\\\cmidrule(lr){2-3}\cmidrule(lr){6-7}
Size & Clinical source & Ridge source & neural & Cox & Clinical source & Ridge source\\\midrule
100 & 0.1713 & 0.1714 & 0.1728 & 0.1816 & 0.0015 (-0.0011, 0.0041) & 0.0013 (-0.0008, 0.0035) \\
200 & 0.1672 & 0.1670 & 0.1669 & 0.1695 & -0.0003 (-0.0021, 0.0015) & -0.00001 (-0.0013, 0.0012) \\
400 & 0.1656 & 0.1655 & 0.1657 & 0.1669 & 0.0001 (-0.0021, 0.0023) & 0.0002 (-0.0018, 0.0022) \\
Full & 0.1661 & 0.1657 & 0.1671 & 0.1669 & 0.0010 (-0.0017, 0.0037) & 0.0014 (-0.0010, 0.0039) \\
CV & 0.1660 & 0.1660 & 0.1669 & 0.1658 & 0.0009 (-0.0016, 0.0033) & 0.0009 (-0.0014, 0.0033) \\
\bottomrule\end{tabular}

\end{table}

Archive Tables~16 and~17 give every prediction comparison and the descriptive differences from clinical Cox. Archive Table~18 gives all learning-curve values, using the method definitions in Appendix~\ref{app:clinicalvalidation}. The clinical historical learner with full correction again reproduces target clinical Cox. Archive Table~19 gives correction selection counts, including small training samples in which the parsimonious score coefficient is frequently retained.

\subsection{Association summaries and sensitivities}
Table~\ref{tab:gbsginference} describes adjusted hormonal therapy associations. Full neural calibration gives HR \GbsgNeuralHr{}, compared with \GbsgClinHr{} for target clinical Cox. Projected neural estimation gives \GbsgProjectedHr{}. These are working-model association summaries conditional on the historical learner. The projected fit is a computational diagnostic outside the randomized treatment theorem. It is not evidence of bias reduction or a causal effect.

\begin{table}[tbp]\centering\small
\caption{GBSG hormonal therapy associations through five years.}\label{tab:gbsginference}
\begin{tabular}{lrrr}
\toprule
Estimator & HR & 95\% interval & Bootstrap SD \\
\midrule
Treatment only & 0.670 & (0.514, 0.859) & 0.130 \\
Target clinical Cox & 0.651 & (0.501, 0.853) & 0.137 \\
Target spline Cox & 0.636 & (0.478, 0.824) & 0.141 \\
Source spline + full correction & 0.633 & (0.485, 0.830) & 0.137 \\
Source neural + full correction & 0.651 & (0.499, 0.851) & 0.136 \\
Projected neural score & 0.651 & (0.488, 0.871) & 0.147 \\
\bottomrule
\end{tabular}

\par\smallskip\noindent Intervals are nominal 95\% percentile intervals from 999 target resamples. The source learner is fixed. Bootstrap SD is on the log hazard ratio scale. Every resample completed. No randomized treatment or causal interpretation is attached to these rows.
\end{table}

Archive Table~20 gives all fixed correction and free rescaling sensitivities. Archive Table~21 changes the projected dictionary and ridge. The primary clinical dictionary has 15 coordinates. Appending the neural score gives 16 columns with full observed rank but substantially worse conditioning. As in OAK, regularized inversion is not a population information guarantee. Across ten partitions for each fold count, the projected association ranges from \GbsgThreePartitionRange{} with three folds and \GbsgFivePartitionRange{} with five folds. All partitions are retained.

Archive Table~22 reports the historical perturbations with the GBSG patients fixed. Free score rescaling is more sensitive to refitting the supplied learner than full clinical calibration. The result is consistent with retaining more historical information through the smaller correction, but is not a comparison of interval coverage.

\subsection{Case mix, calibration, and model limitations}
Archive Figure~3 shows substantial measured case-mix differences. The standardized differences are 0.96 for age and 1.00 for postmenopausal status. These differences help explain why the historical prognosis requires assessment in the target cohort. They do not quantify all unmeasured differences or establish a bounded density ratio. Archive Figure~4 gives group average survival calibration without a causal interpretation of the treatment groups.

The global PH diagnostic for target clinical Cox gives $p=\GbsgGlobalPh{}$. Archive Table~23 gives every term, including departures for age, receptor measurements, and menopausal status. A retrospective sensitivity adds age and estrogen receptor coordinates multiplied by $\log(t/1095.75)$, where $t$ is in days. All original covariates and a constant hormonal therapy coefficient remain in the fit. The resulting association under full neural calibration is \GbsgTimeHr{}, with a sandwich Wald interval. Archive Figure~5 shows the selected residual diagnostics. Other departures remain possible, and this sensitivity is outside the proposed constant nuisance model.

The breast cancer analysis supports prediction and calibration as the relevant use of this public cohort. Its mixed treatment assignment, substantial case-mix shift, missing chemotherapy duration, and PH departures limit treatment interpretation. The censoring weights also require the group specific independence and support conditions described in Appendix~\ref{app:clinicalvalidation}. We retain these limits when interpreting the small training sample gains and the small differences at full training size.

\clearpage
\begingroup
\small
\setlength{\bibsep}{2pt}

\endgroup

\end{document}